\documentclass{article}

\usepackage{PRIMEarxiv}

\usepackage[utf8]{inputenc} 
\usepackage[T1]{fontenc}    
\usepackage{hyperref}       
\usepackage{url}            
\usepackage{booktabs}       
\usepackage{amsfonts}       
\usepackage{nicefrac}       
\usepackage{microtype}      
\usepackage{lipsum}
\usepackage{fancyhdr}       
\usepackage{graphicx}       
\graphicspath{{media/}}     

\usepackage[ruled,vlined,algo2e]{algorithm2e}

\usepackage{float}
\usepackage{subfig}

\usepackage{color}
\usepackage[x11names]{xcolor}
\hypersetup{
	colorlinks = true,
	linkcolor=blue,
	citecolor=SpringGreen4,
}

\usepackage[authoryear,round]{natbib}

\usepackage{amsmath,amssymb,amsfonts,latexsym}

\title{
SPARSE-R: A point-cloud tracer with random forcing
}

\author{
  Daniel Dom\'inguez-V\'azquez\thanks{Electronic email: ddominguezvazquez@sdsu.edu}, \ 
  Gustaaf B. Jacobs\thanks{Author to whom correspondence should be addressed: gjacobs@sdsu.edu}, \\
  Department of Aerospace Engineering \\
  San Diego State University \\
  San Diego, CA 92182, USA\\
}

\begin{document}

\maketitle

\begin{abstract}
A predictive, point-cloud tracer is presented that determines with a quantified uncertainty the Lagrangian motion  of a group of point-particles within a finite region. 
The tracer assumes a random forcing within confidence intervals to account for the empiricism of data-driven force models and stochasticity related to the chaotic nature of the subcloud scale dynamics.
It builds on the closed Subgrid Particle-Averaged Reynolds Stress-Equivalent (SPARSE) formulation presented in Dom\'inguez-V\'azquez~\textit{et al.} [\textit{Int. J. Multiph. Flow.} 161, 104375] that assumes a deterministic forcing.  SPARSE describes the first two moments of particle clouds with moment equations in closed-form, with a theoretical third-order convergence rate with respect to the standard deviations of the cloud variables.
The cloud model alleviates computational cost and enhances the convergence  rate as compared to Monte Carlo (MC) based point-particle methods.
The randomness in the forcing model leads to  virtual stresses that correlate random forcing and field fluctuations. These stresses strain and rotate  the random cloud as compared to a deterministically forced cloud and thus determine to what extent the random forcing propagates into the confidence intervals of the dispersed solution. In symmetric flows the magnitude of the virtual stress is  zero.
Tests in analytical carrier-fields and in a decaying homogeneous isotropic turbulence flow computed with a discontinuous Galerkin (DG) compressible DNS solver are performed to verify and validate the SPARSE method for randomly forced particles.
\end{abstract}

\keywords{Multiphase Flow \and Particle-Laden Flows \and Eulerian-Lagrangian \and Cloud-In-Cell \ Point-Cloud \and Random Forcing}

\section{Introduction}

Multiphase flows in which a dispersed phase composed of particles, droplet or bubbles interacts with a carrier flow are present in many natural and industrial processes.
Examples of applications of these flows are gas and liquid fluidized beds~\citep{deen2007review,sippola2018experimental}, aerosol and spray flows in combustion engines~\citep{ren2019supersonic,jenny2012modeling} and medical devices~\citep{kumar2020perspective}, dispersion of snowflakes in the atmosphere~\citep{nemes2017snowflakes}, transport and mixing phenomena in oceans~\citep{prants2014chaotic}, sea search and rescue algorithms~\citep{serra2020search}, volcanic eruptions~\citep{ongaro2007parallel,dufek2012granular,delannay2017granular}, cavitation in turbomachinery~\citep{bryngelson2019quantitative} or kidney  stone fragmentation~\citep{jamaluddin2011collapse} and sedimentation~\citep{guo2019role}.

The computational simulation of gas and liquid flows laden with liquid or solid particles can be categorized based on the fidelity of the model into firstly a high-resolution particle-resolved (PR) approach, secondly a point-particle approach and thirdly a point-cloud approach.
The highest fidelity approach 
models and   resolves the flow near a particle's surface at scales smaller than the particles' size.
The coupling of the two phases is carried out by imposing no-slip boundary conditions on the particle's surface.
The hydrodynamic force on a particle can be computed by directly integrating the pressure and shear stress distribution along the boundary.
Examples of various implementations include the arbitrary Eulerian-Lagrangian technique~\citep{takashi1992arbitrary,hu2001direct}, the deformable-spatial-domain/stabilized space–time technique~\citep{johnson1999advanced}, the overset grid technique~\citep{burton2005fully}, the constrained interpolation profile method~\citep{yabe2001constrained,lu2023dynamics}, immersed boundary methods~\citep{uhlmann2005immersed,lucci2010modulation,akiki2016immersed,yao2022particle,biegert2017collision}, lattice Boltzmann equations~\citep{ten2004fully,gao2013lattice} or the smoothed profile method~\citep{luo2009smoothed}.
The high computational cost limits the number of particles that can be simulated to hundreds with current computational resources.

In process-scale problems millions of particles must be simulated~\citep{sengupta2009spectral} which for PR methods are not feasible with current day computer infrastructure.
For this problem size, the reduced point-particle approach that models a particle as a volumeless singular point has to be the de facto choice.
The so-called Particle-Source-In-Cell (PSIC) method~\citep{saffman1973settling,crowe1977particle} accounts for the particle's influence introduced in the Eulerian equations that govern the carrier phase through singular source terms.
This model is valid if the smallest hydrodynamical scale of the flow is several orders of magnitudes larger than the particles' size.
The PSIC method requires modeling for the terms in the governing equations that couple the mass (if evaporation and condensation are considered), momentum and heat transfer exchanged between the two phases.
For spherical particles in an steady incompressible, uniform, creeping flow, the drag force is analytically described by the Stokes' law~\citep{stokes1851effect}.
If the flow or the spherical particle motion is unsteady, additional forcing terms have to be considered including the unsteady added-mass and viscous-history forcing effects, which through Newton's second law results in a governing equation for the particle's acceleration known as the Boussinesq-Basset-Oseen (BBO) equation~\citep{boussinesq1885resistance,basset1888treatise,oseen1927hydrodynamik}.
The BBO relation was later modified to include the effects of non-uniform transient ambient flow leading to the widely used Maxey-Riley-Gatignol (MRG) equation~\citep{gatignol1983faxen,maxey1983equation}.
If compressibility effects are considered, additional Fax\'en forces complete the particle equations~\citep{parmar2012equation,annamalai2017faxen,parmar2011generalized}.
Many physical situations of interest however, lead to particle forcings that can not be described analytically.
The default modus operandi to compute these flows relies on the use of empirical and/or data-driven correlations to correct the MRG equation.
Examples of parameters for which empirical correlations are developed include arbitrary Reynolds and Mach numbers~\citep{boiko1997shock,boiko2005drag,loth2008compressibility,parmar2010improved}, particle density and slip coefficients~\citep{tedeschi1999motion} viscosity rations for droplets~\citep{feng2012drag} and inhomogeneities in the particle configurations~\citep{zhao2021inhomogeneous} among others.
For the mass and energy exchange, the use empirical correlations is also the most common practice.

Despite the broad usage of the point-particle approach, the accuracy, convergence and stability of the method are affected by its numerical treatment.
The computational approximation of point-particles requires interpolation between the Lagrangian point tracer and what is usually a grid based approximation of the Eulerian carrier phase model~\citep{balachandar2010turbulent,fox2012large,maxey2017simulation,elghobashi2019direct,brandt2022particle,bec2023statistical}.
Particles move freely through the domain in locations different from the computational grid points of the flow, whose tractability in parallel computing is involved and requires the use of interpolation~\citep{yeung1988algorithm,balachandar1989methods,jacobs2007towards}.
Special attention has to be paid to the treatment of pointwise forcing into the numerical grid to regularize the coupling source terms and its numerical representation in the computational grid~\citep{cortez2001method,suarez2014high,suarez2017regularization,evrard2020euler,gualtieri2015exact,horwitz2022discrete,poustis2019regularization}.
Another known approximation challenge is a nonphysical numerical self-forcing that is connected with this interpolation.
The point-particle description relies on the knowledge of the unperturbed flow velocity which is in principle unknown.
The use of the flow velocity interpolated at the particle location instead of the velocity of the unperturbed flow at that location produces self-motion as a result of a self-induced force that is not physical~\citep{lomholt2003force}.
Some of the earliest work to tackle this issue led to the forcing-coupling method (FCM)~\citep{maxey2001localized,lomholt2003force,dance2003incorporation}, while several corrections to the point-particle method have been recently proposed~\citep{horwitz2018correction,evrard2020euler,horwitz2022discrete,balachandar2023correction}.
The point-particle assumption leads also to convergence issues related to a strong grid dependence because the forcing is modeled by averaging in the volume of the computational cell unless the number of particles per cell exceeds a threshold~\citep{gualtieri2015exact}.
These drawbacks have inspired the development of different alternative approaches for the simulation of particle-laden flows~\citep{koumoutsakos2005multiscale,maxey2017simulation}.
Some of the recent research includes the volume averaged method~\citep{capecelatro2013euler,ireland2017improving,shallcross2020volume}, the modeling of interparticle forcing by the pairwise interaction extended point-particle (PIEP) model~\citep{akiki2017pairwiseFJM,akiki2017pairwiseJCF,moore2019hybrid,balachandar2020toward}, the microstructure-informed probability-driven point-particle (MPP) model~\citep{seyed2020microstructure}, the exact regularized point particle (ERPP) method~\citep{gualtieri2015exact,battista2018application} to tackle convergence issues, the use of discrete Green's functions to find the undisturbed velocity and correct the particle's self-force~\citep{horwitz2022discrete}, and machine learning (ML) models to find closures to reduced descriptions~\citep{he2019supervised,seyed2022physics}.

\begin{figure}[h!]
	\centering
		\includegraphics[width=0.8\textwidth]{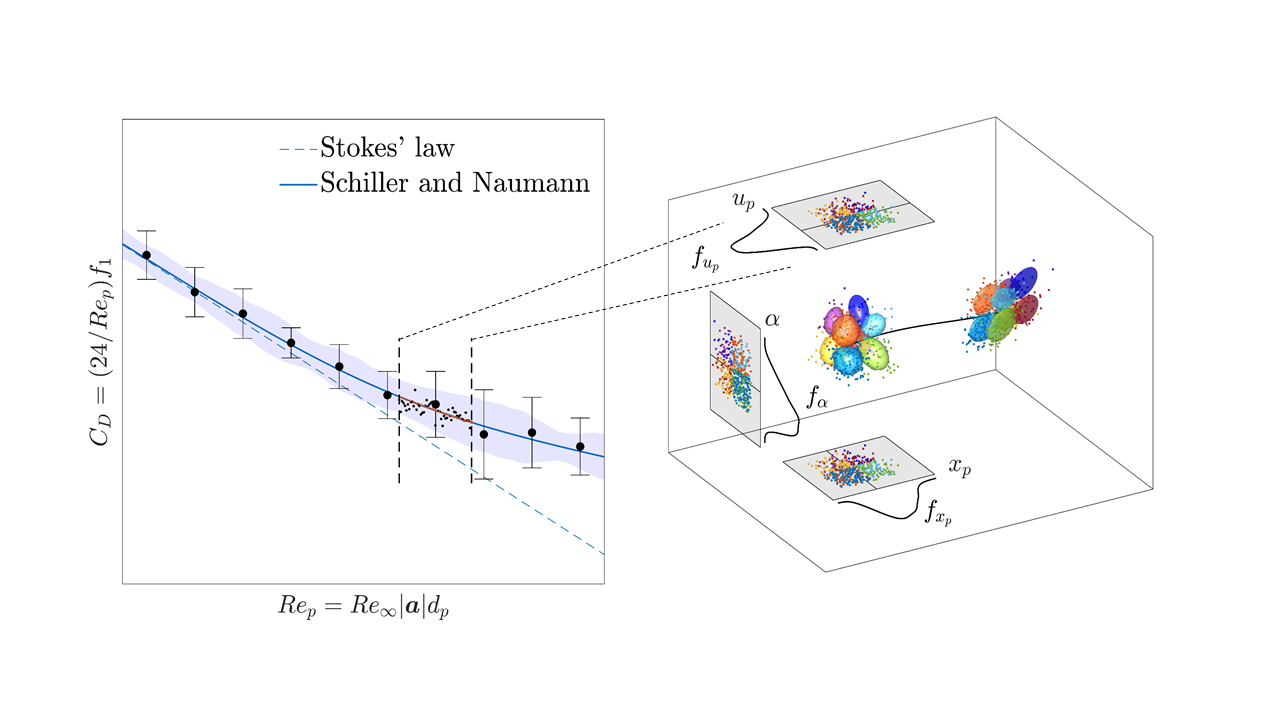} 
	\caption[]{Scheme of a randomly forced cloud described by the SPARSE method. On the left, the drag coefficient $C_D$ as a function of the particle Reynolds number $Re_p=Re_\infty|\boldsymbol{a}|d_p$ with uncertainty bounds where $\boldsymbol{a}$ is the relative velocity, $Re_\infty$ the reference Reynolds number of the flow and $d_p$ the non-dimensional particle diameter. On the right, the representation of a one-dimensional particle cloud in the domain $\alpha-x_p-u_p$. The dots are the point-particles (evolved with the PSIC equations) and the ellipsoids the subclouds or point-clouds (evolved with the SPARSE equations). The PDFs of the particle phase variables are also represented where the point-particles and point-clouds are linked by colors according to a division of the domain in two subclouds per dimension in $\alpha-x_p-u_p$. }
	\label{fig: introSqueme}
\end{figure}

If a forcing function is fit empirically to experimental or computational data, then the resulting expression is analytical and ignores the confidence intervals of the forcing and its effects on the particle trace are discarded.
In that sense, methods to trace (tracers) point-particles based on these analytical models are deterministic.
In an ongoing effort to account for the confidence interval, we are developing disperse phase models that assume the point-particles to be randomly forced (rF).
The probabilistic models propagate the confidence intervals of the forcing into the kinematics, dynamics and heat transfer of the disperse phase~\citep{jacobs2019uncertainty,rutjens2021method,dominguez2021lagrangian}.
The randomness may originate from confidence intervals, i.e., the fitting error of empirical or data-driven forcing functions reconstructed from PR simulations~\citep{sen2015evaluation,sen2017evaluation,sen2018evaluation} as shown in Fig.~\ref{fig: introSqueme} (left). 
The randomness can also be stochastic in nature for sub-scale or subgrid scale model, in which the forcing function is described by a probability density function (PDF) according to the dynamics of the sub-scale~\citep{pandya2003non,mashayek2003analytical}. 
The probabilistic models enable an assessment of sensitivity to uncertain/unknown forcing models in flows where the point-particle assumption is used but analytical descriptions for the particle forcing are not available.
The probabilistic macro-model is a natural complement to the multi-scale data-driven framework proposed in~\cite{sen2015evaluation,sen2017evaluation,sen2018evaluation}.
The multi-scale framework machine learns forcing models from high-resolution PR simulations from homogenized assembles of spheres. 
This procedure yields an approximate forcing function with a computable probability density function.
This probabilistic forcing then propagates into a PDF solution based on the random point-particle dynamics.
In recent efforts~\citep{jacobs2019uncertainty,rutjens2021method,dominguez2021lagrangian}, we have developed point-particle models with random forcing for one- and two-way coupled flows and with different approaches such as Monte Carlo (MC), the method of moments (MoM) and the method of distributions (MoD) in order to characterize the propagation of uncertainty.

Modeling the effect of stochasticity, a form of  randomness, on the particle phase based on a kinetic PDF equation analogous to the Maxwell-Boltzmann equation of classical kinetic theory has its origins on the landmark work of~\cite{buyevich1971statistical1,buyevich1972statistical2,buyevich1972statistical3}.
Since then, a significant body of literature addresses closure analytically~\citep{reeks1980eulerian,reeks1983transport,reeks1991kinetic,reeks1992continuum,swailes1997generalized,bragg2012drift,swailes1998chapman,zaichik2007refinement,alipchenkov2005dispersion,pandya2001probability,pandya2002turbulent,pandya2003non,reeks2021development}.
Numerically, a closure has also been pursued.
Quadrature-based moment methods, for example, have been derived by describing finite moment sets of the joint PDF by an optimal set of quadrature nodes and weights~\citep{marchisio2002comparison,marchisio2002quadratic,raman2003quadrature,marchisio2003quadraturea,marchisio2003quadratureb,heylmun2019quadrature,bryngelson2020gaussian,charalampopoulos2022hybrid,bryngelson2023conditional}.  
To reduce the dimensionality of the PDF formulation, often models based on Langevin dynamics, usually refereed to as generalized Langevin model~\citep{reeks2021development}, use stochastic processes (usually Wiener increments) in the particle equations in an equivalent manner to the case of passive tracers in turbulent flows~\citep{pope1985pdf,haworth1986generalized}.
Stochastic models of the disperse phase based on the use of Wiener increments have also been used in recent years~\citep{iliopoulos2003stochastic,gao2004stochastic_a,gao2004stochastic_b,gao2004stochastic_c,shotorban2005modeling,shotorban2006stochastic_a,shotorban2006stochastic_b,sengupta2009spectral,pozorski2009filtered,pai2012two,tenneti2016stochastic,esteghamatian2018stochastic,lattanzi2020stochastic,knorps2021stochastic,lattanzi2022stochastic,friedrich2022single,pietrzyk2022analysis}.
The computational treatment of Langevin equations however requires sampling techniques, which makes its numerical resolution prohibitively expensive. 
An alternative is to use the corresponding Fokker-Plank equation~\citep{esteghamatian2018stochastic,pietrzyk2022analysis} making use of the It\^o calculus which requires closure and its advection-diffusion character makes its numerical solution a computational challenge that continues inspiring new research~\citep{holubec2019physically,tabandeh2022numerical}.
Only under simplified considerations, the Fokker-Plank equation allows constructive analytical solutions to be found~\citep{lattanzi2020stochastic}, where the temporal evolution of the particle statistics is described analytically.
Because of the complexity of the numerical solution of the Fokker-Plank equation to compute the joint PDF of the particle phase, most studies are restricted to finding the first and second moments of the joint PDF.
The moment equations also require the development of closures that can be based on gradient models or machine learned correlations from PR simulations~\citep{lattanzi2022stochastic}.
{\color{black}
None of these models consider the forcing random, rather the result of a stochastic process, which is a mere subset of randomness.
}

In~\cite{dominguez2021lagrangian}, we proposed a new class of PDF model that starts from an approximation of the random forcing function with a polynomial chaos expansion (PCE), i.e., a polynomial with modes (Chebyshev, for example) weighted by random coefficients~\citep{rutjens2021method}. 
The confidence intervals of the forcing so defined are general and can be caused by either empiricism or stochasticity of the problem.
Following the MoD, the randomly forced Lagrangian point-particle model is governed by a closed hyperbolic PDE for the joint PDF of the particle solution in physical and phase space.
The resulting equation governs a joint PDF that has an augmented dimensionality to include the particle phase variables and the random coefficients.
The moment equations follow naturally from the PDF approach, but they require closure of the higher moments. 
The numerical approximation of the moment model that admits singularities is not trivial~\citep{jacobs2019uncertainty}.

Another approach to account for the effects of random and stochastic forcing is to take a cloud perspective as initially proposed in~\cite{davis2017sparse}. 
By ensemble averaging a group of point-particles within a cloud region and combining that with a Taylor expansion of the forcing function and carrier phase variables around the average values of the particle phase (particle position and relative velocity), we have developed a two-way coupled and closed form point-cloud tracer~\citep{davis2017sparse,taverniers2019two,dominguez2023closed}. 
We coined this cloud tracer Subgrid Particle-Averaged Reynolds Stress-Equivalent (SPARSE), as it naturally accounts for stochasticity in the subcloud scale.
SPARSE was originally developed to account for subgrid effects and pseudo-turbulent kinetic energy (PTKE)~\citep{davis2017sparse}, later extended for two-way coupling regime~\citep{taverniers2019two} and recently a closure and extension to second moments has been developed~\citep{dominguez2023closed}.
Thus far, we have assumed the forcing to be described analytically (deterministically) and the particle cloud to be deterministically forced (dF).

In this article, we present a SPARSE method that accounts  for both randomness of the empirical or data-driven  forcing functions and/or stochasticity of the sub-clouds dynamics.
We expand the random forcing function according to a polynomial choas expansions as proposed in~\cite{rutjens2021method} to trace randomly forced (rF) particle clouds.
The SPARSE method is designed to capture the first two moments of the particle phase by computing a closed set of equations in Lagrangian form combining Taylor expansions around the average magnitudes of the cloud with Reynolds averaging.
The SPARSE method exhibits a third order convergence rate with respect to the standard deviations of the particle phase variables.
By subdividing or splitting the particle cloud into subclouds (see Fig.~\ref{fig: introSqueme} right) the accuracy of the method is ensured.
Using a mixture distribution (MD) of Gaussians, SPARSE approximates higher order moments and enables the reconstruction of the PDF of the of the underlying point-particle population.
The computational savings as compared with Monte Carlo (MC) simulations using the PSIC description (MC-PSIC), is proportional to the number of point-clouds or subclouds divided by the total number of point-particles.
Because of the slow convergence of MC, the number of samples (point-particles) is several order of magnitudes larger than the number of subclouds needed for an accurate SPARSE simulation, ensuring computational savings.
The SPARSE method is a novel approach to compute the moments and the PDF of the particle phase in a closed and computationally efficient manner that makes it scalable to complex flows.
It combines the advantages of the MoM in a closed form, reduction of the computational cost as compared to sampling methods, with the advantages of PDF methods, full statistical description able to capture rare events linked to fat-tailed PDFs in the solution.
Considering the MC-PSIC simulations the ground truth, we compare the SPARSE method with the MC-PSIC method in a variety of verification tests on dilute, one-way coupled particle-laden flow problems with prescribed velocity fields, and a numerical simulation of an isotropic turbulence flow, where the gas is simulated with a compressible discontinuous Galerkin (DG) DNS solver.

In the Section~\ref{sec: formulation}, we summarize the derivation of the SPARSE method for randomly forced particles.
The random forcing model is presented in Section~\ref{sec: rF_model} and the closure model in Section~\ref{sec: closure}, followed by insights on the numerical implementation in Section~\ref{sec: IC_numericalImplementation}.
In Section~\ref{sec: tests} we present the results of particle-laden flow tests.
Concluding remarks and future work are reserved for the Section~\ref{sec: conclusions}.

\section{SPARSE-R: point-cloud model with random forcing} \label{sec: formulation}

\subsection{Point-particle model}

The non-dimensional governing point-particle equations for a small spherical particle immersed in a carrier flow where the inertial effects are dominant is given by
\begin{subequations}\label{eq: xp_up_Tp_nondimensional}
\begin{align}
    \frac{\text{d}\boldsymbol{x}_p}{\text{d}t} &= \boldsymbol{u}_p ,
    \label{eq: dxpdt_nondimensional} \\
    \frac{\text{d}\boldsymbol{u}_p}{\text{d}t} &= \frac{f_1}{St}  \left( \boldsymbol{u} -\boldsymbol{u}_{p}  \right) , 
    \label{eq: dupdt_nondimensional} \\
    \frac{\text{d}T_p}{\text{d}t} &= \frac{2c_r}{3Pr}\frac{f_2}{St}  \left( T -T_p  \right) ,
    \label{eq: dTpdt_nondimensional} 
\end{align}
\end{subequations}
where $\boldsymbol x_p$, $\boldsymbol u_p$ and $T_p$ are the non-dimensional particle location, velocity and temperature and $\boldsymbol u$ and $T$ are the non-dimensional velocity and temperature of the carrier flow evaluated at the particle location.
The Prandtl number $Pr=\tilde{\mu} \tilde{c} / \tilde{k}$ is defined with the dimensional dynamic viscosity $\tilde{\mu}$, specific heat capacity $\tilde{c}$ and conductivity $\tilde{k}$ of the flow. 
The Stokes number $St=\tau_p/\tau_f$ is defined with the characteristic time of the particles $\tau_p={\tilde{\rho}_p \tilde{d}}_p^2 / \left(18\tilde{\mu} \right)$ where $\tilde{\rho}_p$ and $\tilde{d}_p$ are the dimensional density and diameter of the particles and $\tau_f=L_\infty / U_\infty$ is the characteristic time of the flow defined with a reference length and velocity.
Making use of reference values denoted with subscript infinite, we define the reference Prandtl number $Pr_\infty=\mu_\infty c_\infty/k_\infty$ and Reynolds number $Re_\infty=\rho_\infty U_\infty L_\infty/\mu_\infty$, one can rewrite the non-dimensional numbers in terms of the reference values
\begin{align}
    Re_p = Re_\infty\frac{\rho}{\mu} |\boldsymbol{u} -\boldsymbol{u}_p | d_p , \ \ \ \ St   = Re_\infty\frac{\rho_p d_p^2}{18\mu} , \ \ \ \ Pr   = Pr_\infty \frac{\mu c}{k}
    \label{eq: Rep_St_Pr}
\end{align}
using the non-dimensional variables $\rho = {\tilde{\rho}}/{\rho_\infty}$, \ $\mu = {\tilde{\mu}}/{\mu_\infty}$, \ $k = {\tilde{k}}/{k_\infty}, \ c = {\tilde{c}}/{c_\infty}$, \  $\rho_p={\tilde{\rho}_p}/{\rho_\infty}$, \ $c_p={\tilde{c}_p}/{c_\infty}$ and, $ d_p={\tilde{d}_p}/{L_\infty}$.
The specific heat ratio of the two phases is defined as $c_r=c_p/c$.

The correction factors of exchanged momentum and energy are given by the functions $f_1$ and $f_2$ which correct the analytical laminar solution for the drag coefficient $C_D$ and Nusselt number $Nu$ (based on the particle diameter) as
\begin{subequations}\label{eq: f1_f2_definition}
\begin{align}
    C_D&=\frac{24}{Re_p}f_1 , 
    \label{eq: f1_definition} \\
    Nu&=2f_2 .
    \label{eq: f2_definition}
\end{align}
\end{subequations}
These functions, are also denoted the \textit{forcing functions}, typically correct for higher particle Reynolds and Mach numbers and/or other flow parameters~\citep{boiko1997shock,boiko2005drag,loth2008compressibility,parmar2010improved,tedeschi1999motion,feng2012drag}.

\subsection{Point-cloud SPARSE model}

Following the SPARSE approach as described in~\cite{davis2017sparse,taverniers2019two,dominguez2023closed}, we model a cloud of particles with the MoM using a Reynolds decomposition of any instantaneous scalar particle variable $\phi$ into its average and fluctuating component according to $\phi=\overline{\phi}+\phi^\prime$, where the average is defined 
by its ensemble average
\begin{align}
    \overline { \phi  } =\frac { 1 }{ { N }_{ p } } \sum _{ i=1 }^{ { N }_{ p } }{ { \phi  }_{ i } } ,
\end{align}
for $N_{p}$ point-particles within a cloud. 
Using the Reynolds decomposition for the variables of the two phases including the forcing functions $f_1$ and $f_2$, the system of equations for the kinematics, dynamics and thermodynamics of a stochastic cloud of point-particles can be written as
\begin{subequations}\label{eq: x_p_up_Tp_meanPlusPrime}
\begin{align}
\frac {\text{d}\overline { \boldsymbol { x } }_p  }{ \text{d}t }+\frac { \text{d} \boldsymbol { x }_{ p }^\prime   }{ \text{d}t } &= \overline{\boldsymbol u}_p+\boldsymbol u_p^\prime , 
\label{eq: dxpdt_meanPlusPrime} \\
\frac { \text{d}\overline { \boldsymbol{ u} }_p  }{ \text{d}t } +\frac { \text{d} \boldsymbol{ u }_{ p }^\prime }{ \text{d} t } &=\frac{1}{St}\left( 
\overline{f}_1+f_1^\prime \right) \left( \overline{\boldsymbol u} + \boldsymbol u^\prime -\overline{\boldsymbol u} - \boldsymbol u^\prime \right) , 
\label{eq: dupdt_meanPlusPrime} \\
\frac { \text{d}\overline T_p  }{ \text{d} t } +\frac { \text{d} T_p^\prime }{ \text{d}t } &= \frac{2c_r}{3Pr St}\left( \overline{f}_2+f_2^\prime \right) \left( \overline { T } + T^\prime -\overline { T }_p - T_p^\prime\right) , \label{eq: dTpdt_meanPlusPrime}
\end{align}
\end{subequations}
that after manipulation and averaging leads to the equations for the first two moments of the particle phase
\begin{subequations}\label{eq: MoM}
\begin{align}
\frac{\text{d} {\overline{x}_p}_i}{\text{d}t} &= {\overline{u}_p}_i
\label{eq: MoM_mean_xp}, \\
St \frac{\text{d} {\overline{u}_p}_i}{\text{d}t} &= \overline{f}_1\left( \overline{u}_i-{\overline{u}_p}_i \right)+\overline{f_1^\prime u_i^\prime}-\overline{f_1^\prime {u_p^\prime}_i}
\label{eq: MoM_mean_up}, \\
\frac{3Pr St}{2c_r}\frac{\text{d} \overline{T}_p}{\text{d}t} &= \overline{f}_2\left(\overline{T}-\overline{T}_p  \right)+\overline{f_2^\prime T^\prime}-\overline{f_2^\prime {T_p^\prime}}
\label{eq: MoM_mean_Tp}, \\
\frac{\text{d}}{\text{d} t}\left( \overline{{x^\prime_p}_i{x^\prime_p}_j } \right) &= \overline{{x^\prime_p}_i{u_p^\prime}_j} + \overline{{x^\prime_p}_j{u_p^\prime}_i}
\label{eq: MoM_xpxp}, \\
St \frac{\text{d}}{\text{d} t}\left( \overline{{u_p^\prime}_i {u_p^\prime}_j}\right)  &= \overline{f}_1 \left( \overline{u^\prime_i{u^\prime_p}_j}+\overline{u^\prime_j{u^\prime_p}_i} -2\overline{{u^\prime_p}_i{u^\prime_p}_j} \right)+\overline{f^\prime_1{u^\prime_p}_i}\left( \overline{u}_j-{\overline{u}_p}_j \right)+\overline{f^\prime_1{u^\prime_p}_j}\left( \overline{u}_i-{\overline{u}_p}_i \right)
\label{eq: MoM_upup}, \\
\frac{3Pr St}{4c_r} \frac{\text{d} \overline{{T_p^\prime}^2}}{\text{d}t} &= \overline{f}_2\left( \overline{T^\prime T^\prime_p}-\overline{T^\prime_p} \right)+\overline{f^\prime_2 T^\prime_p}\left( \overline{T}-\overline{T}_p \right) 
\label{eq: MoM_TpTp}, \\
\frac{\text{d}}{\text{d}t}\left( \overline{{x_p^\prime}_i {u_p^\prime}_j}\right)  &= \overline{{u_p^\prime}_i{u_p^\prime}_j}+\frac{1}{St}\left[ \overline{f}_1\left( \overline{{x^\prime_p}_iu^\prime_j}-\overline{{x^\prime_p}_i {u^\prime_p}_j} \right)+\overline{f^\prime_1{x^\prime_p}_i}\left( \overline{u}_j-{\overline{u}_p}_j \right) \right]
\label{eq: MoM_xpup}, \\
\frac{\text{d}}{\text{d}t}\left( \overline{{x^\prime_p}_i{T_p^\prime} } \right) &= \overline{{u^\prime_p}_iT^\prime_p}+\frac{2c_r}{3Pr St}\left[ \overline{f}_2\left(\overline{{x^\prime_p}_iT^\prime}-\overline{{x^\prime_p}_iT^\prime_p}  \right) +\overline{f^\prime_2{x^\prime_p}_i} \left( \overline{T}-\overline{T}_p \right) \right]
\label{eq: MoM_xpTp}, \\
\begin{split}
\frac{\text{d}}{\text{d}t}\left( \overline{{u^\prime_p}_i T_p^\prime } \right) &=
\frac{1}{St}\left[ \overline{f}_1\left( \overline{u^\prime_i T^\prime_p}-\overline{{u^\prime_p}_i T^\prime_p} \right)+\overline{f^\prime_1T^\prime_p}\left( \overline{u}_i-{\overline{u}_p}_i\right) \right] \\ &+\frac{2c_r}{3Pr St}\left[ \overline{f}_2\left( \overline{{u^\prime_p}_iT^\prime}-\overline{{x^\prime_p}_iT^\prime_p}\right)+\overline{f^\prime_2{u^\prime_p}_i}\left( \overline{T}-\overline{T}_p \right)   \right] ,
\end{split}
\label{eq: MoM_upTp}
\end{align}
\end{subequations}
where index notation with $i=1,2,3$ and $j=1,2,3$ is used to present a compact version of the equations. 
Note that for any two vectorial variables $\boldsymbol{\eta}$ and $\boldsymbol{\xi}$, the term $\overline{\eta_i^\prime \xi_j^\prime}$ is a component of a three by three tensor, whereas $\overline{\eta}_i$ is a the component of a three-dimensional vector.
For the general three-dimensional case we have $\boldsymbol{x}_p = (x_p \ y_p \ z_p)^\top$ and $\boldsymbol{u}_p = (u_p \ v_p \ w_p)^\top$.
For a combined moment of a scalar $\phi$ and a vectorial magnitude $\boldsymbol{\eta}$, the term $\overline{\phi^\prime \eta_i^\prime}$ represent the $i-$th component of a vector.
Note that the system of equations~\eqref{eq: MoM} truncates moments that are greater than third order while the equations for the first moments~\eqref{eq: MoM_mean_xp}--\eqref{eq: MoM_mean_Tp} are not truncated.
In the second moment equations~\eqref{eq: MoM_xpxp}--\eqref{eq: MoM_upTp} high-order terms are neglected.

\subsection{Random forcing SPARSE-R model} \label{sec: rF_model}

The dependencies of both correction functions $f_1$ and $f_2$ correct for physics that deviate from the case of creeping flow over a spherical particle.
In PSIC, those functions  have generally been used as exact force models  that depend only on the particle phase and the carrier phase.
In practice, however, they can be known only  within confidence intervals as they are  approximate curve fits to experimental and computational data that have  sources of systematic uncertainty or epistemic uncertainty.
Alternatively, the forcing function can be  interpreted in the context of stochastic models to account for stochasticity of subscales.
In other words, the function may be used to account for aleatoricism in the particle forcing.

\begin{figure}[h!]
	\centering
	\subfloat[]{
		\label{fig: focring_stochModes}
		\includegraphics[width=0.45\textwidth]{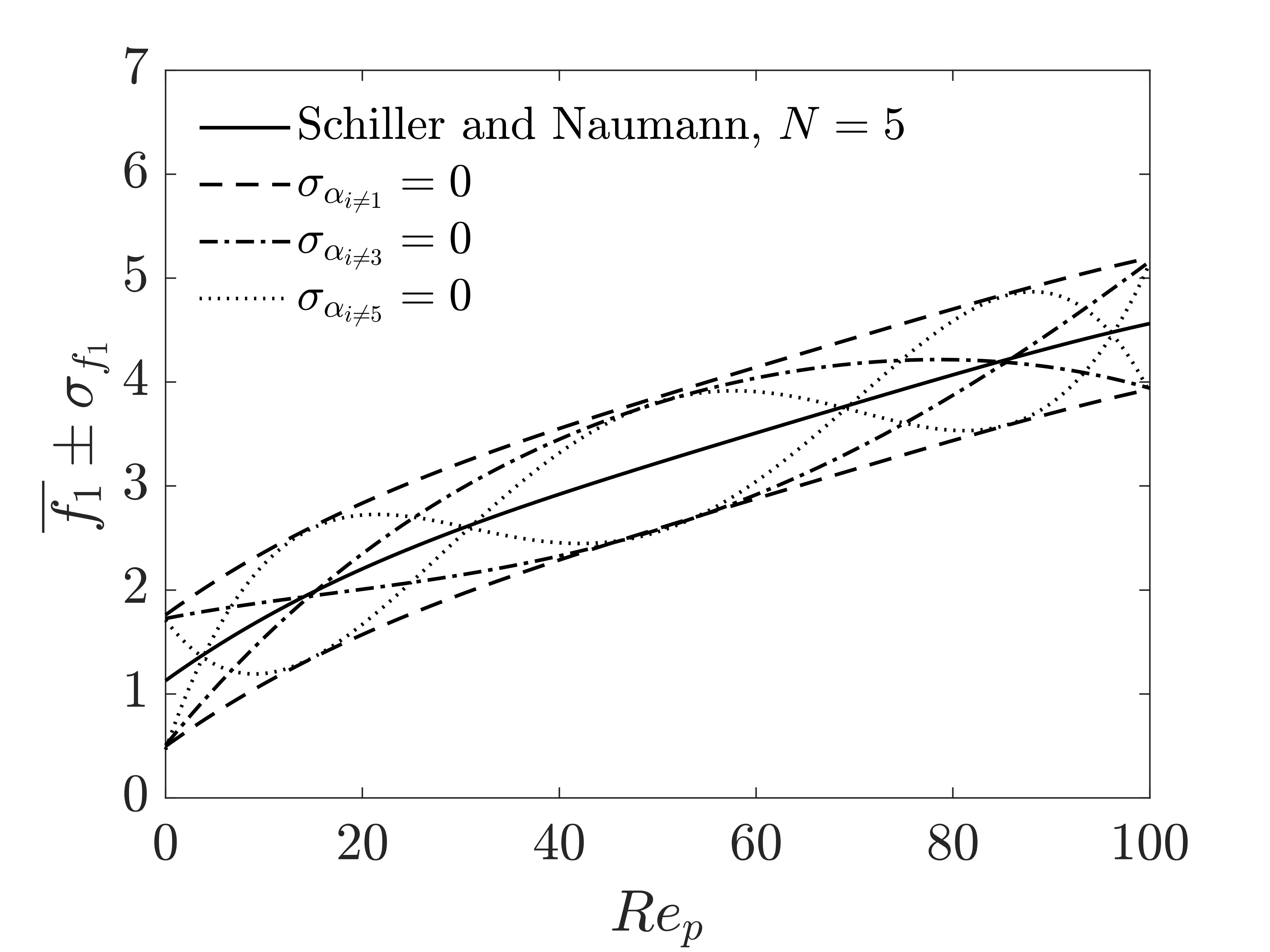}}
        \hfill	
	\subfloat[]{
		\label{fig: focring_stochModes_2test}
		\includegraphics[width=0.45\textwidth]{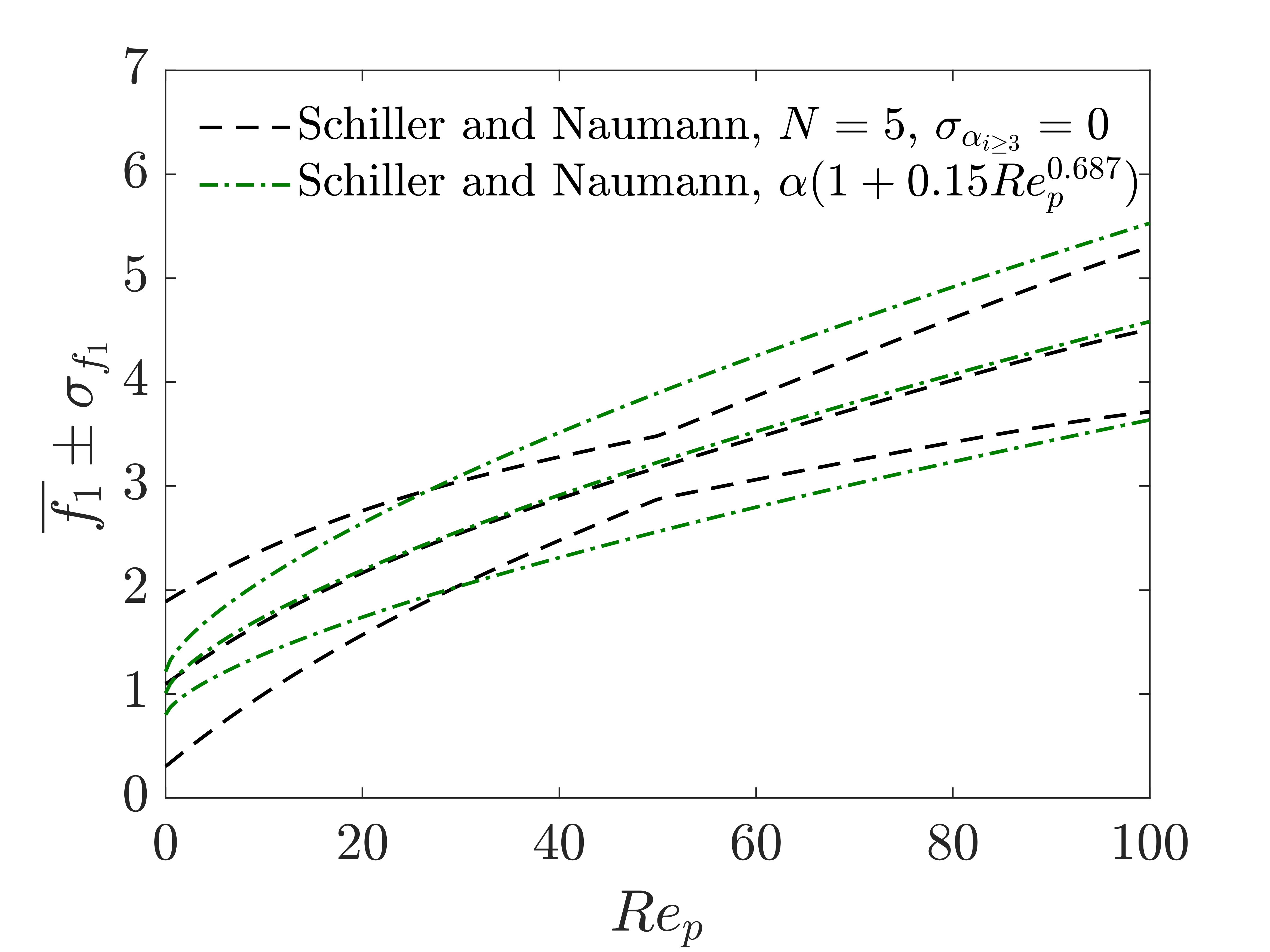}}  
	\caption[]{Two bandwidth representation of the random forcing function $f_1$ when the SN correlation is fitted with the first five modes; (a) considering random only in the first (dashed line), third (dash-dotted line) and fifth (dotted line) modes; (b) combining the random modes first and second (dashed line) and taking $f_1=\alpha(1+0.15Re_p^{0.687})$ (dash-dotted line). The particle Reynolds number is $Re_p=Re_\infty|\boldsymbol{a}|d_p$ according to~\eqref{eq: Rep_St_Pr} where constant density and viscosity are considered.}
	\label{fig: focring_stochModes_figure}
\end{figure}

To model confidence intervals, we follow~\cite{davis2017sparse},~\cite{taverniers2019two} and~\cite{dominguez2023closed}, and consider the functions $f_1$ and $f_2$ to be dependent on the relative velocity $\boldsymbol{a}=\boldsymbol{u}-\boldsymbol{u}_p$ and the random coefficients $\alpha_i$ and $\beta_i$ with $i=1,\dots,N$ where $N$ is the number of modes considered.
As proposed by~\cite{rutjens2021method} and ~\cite{dominguez2022adjoint}, the correction functions with quantified uncertainty can be described with a polynomial chaos expansion as
\begin{align}
    f_1(\boldsymbol{a}) = \sum_{i=1}^{\infty}\alpha_i \psi_i(\boldsymbol{a}) \approx \sum_{i=1}^{N}\alpha_i \psi_i(\boldsymbol{a}), \ \ \ \ \ \ \ \ \ \
    f_2(\boldsymbol{a}) = \sum_{i=1}^{\infty}\beta_i \zeta_i(\boldsymbol{a}) \approx \sum_{i=1}^{N}\beta_i \zeta_i(\boldsymbol{a}),
    \label{eq: f1andf2_chebyshev}
\end{align}
where the variables $\alpha_i$ and $\beta_i$ are correlated or uncorrelated random coefficients that account for the uncertainty in the forcing defined by the joint PDF $f_{\boldsymbol{\alpha \beta}}(\alpha_1,\dots, \alpha_N,\beta_1,\dots, \beta_N)$ 
and the orthogonal basis functions $\psi_i$ and $\zeta_i$.
We assume $f_1$ and $f_2$ to have compact support and consider $\psi_i$ and $\zeta_i$ to be the Chebyshev polynomials of the first kind scaled from $[-1, \ 1]$ to $[\boldsymbol{a}_{min}, \ \boldsymbol{a}_{max}]$.
Note that this polynomial generalizes randomness accounting for both the epistemic and aleatoric character of the forcing.
They can be found using high-resolution simulations, as was done for example in~\cite{sen2015evaluation,sen2017evaluation,sen2018evaluation}.
Using~\eqref{eq: f1andf2_chebyshev} and decomposing the stochastic variables in average plus fluctuation, the mean and variance of the forcing functions are given by
\begin{subequations}\label{eq: f1_mean_and_sigma}
\begin{align}
    \overline{f}_1 &= \sum_{i=1}^{N}\overline{\alpha}_i\psi_i\left(\boldsymbol{a} \right), 
    \label{eq: f1_mean}\\
    \sigma_{f_1}^2 &= \overline{\left(\sum_{i=1}^{N} \alpha_i^\prime \psi_i\left( \boldsymbol{a}\right)\right)^2} = \sum_{i=1}^{N} \overline{{\alpha_i^\prime}^2}\psi_i^2\left(\boldsymbol{a} \right)+\sum_{i\neq j}^{N(N-1)/2}2\overline{\alpha_i^\prime \alpha_j^\prime}\psi_i\left(\boldsymbol{a} \right)\psi_j\left(\boldsymbol{a} \right),
    \label{eq: f1_mean_sigma}
\end{align}
\end{subequations}
and equivalently for $f_2$.
The variance of the forcing function is then described by the second moments of the coefficients and captures the uncertainty in the forcing functions by combining $N$ modes.
As an example consider an empirical forcing that is based on the Schiller and Naumann (SN) correlation~\citep{schiller1933correlation} given by
\begin{align}
    g_1=1+0.15Re_p^{0.687}.
    \label{eq: schiller_and_naumann}
\end{align}
In Figure~\ref{fig: focring_stochModes}, this forcing is plotted with an unknown confidence interval modeled by the first five Chebyshev modes $\psi_1,\dots,\psi_5$ where the random coefficients $\alpha_1,\dots,\alpha_5$ are considered uncorrelated.
Two standard deviation bounds are depicted when considering the standard deviation of all modes zero except for the modes one $\sigma_{i\neq 0}=0$, third $\sigma_{i\neq 3}=0$ and fifth $\sigma_{i\neq 5}=0$ in dashed, dash-dotted and dotted lines respectively.
The number of random modes is chosen such that the polynomial chaos expansion can accurately approximate complex dependencies of the confident interval with the relative velocity.
Considering only the first mode to be random (i.e., only $\alpha_1$ is given by a PDF and the rest $\alpha_2,\dots,\alpha_N$ are deterministic) the standard deviation $\sigma_{f_1}$ does not change with respect to the relative velocity (Fig.~\ref{fig: focring_stochModes} dashed line).
By combining modes, more complex functions and confidence interval can be approximated by~\eqref{eq: f1_mean_sigma}.
In Figure~\ref{fig: focring_stochModes_2test} (dashed line), we show a combination of only two modes that localizes a particle Reynolds number where the forcing is defined with a smaller confidence interval.

We consider a representation of the random function by a single mode as
\begin{align}
    f_1(\boldsymbol{a})=\alpha g_1(\boldsymbol{a}), \ \ \ \ \ \ \ \ \ \
    f_2(\boldsymbol{a})=\beta g_2(\boldsymbol{a}), 
    \label{eq: f1andf2}
\end{align}
where $g_1$ and $g_2$ carry the dependencies with the relative velocity and the random coefficients are reduced to $\alpha$ and $\beta$ with given PDFs $f_{\alpha}(\alpha)$ and $f_\beta(\beta)$.
For the Schiller and Naumann correlation this yields $f_1=\alpha(1+0.15Re_p^{0.687})$ which corresponds to the dash-dotted line in Figure~\ref{fig: focring_stochModes_2test}.
This confidence interval for a single mode representation with the Reynolds number is consistent with trends reported in literature (see for example~\cite{loth2008drag}).

The second moments of the coefficients $\overline{{\alpha^\prime}^2}$, $\overline{{\beta^\prime}^2}$ and $\overline{\alpha^\prime \beta^\prime}$ are probabilistic sources on the SPARSE method related to uncertainty in the forcing function when~\eqref{eq: f1andf2} is considered, i.e., the particles are rF per the description in the introduction.
A deterministic description of the drag force and energy exchanged between phases yields a fine-grained PDF, i.e. a Dirac delta function so that $\overline{{\alpha^\prime}^2}=\overline{{\beta^\prime}^2}=\overline{\alpha^\prime \beta^\prime}=0$, in which case the particles are dF.
Note that the average of the coefficients for rF and dF cases is unity $\overline{\alpha}=\overline{\beta}=1$.

\subsection{Closure model} \label{sec: closure}

The SPARSE equations in~\eqref{eq: MoM} are not yet closed for terms that are on the order of fluctuations squared and that are function of  a combination of particle phase and flow variables, and  correction functions. 
To close~\eqref{eq: MoM}, we follow the procedure in~\cite{dominguez2023closed}. 
The carrier flow velocity $\boldsymbol u$ and temperature $T$ fields are Taylor expanded around the average particle location truncating terms of order greater than two.
By using a Taylor expansion, we obtain an estimate of the carrier flow field at the particle locations within the cloud region and can be interpreted as interpolation within the cloud region.
The forcing functions $f_1$ and $f_2$ are also expanded around the average values of the relative velocity and random coefficients of the particle cloud.
For completeness, we review the closure approach and discuss it for random forcing. 

Starting with the interpolation of the mean carrier flow velocity and temperature, we Taylor expand at the average particle location of the cloud.
Let $\phi$ be a flow variable (for example, a velocity component $u_i$ or the temperature $T$), 
then
\begin{align}
    \overline{\phi} &\approx \overline{\phi \left( \overline{\boldsymbol x}_p \right) + {x_p^\prime}_i\left.\frac{\partial \phi}{\partial x_i} \right|_{\overline{\boldsymbol x}_p}  +\frac{1}{2}{{x_p^\prime}_i {x_p^\prime}_j}\left.\frac{\partial^2 \phi}{\partial x_i \partial x_j} \right|_{\overline{\boldsymbol x}_p}} = \phi \left( \overline{\boldsymbol x}_p \right) +\frac{1}{2}\overline{ {x_p^\prime}_i {x_p^\prime}_j }  \left.\frac{\partial^2 \phi}{\partial x_i \partial x_j} \right|_{\overline{\boldsymbol x}_p},
    \label{eq: closure_means} 
\end{align}
which relates the solution evaluated at the averaged cloud's location, the second derivative of the carrier phase at the averaged cloud's location and the covariances of the locations of the particle cloud $\overline{{x_p^\prime}_i {x_p^\prime}_j}$ which are governed by equation~\eqref{eq: MoM_xpxp}.

To close second moments, we proceed similarly by Taylor expanding the flow variables.
Depending on the type of term to close, the Taylor expansion has to be applied once or twice to find a closure.
If the term involves two variables different from the particle phase variables (particle position, velocity or temperature), it has to be applied twice whereas if it contains a combination of a flow variable or forcing function with a particle variable, it has to be applied only once.
For example, for a term involving a generic particle variable $\xi_p$ and carrier phase variable $\eta$, it follows that
\begin{align}
    \overline{\xi_p^\prime \eta^\prime} = \overline{\xi_p^\prime \left( \eta -\overline{\eta}  \right)} \approx \overline{\xi_p^\prime \left( \eta \left( \overline{\boldsymbol x}_p \right) + {x_p^\prime}_i \left.\frac{\partial \eta}{\partial x_i} \right|_{\overline{\boldsymbol x}_p}-\overline{\eta}   \right) } = \overline{\xi_p^\prime {x_p^\prime}_i}\left.\frac{\partial \eta}{\partial x_i} \right|_{\overline{\boldsymbol x}_p},
    \label{eq: closure_particle_flow}
\end{align}
where $\xi_p$ can be any of the components of the particle location $\boldsymbol{x}_p$, velocity $\boldsymbol{u}_p$ or temperature $T_p$ and $\eta$ can be any component of the flow field $\boldsymbol{u}$ or temperature $T$.
A second moment, however, that combines two flow variables $\xi$ and $\eta$ (components of the flow field or temperature), has to be Taylor expand twice to close it as follows 
\begin{align}
    \overline{\xi^\prime \eta^\prime} \approx  \overline{\xi^\prime {x_p^\prime}_i}\left.\frac{\partial \eta}{\partial x_i} \right|_{\overline{\boldsymbol x}_p} \approx \overline{{x_p^\prime}_i\left(\xi(\overline{\boldsymbol{x}}_p) + {x_p^\prime}_j\left.\frac{\partial \xi}{\partial x_j} \right|_{\overline{\boldsymbol x}_p} -\overline{\xi} \right)  }\left.\frac{\partial \eta}{\partial x_i} \right|_{\overline{\boldsymbol x}_p} = \overline{{x_p^\prime}_i {x_p^\prime}_j}\left.\frac{\partial \xi}{\partial x_j} \right|_{\overline{\boldsymbol x}_p}\left.\frac{\partial \eta}{\partial x_i} \right|_{\overline{\boldsymbol x}_p} ,
    \label{eq: closure_flow_flow}
\end{align}
where the approximation in~\eqref{eq: closure_particle_flow} has been applied first to expand $\eta$ and then to $\xi$.

To account for uncertainties in the forcing functions $f_1$ and $f_2$ we expand as follows
\begin{align}
\begin{split}
    \overline{f}_1 &\approx \overline{f_1 \left(\overline{\alpha},\overline{\boldsymbol a} \right) + {\alpha^\prime}\left.\frac{\partial f_1}{\partial \alpha} \right|_{\overline{\alpha},\overline{\boldsymbol{a}}}+ a_i^\prime\left.\frac{\partial f_1}{\partial {a}_i} \right|_{\overline{\alpha},\overline{\boldsymbol{a}}}+\frac{1}{2}{\alpha^\prime}^2\left.\frac{\partial^2 f_1}{\partial \alpha^2} \right|_{\overline{\alpha},\overline{\boldsymbol{a}}} +\frac{1}{2}{\alpha^\prime}{a_i^\prime}\left.\frac{\partial^2 f_1}{\partial \alpha \partial a_i} \right|_{\overline{\alpha},\overline{\boldsymbol{a}}} +\frac{1}{2}{a_i^\prime}{a_j^\prime}\left.\frac{\partial^2 f_1}{\partial a_i \partial a_j} \right|_{\overline{\alpha},\overline{\boldsymbol{a}}} }  \\ 
    &= f_1 \left(\overline{\alpha},\overline{\boldsymbol a} \right) +\frac{1}{2}\overline{{\alpha^\prime}{a_i^\prime}}\left.\frac{\partial g_1}{\partial a_i} \right|_{\overline{\alpha},\overline{\boldsymbol{a}}}+\frac{1}{2}\left( \overline{{u_i^\prime}{u_j^\prime}}-\overline{{u_i^\prime}{u_p^\prime}_j}-\overline{{u_j^\prime}{u_p^\prime}_i}+\overline{{u_p^\prime}_i{u_p^\prime}_j} \right)\left.\frac{\partial^2 f_1}{\partial a_i \partial a_j} \right|_{\overline{\alpha},\overline{\boldsymbol{a}}},
\end{split}
\label{eq: closure_mean_f1}
\end{align}
where $f_1$ depends linearly on the random coefficient $\alpha$ according to~\eqref{eq: f1andf2} and the identity $\overline{a_i^\prime a_i^\prime}=\overline{u_i^\prime u_j^\prime}-\overline{u_i^\prime{u_p^\prime}_j}-\overline{u_j^\prime{u_p^\prime}_i} +\overline{{u_p^\prime}_i{u_p^\prime}_j} $ has been used.
Note that, the procedure is easily extendable to $N$ randomly weighted modes according to~\eqref{eq: f1andf2_chebyshev} by adding the partials of the functions $\psi_i$ and $\zeta_i$.
The equation above can be also applied for the correction function of the exchanged energy $f_2$ accordingly.
The resulting expression in~\eqref{eq: closure_mean_f1} contains the terms $\overline{u_i^\prime {u_p^\prime}_j}$ and $\overline{{u_i^\prime}u_j^\prime}$ that need to be closed using the relations~\eqref{eq: closure_particle_flow} and~\eqref{eq: closure_flow_flow} respectively.

For the second moments involving the forcing functions with any particle variable $\xi_p$, the closure is applied as follows
\begin{align}
\begin{split}
    \overline{\xi_p^\prime f_1^\prime} &= \overline{\xi_p^\prime \left(f_1-\overline{f}_1 \right)} \approx \overline{\xi_p^\prime\left( f_1 \left(\overline{\alpha},\overline{\boldsymbol a} \right) + {\alpha^\prime}\left.\frac{\partial f_1}{\partial \alpha} \right|_{\overline{\alpha},\overline{\boldsymbol{a}}}+ a_i^\prime\left.\frac{\partial f_1}{\partial {a}_i} \right|_{\overline{\alpha},\overline{\boldsymbol{a}}} -\overline{f}_1 \right)} \\ 
    &=\overline{\alpha^\prime {\xi_p^\prime}} g_1\left(\overline{\alpha},\overline{\boldsymbol a} \right) +\left(\overline{\xi_p^\prime u^\prime_i}-\overline{\xi_p^\prime {u_p^\prime}_i} \right)\left.\frac{\partial f_1}{\partial a_i} \right|_{\overline{\alpha},\overline{\boldsymbol{a}}}
    \label{eq: closure_xip_f1},
\end{split}
\end{align}
where the identities $\partial f_1/\partial \alpha = g_1$, according to~\eqref{eq: f1andf2}, and $a_i=u_i-{u_p}_i$ have been used.
The resulting expression in~\eqref{eq: closure_xip_f1} contains the unclosed term $\overline{\xi_p^\prime u_i^\prime}$ that can be closed using the relation~\eqref{eq: closure_particle_flow}.
For the term $\overline{\alpha^\prime \xi_p^\prime}$ however (combinations of particle variables with random coefficients), we can not apply the closure procedure because the gradients of the random coefficients within the cloud are unknown, preventing its Taylor expansion.
Rather, we use the MoM to derive moment equations that govern these terms
\begin{subequations}\label{eq: closure_axp_system}
\begin{align}
    \frac{\text{d}}{\text{d}t}\left( \overline{\alpha^\prime {x^\prime_p}_i} \right) &= \overline{\alpha^\prime {u^\prime_p}_i}
    \label{eq: closure_axp}, \\
    St\frac{\text{d}}{\text{d}t}\left( \overline{\alpha^\prime {u^\prime_p}_i} \right) &= \overline{f}_1\left(\overline{\alpha^\prime u^\prime_i}-\overline{\alpha^\prime {u^\prime_p}_i} \right)+\overline{\alpha^\prime f^\prime_1}\left(\overline{u}_i-{\overline{u}_p}_i \right)
    \label{eq: closure_aup}, \\
    \frac{3Pr St}{2c_r}\frac{\text{d}}{\text{d}t}\left( \overline{\alpha^\prime {T^\prime_p}} \right) &= \overline{f}_2\left(\overline{\alpha^\prime T^\prime}-\overline{\alpha^\prime {T^\prime_p}} \right)+\overline{\alpha^\prime f^\prime_2}\left(\overline{T}-{\overline{T}_p} \right)
    \label{eq: closure_aTp}, 
\end{align}
\end{subequations}
where $\alpha$ can also be substituted by $\beta$. 
If we considered an expansion with $N$ modes, relations~\eqref{eq: closure_axp_system} have to be solved for each of the $N$ random coefficients for $\overline{\alpha_i {x_p^\prime}_j}$, $\overline{\alpha_i {u_p^\prime}_j}$ and $\overline{\alpha_i {T_p^\prime}}$ with $i=1,\dots,N$.

In equations~\eqref{eq: closure_axp_system}, terms combining the random coefficients and the forcing functions are expressed as
\begin{subequations} \label{eq: closure_af1_af2} 
\begin{align}
    \overline{\alpha^\prime f_1^\prime} &\approx \overline{{\alpha^\prime}^2}g_1\left(\overline{\alpha},\overline{\boldsymbol a} \right)+\left(\overline{\alpha^\prime u^\prime_i}-\overline{\alpha^\prime {u_p^\prime}_i} \right)\left.\frac{\partial f_1}{\partial {a}_i}\right|_{\overline{\alpha},\overline{\boldsymbol{a}}} ,
    \label{eq: closure_af1} \\
    \overline{\alpha^\prime f^\prime_2} &\approx \overline{\alpha^\prime \beta^\prime}g_2\left(\overline{\beta},\overline{\boldsymbol a} \right)+\left(\overline{\alpha^\prime u_i^\prime}-\overline{\alpha^\prime {{u_p^\prime}_i}} \right)\left.\frac{\partial f_2}{\partial {a_i}}\right|_{\overline{\beta},\overline{\boldsymbol{a}}} ,
    \label{eq: closure_af2} 
\end{align}
\end{subequations}
following a similar Taylor expansion.
Finally, the only remaining terms that require closure in~\eqref{eq: closure_af1_af2} combine the random coefficients with the fluid phase (as for example $\overline{\alpha^\prime u_i^\prime}$).
Those terms, generally expressed as $\overline{\alpha^\prime \eta^\prime}$ where $\alpha$ is exchangeable with $\beta$ and $\eta$ is any scalar flow variable, close as follows
\begin{align}
    \overline{\alpha^\prime \eta^\prime} &\approx \overline{\alpha^\prime {x_p^\prime}_i}\left. \dfrac{\partial \eta}{\partial x_i}\right|_{\overline{\boldsymbol x}_p},
    \label{eq: closure_alphaeta} 
\end{align}
that relates again to the system of equations~\eqref{eq: closure_axp_system}.

For closure of  correlations of the forcing functions and a flow variable, the procedure has to be applied twice.
Generalizing, $\eta$ for any flow variable (velocity components or temperature), it follows that
\begin{align}
    \overline{\eta^\prime f_1^\prime} \approx \overline{{x_p^\prime}_if_1^\prime}\left. \frac{\partial \eta}{\partial x_i}\right|_{\overline{\alpha},\overline{\boldsymbol{a}}} \approx \left[\overline{\alpha^\prime {x_p^\prime}_i}g_1\left(\overline{\alpha},\overline{\boldsymbol a} \right)+\left(\overline{{x_p^\prime}_i u^\prime_j}-\overline{{x_p^\prime}_i{u_p^\prime}_j}\right)\left.\frac{\partial f_1}{\partial a_j}\right|_{\overline{\alpha},\overline{\boldsymbol{a}}} \right]\left. \frac{\partial \eta}{\partial x_i}\right|_{\overline{\alpha},\overline{\boldsymbol{a}}},
    \label{eq: closure_etaf1}
\end{align}
where the term $\overline{{x_p^\prime}_iu_j^\prime}$ is closed using the relation~\eqref{eq: closure_particle_flow} and $\overline{\alpha^\prime {x_p^\prime}_i}$ is closed with~\eqref{eq: closure_axp}.

Equations~\eqref{eq: closure_means}--\eqref{eq: closure_etaf1} represent the closed form of the randomly forced SPARSE model.
Its solution depends on inputs of the averages of the carrier flow fields and forcing functions.
The accuracy of the SPARSE was shown to be of third order for the deterministic model consistent with the order of the truncations of the Taylor series and moment equations. 
Similarly here, the accuracy of the randomly forced model depends on: (a) the truncation of terms on the order of statistical correlations of order greater than two and (b) the truncation of the Taylor series terms greater than second order.
The SPARSE method is closed and thus \textit{predictive}.
It converges with a third order expected rate with the size of the initial particle cloud given by the standard deviations in phase space consequently with the retained terms in the SPARSE formulation. 
The analysis of the leading order truncated terms in SPARSE is added in the Appendix~\ref{app: convergence} for completeness.

\subsection{Numerical implementation} \label{sec: IC_numericalImplementation}

\subsubsection{Tracing}

The numerical solution of the governing system of equations~\eqref{eq: MoM} and~\eqref{eq: closure_means}--\eqref{eq: closure_etaf1} requires four stages including (1) locating the host grid or carrier flow domain of the point-cloud, (2) interpolating the flow variable and gradients at the point-cloud's location (because this is a point-cloud, there is only one location, that happens to be the mean location), (3) determining the forcing gradients at the cloud's mean location and (4) integrating in time. 
The first stage requires algorithms that are similar to the PSIC method. 
We locate host cell as described in~\cite{jacobs2007towards} if needed. 
For the second stage, we either use analytical solutions or interpolations~\citep{jacobs2007towards,jacobs2006high}  that are consistent with the carrier phase solver.
A third order total variation diminishing (TVD) Runge-Kutta scheme~\citep{gottlieb1998total} is used for time integration. 

For the third (interpolation) stage we need to compute the first and second derivatives of the forcing functions $f_1$ and $f_2$ with respect to the relative velocity components and similarly the first and second derivatives of the carrier flow velocity and temperature with respect to space.
The derivatives of an analytical forcing function can be precomputed and evaluated at the average values during the time integration in the fourth stage.
A polynomial approximation of the forcing function requires the numerical evaluation of the derivative. 
At each location, this derivative is interpolated at the required average values during the time integration.
In a similar fashion, for analytically prescribed carrier phase flows, the derivatives are precomputed analytically and evaluated at the average cloud's location during the time integration.
If the carrier phase is computed numerically, the derivatives of velocity and temperature field with respect to space are numerically determined and interpolated at the average cloud's location during the fourth (time integration) stage.

\subsubsection{Splitting and convergence}

To compare point-cloud tracer results with MC-PSIC results in the tests below, we define the initial state of a point-particle cloud by its first two moments computed from the point-particles contained in the cloud.
As an example, for any two particle variables $\xi_p$ and $\eta_p$, a first moment $\overline{\eta}_p$ and second moment $\overline{\xi_p^\prime \eta_p^\prime}$ are computed as
\begin{subequations}\label{eq: IC_SPARSEandPSIC}
\begin{align}
    \overline{\eta}_p &= \frac{1}{N_p}\sum_{i=1}^{N_p}{\eta_p}_i, 
    \label{eq: IC_SPARSEandPSIC_means} \\
    \overline{\xi_p^\prime \eta_p^\prime} &= \frac{1}{N_p}\sum_{i=1}^{N_p}\left({\xi_p}_i-\overline{\xi}_p\right) \left({\eta_p}_i-\overline{\eta}_p \right),
    \label{eq: IC_SPARSEandPSIC_cm2} 
\end{align}
\end{subequations}
for a cloud containing $N_p$ point-particles.
\begin{figure}[h!]
	\centering
	\includegraphics[width=0.9\textwidth]{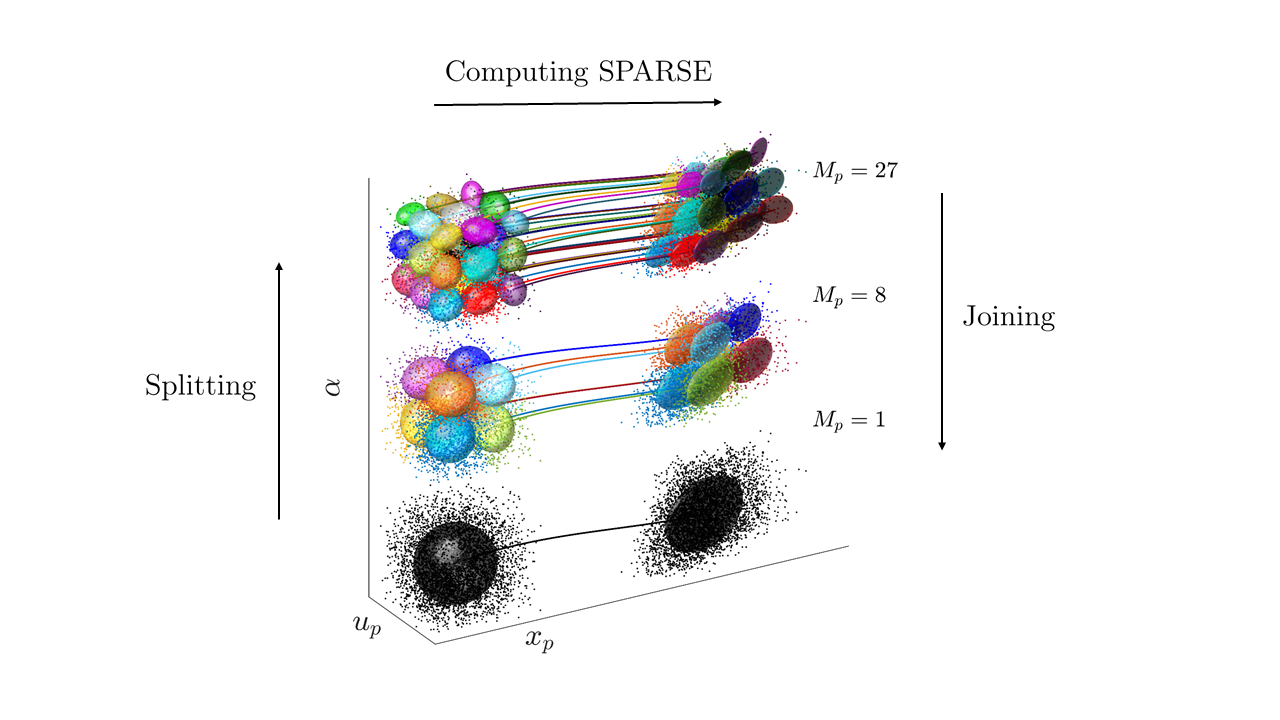}
	\caption[]{Illustration of a cloud of $N_p=10^3$ point-particles represented by its point-particles depicted as points and subclouds (point-clouds) as ellipses at the initial time $t=0$ and a later time $t>0$. The initial cloud is split into $M_p$ subclouds whose moments are computed from the PSIC particles according to the Algorithm~\ref{alg: spliting} described in the Appendix~\ref{app: splitting_algorithm}.}
	\label{fig: IC}
\end{figure}  

 SPARSE was shown to be third-order accurate for a union of computational clouds  with respect to the standard deviations in  each independent (physical  and phase space)  dimension in \cite{dominguez2023closed}.  To improve accuracy,  a splitting algorithm was proposed that converges the solution according to this rate by  reducing the cloud's sizing and  increasing the number of clouds along each independent dimension. 
To split, we divide the cloud of point-particles at the initial time into a union of uniform sets.
For example for a one-dimensional case we divide the cloud in $\alpha-x_p-u_p$, which leads to $M_p=M_p^{\alpha}M_p^{x_p}M_p^{u_p}$ subclouds with $M_p^{\alpha}$, $M_p^{x_p}$ and $M_p^{u_p}$ the number of divisions along $\alpha$, $x_p$ and $u_p$ respectively.
Considering uniform splitting along all dimensions, in the general three-dimensional non-isothermal case one has $M_p=M^{N+d+e}$ where $N$ is the number of modes considered in the random forcing, $d=1,2,3$ is the dimension of the problem and $e=1$ for non-isothermal and $e=0$ for isothermal flow.
We identify $M$ as the level of splitting of the cloud.
The splitting algorithm  is  schematically illustrated in Figure~\ref{fig: IC}, which visualizes the sampling of a group of point particles and its division into $M_p=1$, $M_p=2^3$ and $M_p=3^3$ subclouds, corresponding to the levels of splitting $M=[1, \ 2, \ 3]$, in the three dimensional space spanned by the particle's coordinate $x_p$, its velocity $u_p$ and the random forcing coefficient $\alpha$.
Once the cloud is split, the moments of each subcloud are computed using the relations in~\eqref{eq: IC_SPARSEandPSIC}. For further details of the algorithm, we refer to \cite{dominguez2023closed} and the description in Appendix~\ref{alg: spliting}. 


The moments of the cloud composed by the total number of point-particles, i.e., without splitting, are post-processed by joining each subcloud $k$, with $k=1,\dots, M_p $ as follows
\begin{subequations}\label{eq: join_moments}
\begin{align}
\overline{\phi} &= \sum_{k=1}^{M_p}w_k \overline{\phi}_k, 
\label{eq: join_moments_mean} \\
\overline{\xi^\prime \eta^\prime} &= \sum_{k=1}^{M_p}w_k \overline{\xi^\prime \eta^\prime}_k+\sum_{k=1}^{M_p}w_k(\overline{\xi}_k-\overline{\xi})(\overline{\eta}_k-\overline{\eta}), 
\label{eq: join_moments_cm2}
\end{align}
\end{subequations}
for any three solution variables $\phi$, $\eta$ and $\xi$.
We will refer to this as the \textit{Cumulative Cloud} throughout the paper, i.e., a multivariate Gaussian description of the cumulation of the subclouds where the moments of each of them have been computed with SPARSE and then joined with~\eqref{eq: join_moments}.
The moments of the Cumulative Cloud converge to the Monte Carlo results using~\eqref{eq: IC_SPARSEandPSIC} with the level of splitting.
The relations~\eqref{eq: join_moments} are exact and do not add any approximation error to the computation.

The Cumulative Cloud can be visualized as an ellipsoid or a prolate spheroid for any three variables of the particle phase and ellipse for any two variables by scaling the principle axis, in the directions of the eigenvectors of the covariance matrix, with the eigenvalues of the covariance matrix in the $N+d+e$ dimensional space. 
For example, for three particle variables $\phi$, $\xi$ and $\eta$, one has the following covariance matrices
\begin{align}
{K}_{\phi \xi \eta} = 
\begin{bmatrix}
    \overline{{\phi^\prime}^2} & \overline{\phi^\prime \xi^\prime} & \overline{\phi^\prime \eta^\prime}  \\
    \overline{\phi^\prime \xi^\prime} & \overline{{\xi^\prime}^2} & \overline{\xi^\prime \eta^\prime}  \\
    \overline{\phi^\prime \eta^\prime} & \overline{\xi^\prime \eta^\prime} & \overline{{\eta^\prime}^2}
\end{bmatrix}, \ \ 
{K}_{\phi \xi} = 
\begin{bmatrix}
    \overline{{\phi^\prime}^2} & \overline{\phi^\prime \xi^\prime} \\
    \overline{\phi^\prime \xi^\prime} & \overline{{\xi^\prime}^2} 
\end{bmatrix}, \ \ 
{K}_{\phi \eta} = 
\begin{bmatrix}
    \overline{{\phi^\prime}^2} & \overline{\phi^\prime \eta^\prime} \\
    \overline{\phi^\prime \eta^\prime} & \overline{{\eta^\prime}^2} 
\end{bmatrix}, \ \ 
{K}_{\xi \eta} = 
\begin{bmatrix}
    \overline{{\xi^\prime}^2} & \overline{\xi^\prime \eta^\prime} \\
    \overline{\xi^\prime \eta^\prime} & \overline{{\eta^\prime}^2} 
\end{bmatrix}.
\end{align}
Note that this representation may be also performed for a set of samples of the variables $\phi$, $\xi$ and $\eta$ in a discrete manner from MC-PSIC results.

In approximation, statically moments greater than second of the Cumulative Cloud can also be computed as follows
\begin{align}
    \overline{\phi^\prime \xi^\prime \eta^\prime} &= \sum_{k=1}^{M_p}w_k\left(\overline{\phi}_k-\overline{\phi}\right)
\left(\overline{\xi}_k-\overline{\xi}\right)\left(\overline{\eta}_k-\overline{\eta}\right), 
\label{eq: join_moments_cm3}
\end{align}
which is the equivalent of a Monte Carlo sampling of point-particles, but for the SPARSE subclouds.
In~\eqref{eq: join_moments_cm3} $\overline{\phi}$, $\overline{\xi}$ and $\overline{\eta}$ are the average values of the Cumulative Cloud and $k$ is an index that loops over all subclouds. 
The PDF of the Cumulative Cloud is non-Gaussian and can be reconstructed using a mixture distribution by adding weighted Gaussians that are defined with the two moments of each subcloud for a given level of splitting.
The PDF of any particle variable $\phi$ of a cloud split into $M_p$ subclouds can be computed as
\begin{align}
    f_{\phi}(\phi) &= \sum_{k=1}^{M_p}\frac{w_k}{\sqrt{2\pi \overline{{\phi_k^\prime}^2} } }\exp{\left(-\frac{\left(\phi-\overline{\phi}_k \right)^2}{2\overline{{\phi_k^\prime}^2}}\right)}. 
\label{eq: join_PDF}
\end{align}

\subsubsection{Error measurement}
The computational savings of using SPARSE can be estimated by determining the reduction of degrees of freedom as compared with MC-PSIC, i.e., the number of variables to solve along time.
The ratio of computational cost of SPARSE as compared to a MC-PSIC when considering $N$ modes to define the forcing functions is  
\begin{align}
    r =  \frac{ 2 N \left( d+ e\left( d+1\right) \right)+2 d^2+3 d +2e\left( d+1\right)}{2d+e} \left( \frac{M_p}{N_p} \right).
    \label{eq: savings}
\end{align}
The factor multiplying $M_p/N_p$ in the above expression is equal to $7$ for a three-dimensional $d=3$ non-isothermal $e=1$ simulation or smaller if simplified to less dimensions or isothermal when considering a single mode $N=1$.
Generally, we find that $N_p\gg M_p$ to reproduce accurate results by the SPARSE method, ensuring computational savings.

To measure the errors of the SPARSE method, we normalize the $L_2$ norm of the difference between the MC-PSIC and SPARSE results of a given variable with the $L_\infty$ norm of the MC-PSIC result as follows
\begin{align}
    \varepsilon(\cdot) = \frac{\| (\cdot)^{\mbox{SPARSE}}-(\cdot)^{\mbox{MC-PSIC}} \|_2}{\| (\cdot)^{\mbox{MC-PSIC}}\|_{\infty} }.
    \label{eq: error}
\end{align}
To measure the SPARSE error, we define averaged magnitudes of the first and second moments $\mu_1$ and $\mu_2$ respectively that for a SPARSE computation with $N_{var}$ number of variables ${\phi}_j$ with $j=1,\dots,N_{var}$, are given by
\begin{align}
    \mu_1 = \left(\sum_{j=1}^{N_{var}} \overline{\phi}_j^2 \right)^{1/2}, \ \ \ \ \ \
    \mu_2 =   \begin{vmatrix}
        \overline{{\phi_1^\prime}^2} & \dots & \overline{\phi_{N_{var}}^\prime \phi_1^\prime} \\
        \vdots & \ddots  & \vdots \\
        \overline{\phi_1^\prime \phi_{N_{var}}^\prime} & \dots  & \overline{{\phi_{N_{var}}^\prime}^2} \\
\end{vmatrix}  .
\label{eq: mu1_mu2}
\end{align}
Here $N_{var}$ is the total number of variables in the left hand side of~\eqref{eq: MoM}.
This way we take into account all variables that SPARSE computes when analyzing the errors $\varepsilon(\mu_1)$ and $\varepsilon(\mu_2)$ with the definition in~\eqref{eq: error}.

\section{Numerical experiments} \label{sec: tests}

\subsection{One-dimensional sinusoidal velocity field}\label{sec: test_sine1D}

To test the accuracy of SPARSE as compared to MC-PSIC, we consider a cloud of point-particles carried by an isothermal one-dimensional sinusoidal flow velocity field given by
\begin{align}
    u(x) = 1 +\frac{1}{2}\sin{\left( 2x \right)},
    \label{eq: sine1D_flow}
\end{align}
where the forcing $f_1$ is defined according to~\eqref{eq: f1andf2} and $g_1$ according to the the Schiller and Naumann correlation such that
\begin{align}
    f_1 = \alpha \left(1+0.15Re_p^{0.687}  \right).
    \label{eq: sine1D_correlation}
\end{align}
This test is not necessarily physical, but the sinusoidal velocity field is  typical and representative of a  modal (Fourier) analysis in chaotic flows. 
The simple one-dimensional flow enables testing and verification of accuracy.
Moreover, it provides fundamental insights into the behavior of cloud dynamics that can serve as a reference for physically relevant simulations.

The initial condition is set according to a uniform distribution $\mathcal{U}$ to sample the initial locations $x_{p_0}$ and velocities $u_{p_0}$ such that
\begin{align}
{x_p}_0 \sim \mathcal{U}\left[{{x_p}_0}_{min}, \ {{x_p}_0}_{max}\right], \ \ \
    {u_p}_0 \sim \mathcal{U}\left[{{u_p}_0}_{min}, \ {{u_p}_0}_{max}\right].
\label{eq: sine1D_IC_xp_up}
\end{align}
We test the formulation for both, dF and rF particles with $\alpha=1$ and 
\begin{align}
    \alpha \sim \mathcal{U}\left[\alpha_{min}, \ \alpha_{max}\right], \label{eq: sine1D_IC_alpha}
\end{align}
respectively.
The uniform distributions $\mathcal{U}$ are defined by the limit values $\alpha_{min}$ and $\alpha_{max}$, which are directly related to its average and standard deviation as $\alpha_{min}=\overline{\alpha}-\sqrt{3}\sigma_{\alpha}$ and $\alpha_{max}=\overline{\alpha}+\sqrt{3}\sigma_{\alpha}$ with $\overline{\alpha}=1$ and $\sigma_{\alpha}=0.3$.
We also initialize the particle locations and velocities, according to the uniform distributions~\eqref{eq: sine1D_IC_xp_up}, with ${\overline{x}_p}_0={\overline{u}_p}_0=0$, $\sigma_{{x_p}_0}=0.2$ and $\sigma_{{u_p}_0}=0.1$.
Initially, the variables $\alpha$, ${x_p}_0$ and ${u_p}_0$ are statistically independent and uncorrelated and therefore, the following moments are zero
\begin{align}
    \overline{\alpha^\prime {x_p^\prime}_0} = \overline{\alpha^\prime {u_p^\prime}_0} = \overline{{x_p^\prime}_0 {u_p^\prime}_0} = 0.
\end{align}
The particle response time is set to $St=0.5$, i.e., the particle time scale is similar to the carrier flow convective time scale,  such that the inertial effects are significant and of influence to test.
We also set the reference Reynolds number to $Re_\infty=10^4$ so that the relative particle Reynolds number is greater than unity and the forcing of the particles lies in a regime beyond the Stokes drag.
The non-dimensional particle density is $\rho_p=250$ and the non-dimensional particle diameter $d_p=2\times10^{-3}$.

\begin{figure}[h!]
	\centering
	\subfloat[]{
		\label{fig: sine1D_phaseSpace_deter_2x2}
		\includegraphics[width=0.4\textwidth]{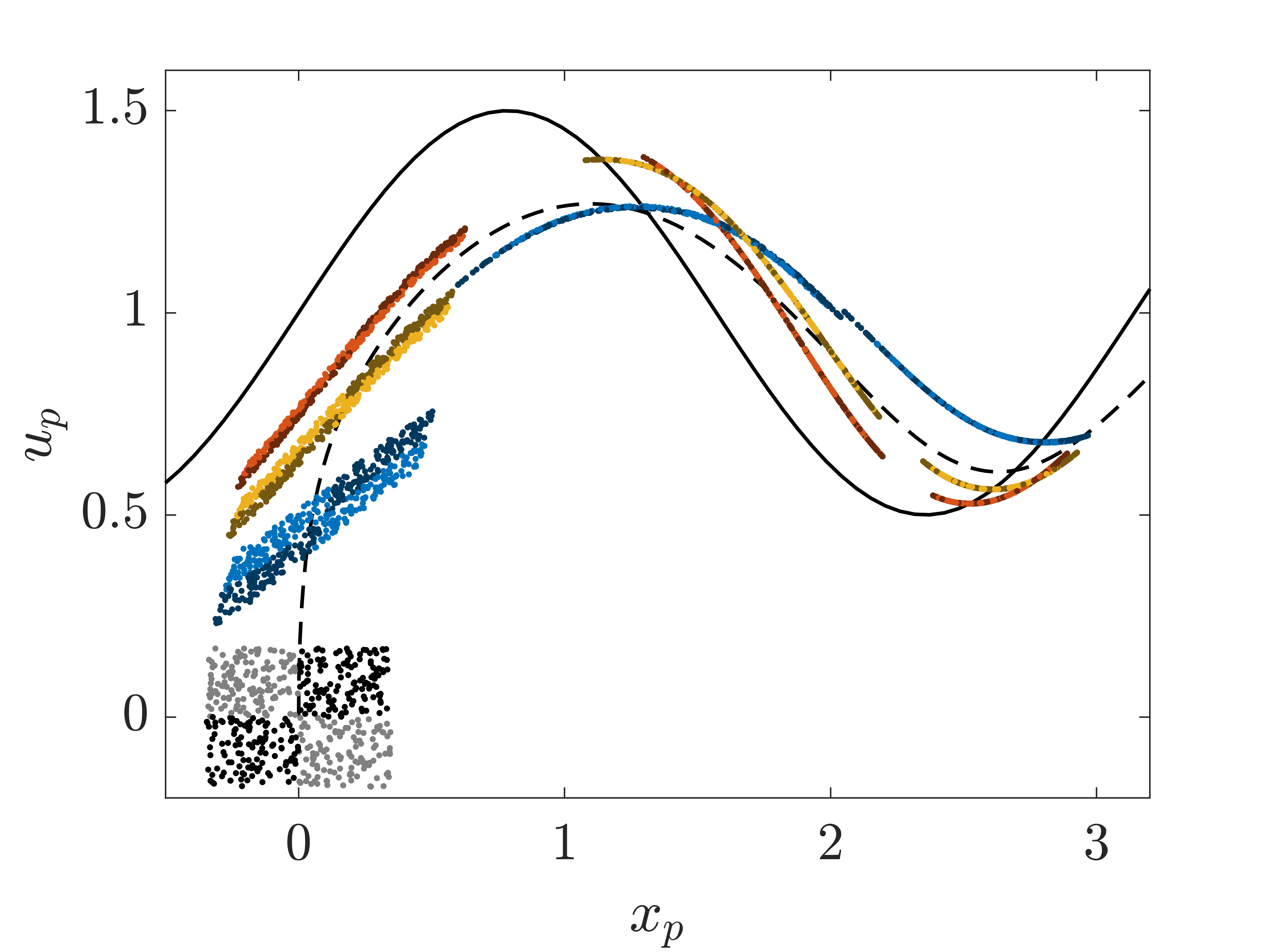}}
	\subfloat[]{
		\label{fig: sine1D_phaseSpace_PSIC_detVSrand}
		\includegraphics[width=0.4\textwidth]{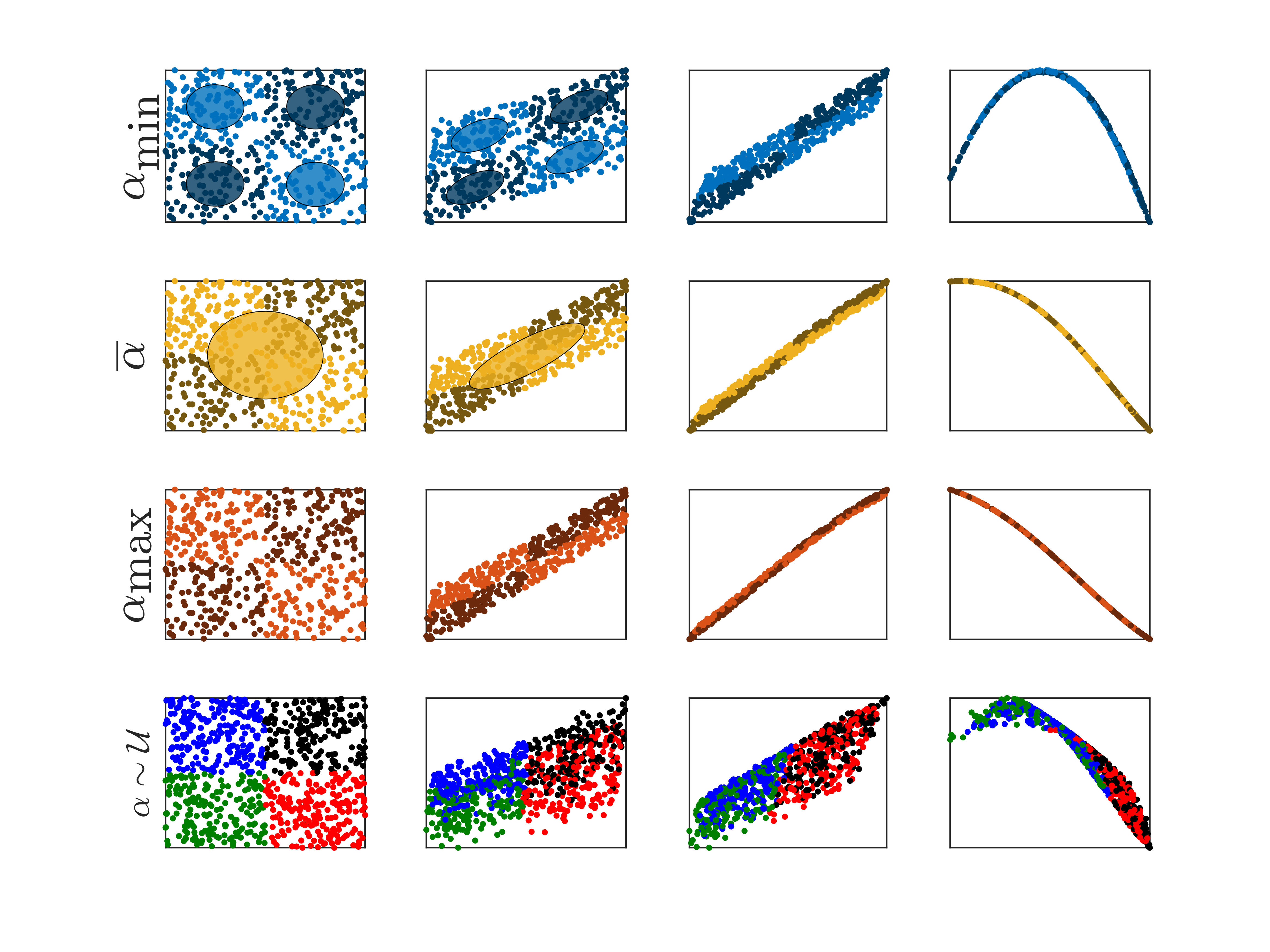}} \\
    \subfloat[]{
		\label{fig: sine1D_phaseSpace_random_2x2x2}
		\includegraphics[width=0.4\textwidth]{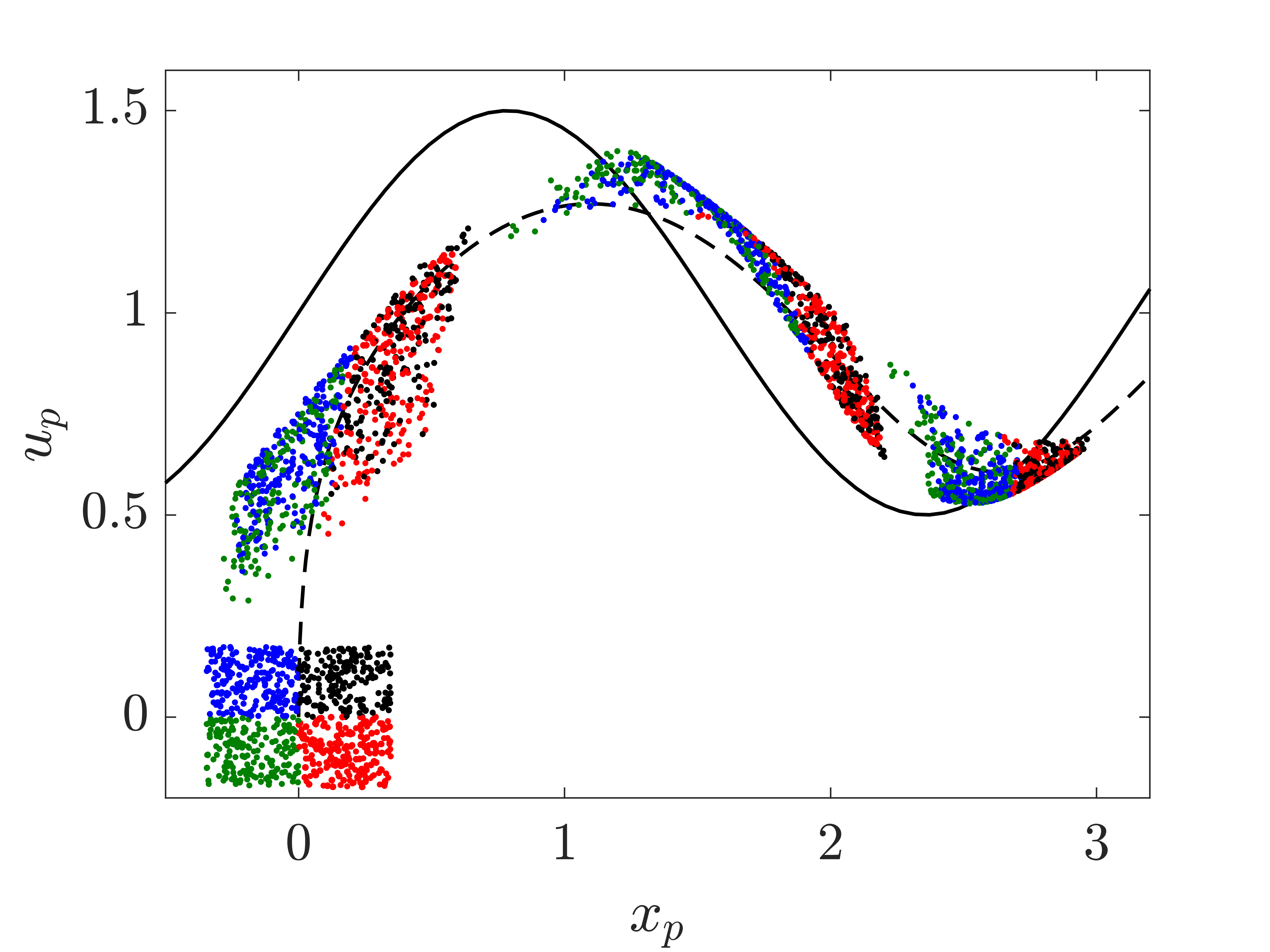}} 
	\subfloat[]{
		\label{fig: sine1D_phaseSpace_random_2x2x2_SPARSE}
		\includegraphics[width=0.4\textwidth]{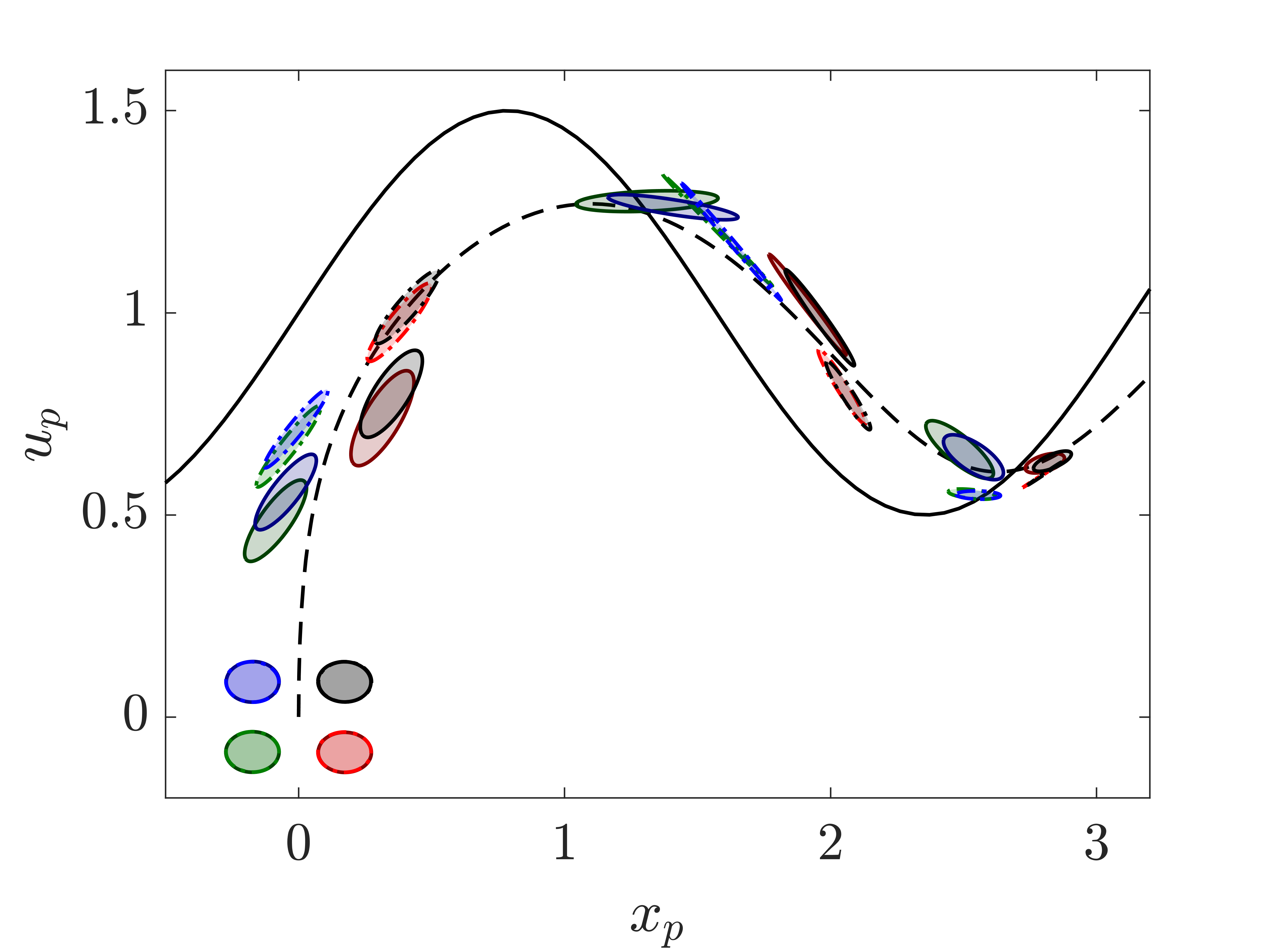}}
	\caption[]{Particle phase solution of the sinusoidal one-dimensional test case for: (a) dF particle clouds for three deterministic values of the random coefficient $\alpha_{min}$, $\overline{\alpha}$ and $\alpha_{max}$ for four instants of time computed with MC-PSIC divided in quadrants by colors; (b) point-particle clouds at four instants of time for the cases in (a) and (c) in zoomed $x_p-u_p$ axis divided in quadrants by colors and with ellipses representing the total set of particles (second row) or subsets (first row); (c) rF cloud considering $\alpha \sim \mathcal{U}$ for four instants of time with coloring according to quadrants in $x_p-u_p$, i.e., each quadrant has a distribution of values of $\alpha$; and (d) SPARSE solution with $M=2$ of the same case than in (c) with ellipses with continuous border line for the lower range of $\alpha$ and dot-dashed border line for the larger range of $\alpha$. The times in (a), (c) and (d) are $t=[0, \ 0.35, \ 1.75, \ 2.98]$ and for (b) $t=[0, \ 0.1, \ 0.35, \  1.75]$. The flow is depicted with a black continuous line in (a), (c) and (d). For a better visualization, $500$ point-particles are depicted from the $N_p=10^5$ used for the computations. }
\end{figure}

{To develop a general understanding of the effect of the deterministic and random forcing functions  on the  solution behavior of clouds of particles in a sinusoidal carrier field, we first discuss the traces of three groups of PSIC point-particles that are computed with  a deterministic forcing function according to  three values of $\alpha$, including $\alpha_{min}$, $\alpha_{max}$, and $\overline{\alpha}=(\alpha_{min}+\alpha_{max})/2=1$.
We will refer to these determinstic forced cloud traces as  dF clouds throughout the rest of the paper.
The groups for each $\alpha$ contain $N_p=10^5$ particles to ensure a converged MC moment error that is approximately $1/\sqrt{N_p}$.

The locations of the  point-particles so computed are visualized in Fig.~\ref{fig: sine1D_phaseSpace_deter_2x2} for different instants of time in the phase space $x_p-u_p$. Initially, all clouds coincide. As time progresses, 
 each group  accelerates in positive direction   because the carrier velocity   (black line) is positive, which combined with  negligible particle velocities at early times yields a positive relative particle velocity. The relative velocity has the same sign as the acceleration, and thus is  positive  at each particle location also.  
Each point-particle cloud accelerates proportionally to the magnitude of its forcing coefficient, $\alpha$. 
For $\alpha \rightarrow \infty$ the particle time response $St / \alpha \rightarrow 0$ and the trajectory tends to that of a particle without inertia (a tracer).
For small $\alpha$, the time response increases.
In the limit $\alpha \rightarrow 0$, the time response tends to infinity $St/ \alpha \rightarrow \infty$ and its movement is asymptotically zero.
At later times,
the clouds accelerate and decelerate successively around the average unity value of the carrier phase velocity.
Because of its inertia, each particle's trajectory  has a phase delay and decrease of amplitude as compared to those of the carrier flow velocity field~\eqref{eq: sine1D_flow} .
Initially uniformly distributed around the average location $({x_p}_0,{u_p}_0)=(0,0)$, the particle clouds stretch and rotate  in phase space while they are simultaneously advected by the flow.

A more detailed understanding of the groups' rotation and deformation characteristics can be obtained by the four quadrant depiction of the three tracers (first three rows) in time (four collumns) in  Fig.~\ref{fig:  sine1D_phaseSpace_PSIC_detVSrand}.
In each quadrant the particles are colored according to a shade of a color scheme. Moreover, the mean location, the principle stretching and rotation  of the clouds in each quadrant  and of a Cummulative Cloud can be 
represented by the mean location, the principle axes and the rotation  of an  ellipse,
 based on  the average $(\overline{x}_p,\overline{u}_p)$ and the eigenvalues and eigenvectors of the covariance matrix $K_{x_p u_p}$.
These  moments are of course precisely the moments that are modeled by SPARSE  and the MC-PSIC moment results can thus be used  for comparison and assessment of the predictive accuracy of SPARSE. 
The ellipses and groups of  point-particles show that particles in different quadrants align naturally in phase space over a time interval that is on  the order of $St/\alpha$ as the velocities of all particles are damped towards the carrier velocity.
The particle clouds are straining  in the $x_p-u_p$ plane consistently with the velocity gradient of the carrier phase and the particles' inertia. At the initial time this velocity gradient is positive and thus the particle cloud widens in the physical space. A particle group with a smaller response time, $St/\alpha$ stretches more along the velocity gradient. In phase space the cloud compresses as it settles to the carrier velocity. }


A randomly forced cloud tracer with $\alpha ~\sim \mathcal{U}$ does not concentrate along a line  like the dF cloud, but     distributes over a region according to the probability of the forcing coefficient and of the associated deterministic tracers  (Fig.~\ref{fig: sine1D_phaseSpace_random_2x2x2}).This distribution is a measure of the uncertainty in the solution. The footprint of the distributed particles in phase space for an rF cloud  is the  confidence interval that concisely visualizes this uncertainty. The rF cloud's dispersion can be understood qualitatively as a summation of the dF tracers that are determined with  forcing coefficients that randomly span the range of the PDF forcing coefficient. Roughly this  can be seen as  a summation of the locations of the dF clouds in rows 1 to 3 in Fig.~\ref{fig: sine1D_phaseSpace_PSIC_detVSrand}. In addition to this increased dispersion, the distributions of  rF clouds in each quadrant overlap and  can  thus be said to be more mixed at later times.  The driving mechanisms for this enhanced mixing and increase in the associated confidence intervals is the virtual stress correlation between the random forcing and the flow field fluctuations that appears in the SPARSE formulation and that drives the uncertainty in the rF cloud's solution.
The dF cloud that is forced according to the average value of $\bar{\alpha}$ compares closely with  the mean solution of a rF cloud  (compare second and fourth row in Fig.~\ref{fig: sine1D_phaseSpace_PSIC_detVSrand}). 

Using the SPARSE point-cloud tracer we can compute the dispersion of the clouds of point-particles depicted in Figs.~\ref{fig: sine1D_phaseSpace_deter_2x2}--~\ref{fig: sine1D_phaseSpace_random_2x2x2}   at a reduced computational cost.
Splitting  the cloud in quadrants so that $M=2$, i.e., two divisions along $\alpha$, $x_p$ and $u_p$, we specify  the first two moments of each subdivision (subset of point-particles)  as initial conditions for the point-cloud simulation (see Fig.~\ref{fig: IC}).
The resulting phase space depicted with an ellipse for each subcloud is represented in Fig~\ref{fig: sine1D_phaseSpace_random_2x2x2_SPARSE} for different instants of time.
The solution in each quadrant is now represented by two ellipses along the $\alpha$ dimension that coincide initially in phase space. At later times the two ellipses are located at different positions and are deformed differently because they are forced differently.
 The point clouds show the same trends in terms of advection, straining and rotation as compared to the groups of  MC-PSIC particles. 

\begin{figure}[h!]
	\centering
	\subfloat[]{
		\label{fig: sine1D_phaseSpace}
		\includegraphics[width=0.4\textwidth]{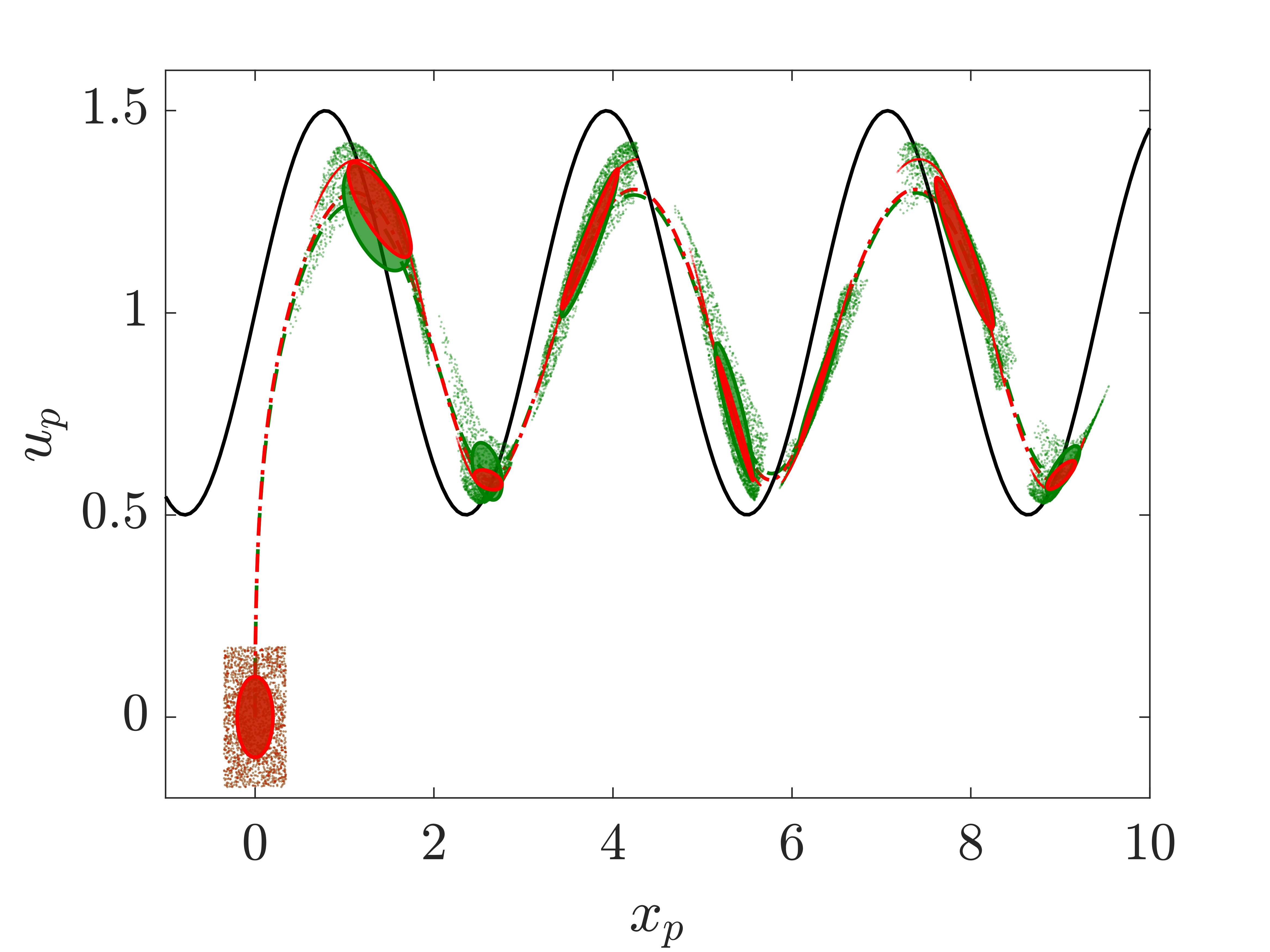}}  
	\caption[]{Particle phase solution of the sinusoidal one-dimensional test case for the rF (green) and dF (red) computed with SPARSE with $M=7$ as ellipses for the Cumulative Cloud at different instants of time along the average particle phase (green dashed line for rF and red dot-dashed line for dF). The point-particles are depicted as points with the corresponding coloring. The flow is depicted with a black continuous line. The time instants are $t=[0, \ 1.43, \  2.85, \ 4.28, \ 5.73, \  7.15, \  8.5, \ 10]$. For a better visualization, $500$ point-particles are depicted from the $N_p=10^5$ used for the computations.}
	\label{fig: sine1D_phaseSpace_and_splitting}
\end{figure}

\begin{figure}[h!]
	\centering
	\subfloat[]{
		\label{fig: sine1D_mean_xpup}
		\includegraphics[width=0.32\textwidth]{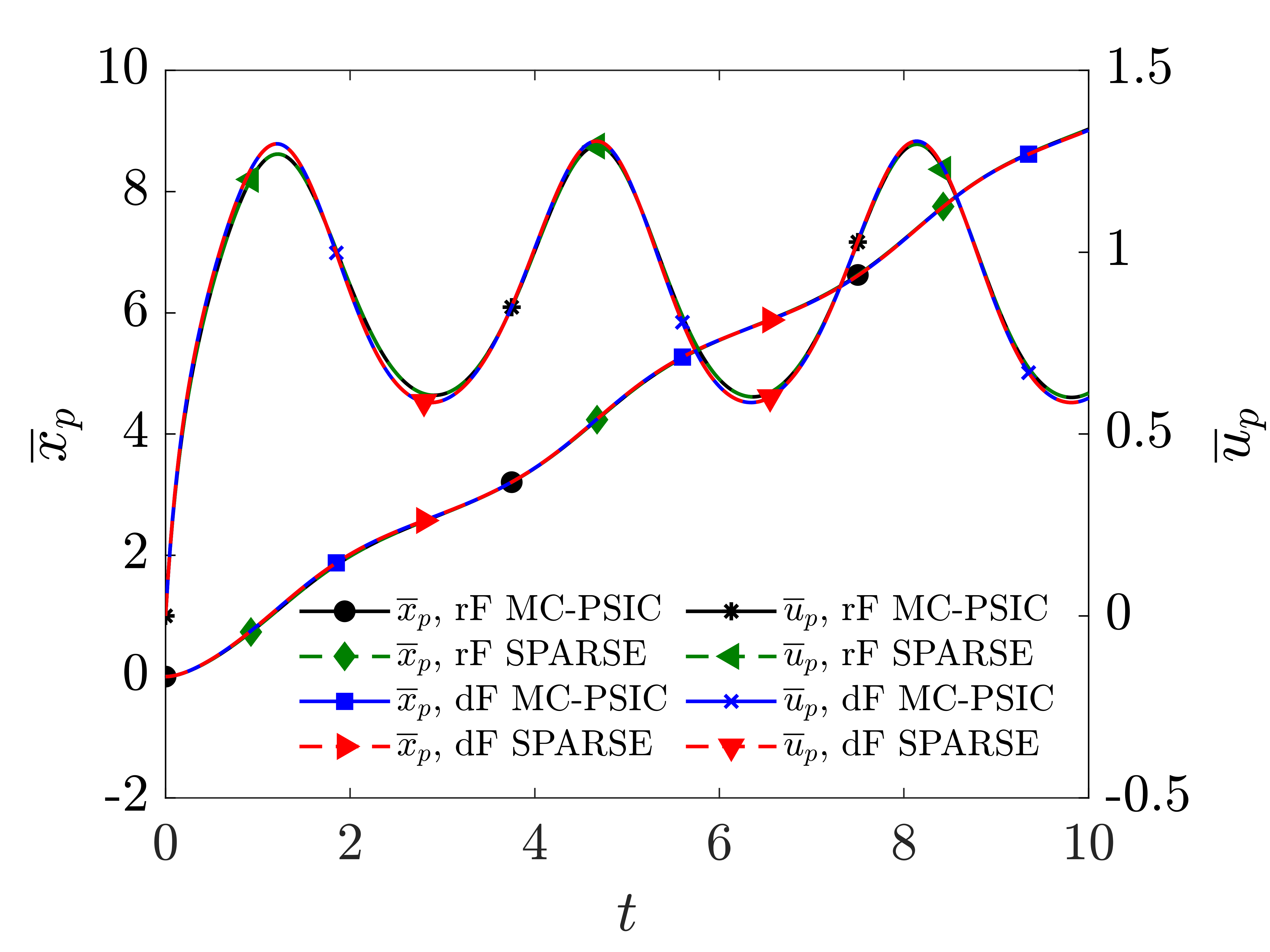}}
        \hfill	
	\subfloat[]{
		\label{fig: sine1D_sigma_xp}
		\includegraphics[width=0.32\textwidth]{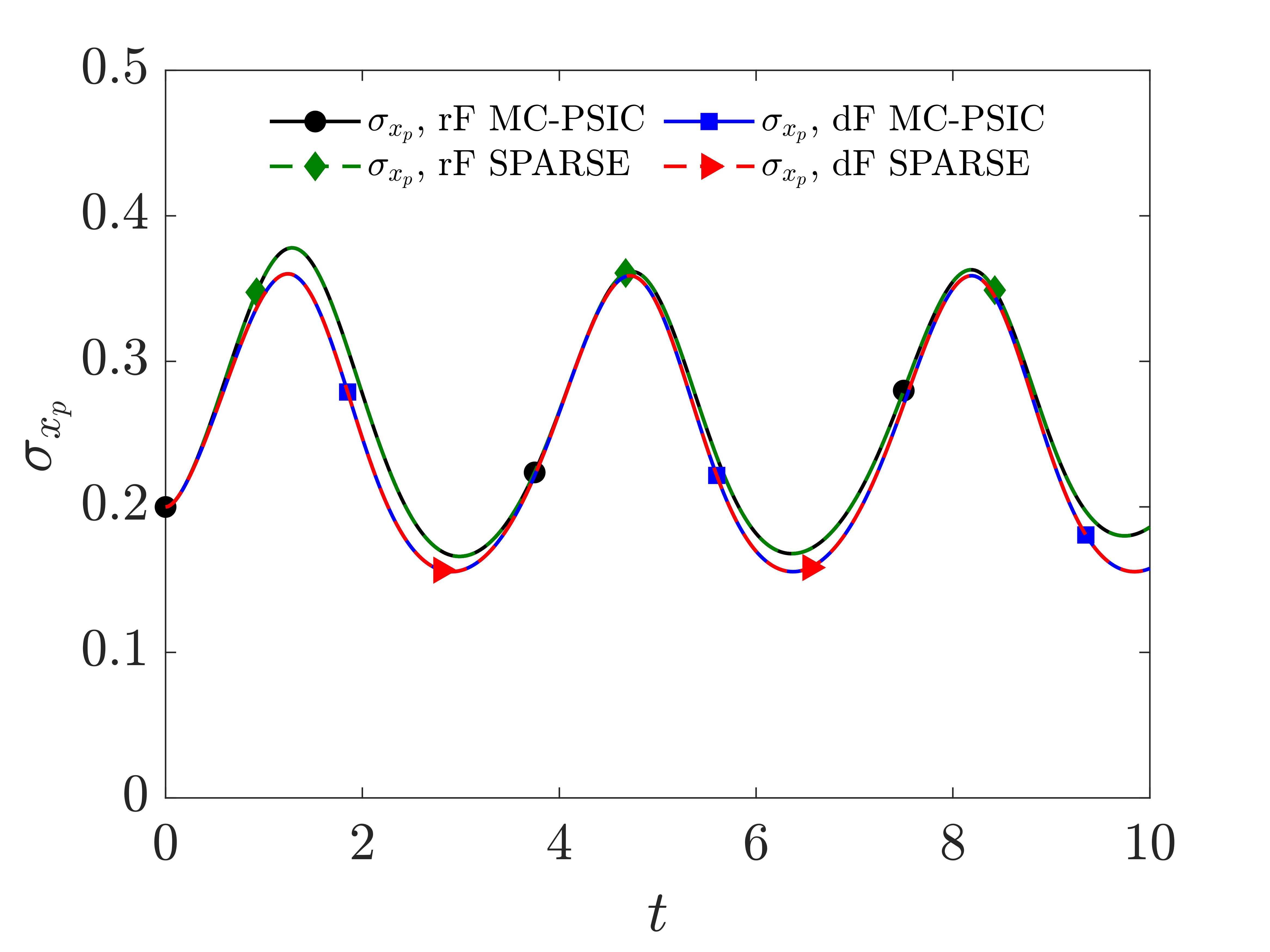}} 
		\hfill
	\subfloat[]{
		\label{fig: sine1D_sigma_up}
		\includegraphics[width=0.32\textwidth]{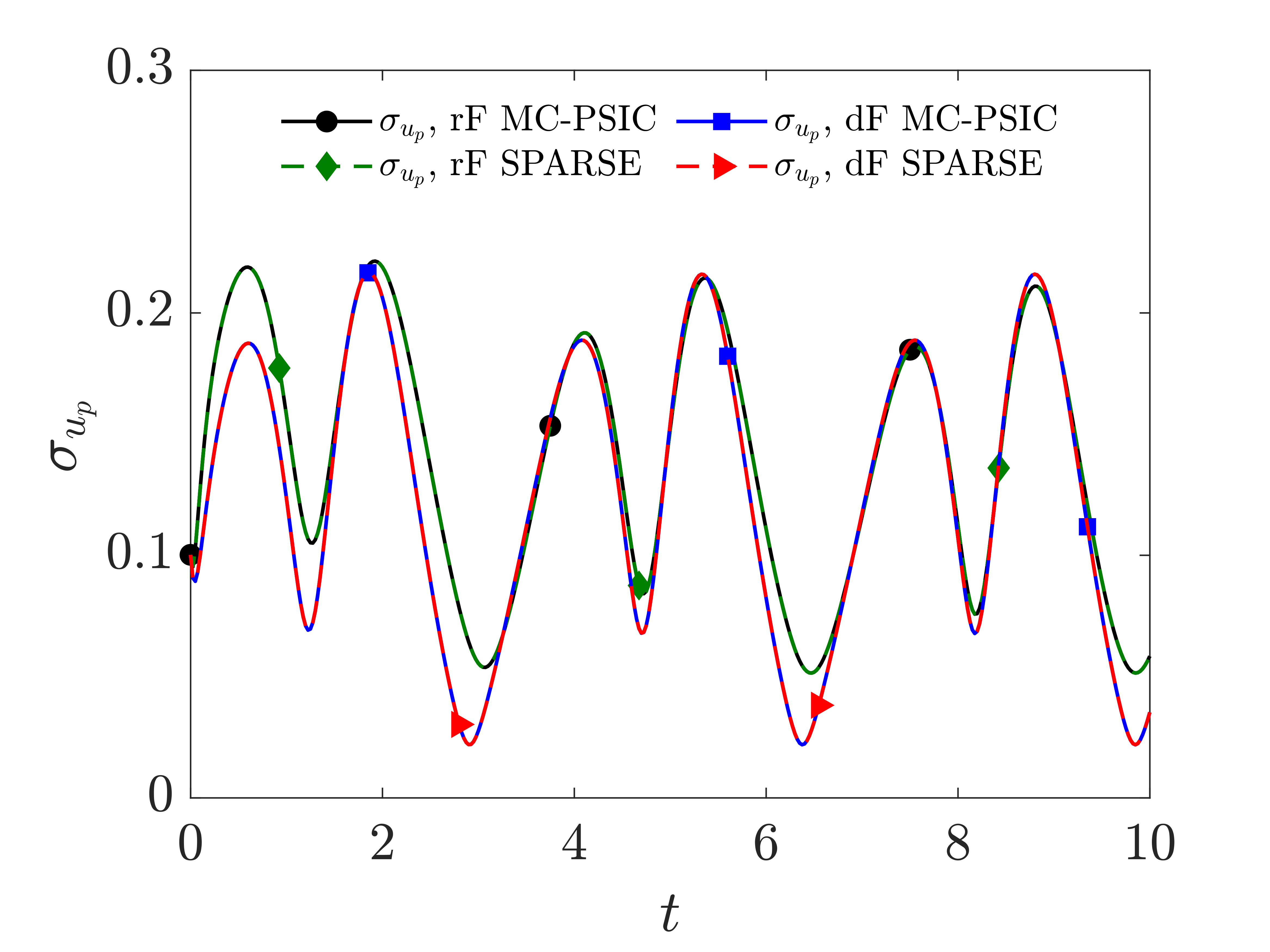}}  \\
	\subfloat[]{
		\label{fig: sine1D_xpup}
		\includegraphics[width=0.32\textwidth]{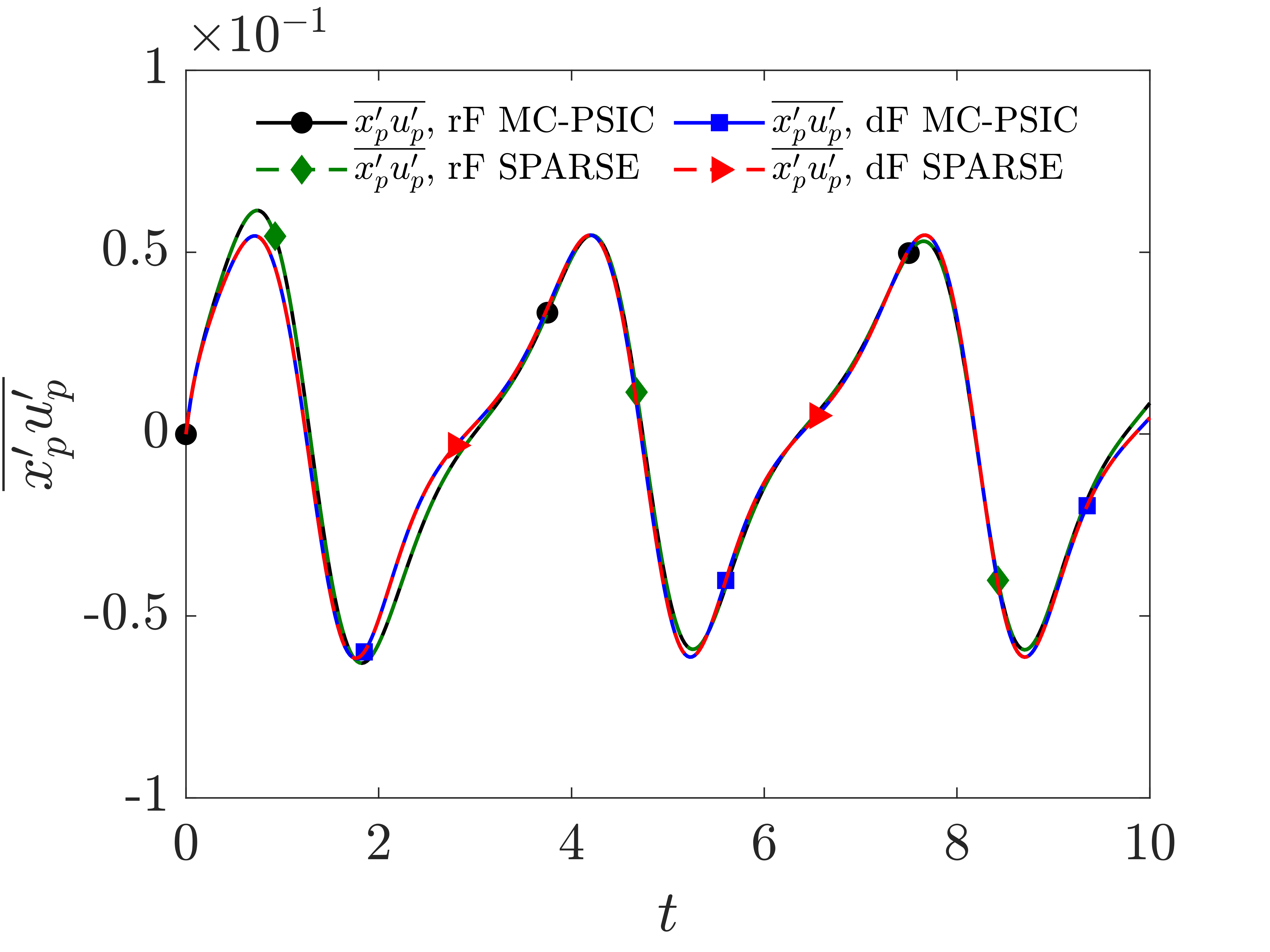}} 
		\hfill
	\subfloat[]{
		\label{fig: sine1D_axpaup}
		\includegraphics[width=0.32\textwidth]{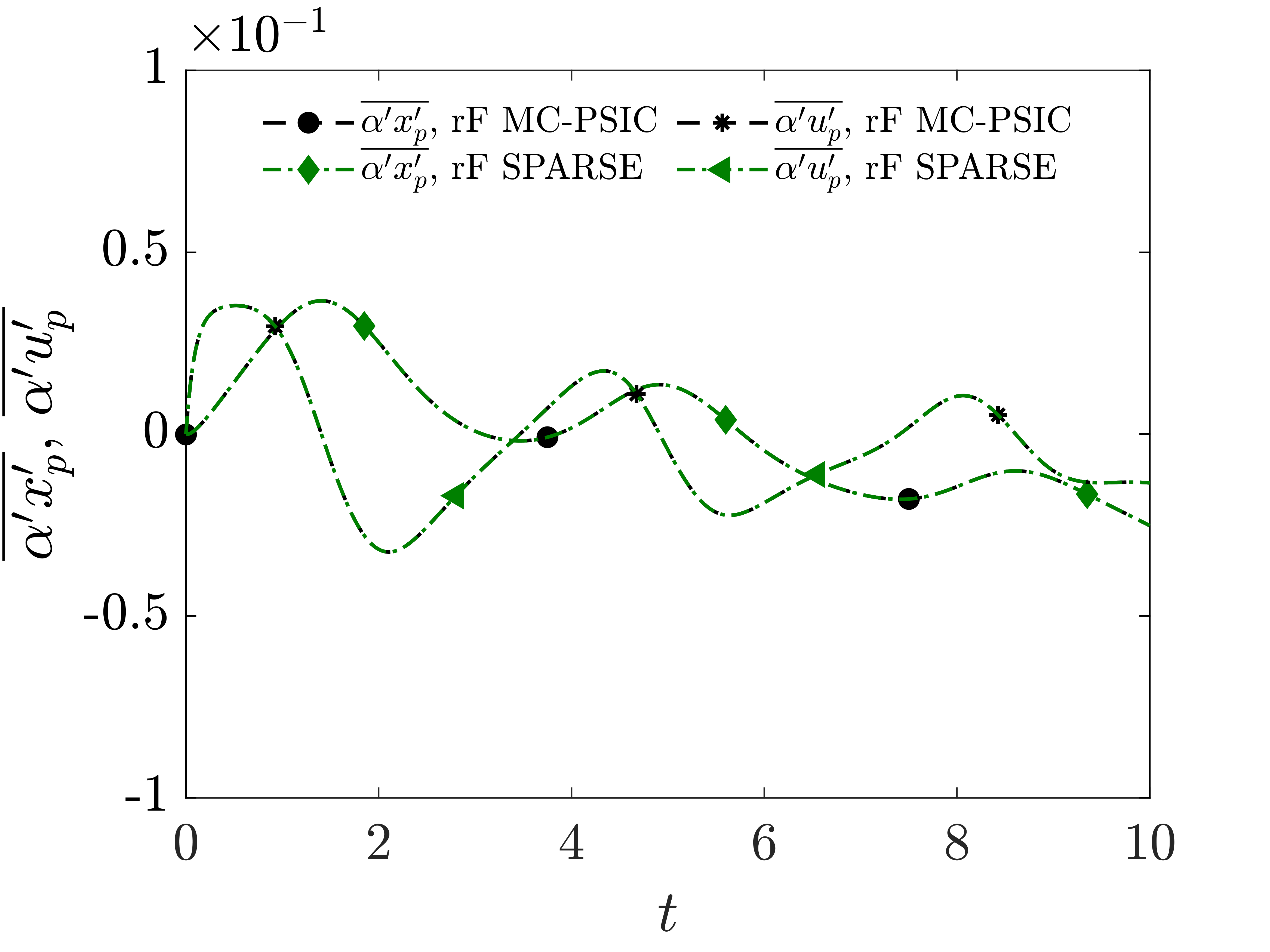}} 
		\hfill
	\subfloat[]{
		\label{fig: sine1D_errors}
		\includegraphics[width=0.32\textwidth]{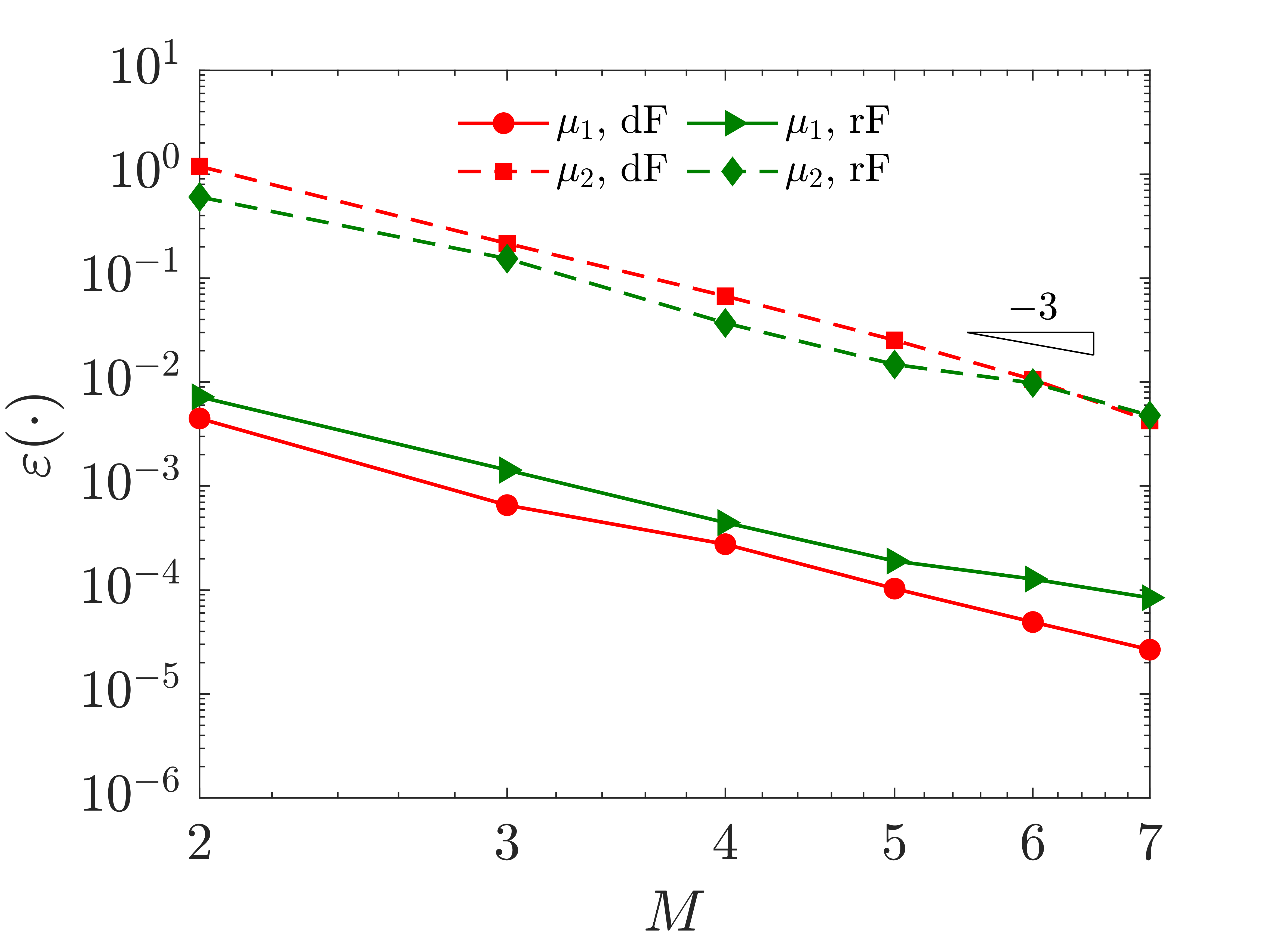}} 
	\caption[]{First and second moments of the one-dimensional sinusoidal velocity field test case for stochastic and deterministic forcing computed with SPARSE and MC-PSIC. Figure (a) shows the average particle location and velocity, (b) standard deviation of the particle location, (c) standard deviation of the particle velocity, (d) combined second moment of the particle phase, (e) combined second moments of the random coefficient and particle phase variables and (f) expected convergence of SPARSE.}
	\label{fig: sine1D}
\end{figure}
\begin{figure}[h!]
	\centering
	\subfloat[]{
		\label{fig: sine1D_cm3_xp}
		\includegraphics[width=0.32\textwidth]{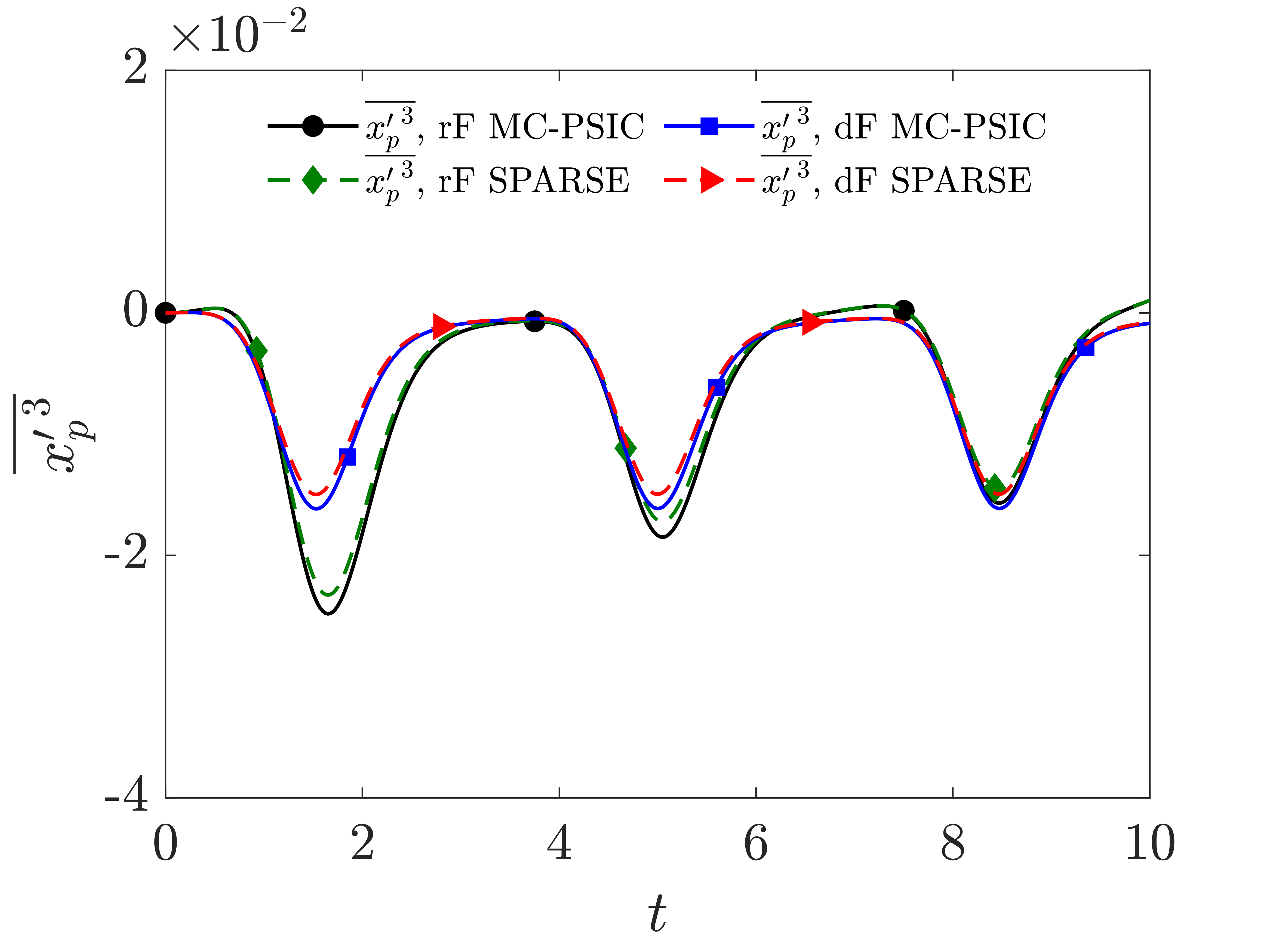}}
        \hfill	
	\subfloat[]{
		\label{fig: sine1D_cm3_up}
		\includegraphics[width=0.32\textwidth]{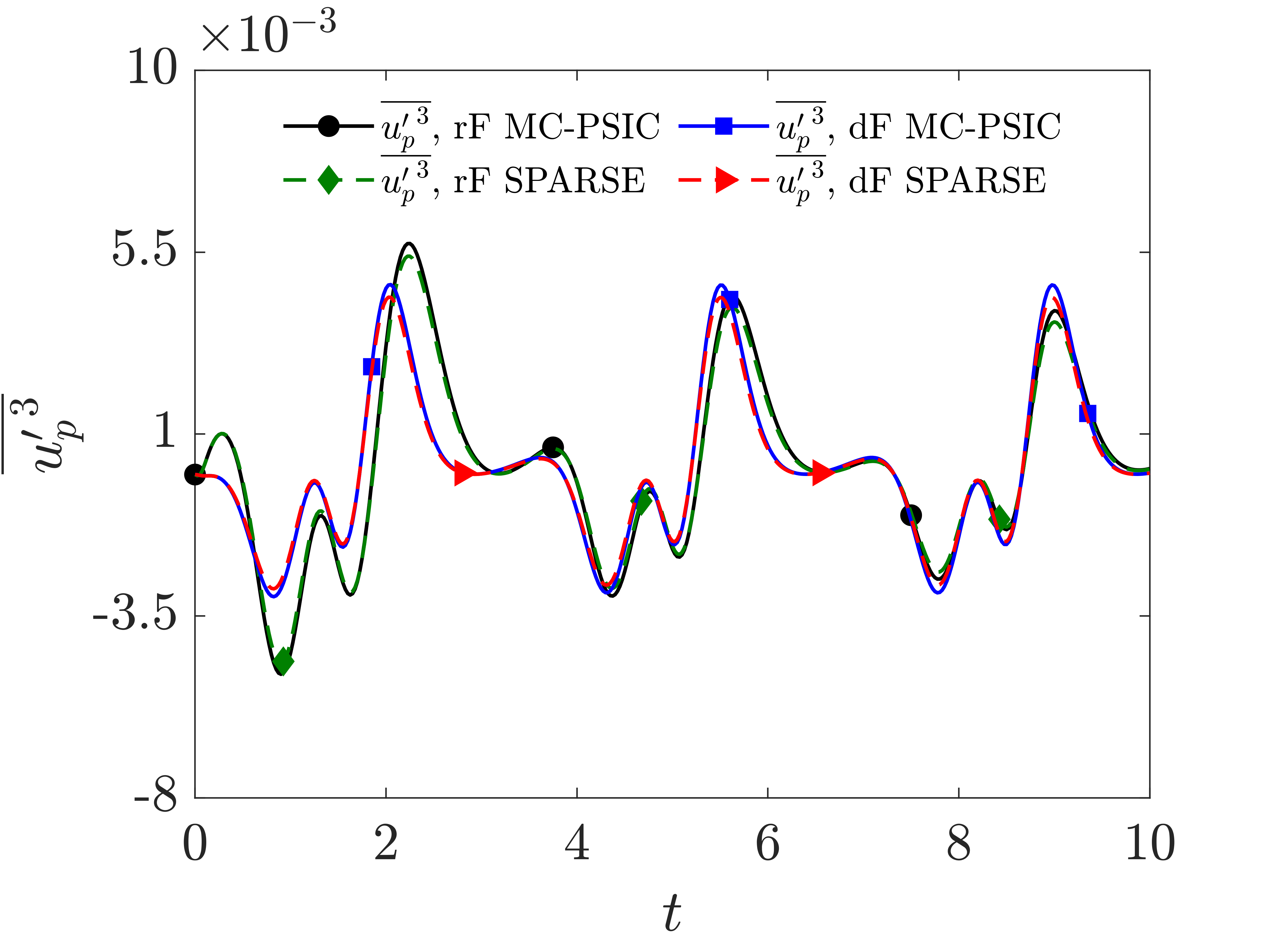}} 
		\hfill
	\subfloat[]{
		\label{fig: sine1D_xpxpup}
		\includegraphics[width=0.32\textwidth]{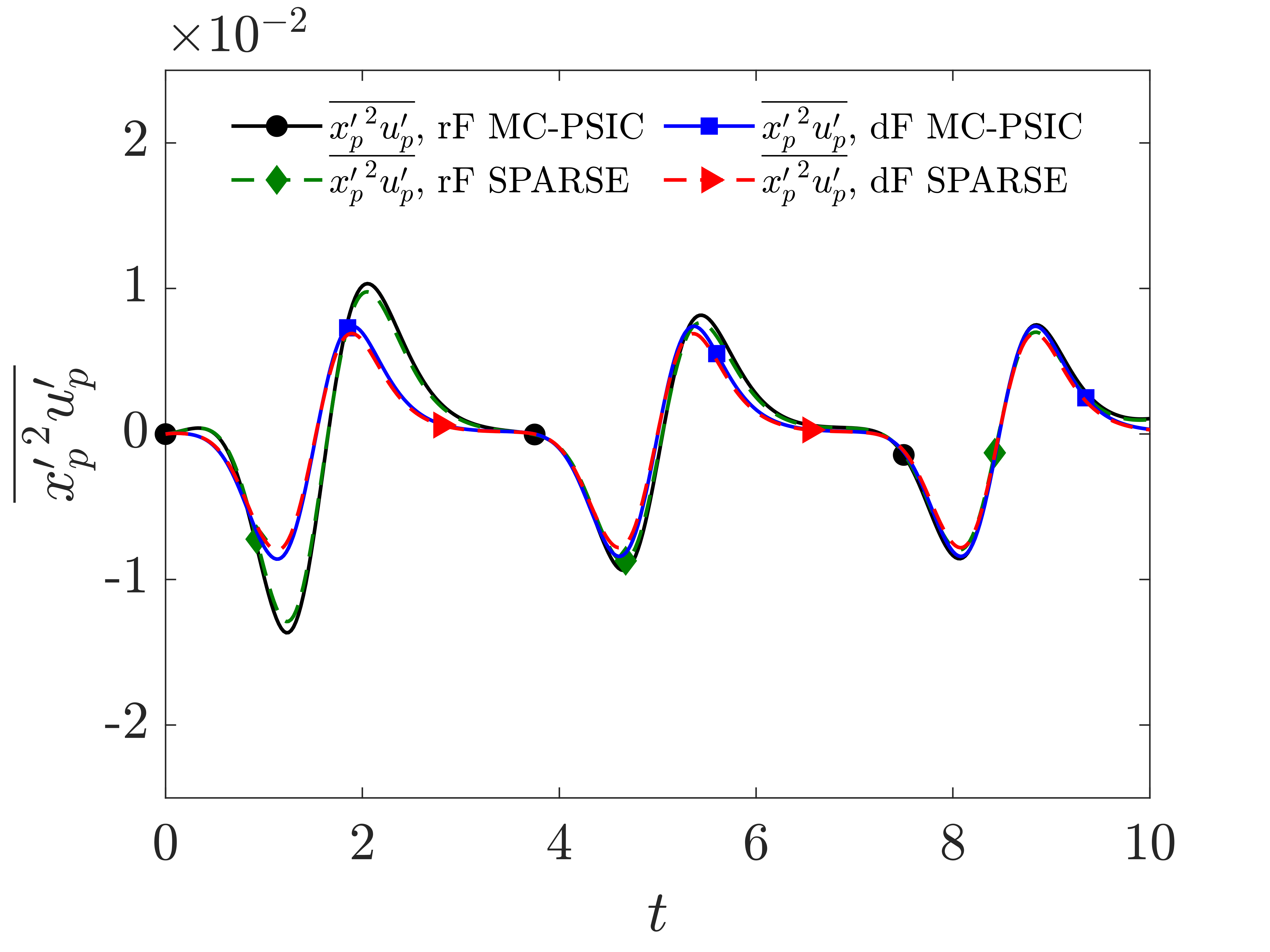}}  \\
	\subfloat[]{
		\label{fig: sine1D_xpupup}
		\includegraphics[width=0.32\textwidth]{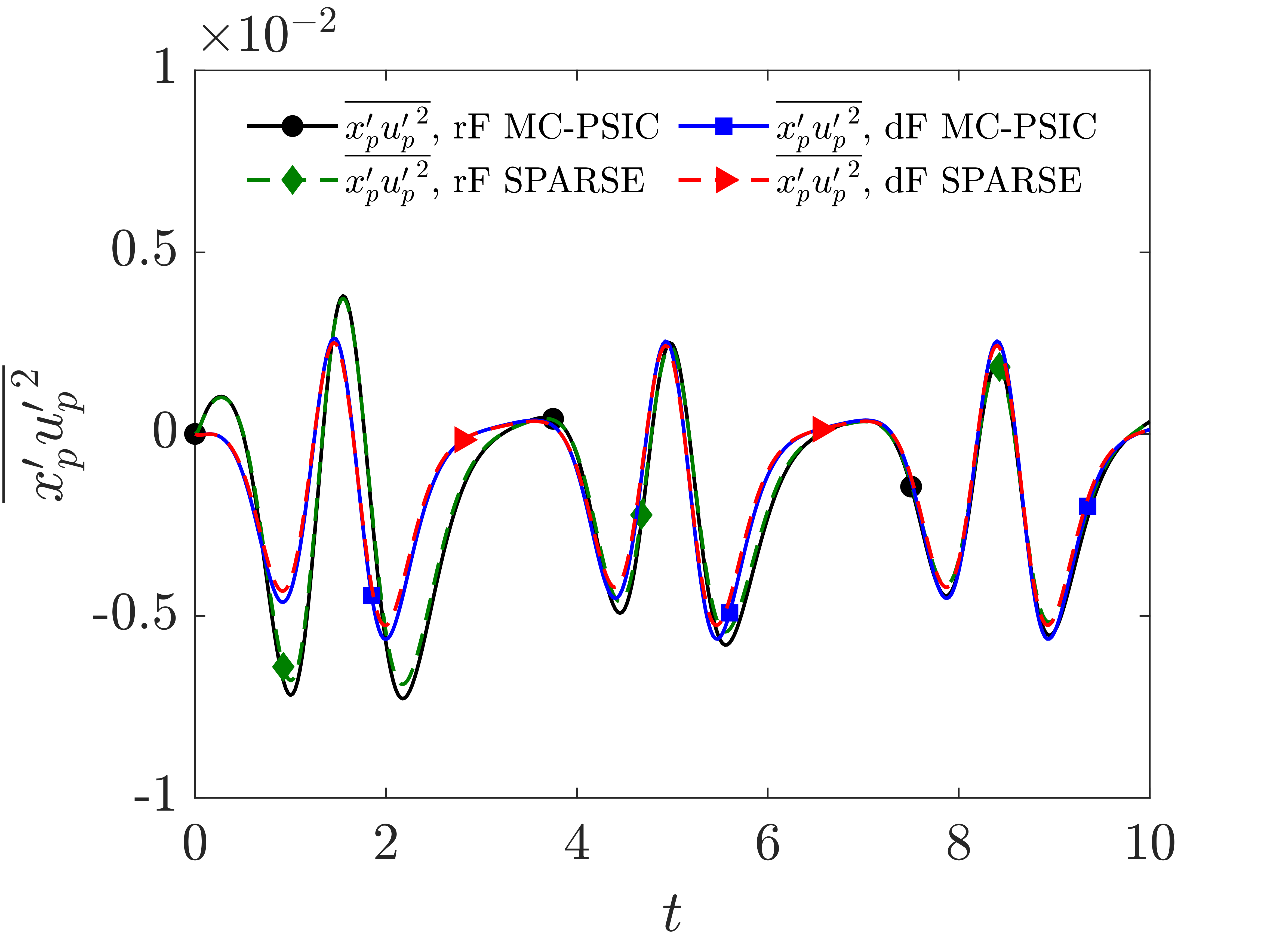}} 
		\hfill
	\subfloat[]{
		\label{fig: sine1D_alphaxpup}
		\includegraphics[width=0.32\textwidth]{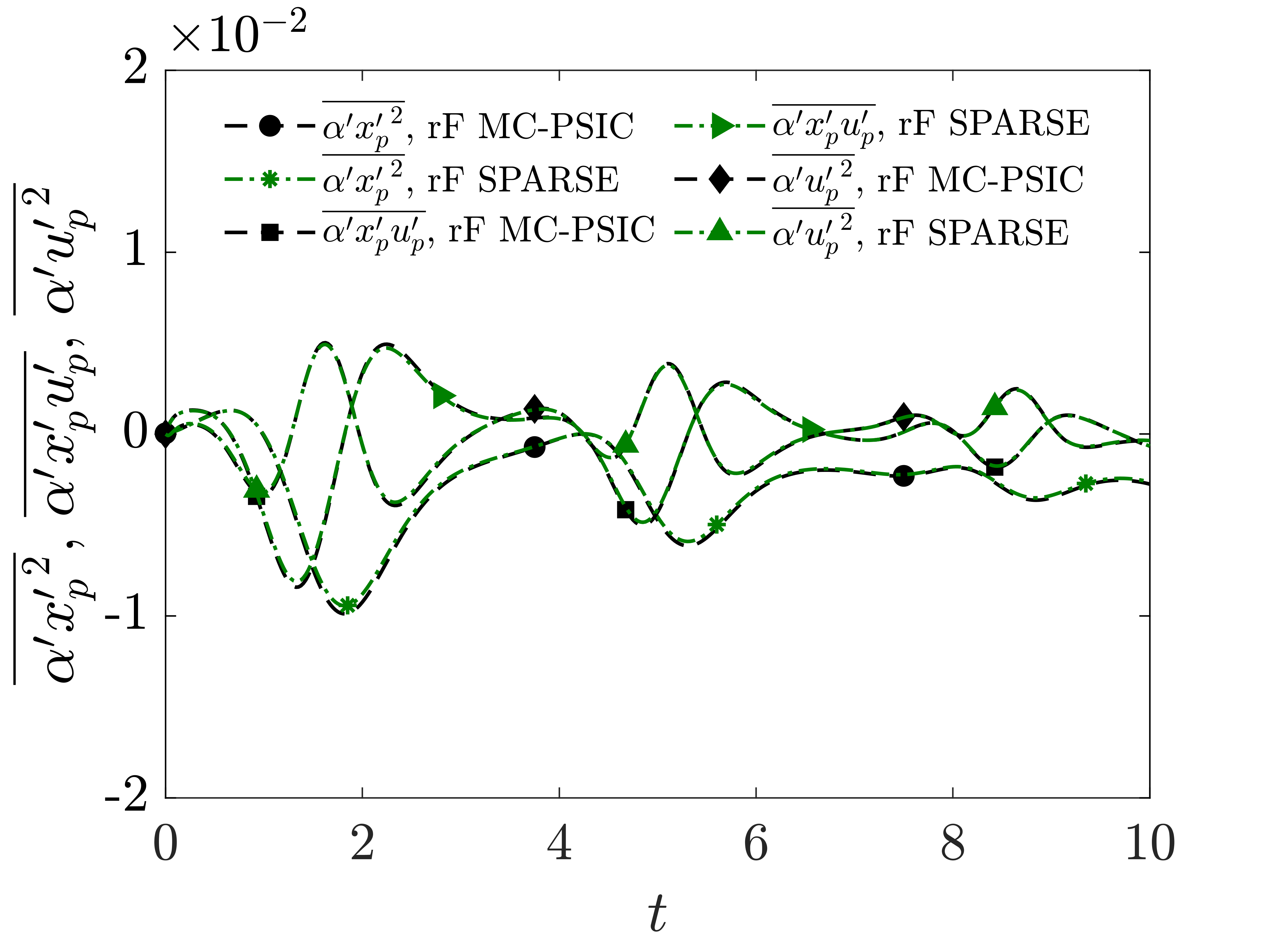}} 
		\hfill
	\subfloat[]{
		\label{fig: sine1D_alphaalphaxp}
		\includegraphics[width=0.32\textwidth]{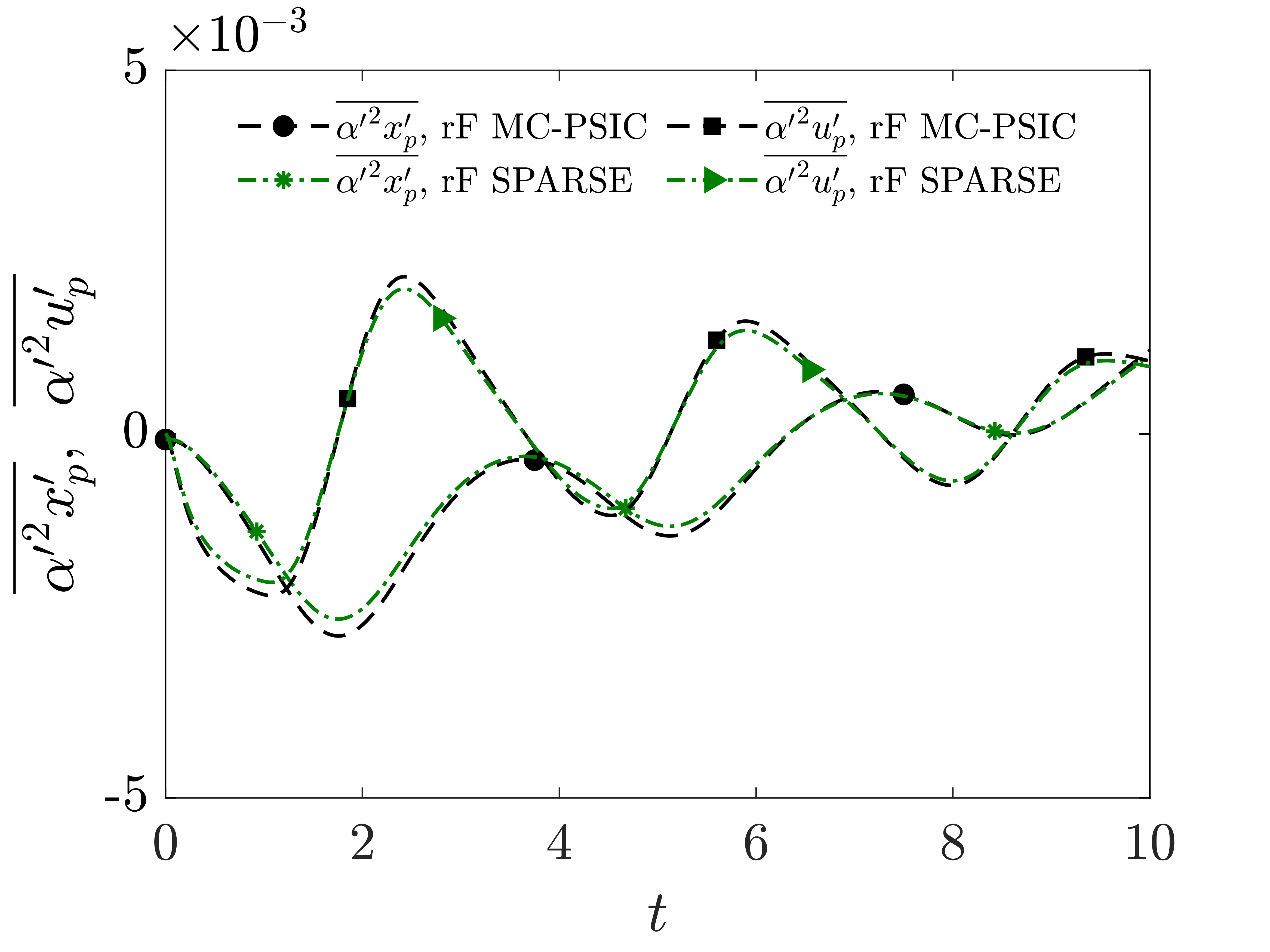}} 
	\caption[]{Third moments of the one-dimensional sinusoidal velocity field test case for rF and dF cases computed with SPARSE and MC-PSIC using the relation~\eqref{eq: join_moments_cm3}; (a) shows the skewness of the particle position; (b) the particle velocity; (c) and (d) combined third moments of the particle phase; and (e) and (f) combined third moments of the random coefficient and particle phase variables.}
	\label{fig: sine1D_cm3}
\end{figure}

We assess the accuracy of the point-cloud tracer by comparing computational results for different levels of splitting $M=1,\dots,7$, such that the total number of subclouds is $M_p=M^2$ for the dF case (along $x_p$ and $u_p$) and $M_p=M^3$ for the rF case (along $\alpha$, $x_p$ and $u_p$) respectively. 
Figure~\ref{fig: sine1D_phaseSpace} compares   the Cumulative Cloud solution for  the rF and dF case with $M=7$   and shows that the deterministic and random solutions while different are strongly correlated.
In fact, a comparison of their average  trajectories in Fig.~\ref{fig: sine1D_mean_xpup} in phase space shows that   they are within $3\%$ of each other.
They match so surprisingly well, because
the second moments on the right hand side of the  first moment equations are  negligible as compared to the dominant  acceleration term $\overline{f}_1(\overline{u}-\overline{u}_p)$.
This is however not the case for the second moment equations and their solution that depend on the second and third moments~\citep{dominguez2021lagrangian} as plotted
in Figures~\ref{fig: sine1D_mean_xpup}--~\ref{fig: sine1D_alphaalphaxp}.
The maximum difference between the rF and dF solution for the second moments is  $20\%$ in $\sigma_{u_p}$ (see Fig.~\ref{fig: sine1D_sigma_up}) in the initial acceleration stages.
The standard deviations and correlations  show an oscillatory trend. The  solutions for the rF cloud are mostly different from the dF cloud near its maxima.

The second moments  $\overline{\alpha^\prime x_p^\prime}$ and $\overline{\alpha^\prime u_p^\prime}$ can
be interpreted as "virtual" stresses 
 that cause the mixing of the random cloud. We dub these stresses virtual, because they are not physical and only affect the random sample space.
  The two correlations of the force coefficients  with  location and velocity are oscillatory and  in phase with their respective position and velocity standard deviation  trends. This phase-locked behavior indicates that the virtual stress cross-correlations  are greater if the  principle strain of the random cloud  is greater in the related physical or phase space dimension (in general physics, the stress-strain relation is well-known~\citep{pope2000turbulent}).
Note that these terms have a zero value  for the dF case because $\alpha^\prime=0$ and they are thus not included the figure.

\begin{figure}[h!]
	\centering
    \subfloat[]{
		\label{fig: sine1D_f1_Rep}
		\includegraphics[width=0.4\textwidth]{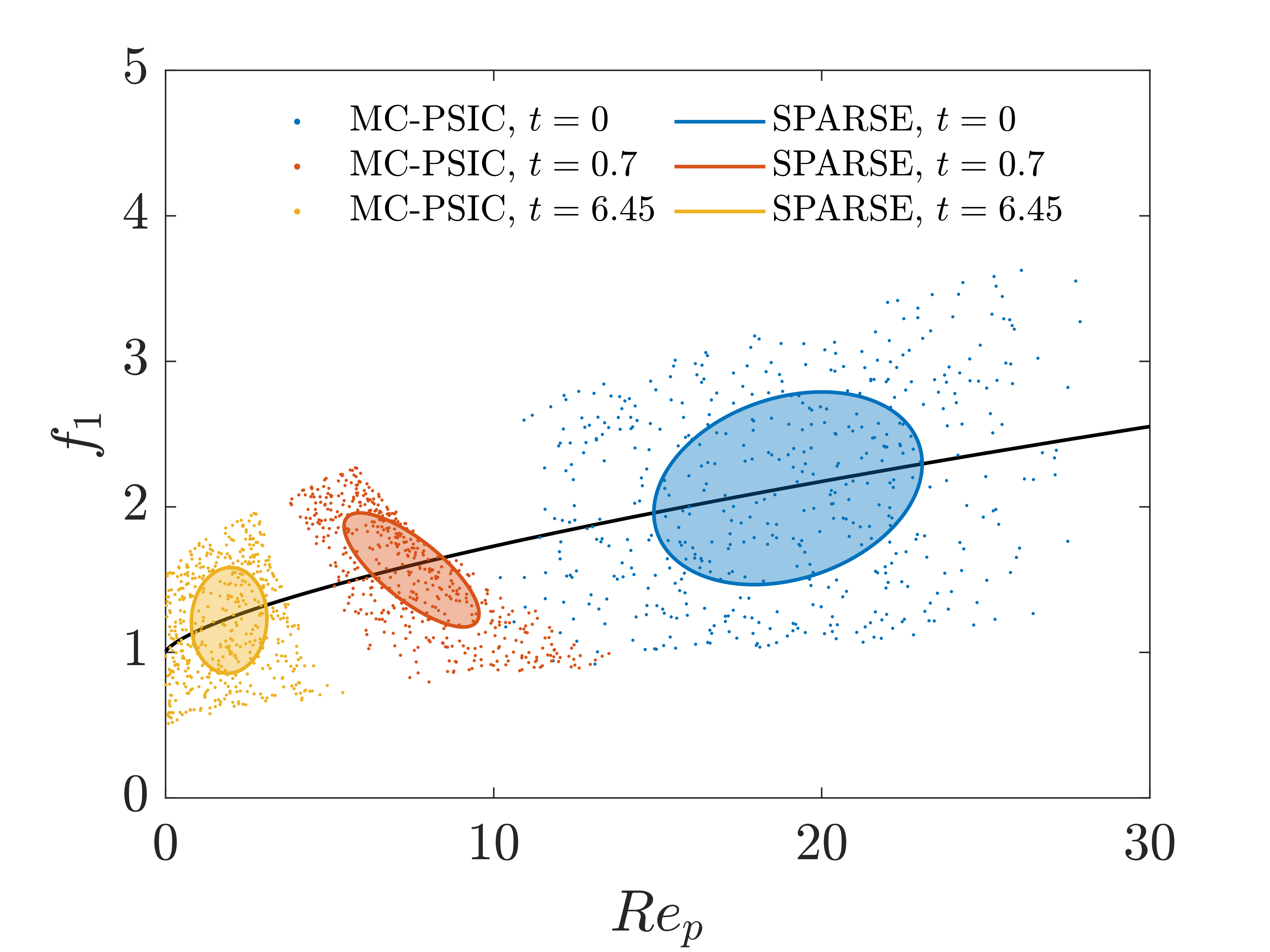}}
	\caption[]{Forcing of the Cumulative Cloud of the sinusoidal one-dimensional test case for three instants of times computed with SPARSE ($M=7$) depicted as ellipses in $f_1-Re_p$ space and point-particles computed with MC-PSIC depicted as points. For a better visualization, $500$ point-particles are depicted from the $N_p=10^5$ used for the computations.}
	\label{fig: sine1D_phaseSpace_and_splitting}
\end{figure}

Figure~\ref{fig: sine1D_f1_Rep} shows  the MC-PSIC and Cumulative Cloud solution from a different perspective in the  $f_1-Re_p$ plane. The dF cloud follows the line described by the deterministic forcing function. 
The rF cloud is distributed in the space spanned by the forcing and the relative velocity and can be visualized and computed with SPARSE using the ellipsoidal approximation. The results show that  the particle's relative Reynolds number  is initially  high (on the order of $20$) when the clouds  accelerate.
Later, the particle cloud oscillates around moderate values of $Re_p$ associated with a lower correction of the Stokes drag and a forcing function closer to unity despite the randomness in $\alpha$.

Figure~\ref{fig: sine1D_errors} confirms the theoretical third-order convergence rate of SPARSE with respect to the level of splitting $M$.
We find that for $M=7$ the maximum relative error of all first and second moments of the rF particle cloud computed with SPARSE as compared to MC-PSIC is $1\%$ for the variable $\sigma_{u_p}^2$.
For the rest of variables of the particle phase the error is lower than that, which denotes good agreement between the SPARSE and MC-PSIC results.
The relative error for the other second moments involving $\alpha$  is at most $1.5\%$ 
(for $\overline{\alpha^\prime x_p^\prime})$.
For the dF case, the maximum relative error occurs also in the variance of the particle velocity with a value of $0.1\%$.
The computational savings are significant even for $M=7$.
The ratio of computational cost of SPARSE as compared to MC-PSIC defined as in~\eqref{eq: savings} that takes into account the reduction of degrees of freedom is $r=1.7\cdot10^{-3}$ for the dF case and $r=1.2\cdot10^{-2}$ for the rF case, that translates to a $0.17\%$ and a $1.2\%$ of the computational cost respectively as compared to MC-PSIC.

\begin{figure}[h!]
	\centering
	\subfloat[]{
		\label{fig: sine1D_f_xp_stoch_MC}
		\includegraphics[width=0.32\textwidth]{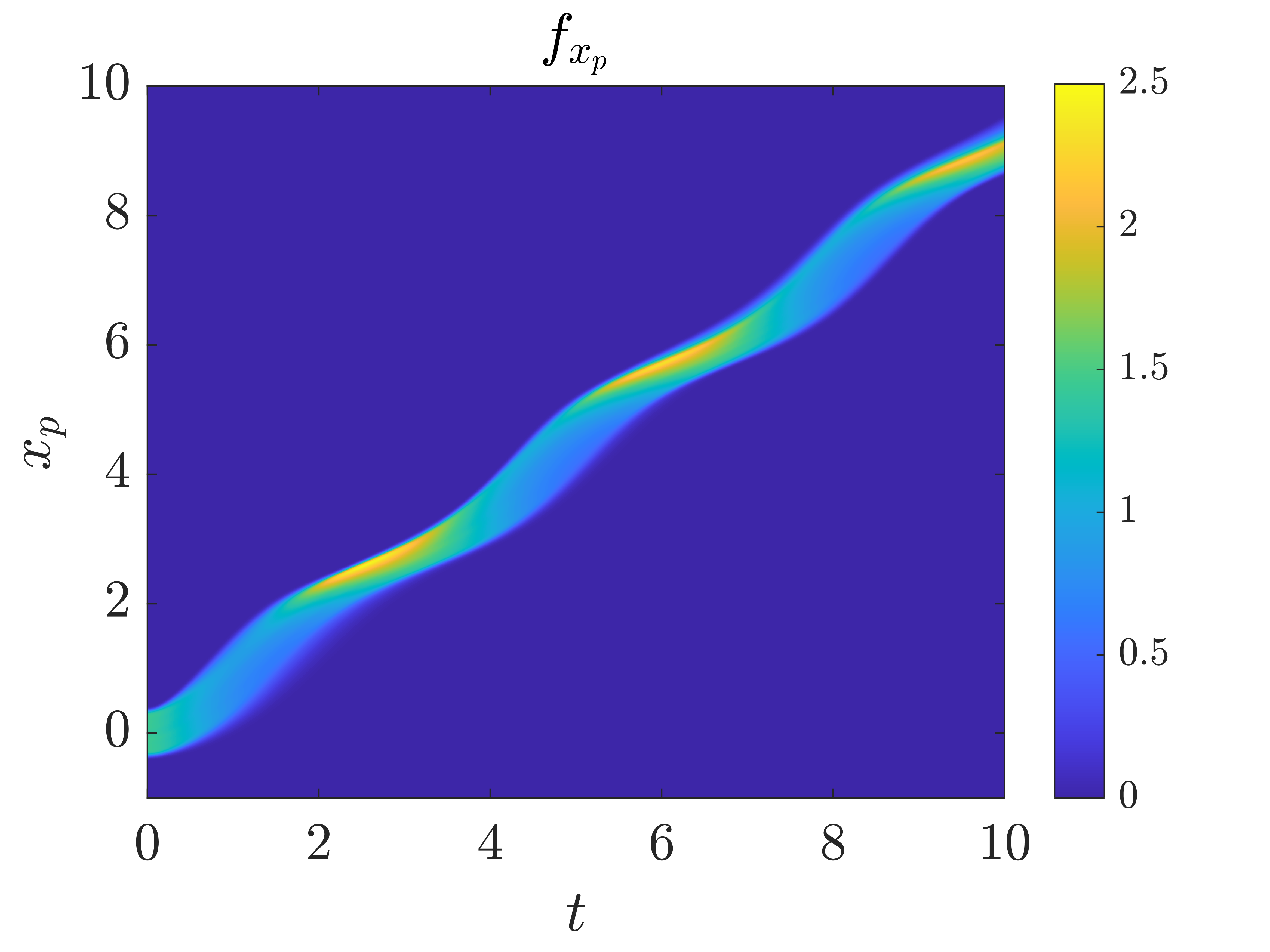}}
        \hfill	
	\subfloat[]{
		\label{fig: sine1D_f_xp_stoch_SPARSE}
		\includegraphics[width=0.32\textwidth]{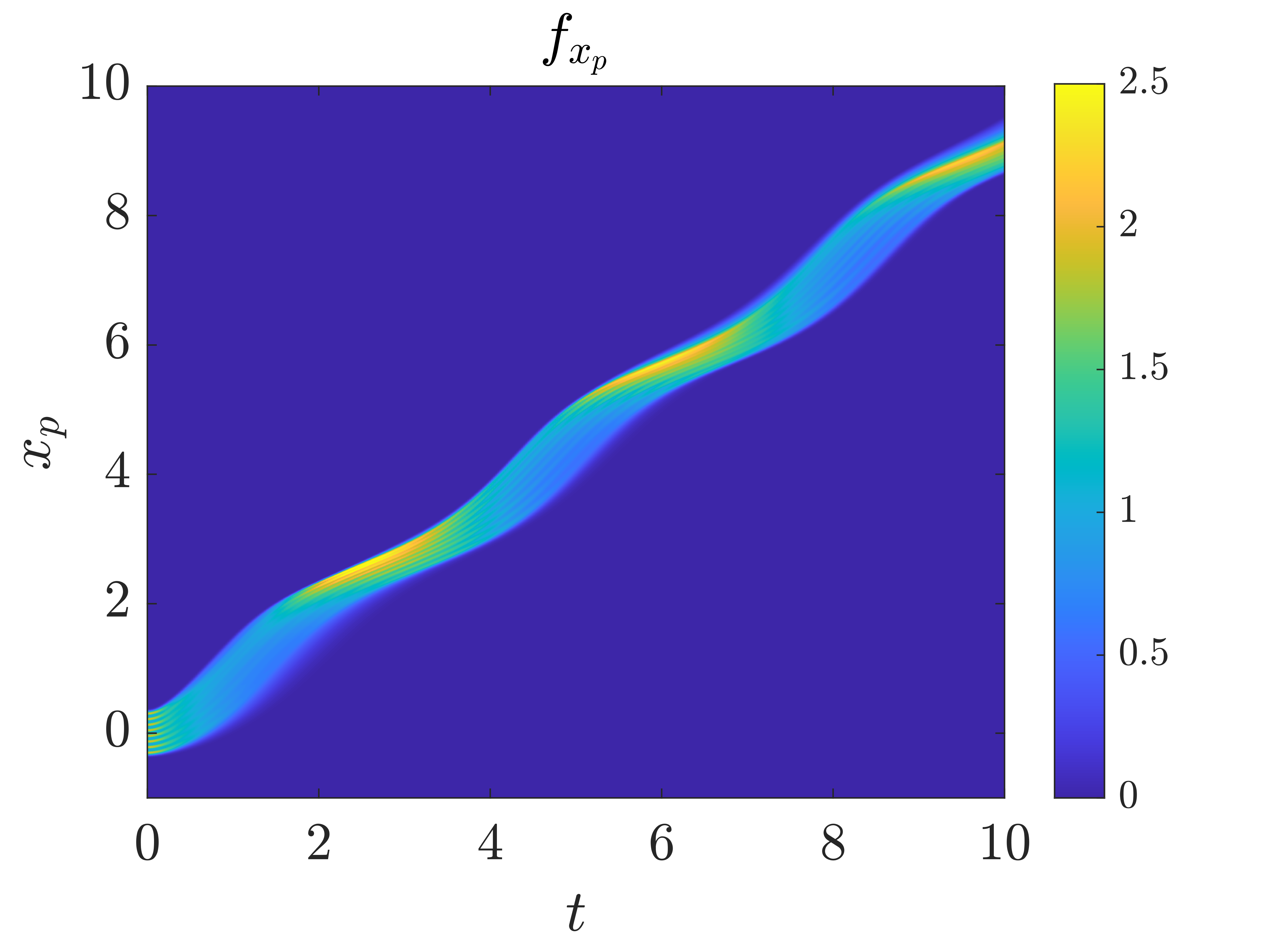}}  
        \hfill	
    \subfloat[]{
		\label{fig: sine1D_f_xp_stoch}
		\includegraphics[width=0.32\textwidth]{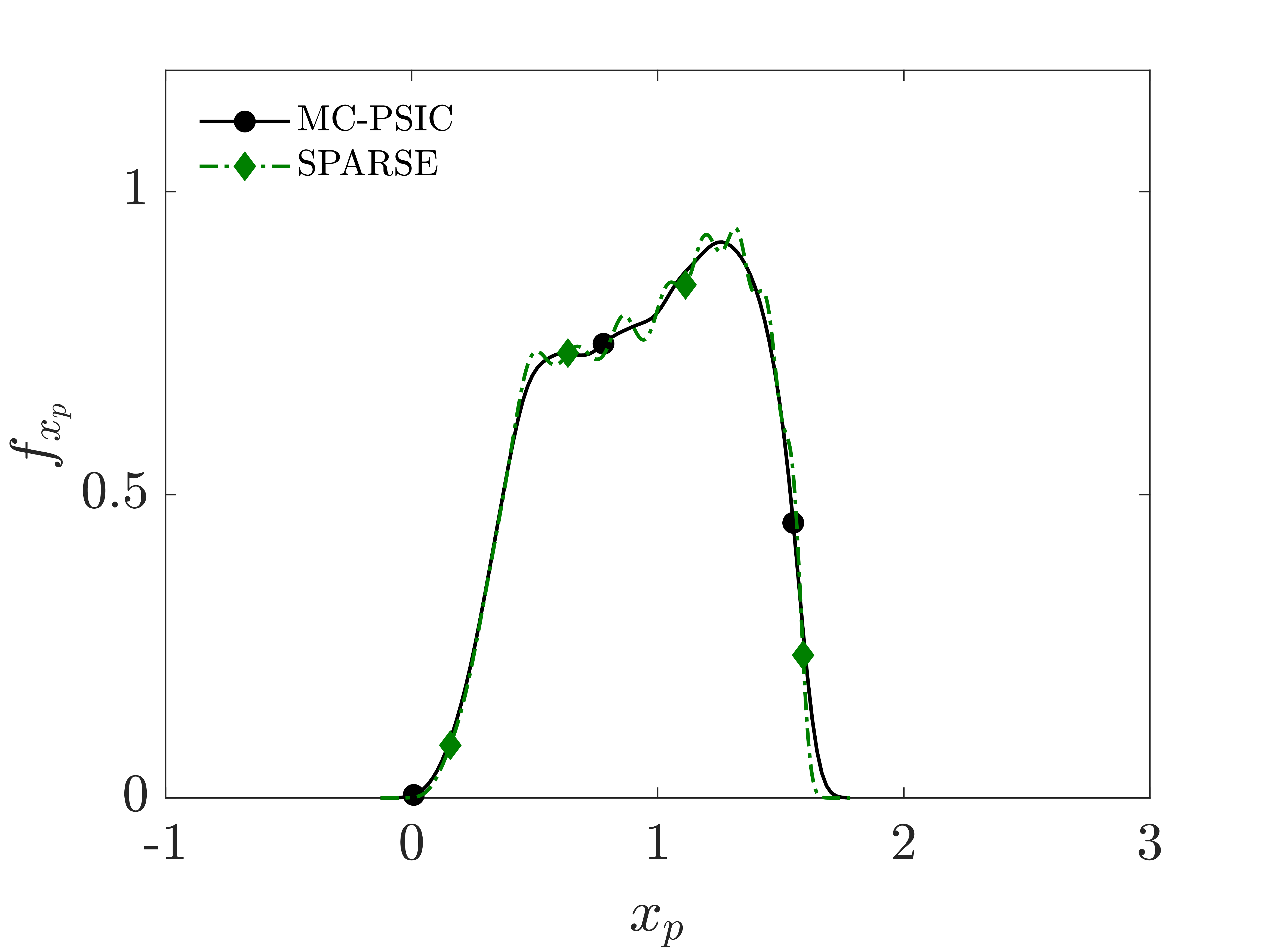}}   \\
	\subfloat[]{
		\label{fig: sine1D_f_up_stoch_MC}
		\includegraphics[width=0.32\textwidth]{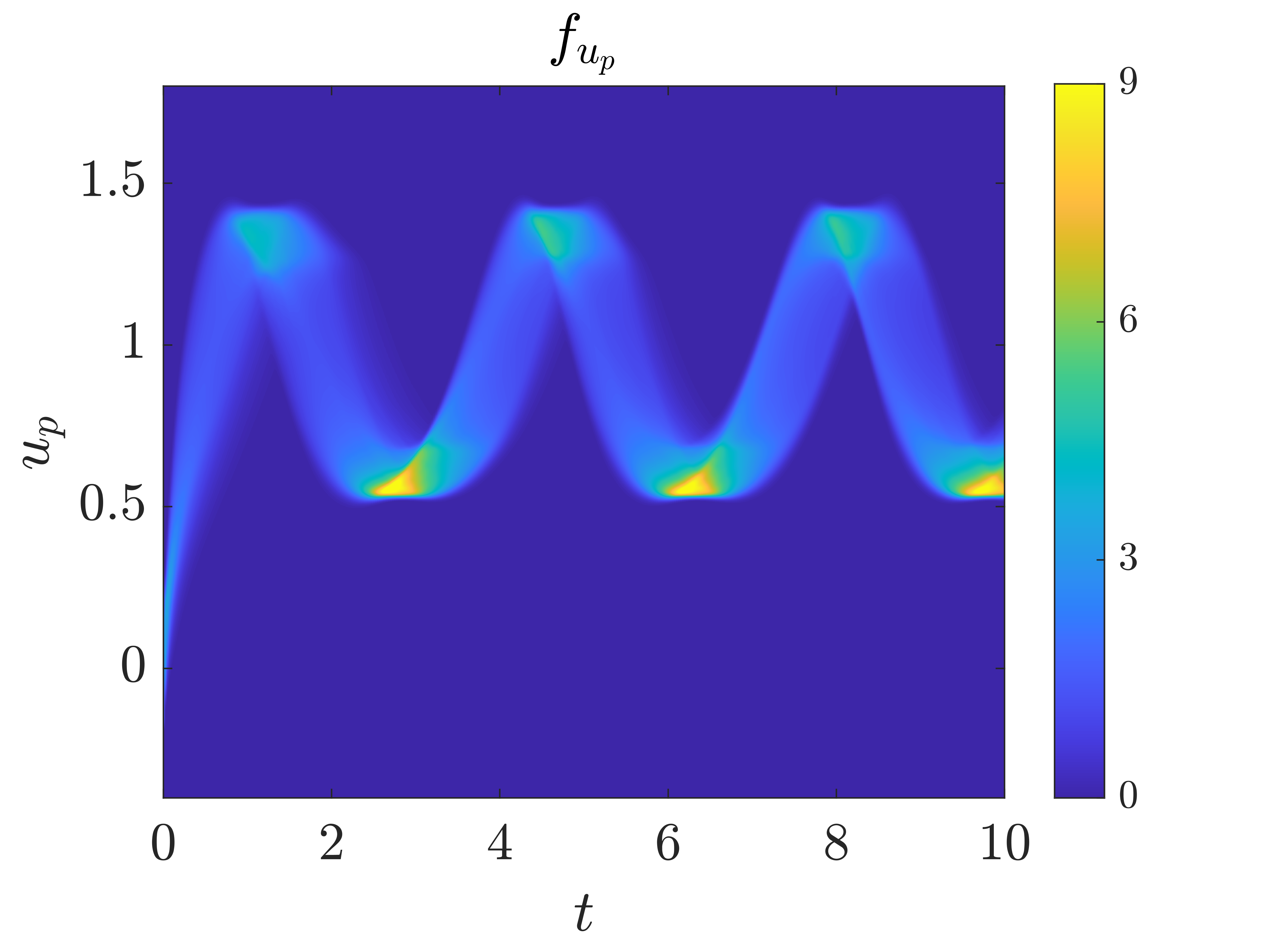}}
        \hfill	
	\subfloat[]{
		\label{fig: sine1D_f_up_stoch_SPARSE}
		\includegraphics[width=0.32\textwidth]{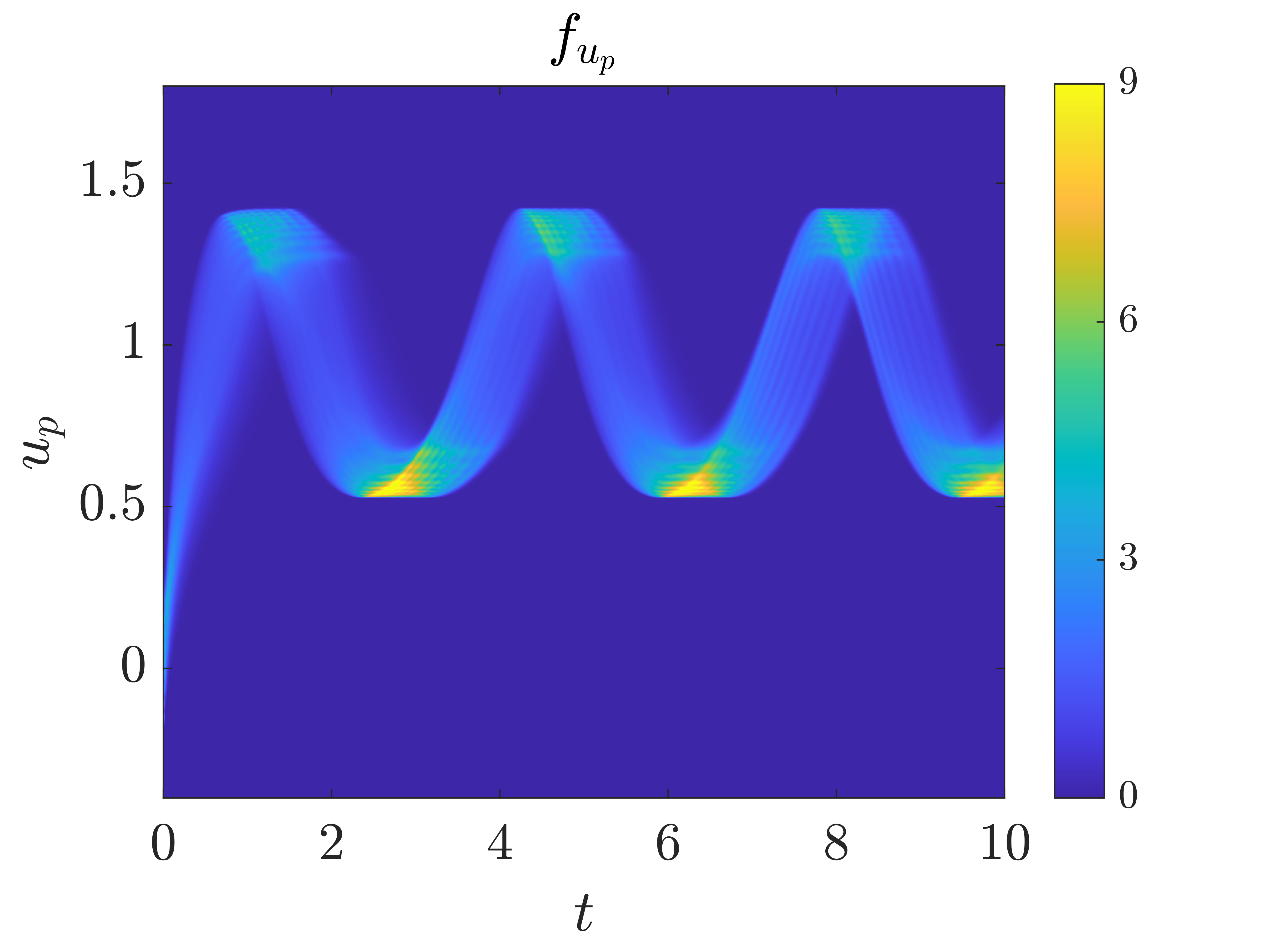}}  
        \hfill	
    \subfloat[]{
		\label{fig: sine1D_f_up_stoch}
		\includegraphics[width=0.32\textwidth]{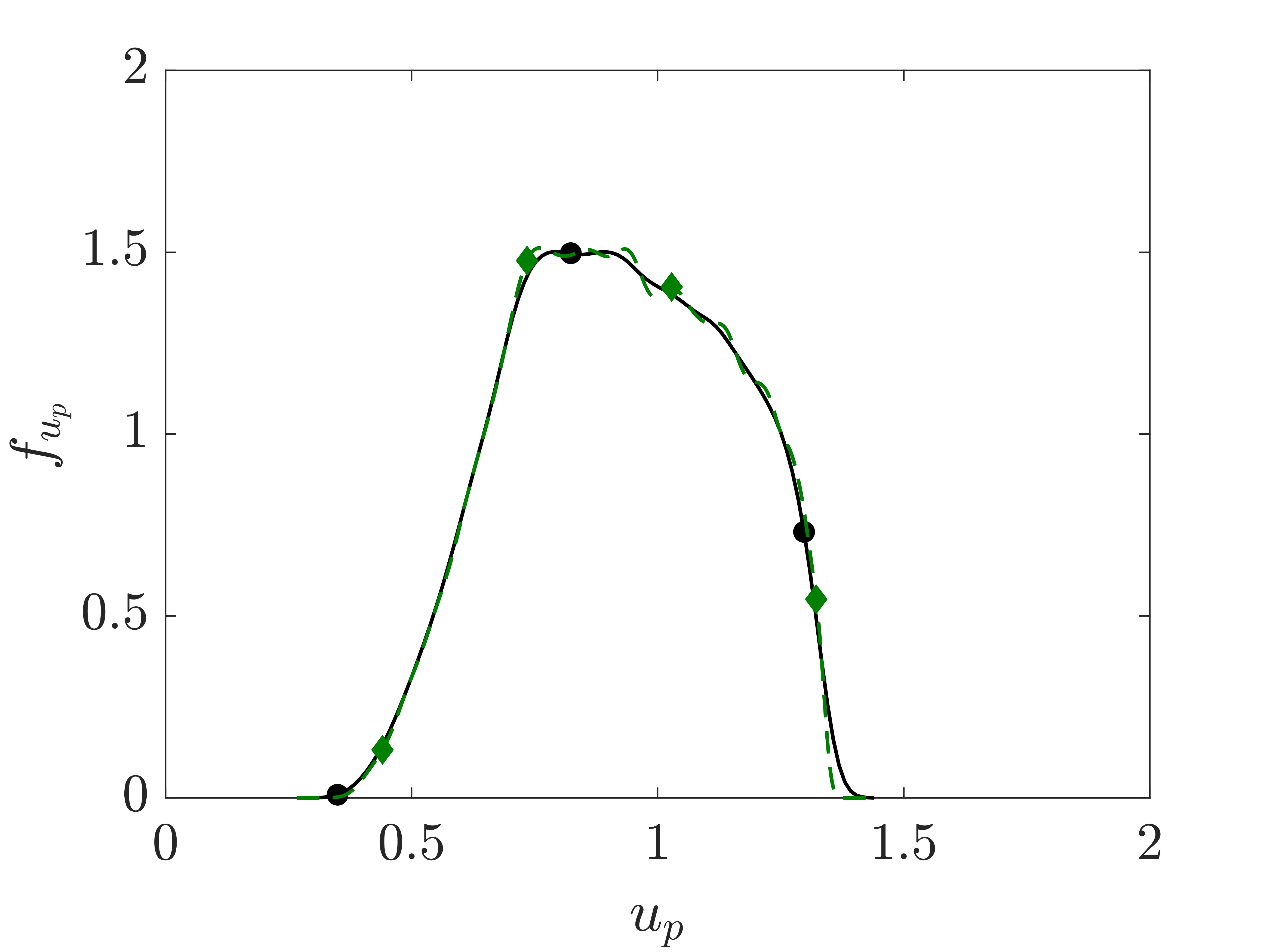}} 
	\caption[]{PDFs of the components of the particle location $f_{x_p}$ and velocity $f_{u_p}$ for the rF case computed with MC-PSIC and SPARSE methods for the one-dimensional sinusoidal velocity field test case. The Figure shows: PDF of $x_p$ computed with (a) MC-PSIC and (b) SPARSE and (c) comparison of both methods at time $t=0.3$ and the corresponding results for $u_p$ in (d)--(f). The legend in (c) also corresponds to (f).}
	\label{fig: sine1D_PDFs}
\end{figure}

Using multiple clouds, we can compose the PDF of the Cumulative Cloud with the sum of weighted Gaussians as defined in~\eqref{eq: join_PDF}.
In Figure~\ref{fig: sine1D_PDFs} we show these PDFs and compare them to the MC-PSIC method for the rF case. Clearly, the PDF is not trivial, i.e. not symmetric or Gaussian and SPARSE accurately predicts it. 
The evolution of PDF of the particle location in time computed with MC-PSIC (Fig.~\ref{fig: sine1D_f_xp_stoch_MC}) compares well with SPARSE (Fig.~\ref{fig: sine1D_f_xp_stoch_SPARSE}) .
In a similar manner, for the PDF of the particle velocity, we show in Figures~\ref{fig: sine1D_f_up_stoch_MC}--\ref{fig: sine1D_f_up_stoch} contour maps of the PDF along time computed with MC-PSIC (Fig.~\ref{fig: sine1D_f_up_stoch_MC}) and SPARSE (Fig.~\ref{fig: sine1D_f_up_stoch_SPARSE}) and compare both at the same time instant in Figure~\ref{fig: sine1D_f_up_stoch}.
The mixture distribution composed by Gaussians shows oscillations related to the underlying Gaussians of each subcloud.
The results are within a $5\%$ agreement and SPARSE uses two order of magnitudes fewer degrees of freedom than the MC-PSIC approach.
Additionally, third order moments computed according to~\eqref{eq: join_moments_cm3} are depicted in Fig.~\ref{fig: sine1D_cm3} for the dF and rF cases computed with SPARSE and MC-PSIC showing also agreement.

\subsection{Stagnation flow} \label{sec: test_SF}


The two-dimensional stagnation flow  is analytically described 
according to~\cite{hiemenz1911grenzschicht} for an inviscid irrotational fluid, in the domain $(x,y) \in [-\infty,0]\times[-\infty,\infty]$ by  
\begin{subequations}\label{eq: SF_flow}
\begin{align}
    u &= -k x, 
    \label{eq: SF_flow_u} \\ 
    v &= k y,
    \label{eq: SF_flow_v}
\end{align}
\end{subequations}
where $y$ is the coordinate perpendicular to the flow direction, and $k$ is a constant.
Point-particles carried by this flow admit an analytical solution for their trajectories as well~\citep{dominguez2021lagrangian}.
These analytical descriptions are helpful for testing and understanding of the basic  characteristics and canonical behaviour of the solutions of dF and rF clouds of particles.
Moreover, because there is symmetry with respect to the horizontal axes, the equations decouple by coordinates and similarities between rF and dF clouds under symmetry conditions can be analyzed.
Closure terms that depend on the linear flow velocity field are exact because the Taylor expansion of the linear field is exact.



Following \cite{dominguez2021lagrangian}, we reduce the complexity of the test  further by assuming a  forcing function that sets  $f_1=\alpha$ with $g_1=1$ in~\eqref{eq: f1andf2}, i.e. the forcing function is random but not dependent on the relative velocity, so that the only the non-zero truncation are the third moments in the second moment equations~\eqref{eq: MoM_xpup} and~\eqref{eq: MoM_upup} and~\eqref{eq: closure_axp_system}. For reference, we  present the complete closed system of equations for this form of the  forcing function in Appendix~\ref{app: rF_equations}.

We specify initial conditions for the cloud's location and velocity by sampling from uniform distributions according to
\begin{subequations}\label{eq: SF_IC}
\begin{align}
{x_p}_0 \sim \mathcal{U}\left[{{x_p}_0}_{min}, \ {{x_p}_0}_{max}\right], \ \ \
    {y_p}_0 \sim \mathcal{U}\left[{{y_p}_0}_{min}, \ {{y_p}_0}_{max}\right],
\label{eq: SF_IC_xp_yp} \\
{u_p}_0 \sim \mathcal{U}\left[{{u_p}_0}_{min}, \ {{u_p}_0}_{max}\right], \ \ \
    {v_p}_0 \sim \mathcal{U}\left[{{v_p}_0}_{min}, \ {{v_p}_0}_{max}\right].
\label{eq: SF_IC_up_vp}
\end{align}
\end{subequations}
With the particle cloud initially at rest, both components of the velocity $u_p$ and $v_p$ are zero for all point-particles contained in the cloud.
For the rF case the random coefficient $\alpha$ is also sampled from an uniform distribution
\begin{align}
    \alpha \sim \mathcal{U}\left[\alpha_{min}, \ \alpha_{max}\right],
    \label{eq: SF_IC_alpha}
\end{align}
whereas for the dF case the coefficient $\alpha$ is deterministic with $\alpha^\prime=0$.
The averages of the distribution functions are $\overline{x}_p=-1$, ${\overline{y}_p}_0={\overline{u}_p}_0={\overline{v}_p}_0=0$, and the standard deviations $\sigma_{{x_p}_0}=\sigma_{{y_p}_0}=\sigma_{{u_p}_0}=\sigma_{{v_p}_0}=0.08$.
For the rF case, we also define $\overline{\alpha}=1$ and $\sigma_{\alpha} = 0.3$ whereas for the dF all point-particles have the same value of $\alpha$ with zero fluctuations so that $\alpha=\overline{\alpha}=1$, $\sigma_{\alpha}=0$.
Second moments combining any of the variables $\alpha$, ${x_p}_0$, ${y_p}_0$, ${u_p}_0$ and ${v_p}_0$ are zero initially because the variables are considered uncorrelated at $t=0$.
However, because in computational practice we can only use  the limited number of samples, $N_p=10^5$, these moments include sampling errors that converge at the rate $1/\sqrt{N_p}$ with respect to the averages and second moments described above.
The initial condition for the SPARSE method is determined by sampling the point-particles as described in Section~\ref{sec: IC_numericalImplementation}.
We set the Stokes number and the constant $k$ in~\eqref{eq: SF_flow} to unity so that $St=k=1$.

\begin{figure}[h!]
	\centering
	\subfloat[]{
		\label{fig: sFcenter_3Dplot_xpyp}
		\includegraphics[width=0.49\textwidth]{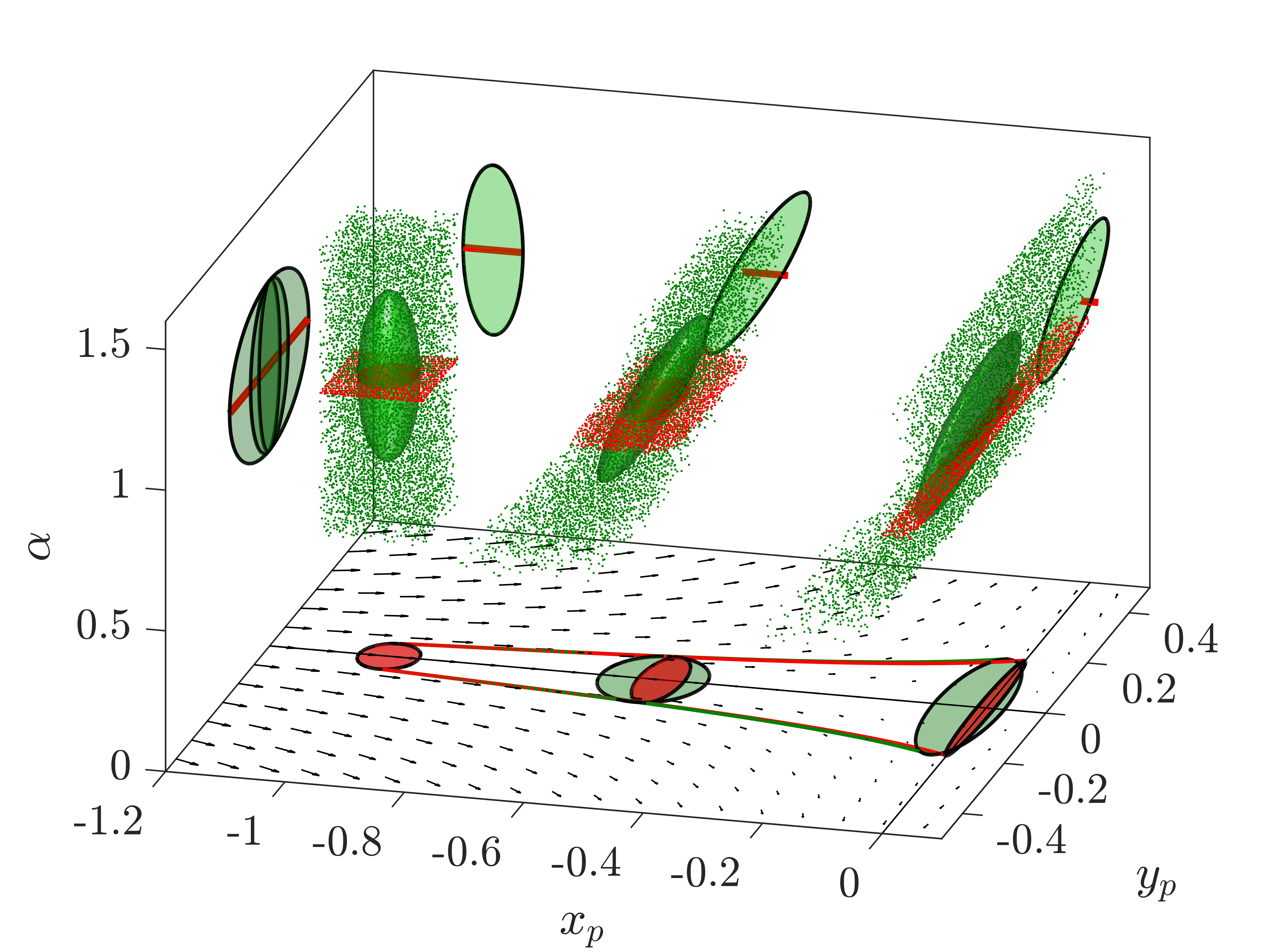}} 
		\hfill
	\subfloat[]{
		\label{fig: sFcenter_3Dplot_xpup}
		\includegraphics[width=0.49\textwidth]{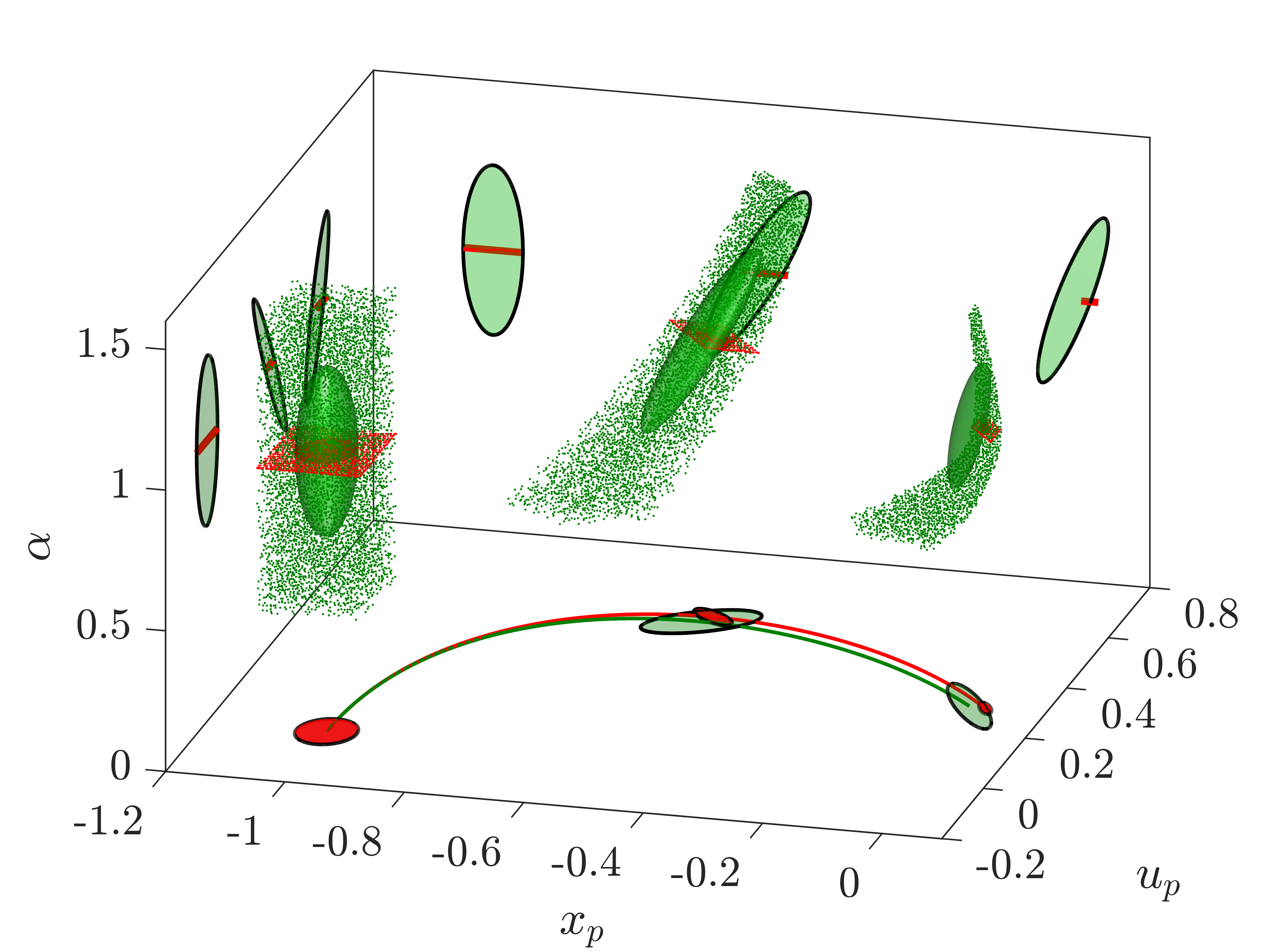}}
	\caption[]{Evolution of the particle cloud immersed in the stagnation flow for different instants of time $t=[0, \ 0.5, \ 1.65, \ 2.2]$. The dots represent the point-particles traced by the MC-PSIC method and the Cumulative Cloud traced with the SPARSE method is represented by an ellipsoid for the rF case (red) and dF case (green). The projections of the SPARSE solution are represented in the background planes.}
	\label{fig: sFcenter_3Dplot}
\end{figure}  

\begin{figure}[h!]
	\centering
	\subfloat[]{
		\label{fig: sFcenter_mean_xpup}
		\includegraphics[width=0.32\textwidth]{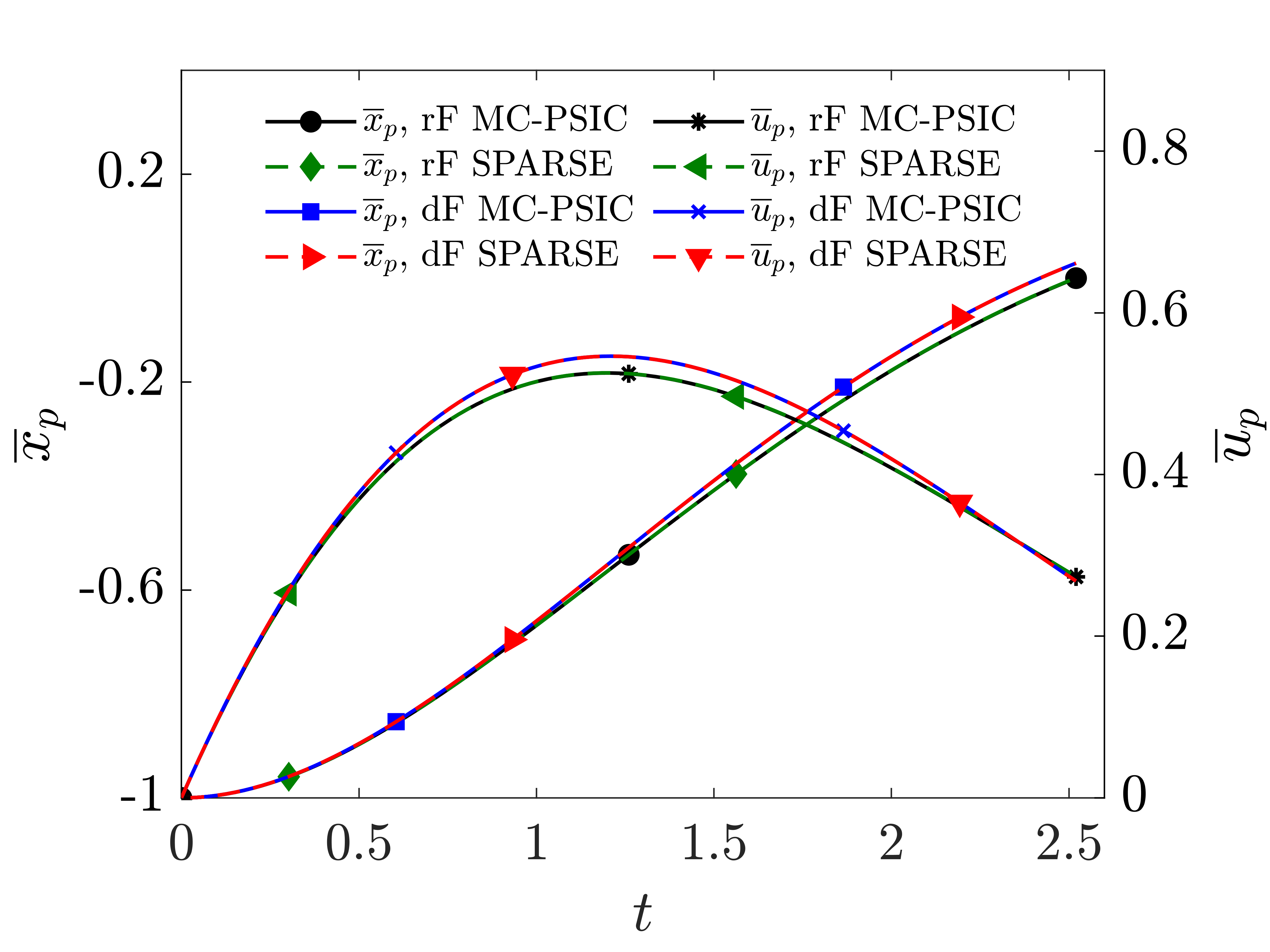}}
        \hfill	
	\subfloat[]{
		\label{fig: sFcenter_mean_ypvp}
		\includegraphics[width=0.32\textwidth]{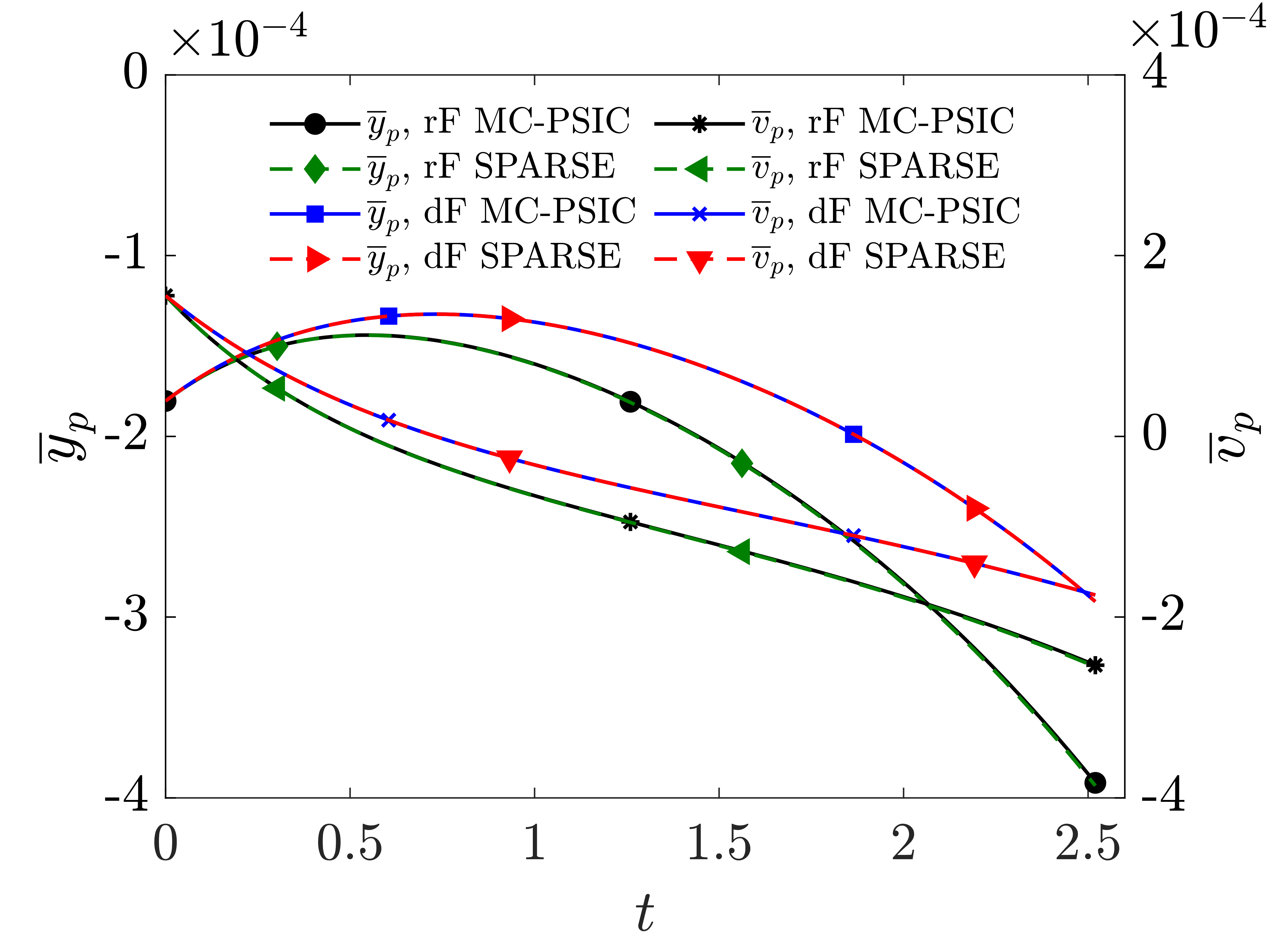}} 
	\subfloat[]{
		\label{fig: sFcenter_sigma_xpyp}
		\includegraphics[width=0.32\textwidth]{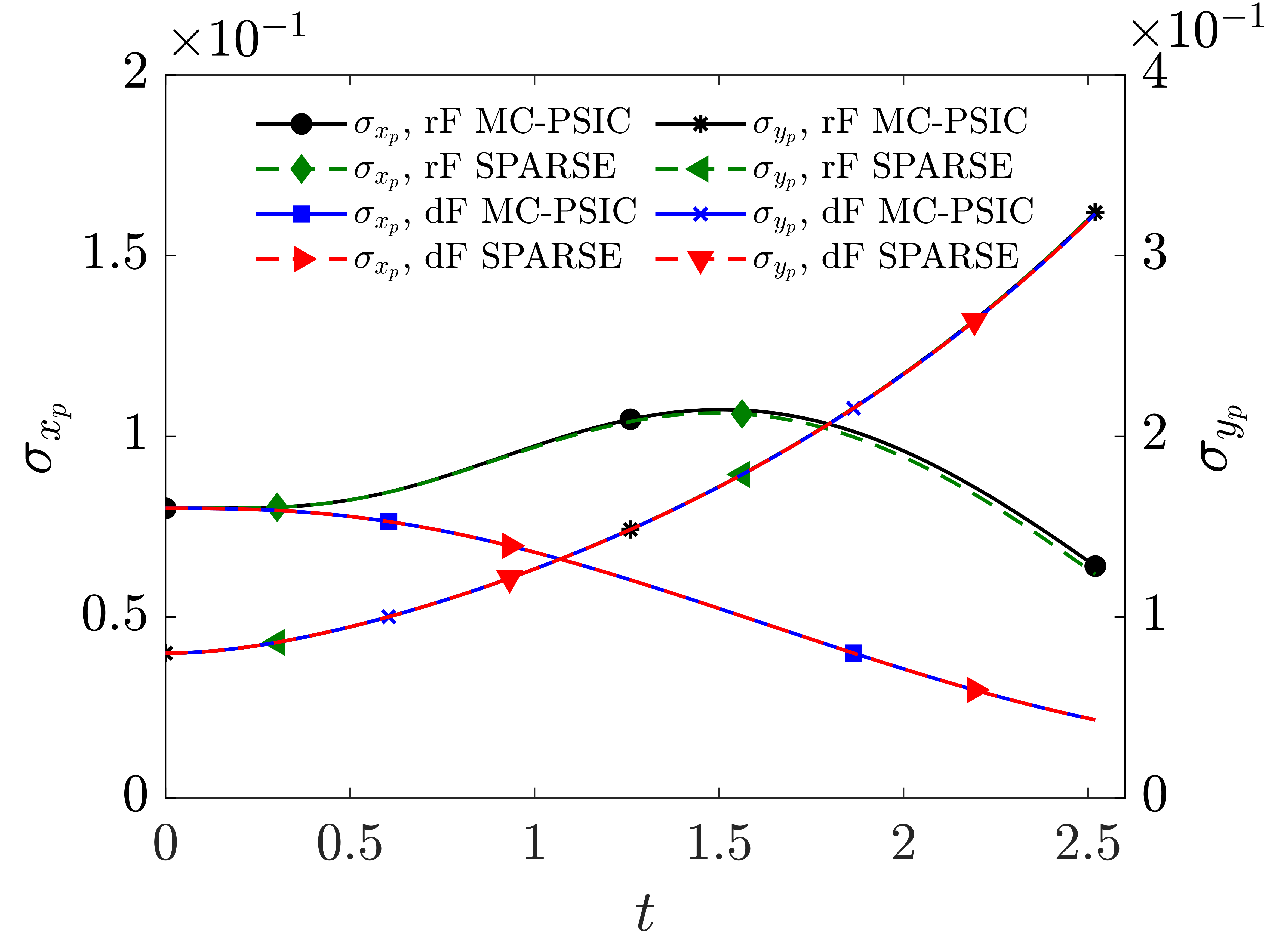}} \\ 
	\subfloat[]{
		\label{fig: sFcenter_sigma_upvp}
		\includegraphics[width=0.32\textwidth]{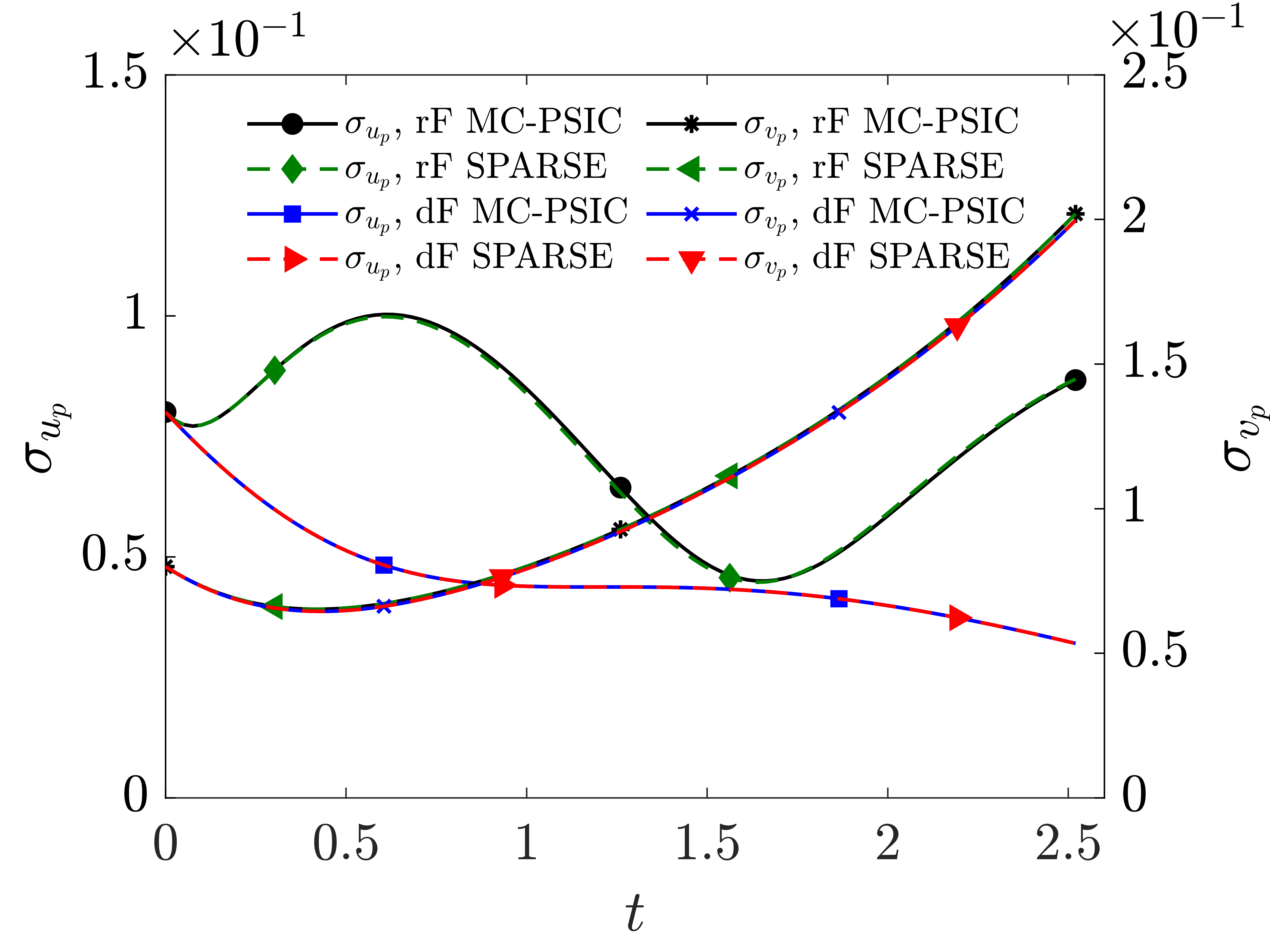}} 
		\hfill
	\subfloat[]{
		\label{fig: sFcenter_xpup_ypvp}
		\includegraphics[width=0.32\textwidth]{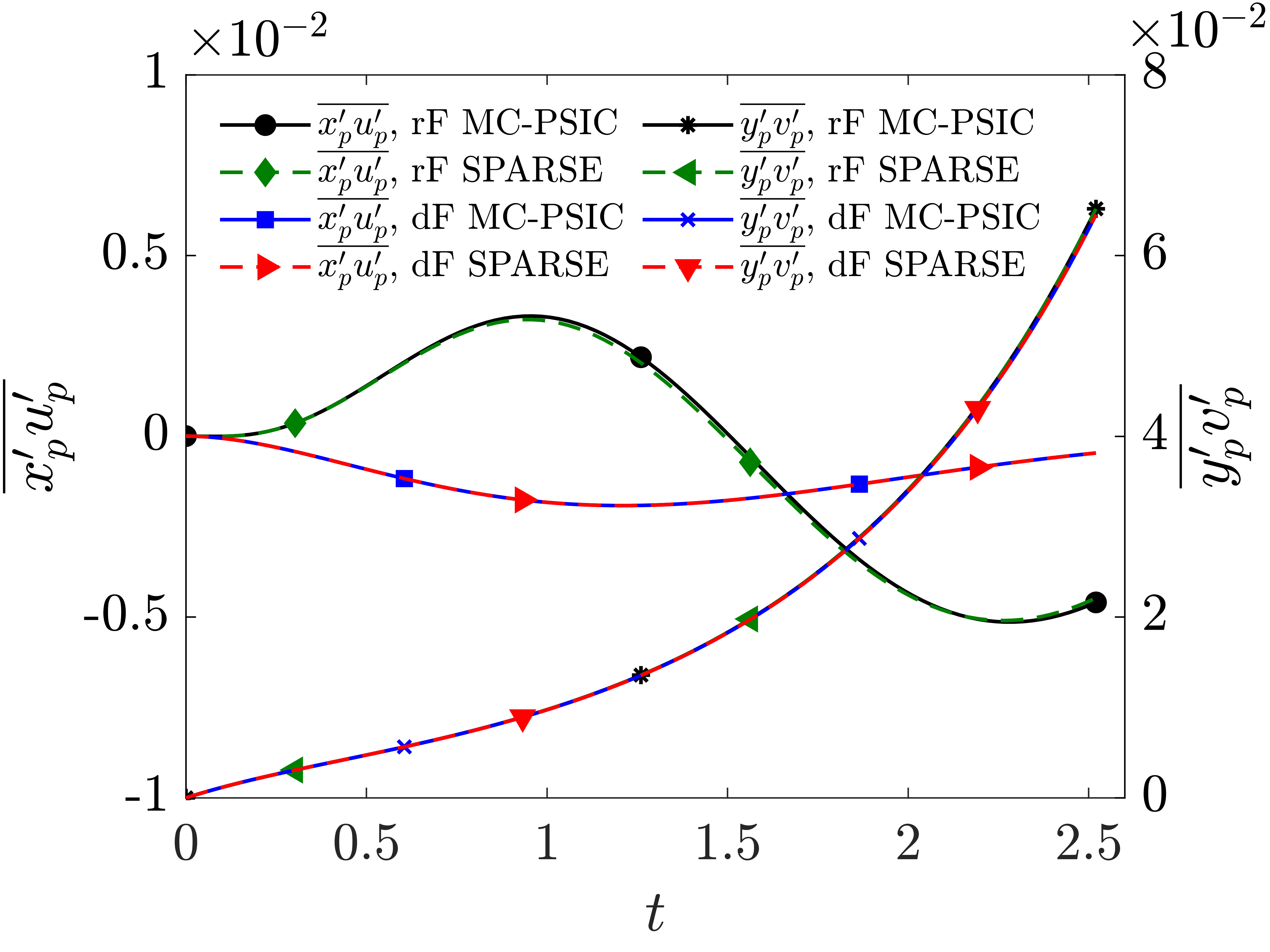}}
		\hfill
	\subfloat[]{
		\label{fig: sFcenter_xpyp_upvp}
		\includegraphics[width=0.32\textwidth]{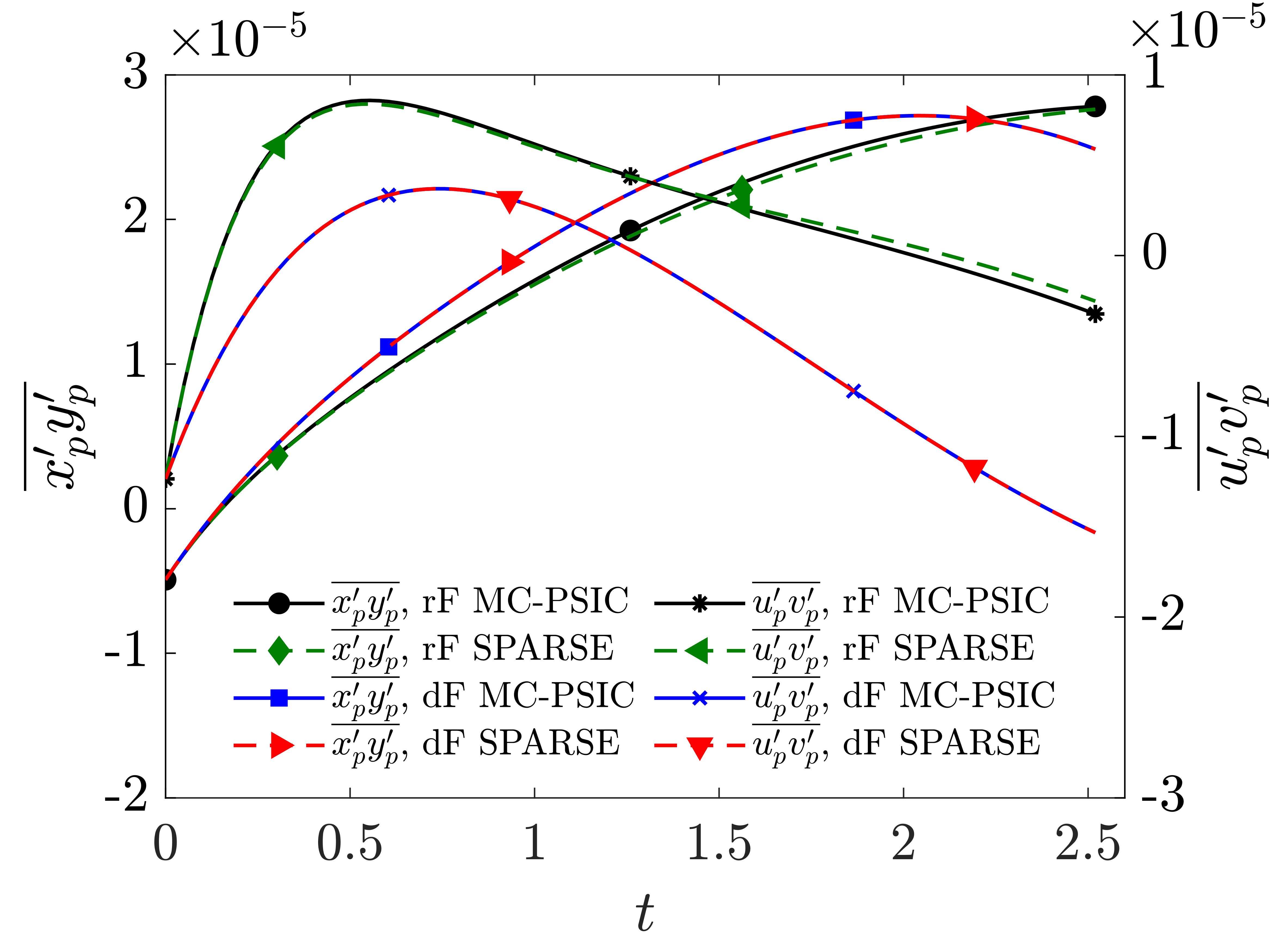}} \\
	    \hfill
	\subfloat[]{
		\label{fig: sFcenter_xpvp_ypup}
		\includegraphics[width=0.32\textwidth]{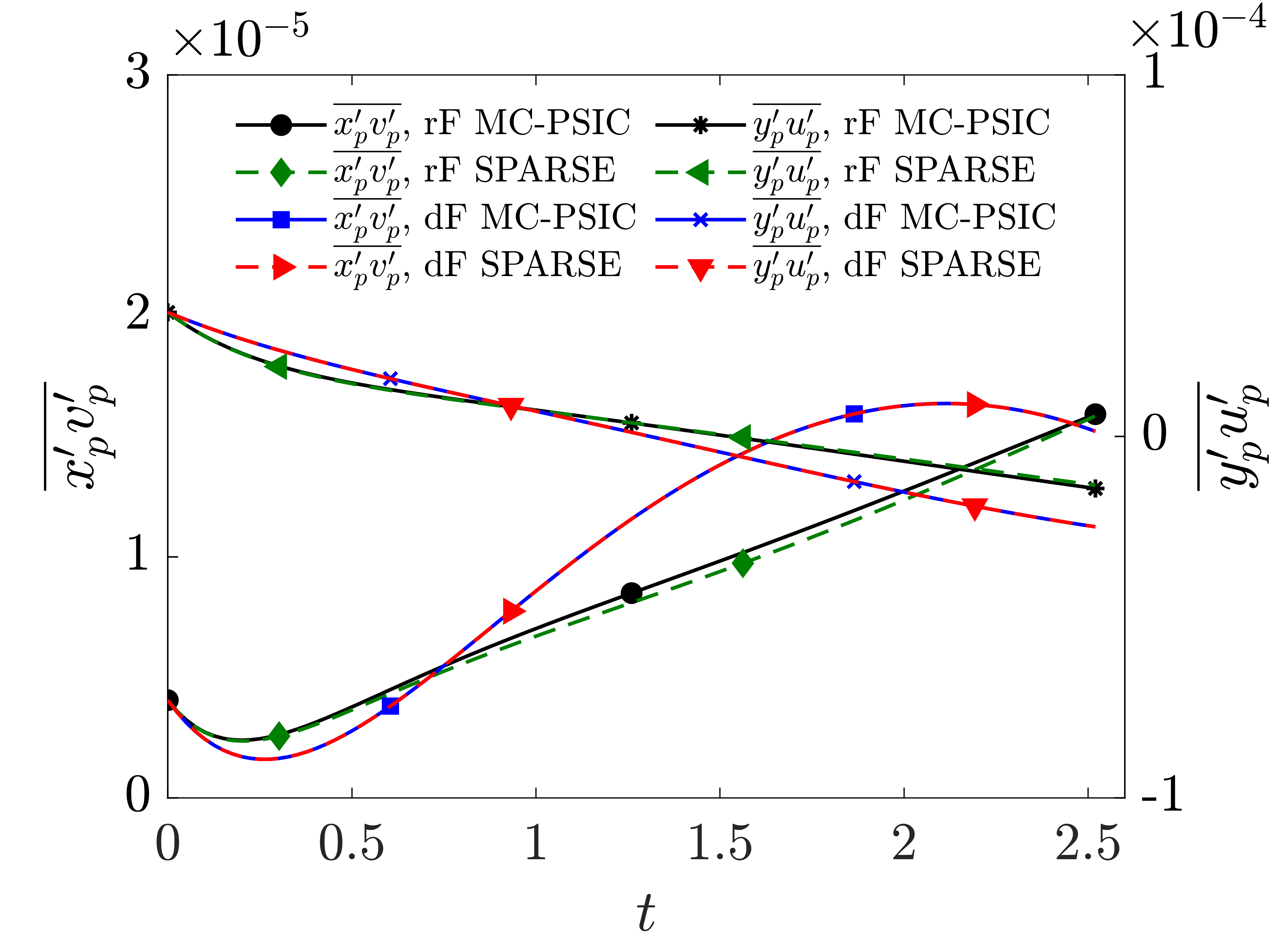}} 
		\hfill
	\subfloat[]{
		\label{fig: sFcenter_xpa_ypa}
		\includegraphics[width=0.32\textwidth]{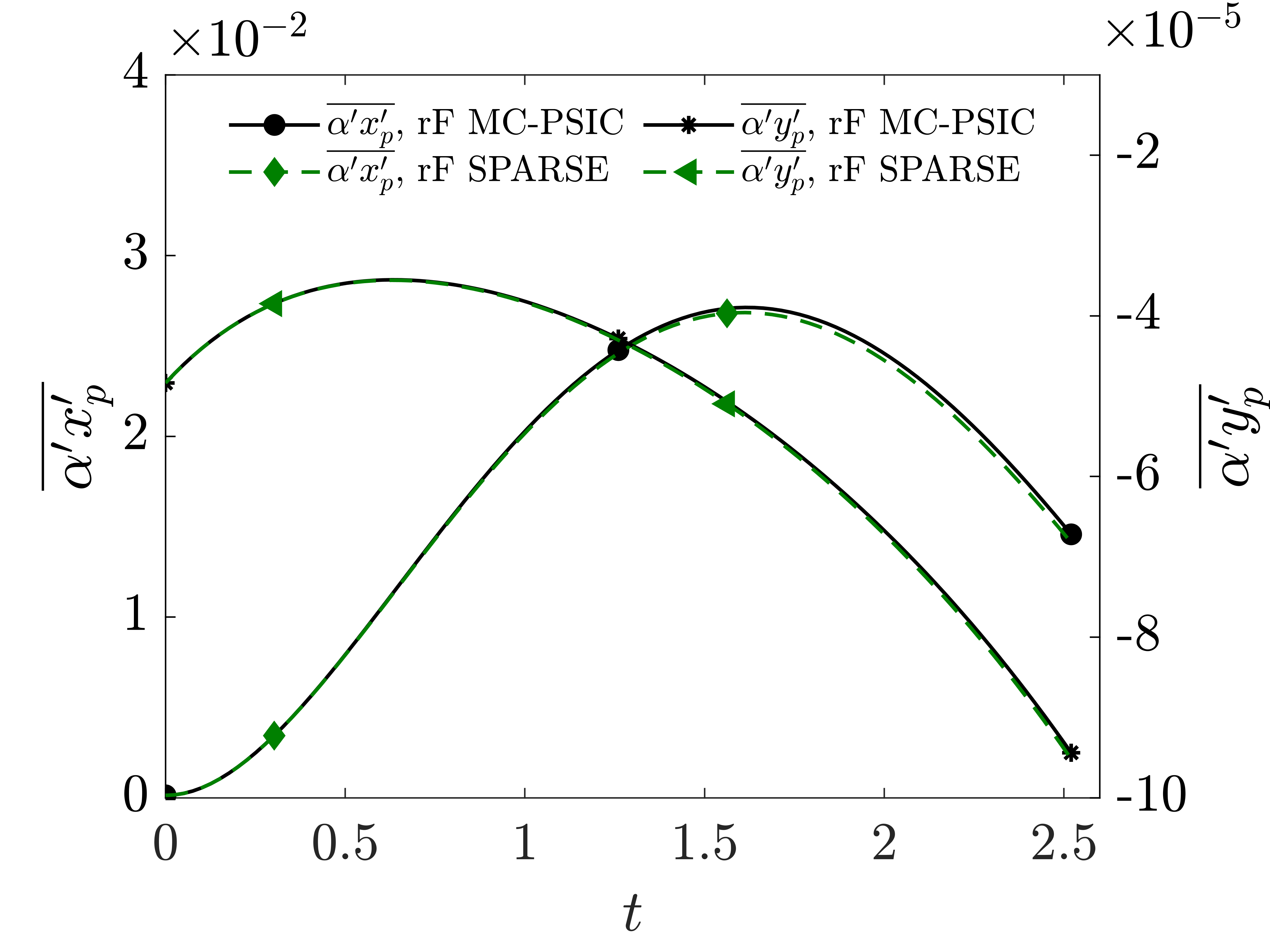}}
		\hfill
	\subfloat[]{
		\label{fig: sFcenter_upa_vpa}
		\includegraphics[width=0.32\textwidth]{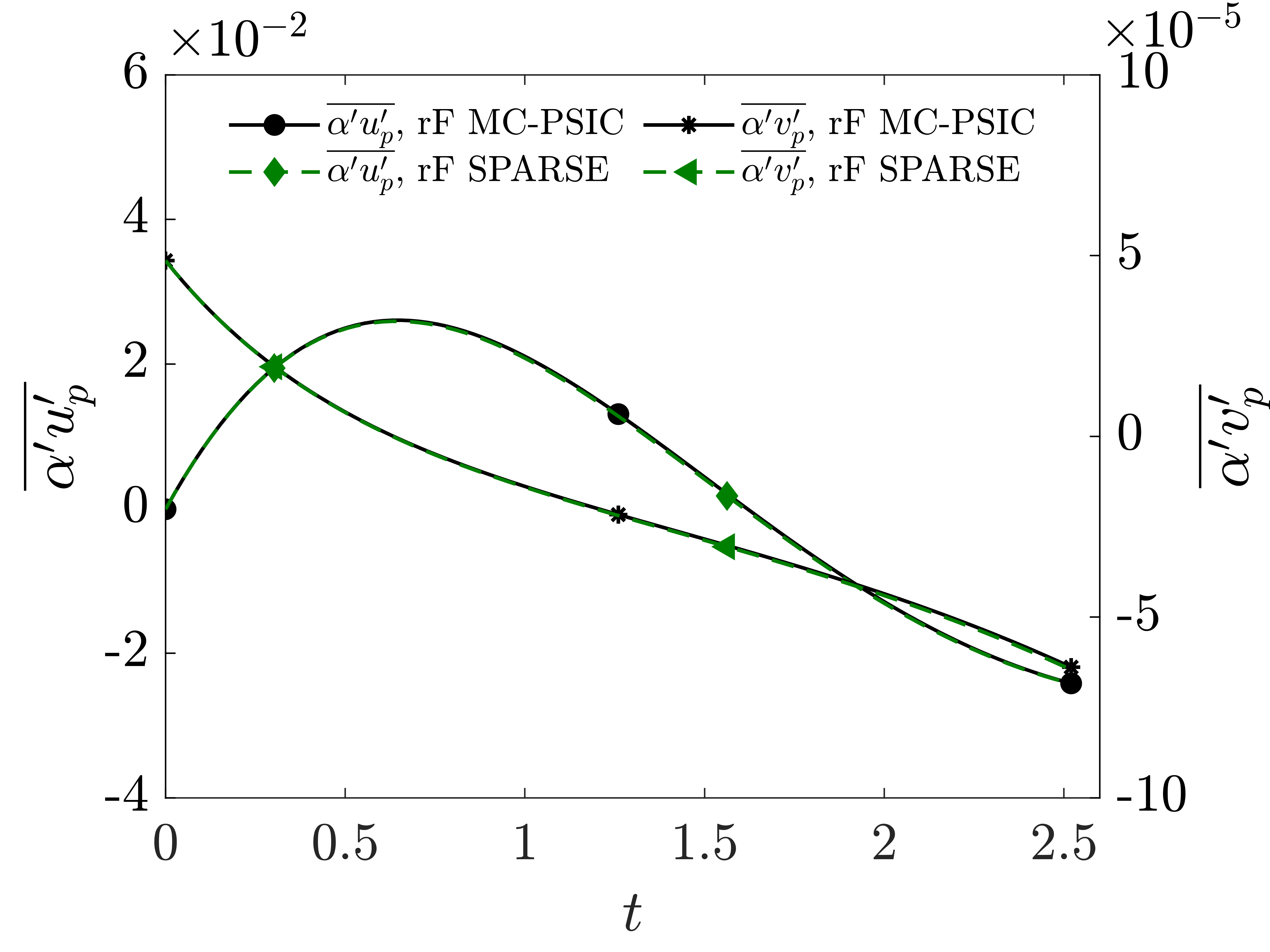}} 
	\caption[]{First two moments for of the stagnation flow test case for the rF and dF particles computed with the SPARSE and MC-PSIC methods. Including (a) horizontal and (b) vertical averages of the phase, (c) and (d) deviations of the particle phase, (e)--(g)combined moments of the particle variables and (h)--(i) combined moments of the particle variables and the random coefficient $\alpha$. The superscript `S' indicates rF and `D' dF.}
	\label{fig: sFcenter_cm2}
\end{figure}

Figure~\ref{fig: sFcenter_3Dplot} shows the solutions of the rF and dF cases identified with green and red colors, respectively.
The particle cloud is advected by the flow field in the time interval $t\in[0,2.52]$ in which the average location of the dF cloud reaches the wall located at $x=0$.
The solution is presented in the three-dimensional space $\alpha-x_p-y_p$ in Figure~\ref{fig: sFcenter_3Dplot_xpyp} for three instants of time $t=[0,1.26,2.52]$.
The point-particles (computed with MC-PSIC) are depicted as points and the point-cloud (computed with SPARSE) is represented by either a ellipse in the dF case (because $\alpha=1$ for all particles then the $\alpha$ collapses in a single point) or a prolate spheroid for the rF case with a magnitude and direction of the principle axes that are equal to  eigenvalues and direction of the eigenvectors of the covariance matrices.
The two covariance matrices used to represent the SPARSE solution for the rF case in Figure~\ref{fig: sFcenter_3Dplot} are $K_{\alpha x_p y_p}$ and $K_{\alpha x_p u_p}$ whose eigenvalues are given by the following Characteristic polynomials
\begin{align}
\begin{split}
\lambda^3 
&- \left( \overline{{x_p^\prime}^2} + \overline{{y_p^\prime}^2}  + \overline{{\alpha^\prime}^2}\right) \lambda^2 
+ \left( \overline{{\alpha^\prime}^2} \left( \overline{{x_p^\prime}^2} + \overline{{y_p^\prime}^2} \right) + \overline{{x_p^\prime}^2} \   \overline{{y_p^\prime}^2} - \overline{{x_p^\prime y_p^\prime}}^2 -\overline{{\alpha^\prime x_p^\prime}}^2 -\overline{{\alpha^\prime y_p^\prime}}^2  \right) \lambda \\
&+ \overline{{x_p^\prime}^2}\overline{\alpha^\prime{y_p^\prime}}^2 +\overline{{y_p^\prime}^2}\overline{\alpha^\prime{x_p^\prime}}^2 +\overline{{\alpha^\prime}^2}\overline{{x_p^\prime}{y_p}^\prime}^2 - \overline{{x_p^\prime}^2} \ \overline{{y_p^\prime}^2} \ \overline{{\alpha^\prime}^2} - 2\overline{{x_p^\prime}{y_p^\prime}} \ \overline{{\alpha^\prime}{x_p^\prime}} \ \overline{{\alpha^\prime}{y_p^\prime}}
= 0,
\end{split}
\end{align}
and 
\begin{align}
\begin{split}
\lambda^3 
&- \left( \overline{{x_p^\prime}^2} + \overline{{u_p^\prime}^2}  + \overline{{\alpha^\prime}^2}\right) \lambda^2 
+ \left( \overline{{\alpha^\prime}^2} \left( \overline{{x_p^\prime}^2} + \overline{{u_p^\prime}^2} \right) + \overline{{x_p^\prime}^2} \   \overline{{u_p^\prime}^2} - \overline{{x_p^\prime u_p^\prime}}^2 -\overline{{\alpha^\prime x_p^\prime}}^2 -\overline{{\alpha^\prime u_p^\prime}}^2  \right) \lambda \\
&+ \overline{{x_p^\prime}^2}\overline{\alpha^\prime{u_p^\prime}}^2 +\overline{{u_p^\prime}^2}\overline{\alpha^\prime{x_p^\prime}}^2 +\overline{{\alpha^\prime}^2}\overline{{x_p^\prime}{u_p}^\prime}^2 - \overline{{x_p^\prime}^2} \ \overline{{u_p^\prime}^2} \ \overline{{\alpha^\prime}^2} - 2\overline{{x_p^\prime}{u_p^\prime}} \ \overline{{\alpha^\prime}{x_p^\prime}} \ \overline{{\alpha^\prime}{u_p^\prime}}
= 0,
\end{split}
\end{align}
respectively.
The projections of the solution onto the planes $x_p-y_p$, $\alpha-x_p$ and $\alpha-y_p$ are also plotted in Figure~\ref{fig: sFcenter_3Dplot_xpyp}.
Accordingly, in Figure~\ref{fig: sFcenter_3Dplot_xpup} the projections in planes $x_p-u_p$, $\alpha-x_p$ and $\alpha-u_p$ are also plotted.
For the rF case, the projections are represented by ellipses in all those planes.
For the dF case the projections simply lead to lines.
We also show the first two moments of both dF and rF clouds computed with the MC-PSIC and SPARSE methods in Figure~\ref{fig: sFcenter_cm2} for comparison.
The solution shown in Figures~\ref{fig: sFcenter_3Dplot} and~\ref{fig: sFcenter_cm2} corresponds to a level of splitting $M=5$.
Note that the initial condition is split along the random parameters $\alpha$ and the two components of the particle velocity and location.
According to this, for the dF case the total number of subclouds is related to the level of splitting $M_p=M^{4}$ and for rF $M_p=M^5$.


The clouds approach the stagnation point   along the $x-$axis as depicted in Figure~\ref{fig: sFcenter_3Dplot_xpyp} in the $x_p-u_p$ plane.
 The average horizontal location and velocity trends show three different stages in time, including a first stage of acceleration, a second transitional stage and a third deceleration stage.
These stages are observed for both the rF and dF clouds.
Because the front of the cloud decelerates while the tail is still increasing its velocity, 
the standard deviations of the horizontal velocity $\sigma_{u_p}$ in the second stage show  a minimum at $t\approx1.6$ (see Figure~\ref{fig: sFcenter_sigma_upvp}) for the rF case.
The dF case does not exhibit this minimum in  $\sigma_{u_p}$, but decreases monotonically over the entire time interval and  plateaus only at that time instant.
In the third stage all point-particles  in the cloud region decelerate towards the stagnation point.
Because of the symmetry of the problem, the mean vertical location and velocity should be zero. The results, however, show  a slight initial motion in the negative $y$ direction as a result of the Monte Carlo sampling error
that is on the order of $10^{-4}$ 
for $N_p=10^5$  (see Figure~\ref{fig: sFcenter_mean_ypvp}). Over time this non-zero initial condition leads to vertical motion that is consistent with the three stages.



To assess the uncertainty in the rF point-cloud tracer,  we compare the dF and the rF SPARSE cloud prediction in the phase space $x_p-u_p$ in Figure~\ref{fig: sFcenter_3Dplot_xpup}. 
The three stages  can once again be observed in the evolution  of $\sigma_{x_p}$, but are notably different for dF and rF as follows:
in the first stage  $\sigma_{x_p}$ decreases for the dF cloud (see also Fig. ~\ref{fig: sFcenter_sigma_xpyp}) whereas for the rF cloud it increases.
The latter increase is a result of the mix of fast and slow responses  to the fluid velocity of the point-particles in the rF cloud. This mix is consistent with the range of the random force coefficients, $\alpha$, and the resulting range in  the time responses of the point-particles as given by $St/\alpha$.
In fact, the  enhanced mixing of the rF cloud in space is a result of the virtual stress per the correlation of the coefficient $\alpha$ and the particle phase variables $x_p$ and $u_p$ (as was also the case for the sine field test discussed above) 
as it appears in the equations \eqref{eq: SFeqs_xpup} that govern
$\overline{x_p^\prime u_p^\prime}$. This term  in turn is the only driving source term in the dynamics of 
$\sigma_{x_p}$ in ~\eqref{eq: SFeqs_cm2_xp}.
The magnitude of the term $\overline{\alpha}\left( k\overline{{x_p^\prime}^2}+\overline{x_p^\prime u_p^\prime } \right)$ in equation~\eqref{eq: SFeqs_xpup} is approximately the same for rF and dF with $\overline{\alpha}=1$, whereas  the second term on the right hand side, $\overline{\alpha^\prime x_p^\prime}\left(k\overline{x}_p+\overline{u}_p \right)$, is  initially on the order of  $\overline{\alpha^\prime x_p^\prime}$, which in turn is governed by equations~\eqref{eq: SFeqs_xpa}--\eqref{eq: SFeqs_upa}. In ~\eqref{eq: SFeqs_xpa}--\eqref{eq: SFeqs_upa}  the only term that is initially different from zero is $\overline{{\alpha^\prime}^2}\left(k\overline{x}_p+ \overline{u}_p\right)$ 
and it is therefor
the root cause for the increase of $\sigma_{x_p}$ in the rF cloud.

In vertical direction the solution is symmetric and  it turns out that because of this symmetry that $\sigma_{y_p}$ evolves identically for rF and dF.
 This can be understood using the simplification of the point-cloud equations in  the $y$-direction  for the $y$-symmetric Hiemenz flow  (Appendix~\ref{app: rF_equations}), which shows that the term $-\overline{\alpha^\prime y_p^\prime}$ in equation~\eqref{eq: SFeqs_ypvp} is multiplied by the average vertical relative velocity which is zero over the simulated time interval (assuming no sampling errors).
 Ergo, the random forcing does not  change the vertical stress or strain 
 (Figure~\ref{fig: sFcenter_3Dplot_xpyp}), which  suggests that the uncertainty in solution  is zero in vertical direction. This results generalizes to any particle-laden that is symmetric and is randomly forced according to a symmetric PDF. We conclude  that in these case the uncertainty in the forcing does not propagate into the solution. 
 A closer inspection 
of $\sigma_{y_p}$in Figure~\ref{fig: sFcenter_sigma_xpyp} shows trends on the order of the sampling errors. These trends as in $x$-direction are governed by the intrinsic interaction of correlation terms in the governing system in $y$-direction.

\begin{figure}[h!]
		\centering
	\subfloat[]{
		\includegraphics[width=0.4\textwidth]{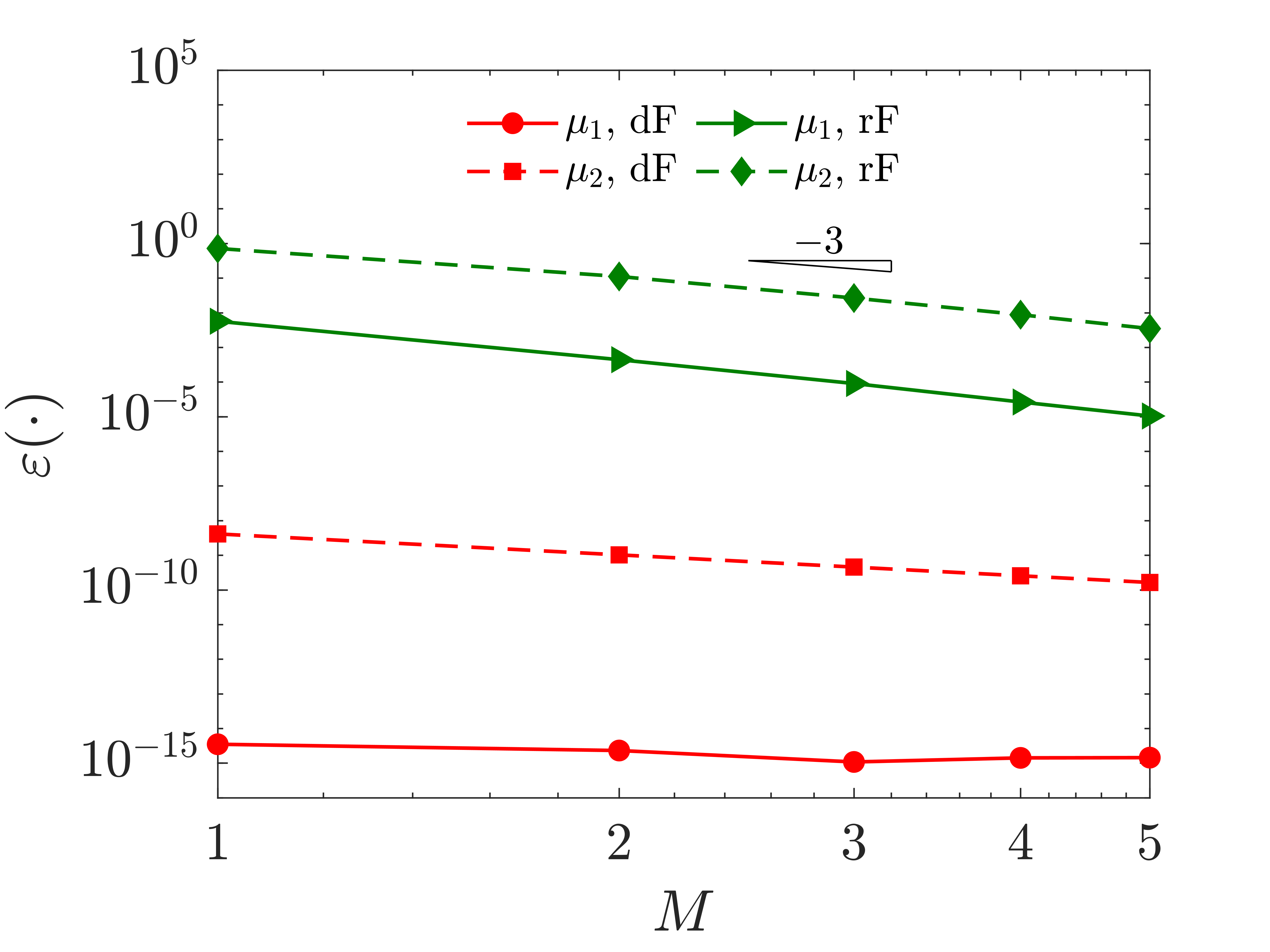}} 
	\caption[]{Convergence of the errors of the SPARSE method as compared to the MC-PSIC method when using splitting for the stagnation flow case. The superscript `S' indicates rF and `D' dF.}
	\label{fig: sFcenter_errors}
\end{figure}


In Figure~\ref{fig: sFcenter_errors} we plot the relative error versus splitting levels in the range $M=1,\dots,5$ .
The third order convergence rate is once more validated.
The dF test cases reaches machine precision and does not require splitting to reduce the errors.
We find that for the maximum level of splitting ($M=5$) considered which leads to $M_p=3,125$ and $M_p=625$ for the rF and dF cases respectively, the greatest relative error of all first and second moments of the particle phase computed with SPARSE as compared to MC-PSIC is $3 \%$ and it occurs
$\sigma_{u_p}^2$.
The remaining variables of the cloud show smaller relative errors as compared to MC-PSIC, validating the SPARSE method.
The ratio of computational cost of SPARSE as compared to MC-PSIC defined as in~\eqref{eq: savings} is $r=0.022$ for the dF case and $r=0.141$ for the rF case. 
This shows that using SPARSE implies only a $14.1\%$ of the computational effort that takes to solve the problem with the MC-PSIC method for the rF case and a $2.2\%$ for the dF case.

\subsection{Isotropic turbulence} \label{sec: test_isoTurb}



Following the validation of the dF clouds  as presented in \cite{dominguez2023closed}, we test the randomly forced SPARSE formulation in a decaying isotropic turbulence velocity field. For a detailed description of the test setup we refer to that article. Summarizing for completeness, the isotropic turbulence simulation is performed in a cube with periodic boundary conditions with the validated discontinuous Galerkin code as described in~\cite{klose2020assessing} and references therein. Initial conditions  are adopted from~\cite{jacobs2005validation}.
The simulation is performed on a domain $\Omega$ spanned by coordinates $(x,y,z)$, defining a cube of size $2\pi$ so that $\Omega=[0, 2\pi]\times[0, 2\pi]\times[0, 2\pi]$.

As in the previous test cases, we seed  a cloud of $N_p=10^5$  Monte Carlo point-particles in the flow.
The particles in the cloud are initially at rest 
and 
therefore the cloud's average velocity is $\overline{u}_p=\overline{v}_p=\overline{w}_p=0$. The cloud's temperature is set constant such that $T_p^\prime=0$ for the cloud with $\overline{T}_p=1$.
The locations of point-particles are sampled from the uniform distribution functions
\begin{align}
{x_p}_0 &\sim \mathcal{U}\left[{{x_p}_0}_{min},{{x_p}_0}_{max}\right], \ \ \
    {y_p}_0 \sim \mathcal{U}\left[{{y_p}_0}_{min},{{y_p}_0}_{max}\right], \ \ \ {z_p}_0 \sim \mathcal{U}\left[{{z_p}_0}_{min},{{z_p}_0}_{max}\right].
\label{eq: isoTurb_IC_xp_yp_zp} 
\end{align}
The random momentum and energy coefficients, $\alpha$ and $\beta$ are also sampled from an uniform distribution
\begin{align}
    \alpha = \beta \sim \mathcal{U}\left[\alpha_{min},\alpha_{max}\right].
    \label{eq: isoTurb_IC_alpha}
\end{align}
The average and standard deviation values  are $\overline{x}_p=\overline{y}_p=\overline{z}_p=\pi$ and $\overline{\alpha}=1$ and $\sigma_{{x_p}_0}=\sigma_{{y_p}_0}=\sigma_{{z_p}_0}=0.05$, and $\sigma_{\alpha}=0.3$. 
For the dF case $\sigma_{\alpha}=0$.

The drag and heat transfer correction factors for this case are adopted from~\cite{boiko1997shock} and~\cite{michaelides2016multiphase} respectively and read as
\begin{subequations} \label{eq: isoTurb_f1f2}
\begin{align}
f_1 &= \alpha \left(1+0.38\frac{Re_p}{24}+\frac{Re_p^{0.5}}{6}   \right)\left[ 1+\exp \left( \dfrac{-0.43}{M_p^{4.67}} \right)    \right],
\label{eq: isoTurb_f1} \\
f_2 &= \alpha \left( 1+0.3Re_p^{0.5}Pr^{0.33} \right).
\label{eq: isoTurb_f2}
\end{align}
\end{subequations}
The computed carrier phase velocities are used to determine the particle Reynolds as defined in~\eqref{eq: Rep_St_Pr} and the particle Mach number $M_{p}=|\boldsymbol{u}-\boldsymbol{u}_p|/\sqrt{T_{f}}$, in the forcing correction factors~\eqref{eq: isoTurb_f1f2}. 
The Prandtl number is $Pr=0.7$, the relative heat capacity is set to unity $c_r=1$ and the density ratio of the two phases is $\rho_p=250$.
We also set the Stokes number to be $St=0.5$ such that the inertia is dominant in the dynamics of the point-particles.
The non-dimensional particle diameter is $d_p=4\cdot 10^{-3}$ according to~\eqref{eq: Rep_St_Pr}.
The isotropic turbulence case is set up with a reference Reynolds $Re_\infty=2,357$ and Mach number $M_\infty=0.05$~\citep{jacobs2005validation}.


Using  splitting along the particle locations and the random parameter $\alpha$ for the point-cloud simulation, 
it follows that for the rF case the total number of subclouds is $M_p=M_p^{x_p}M_p^{y_p}M_p^{z_p}M_p^{\alpha} = M^4$ with superscripts indicating the dimension where the divisions are performed.
For the dF case then we have $M_p=M^3$.
We consider several levels of splitting $M=1,\dots,8$ that correspond to a maximum number of subclouds of $M_p=4,096$ and $M_p=512$ for rF and dF cases respectively. 
Note that in this test case, all relative errors should be expected to be non-zero  because the flow is non-linear and the correction functions of the forcing depend on the relative velocity of the particles through the particle Reynolds and Mach numbers and so all Taylor and moment truncation affect the accuracy of SPARSE.

\begin{figure}[h!]
	\centering
	\subfloat[]{
		\label{fig: isoTurb_contour1}
		\includegraphics[width=0.32\textwidth]{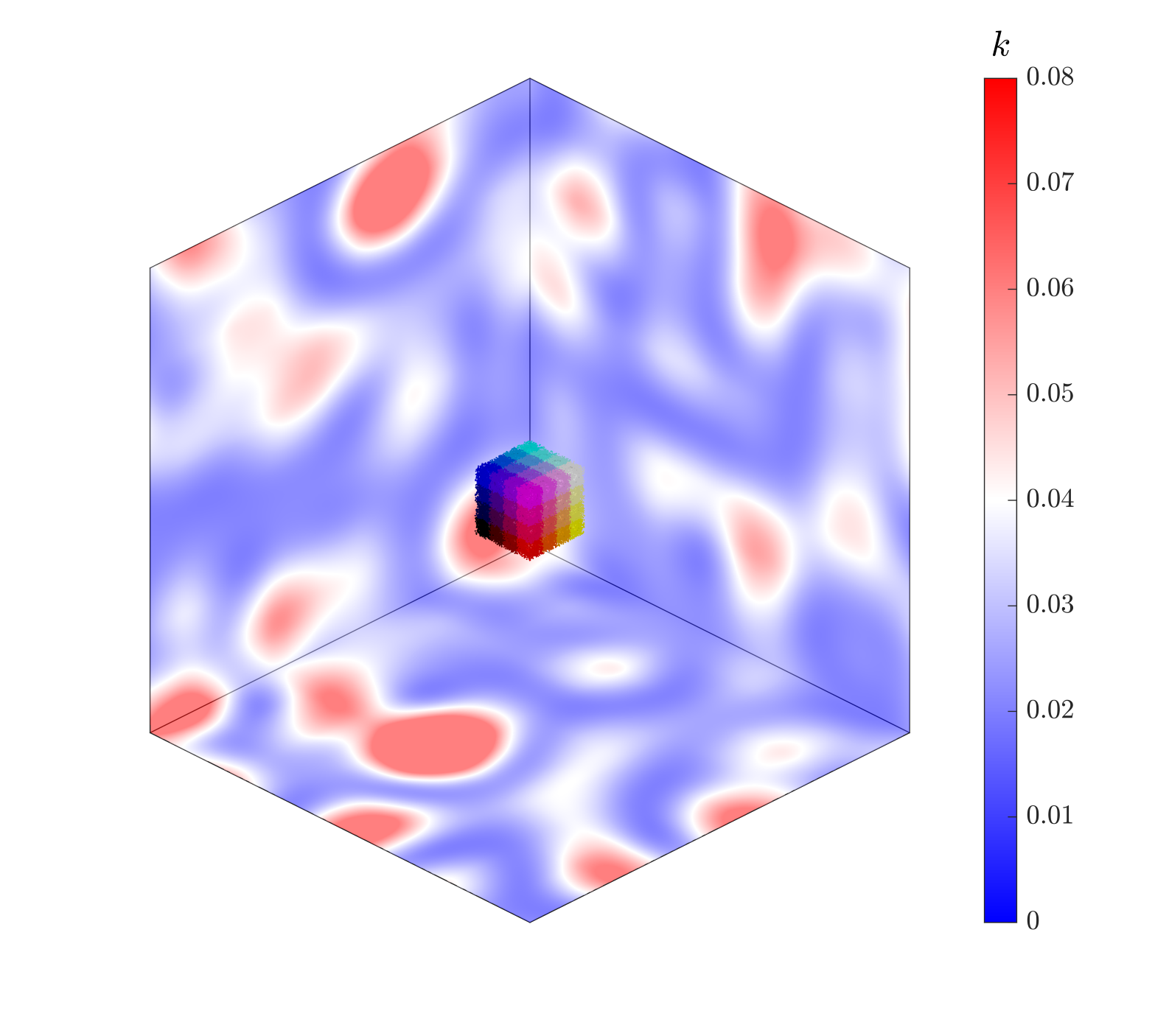}}
        \hfill	
	\subfloat[]{
		\label{fig: isoTurb_contour2}
		\includegraphics[width=0.32\textwidth]{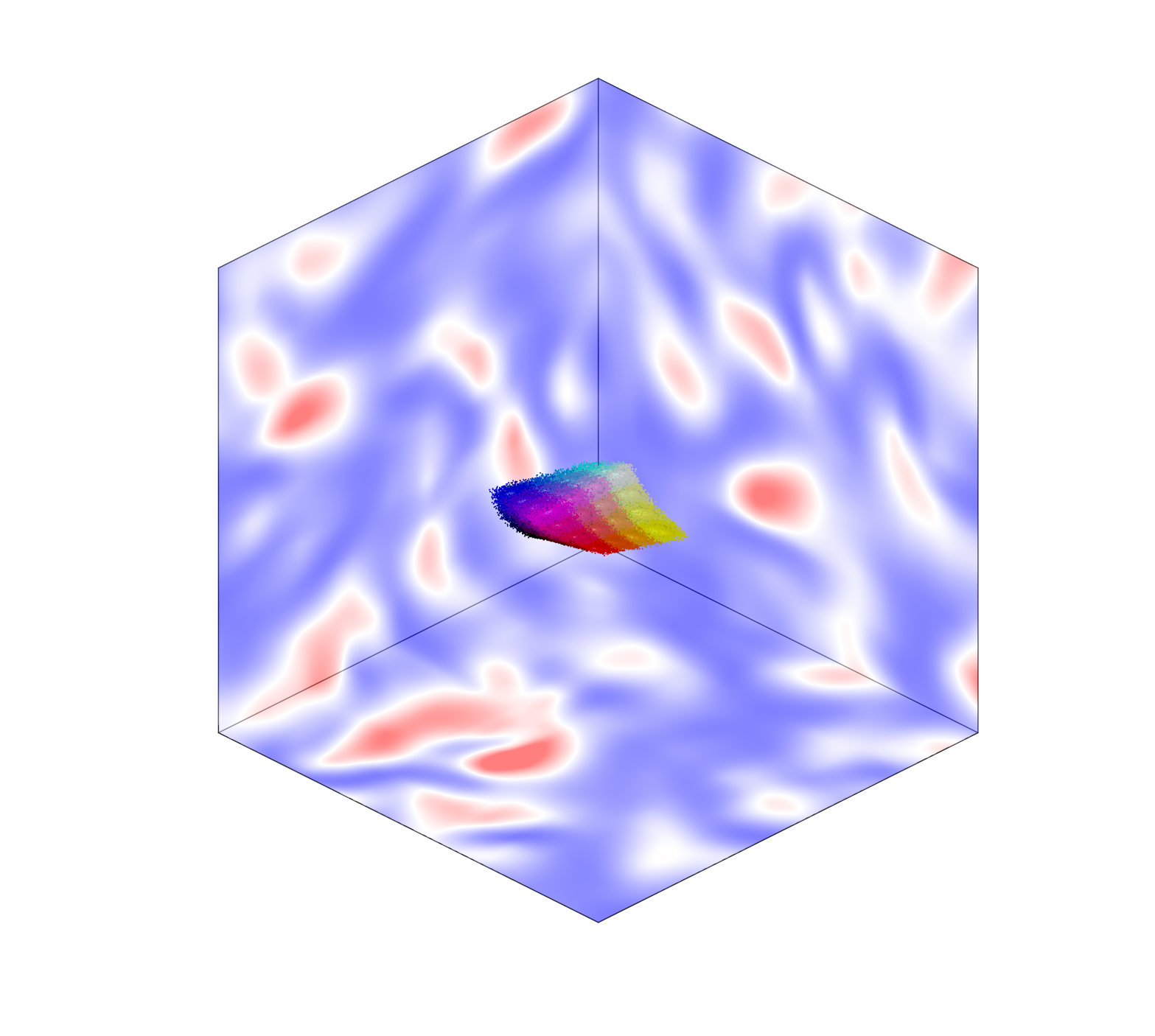}} 
		\hfill
	\subfloat[]{
		\label{fig: isoTurb_contour3}
		\includegraphics[width=0.32\textwidth]{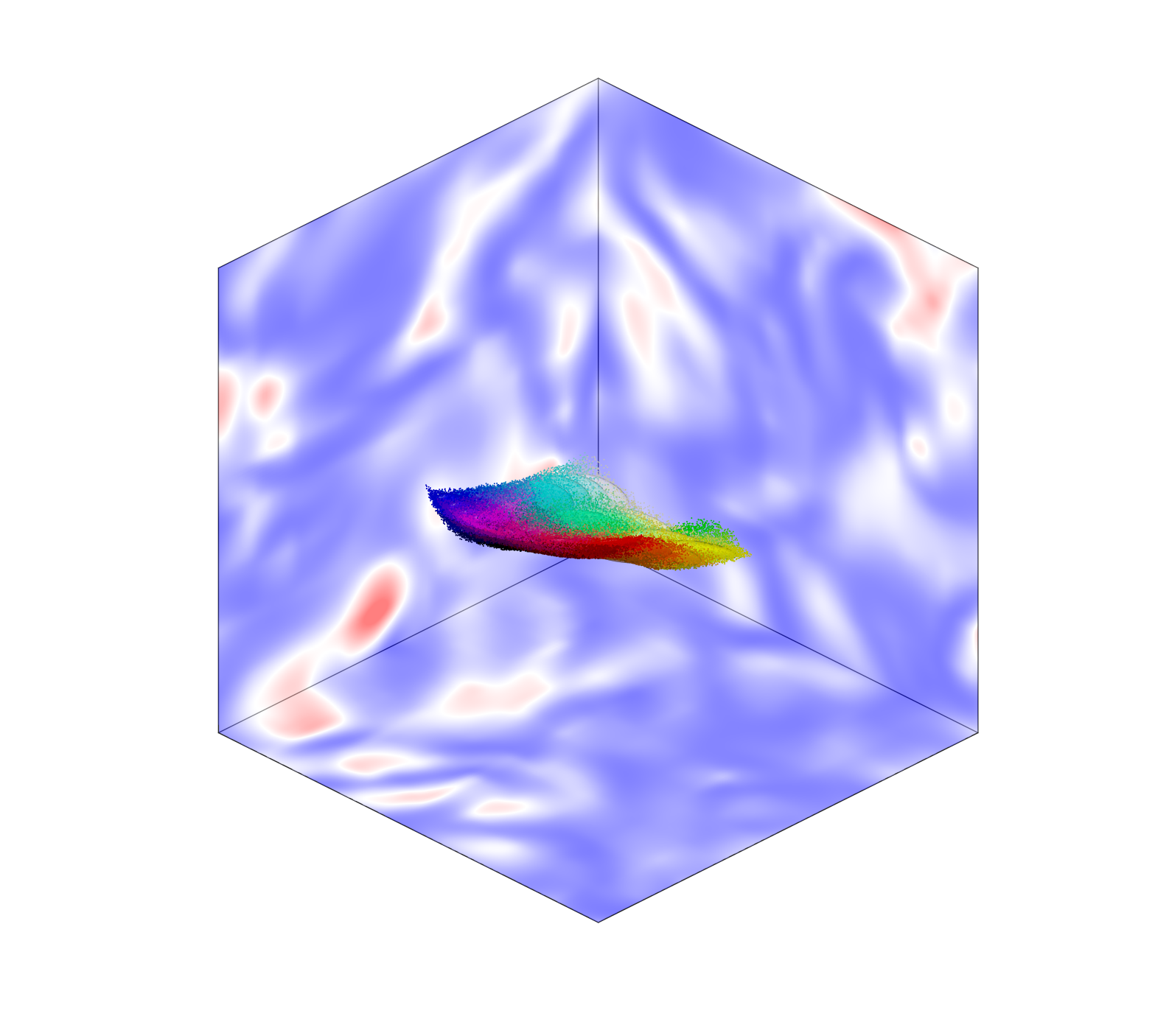}} \\
	\subfloat[]{
		\label{fig: isoTurb_contour4}
		\includegraphics[width=0.32\textwidth]{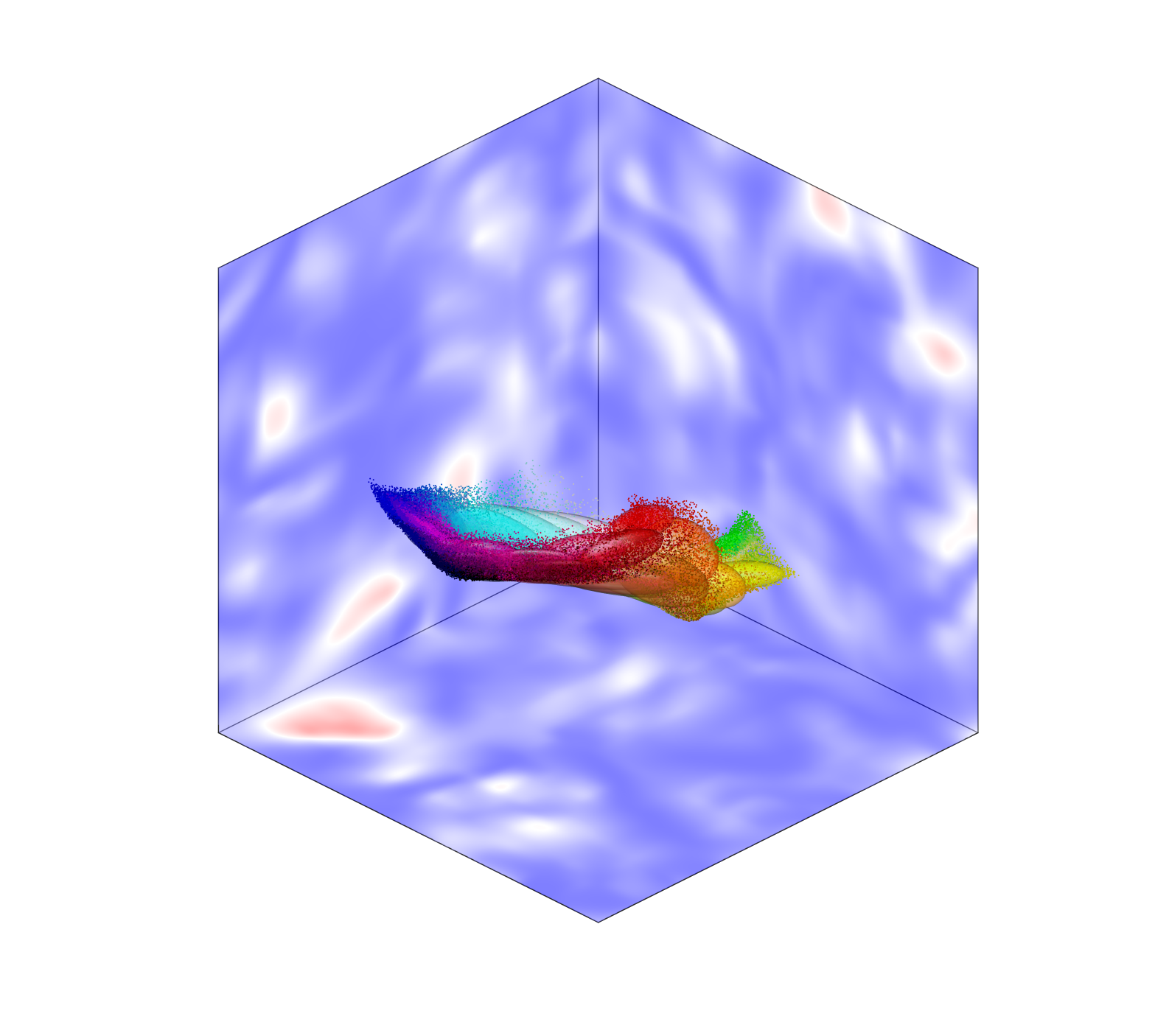}}
        \hfill	
	\subfloat[]{
		\label{fig: isoTurb_contour5}
		\includegraphics[width=0.32\textwidth]{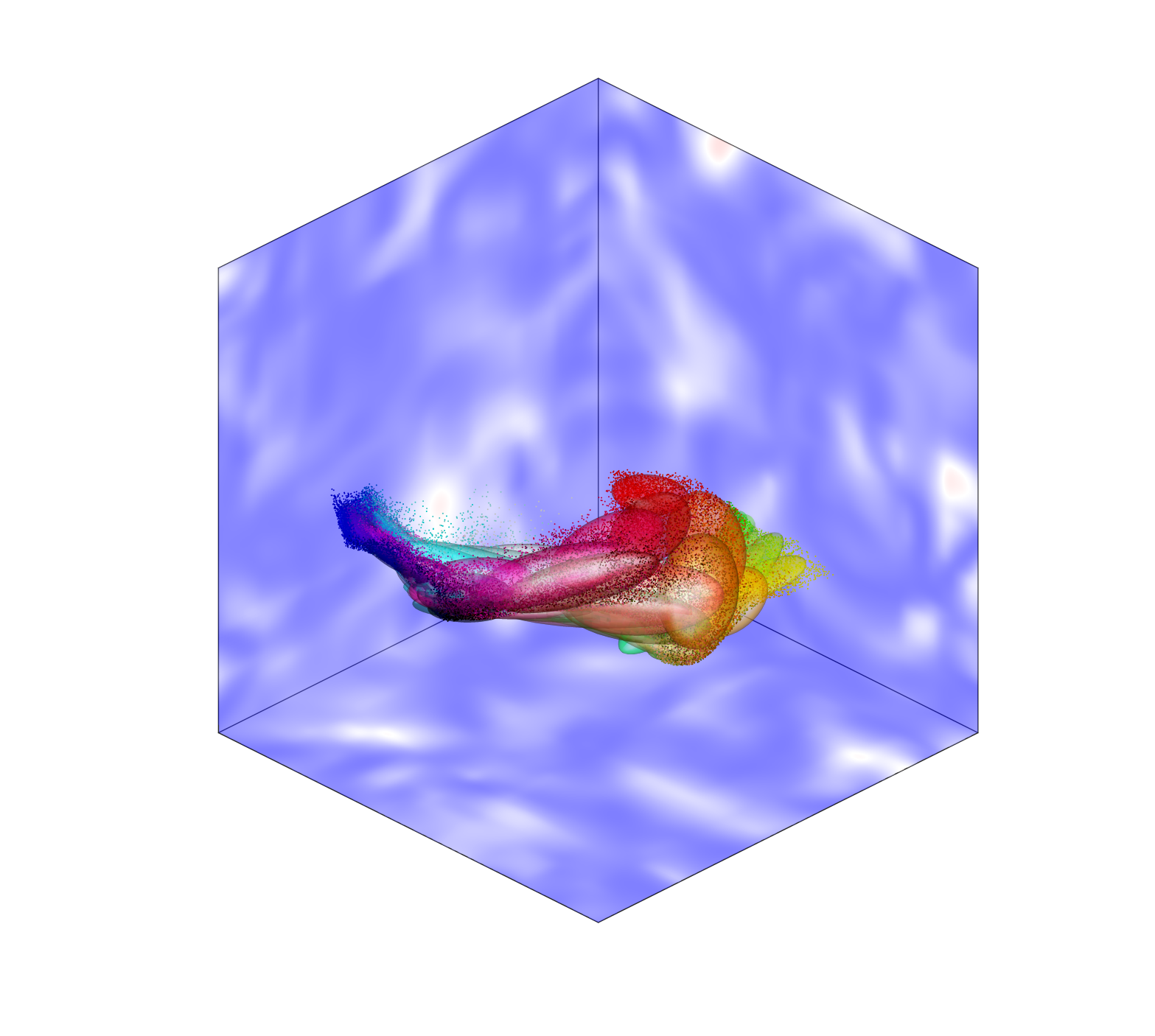}} 
		\hfill
	\subfloat[]{
		\label{fig: isoTurb_contour6}
		\includegraphics[width=0.32\textwidth]{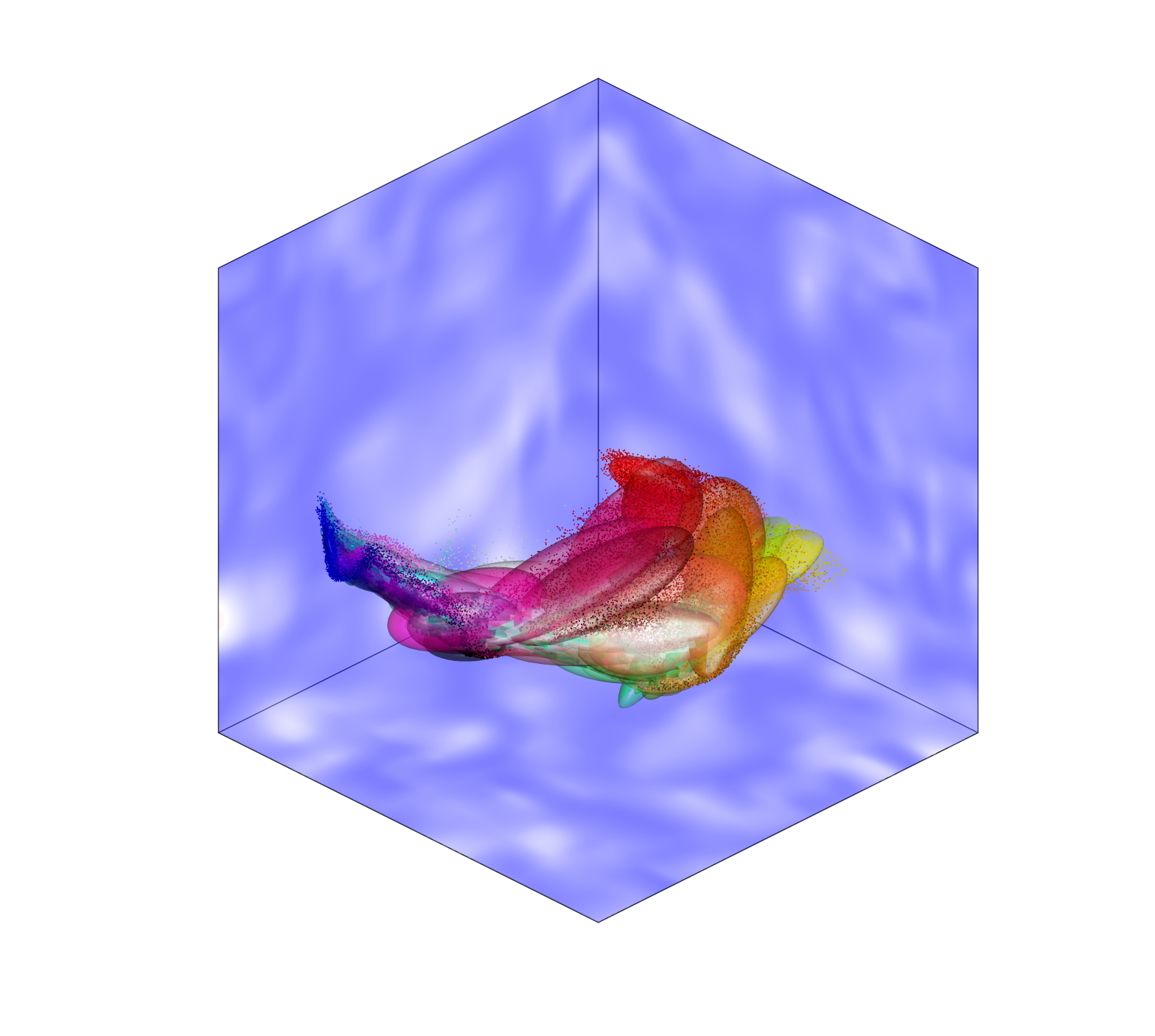}} 
	\caption[]{Locations of the point-particles and point-clouds computed with MC-PSIC and SPARSE methods respectively for the isotropic turbulence case for a level of splitting of $M=4$ at times (a) $t=0$, (b) $t=0.8$, (c) $t=1.6$, (d) $t=2.4$, (e) $t=3.2$ and (f) $t=4$. The contours show the turbulent kinetic energy of the flow in the boundaries of the domain. The PSIC particles are represented with dots and the SPARSE clouds with ellipsoids.}
	\label{fig: isoTurb_contour}
\end{figure}

In Figure~\ref{fig: isoTurb_contour} we show the rF test case results for six equispaced instants of times in the interval $t\in[0, 4]$.
The MC-PSIC particles are depicted as points and the SPARSE clouds as prolate spheroids for a level of splitting with $M=4$ which defines four subclouds along the variables $x_p$, $y_p$, $z_p$ and $\alpha$.
We select this somewhat coarser level of splitting for clarity of the visualization in Figure~\ref{fig: isoTurb_contour}.
The SPARSE subclouds 
are colored according to the spatial four divisions along each coordinate to identify the point-particles that correspond to each point-cloud.
The same color has been used for a given subcloud and its corresponding point-particles in the Figures~\ref{fig: isoTurb_contour1}--\ref{fig: isoTurb_contour6}.
The background planes represents contours of the turbulent kinetic energy of the carrier phase’s turbulent structures for each instant of time.
The ellipsoids for each subcloud computed with SPARSE are represented using the eigenvalues and eigenvectors of the covariance matrix of the locations $x_p$, $y_p$ and $z_p$.
The particles initially at rest respond to the carrier flow producing a deformation of the initial cube where the cloud of point-particles is defined. 
Governed by the inertial effects, the response is smoother as compared with the changes in the carrier flow.
As time evolves, the particle cloud is dispersed occupying an increasingly larger domain.


\begin{figure}[h!]
	\centering
	\subfloat[]{
		\label{fig: isoTurb_tend4_IC0_mean_xpypzp}
		\includegraphics[width=0.32\textwidth]{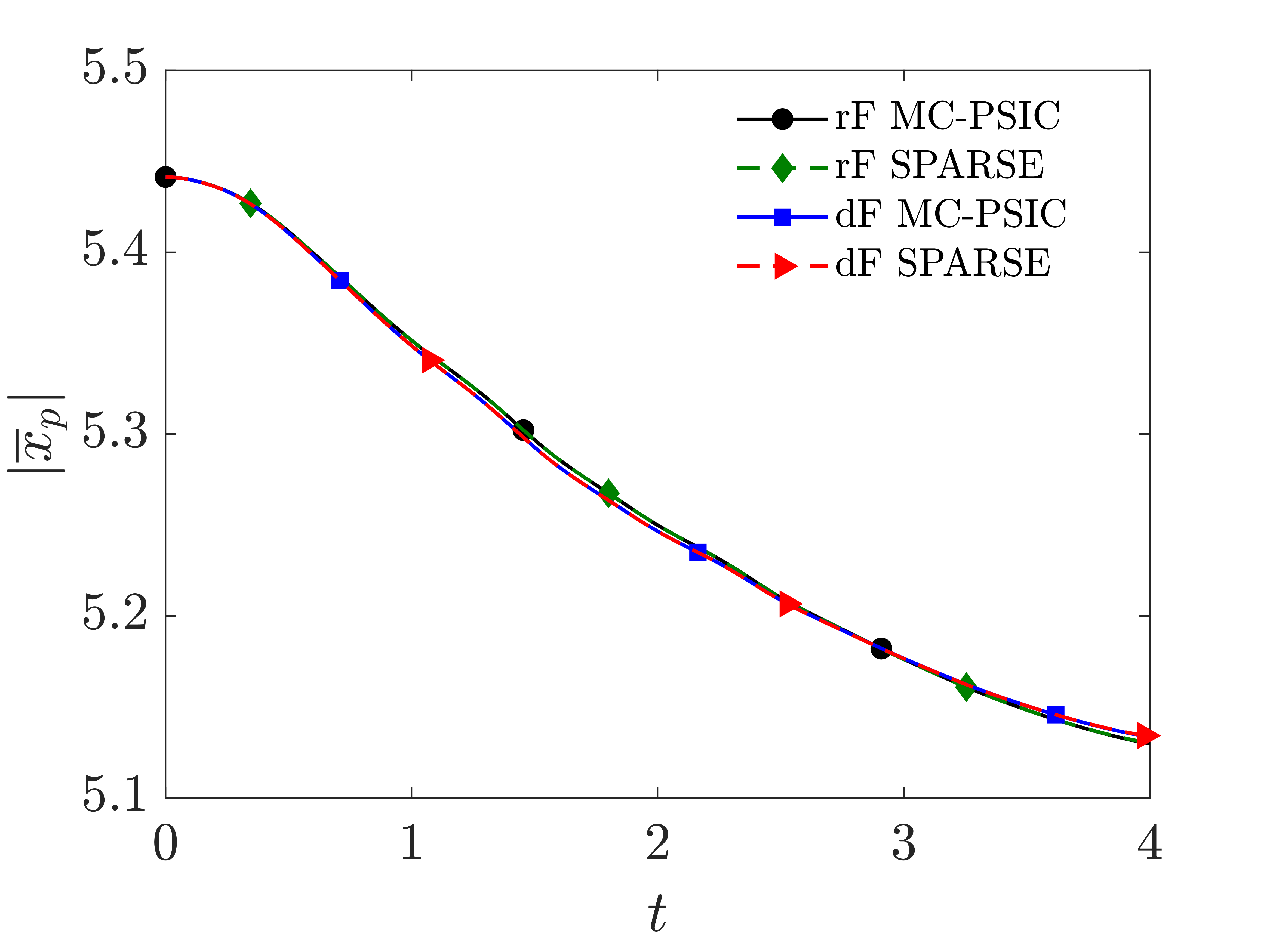}}
        \hfill	
	\subfloat[]{
		\label{fig: isoTurb_tend4_IC0_mean_upvpwp}
		\includegraphics[width=0.32\textwidth]{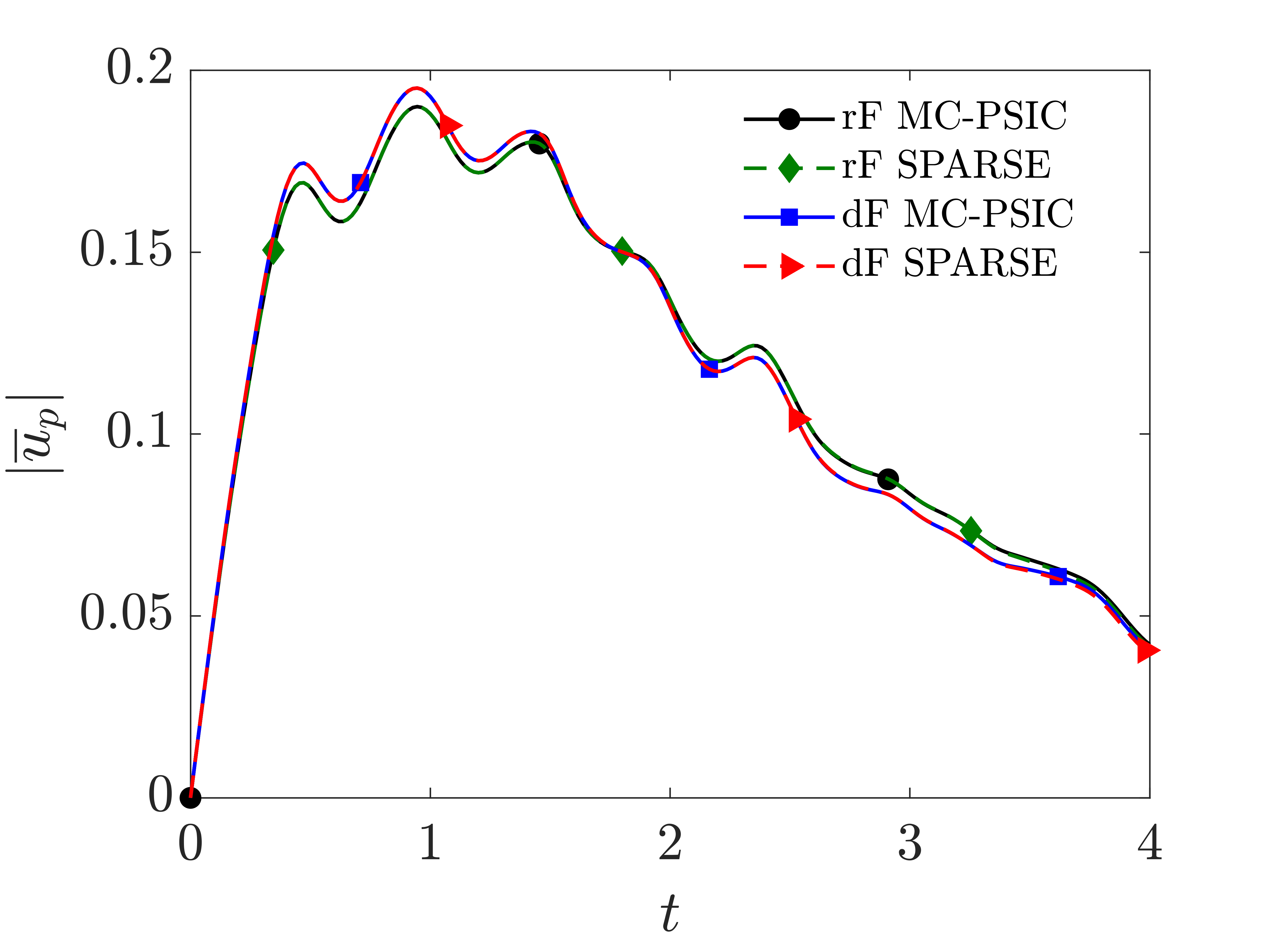}} 
		\hfill
	\subfloat[]{
		\label{fig: isoTurb_tend4_IC0_mean_Tp}
		\includegraphics[width=0.32\textwidth]{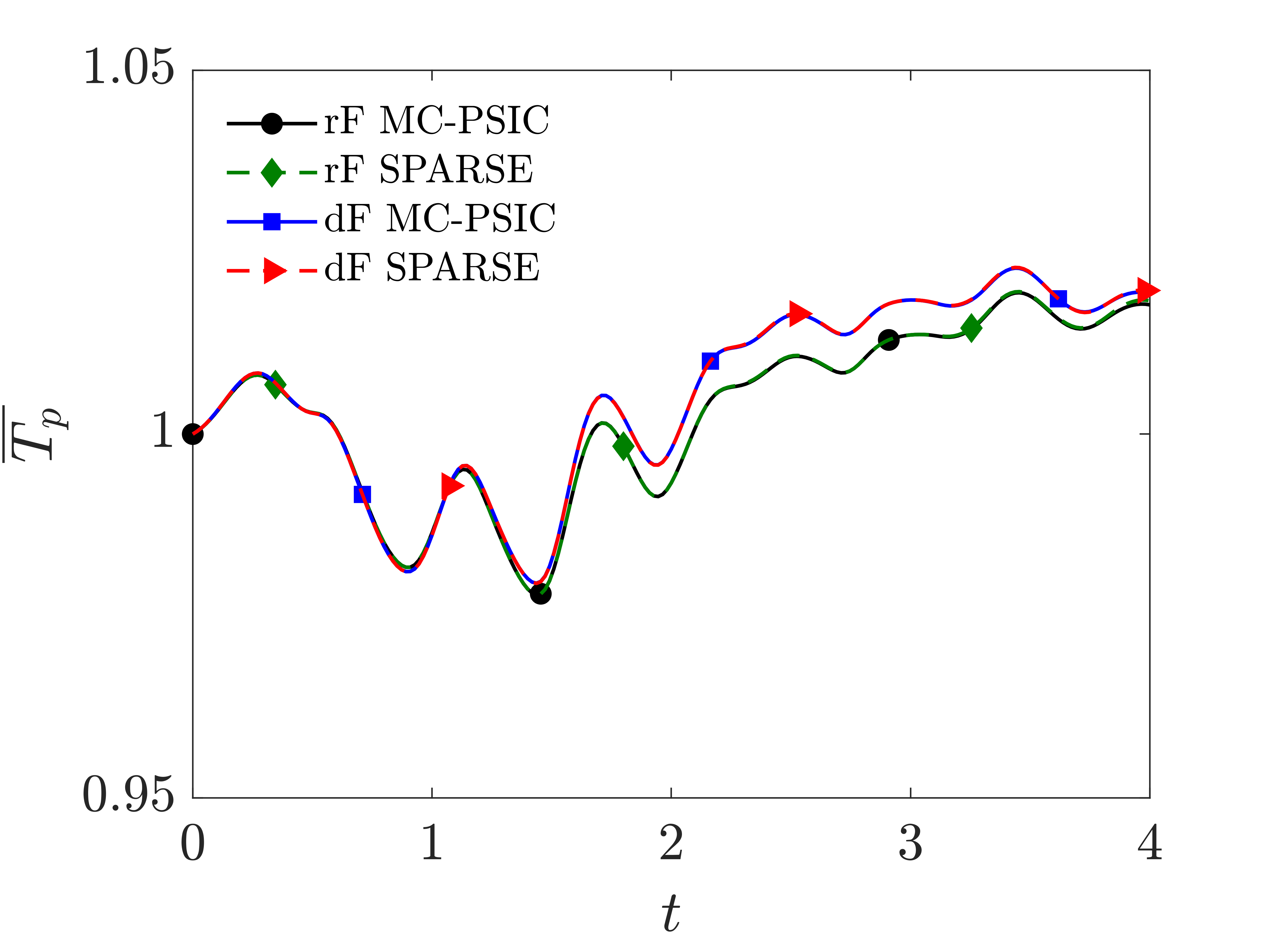}} \\
	\subfloat[]{
		\label{fig: isoTurb_tend4_IC0_detK_xp}
		\includegraphics[width=0.32\textwidth]{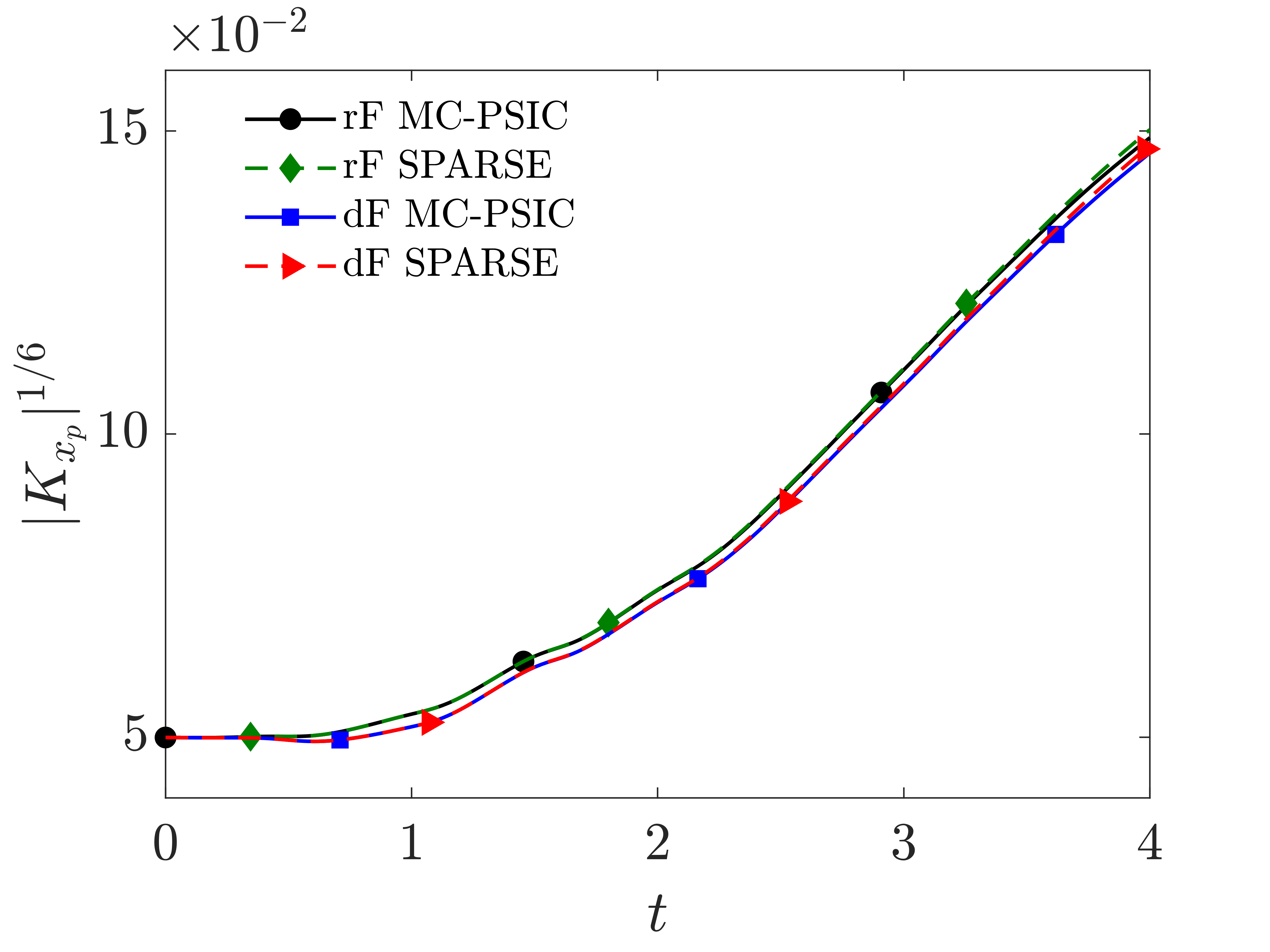}}
        \hfill	
	\subfloat[]{
		\label{fig: isoTurb_tend4_IC0_detK_xpup}
		\includegraphics[width=0.32\textwidth]{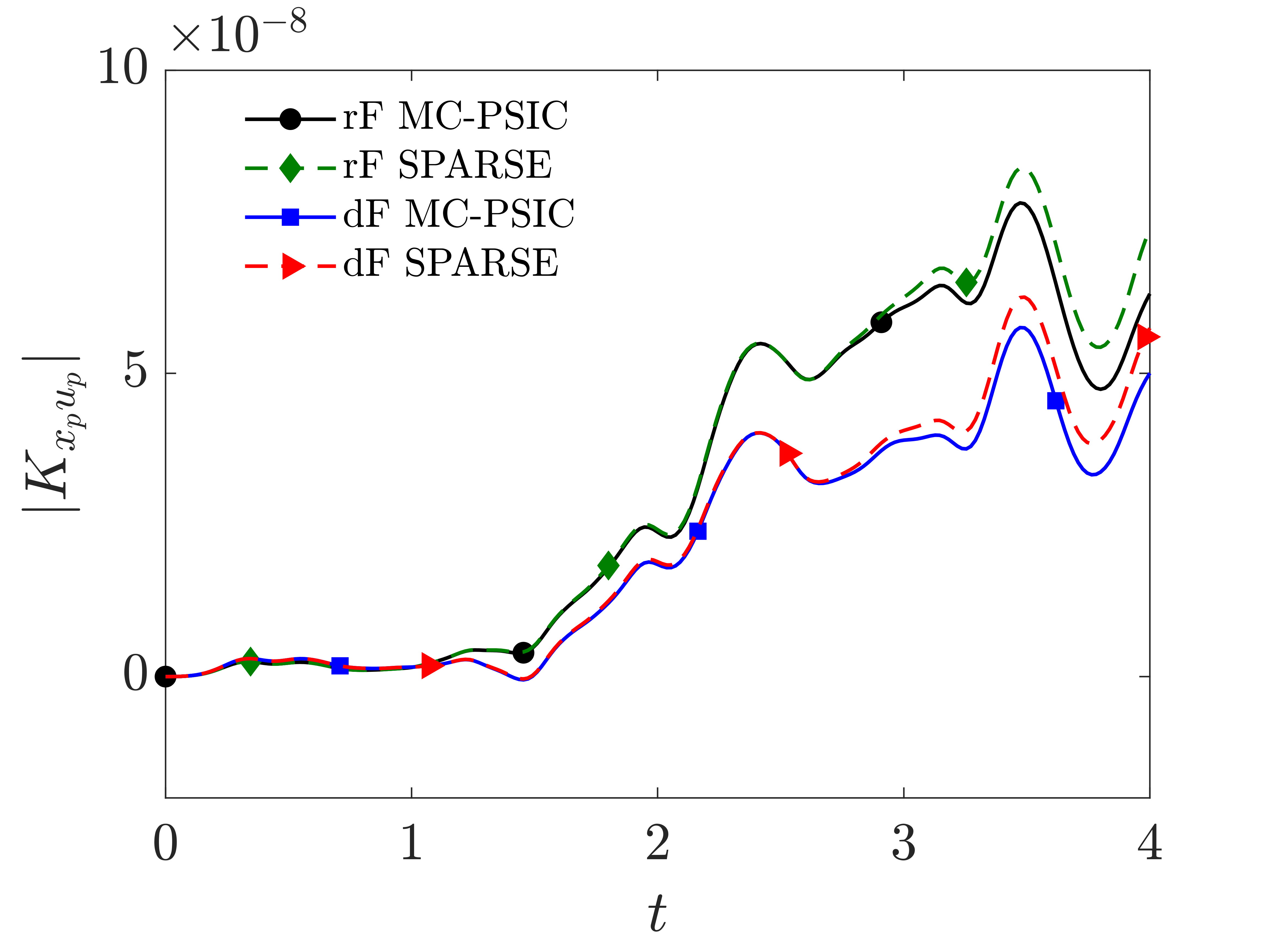}} 
		\hfill
	\subfloat[]{
		\label{fig: isoTurb_tend4_IC0_modK_xpTp}
		\includegraphics[width=0.32\textwidth]{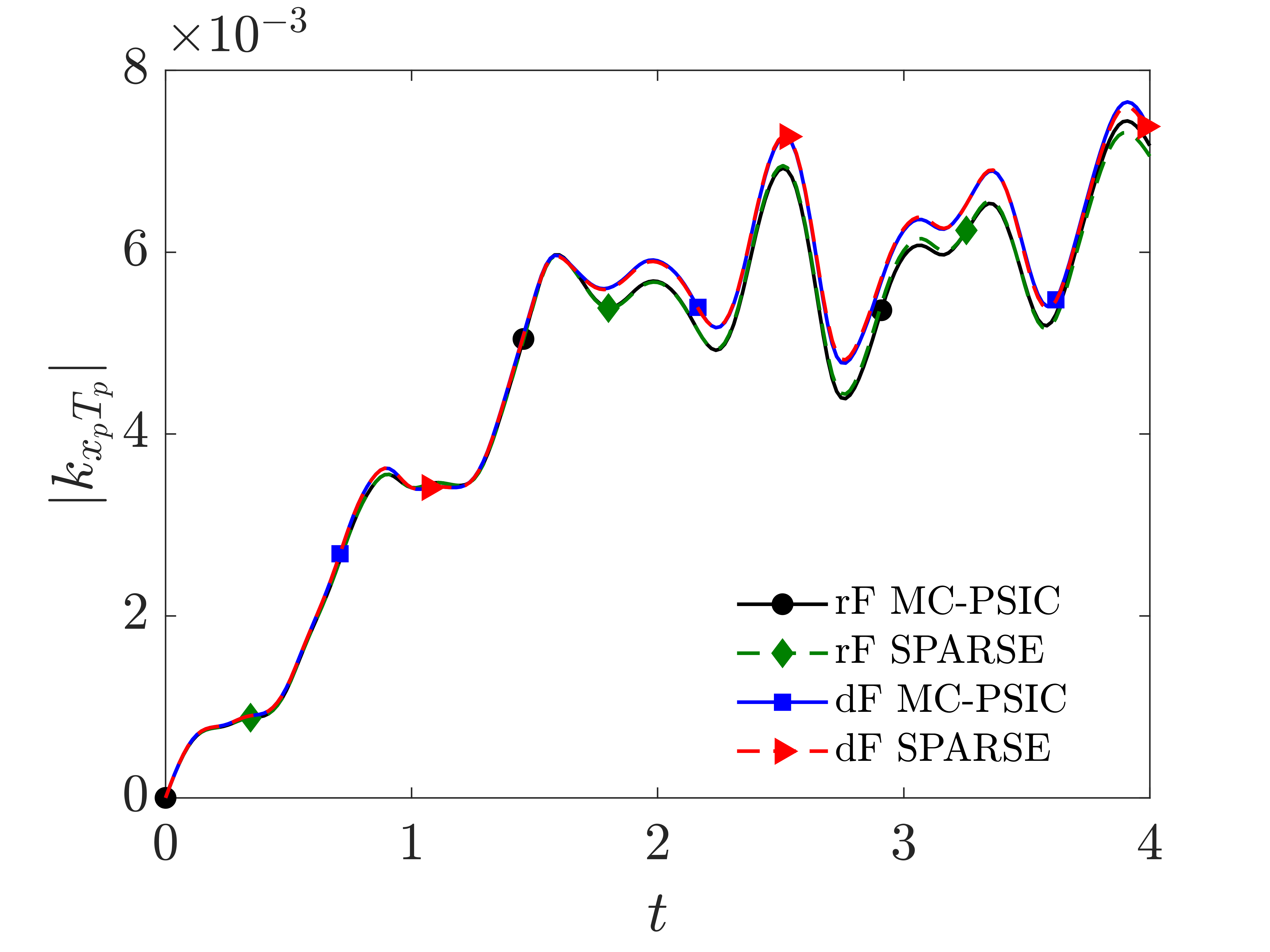}} \\
    \subfloat[]{
		\label{fig: isoTurb_tend4_IC0_detK_up}
		\includegraphics[width=0.32\textwidth]{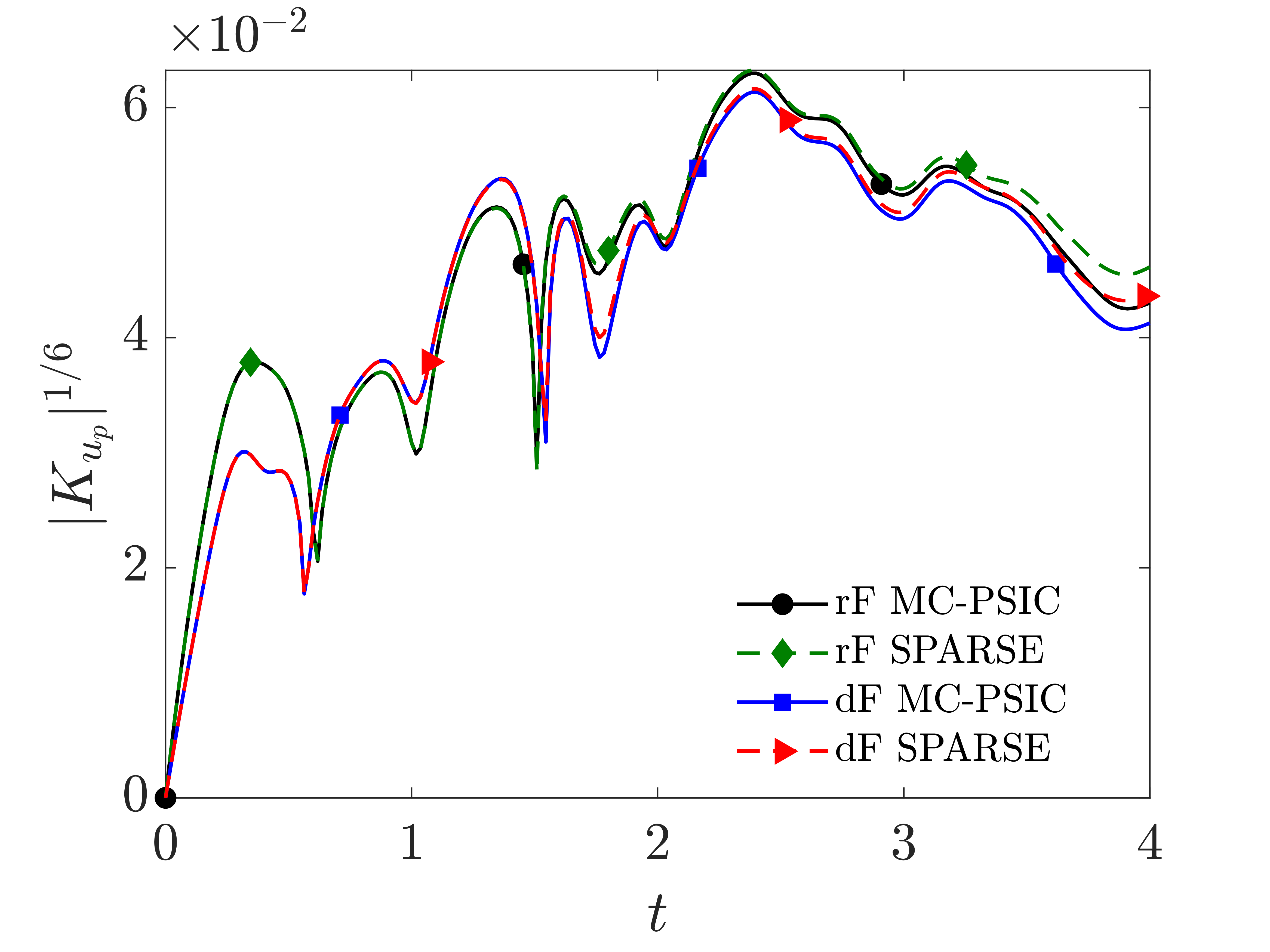}}
        \hfill	
	\subfloat[]{
		\label{fig: isoTurb_tend4_IC0_modK_upTp}
		\includegraphics[width=0.32\textwidth]{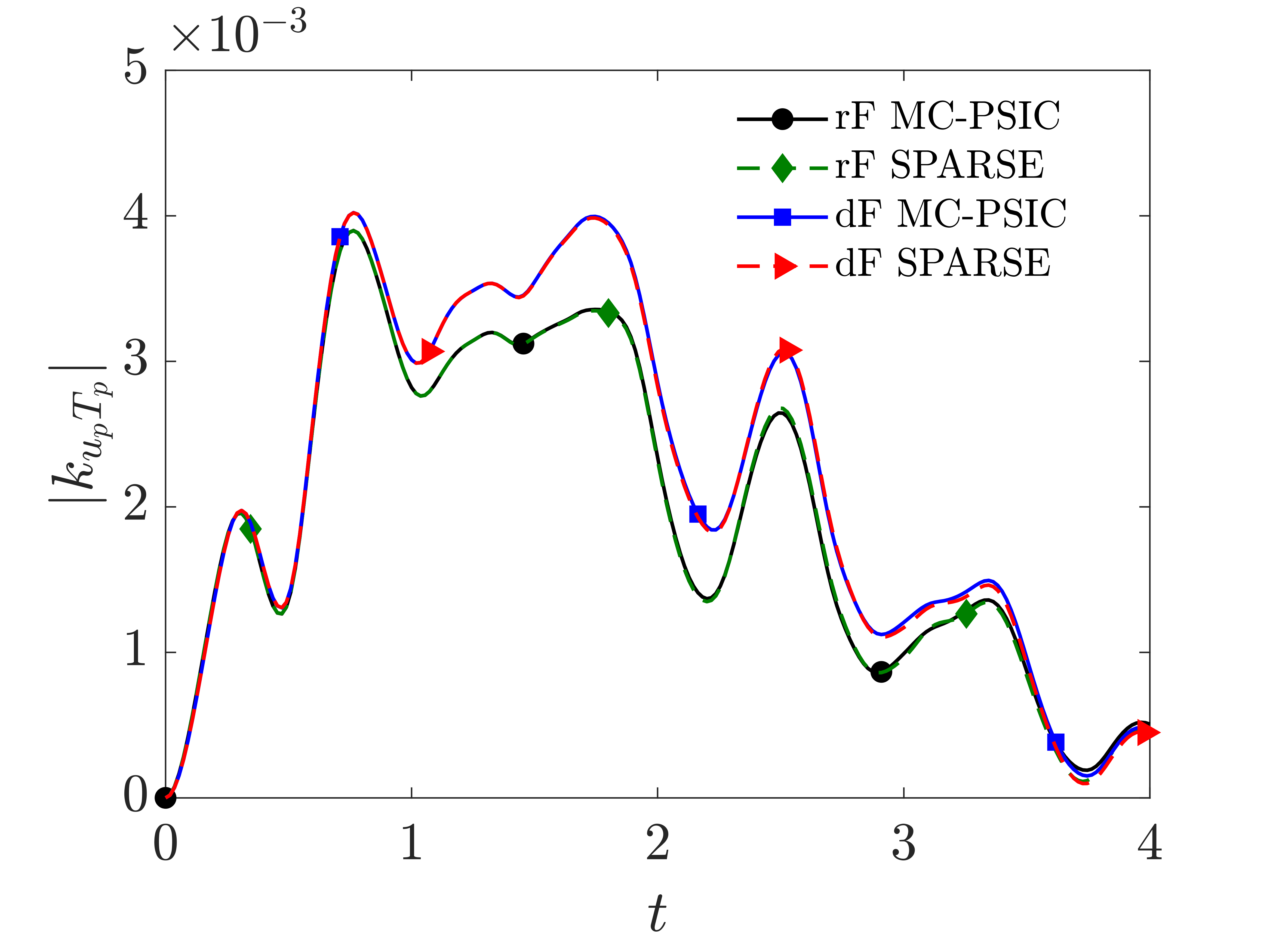}} 
		\hfill
	\subfloat[]{
		\label{fig: isoTurb_tend4_IC0_sigma_Tp}
		\includegraphics[width=0.32\textwidth]{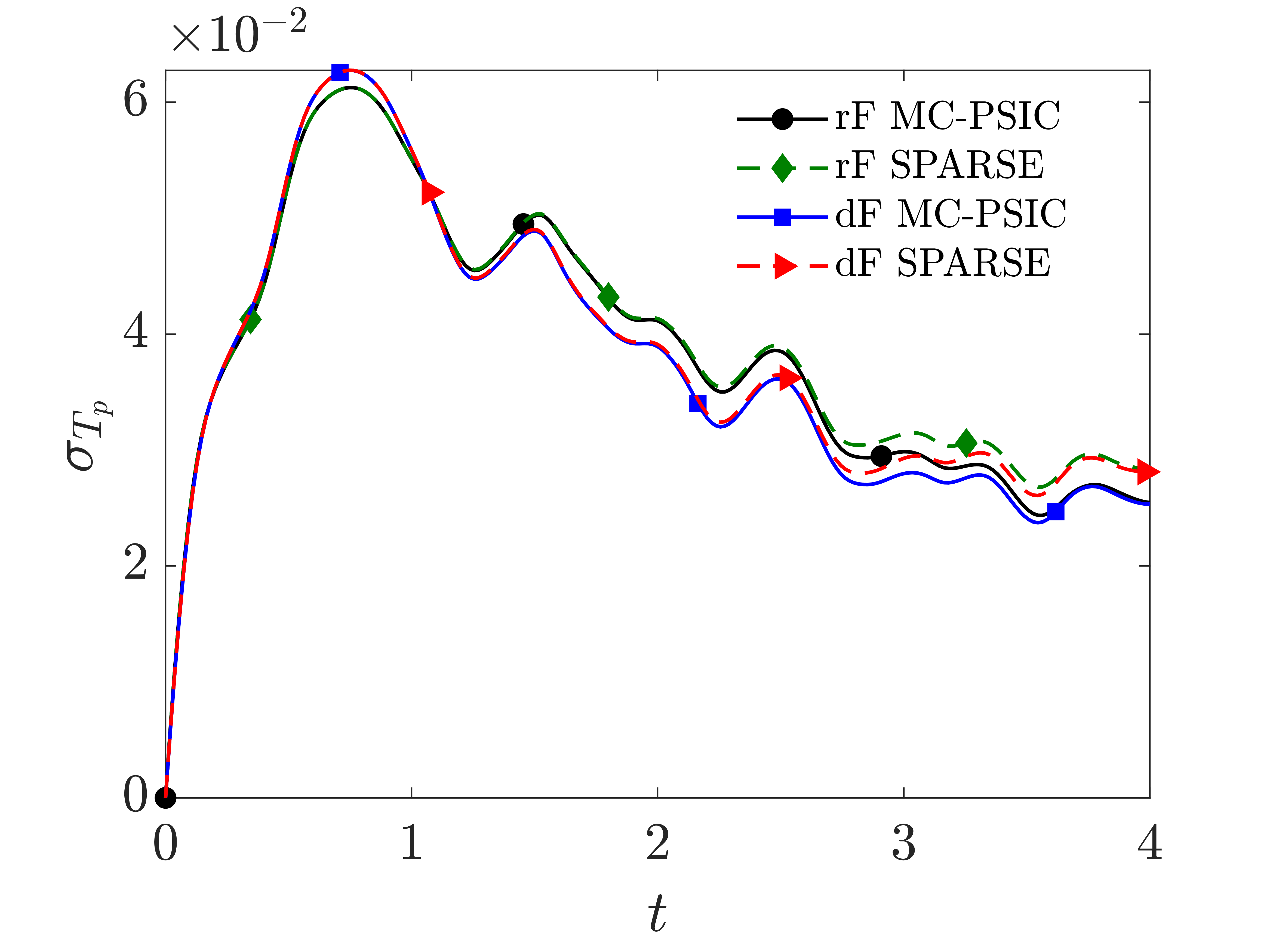}} \\		
	\subfloat[]{
		\label{fig: isoTurb_tend4_IC0_modK_xpalpha}
		\includegraphics[width=0.32\textwidth]{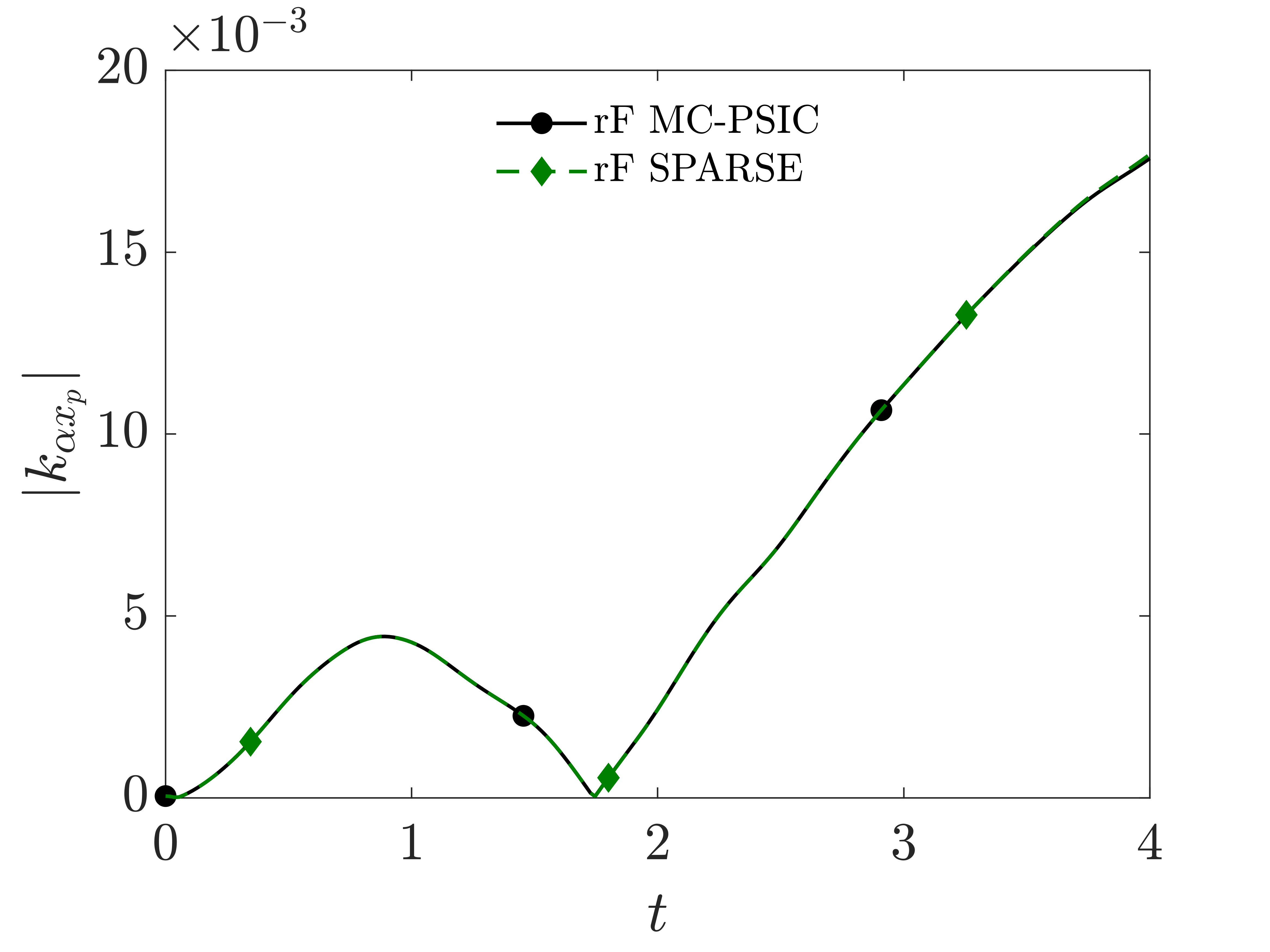}}
        \hfill	
	\subfloat[]{
		\label{fig: isoTurb_tend4_IC0_modK_upalpha}
		\includegraphics[width=0.32\textwidth]{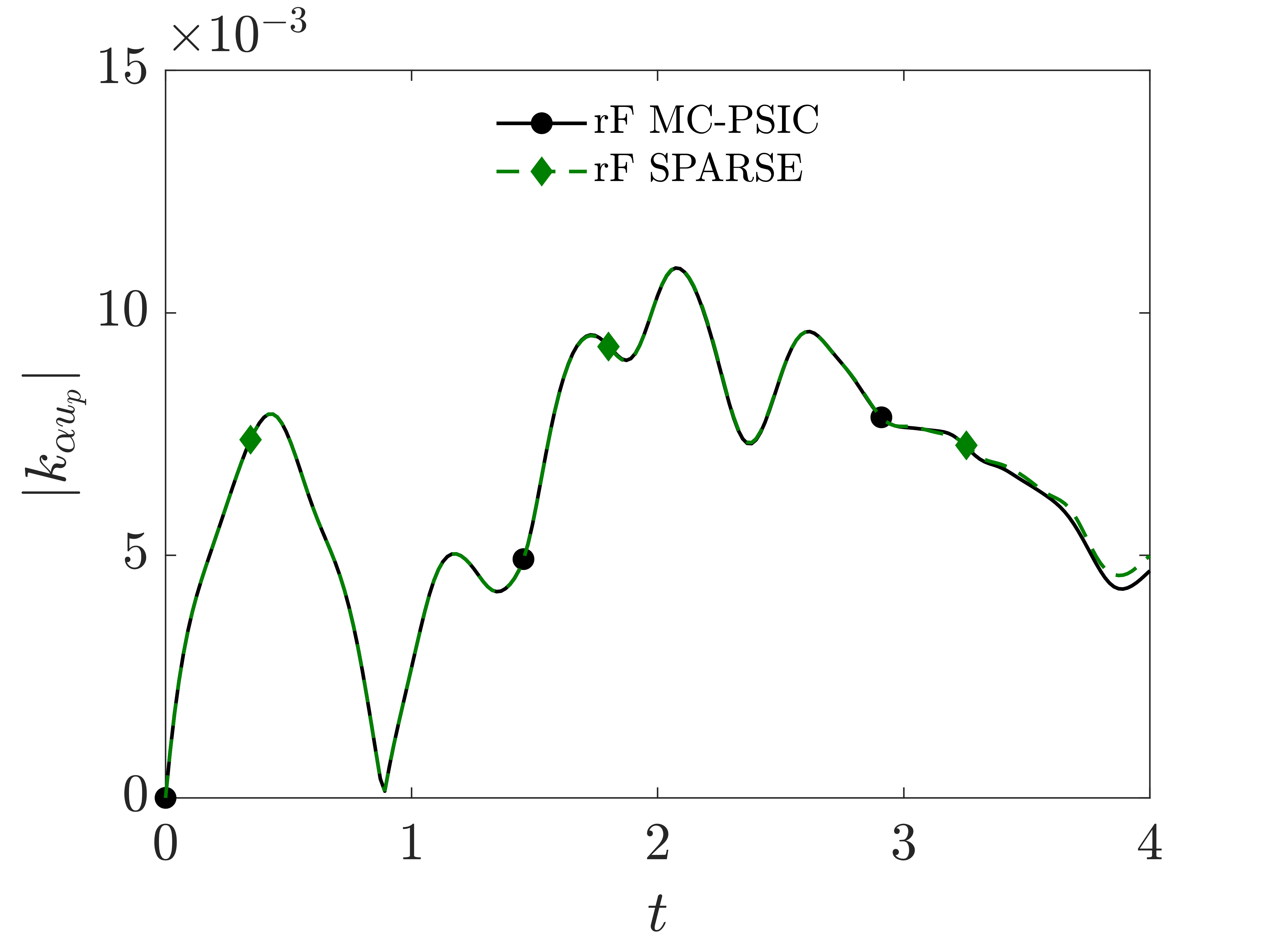}} 
		\hfill
	\subfloat[]{
		\label{fig: isoTurb_tend4_IC0_Tpalpha}
		\includegraphics[width=0.32\textwidth]{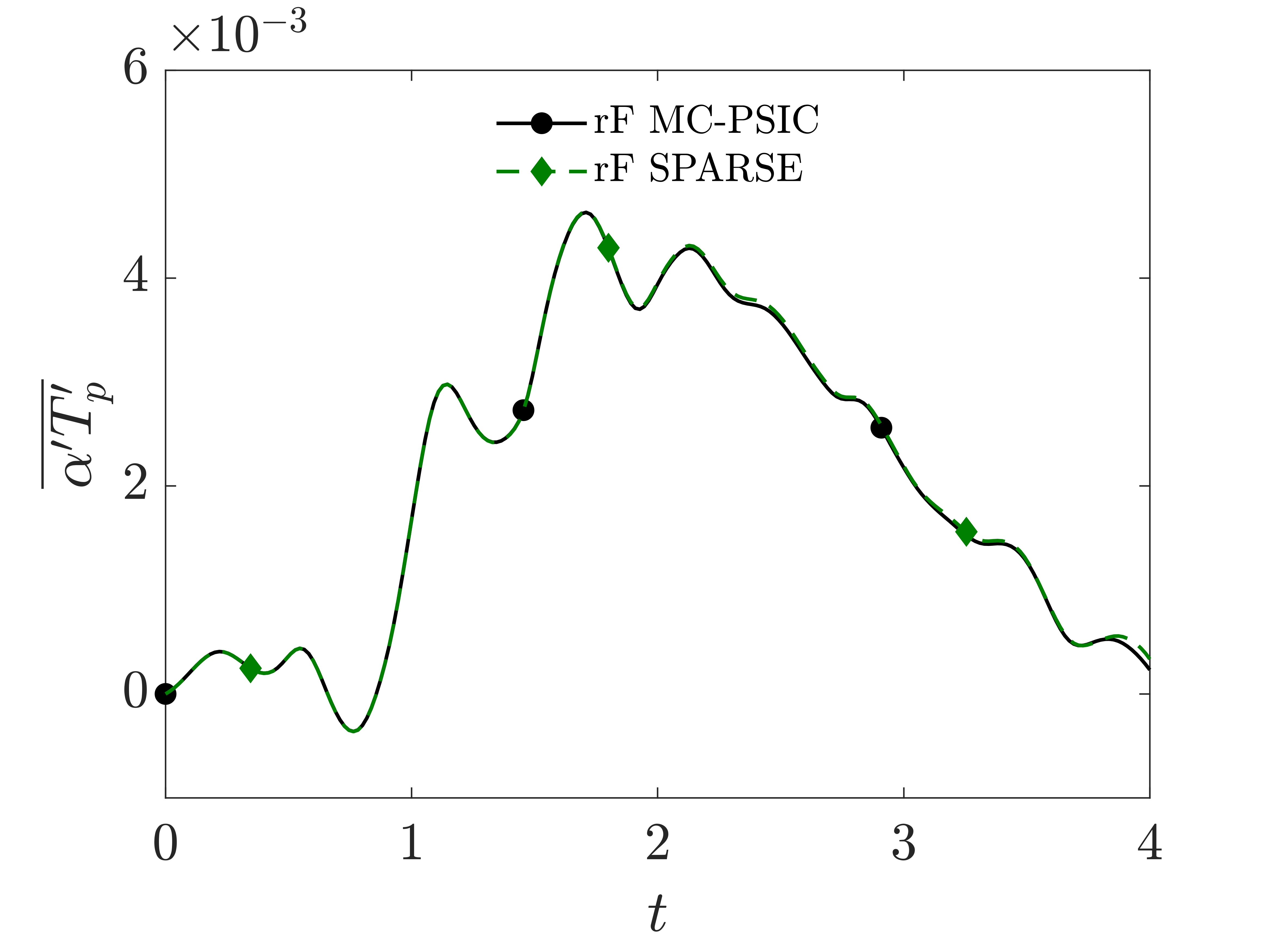}}
	\caption[]{Statistics of the rF and dF inertial particle clouds in the three-dimensional decaying isotropic turbulence case computed with MC-PSIC and SPARSE model; (a)--(c) average values of the first moments of the particle phase, (d)--(i) average values of second moments of the particle phase and (j)--(l) average values of second moments combining the particle phase and random coefficient $\alpha$.}
	\label{fig: isoTurb_moments}
\end{figure}

\begin{figure}[h!]
		\centering
		\includegraphics[width=0.4\textwidth]{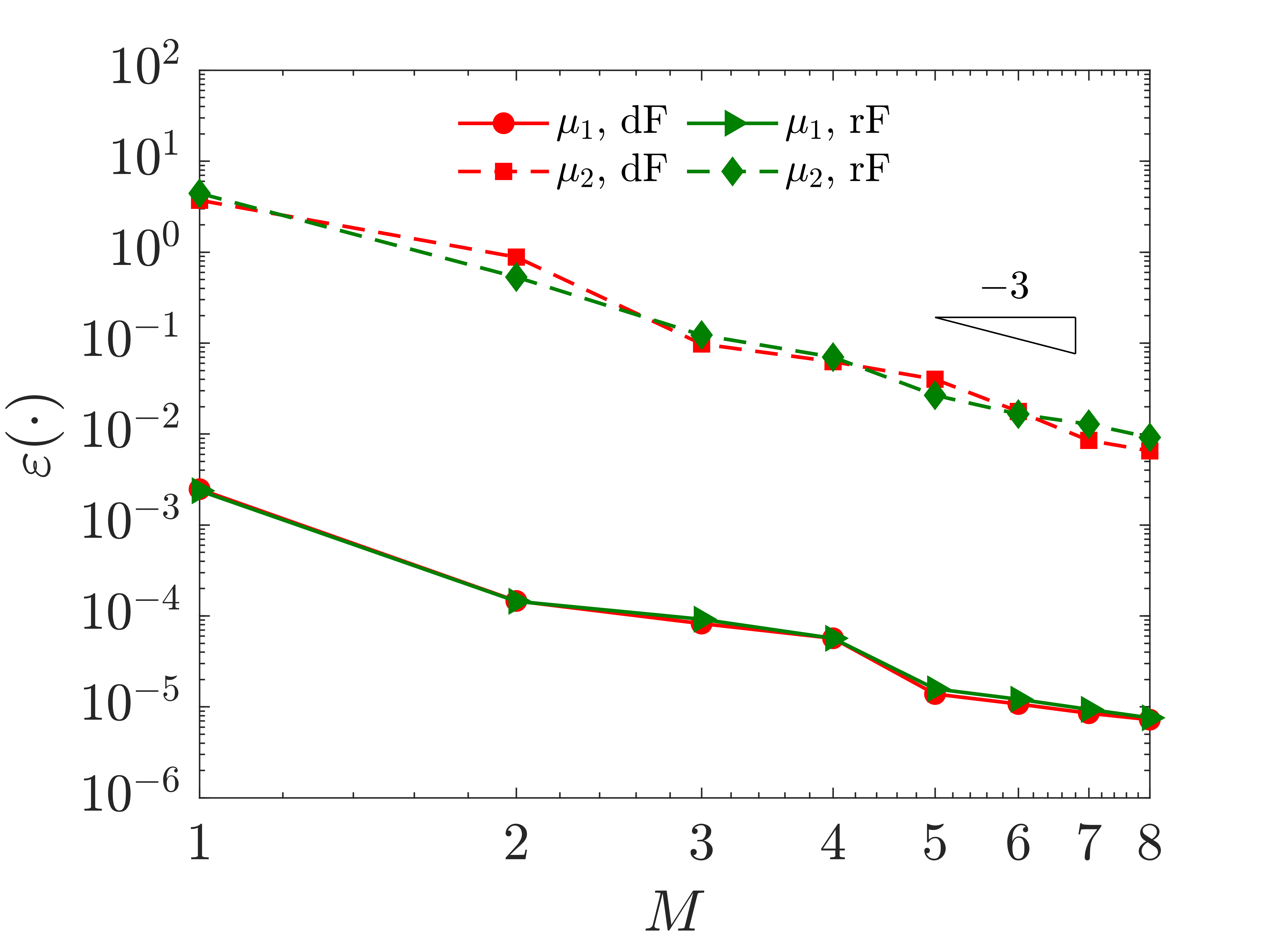} 
	\caption[]{Convergence of the errors of the SPARSE method as compared to the MC-PSIC method when using splitting for the isotropic turbulence test case.}
	\label{fig: isoTurb_errors_mu1mu2}
\end{figure}

In Figure~\ref{fig: isoTurb_moments}, we show the results of first and second moments for the rF and dF where the level of splitting corresponds to the maximum division in the convergence study $M=8$ (see Figure~\ref{fig: isoTurb_errors_mu1mu2}).
In particular, to concisely show the agreement between the MC-PSIC and the SPARSE results we group moments into vectors and matrices and compute their modulus and determinant, respectively.
The first moments of  the average location and velocity are entries  in the vectors  with the following moduli 
\begin{align}
\begin{split}
    |\overline{\boldsymbol{x}}_p| = \left(\overline{x}_p^2+\overline{y}_p^2+\overline{z}_p^2 \right)^{1/2}, \ \ \ \ \ \
    |\overline{\boldsymbol{u}}_p| = \left(\overline{u}_p^2+\overline{v}_p^2+\overline{w}_p^2 \right)^{1/2},
    \label{eq: isoTurb_mean_xp_and_up} 
\end{split}
\end{align}
which are plotted in Figures~\ref{fig: isoTurb_tend4_IC0_mean_xpypzp} and~\ref{fig: isoTurb_tend4_IC0_mean_upvpwp}.
The average temperature $\overline{T}_p$ is depicted in Figure~\ref{fig: isoTurb_tend4_IC0_mean_Tp}.
For the second moments of the particle phase, we define the following matrices 
\begin{align}
        K_{x_p} &= 
\begin{bmatrix}
    \overline{{x_p^\prime}^2} & \overline{{x_p^\prime}{y_p^\prime}} & \overline{{x_p^\prime}{z_p^\prime}}  \\
    \overline{{x_p^\prime}{y_p^\prime}} & \overline{{y_p^\prime}^2} & \overline{{y_p^\prime}{z_p^\prime}}  \\
    \overline{{x_p^\prime}{z_p^\prime}} & \overline{{y_p^\prime}{z_p^\prime}} & \overline{{z_p^\prime}^2}
\end{bmatrix}, \ \ \ 
    K_{x_p u_p} = 
\begin{bmatrix}
    \overline{{x_p^\prime}{u_p^\prime}} & \overline{{x_p^\prime}{v_p^\prime}} & \overline{{x_p^\prime}{w_p^\prime}}  \\
    \overline{{y_p^\prime}{u_p^\prime}} & \overline{{y_p^\prime}{v_p^\prime}} & \overline{{y_p^\prime}{w_p^\prime}}  \\
    \overline{{z_p^\prime}{u_p^\prime}} & \overline{{z_p^\prime}{v_p^\prime}} & \overline{{z_p^\prime}{w_p^\prime}}
\end{bmatrix}, \ \ \ 
    K_{u_p} = 
\begin{bmatrix}
    \overline{{u_p^\prime}^2} & \overline{{u_p^\prime}{v_p^\prime}} & \overline{{u_p^\prime}{z_p^\prime}}  \\
    \overline{{u_p^\prime}{v_p^\prime}} & \overline{{v_p^\prime}^2} & \overline{{v_p^\prime}{w_p^\prime}}  \\
    \overline{{u_p^\prime}{w_p^\prime}} & \overline{{v_p^\prime}{w_p^\prime}} & \overline{{w_p^\prime}^2}
\end{bmatrix},  \label{eq: isoTurb_Ks} 
\end{align}
whose determinants are depicted in Figures~\ref{fig: isoTurb_tend4_IC0_detK_xp},~\ref{fig: isoTurb_tend4_IC0_detK_xpup} and~\ref{fig: isoTurb_tend4_IC0_detK_up}. 
The following vectors are defined based on  correlations between particle locations, velocities, temperatures and the random coefficient $\alpha$ 
\begin{align}
k_{x_p T_p} = 
\begin{bmatrix}
    \overline{{x_p^\prime} T_p^\prime}   \\
    \overline{{y_p^\prime} T_p^\prime}  \\
    \overline{{z_p^\prime} T_p^\prime}
\end{bmatrix}, \ \ \ 
k_{u_p T_p} = 
\begin{bmatrix}
    \overline{{u_p^\prime} T_p^\prime}   \\
    \overline{{v_p^\prime} T_p^\prime}  \\
    \overline{{w_p^\prime} T_p^\prime}
\end{bmatrix}, \ \ \ 
k_{\alpha x_p} = 
\begin{bmatrix}
    \overline{\alpha^\prime {x_p^\prime}}  \\
    \overline{\alpha^\prime {y_p^\prime}}  \\
    \overline{\alpha^\prime {z_p^\prime}}
\end{bmatrix}, \ \ \ 
k_{\alpha u_p} = 
\begin{bmatrix}
    \overline{\alpha^\prime {u_p^\prime}}  \\
    \overline{\alpha^\prime {v_p^\prime}}  \\
    \overline{\alpha^\prime {w_p^\prime}}
\end{bmatrix}.\label{eq: isoTurb_ks} 
\end{align}
The moduli are shown in Figures~\ref{fig: isoTurb_tend4_IC0_modK_xpTp},~\ref{fig: isoTurb_tend4_IC0_modK_upTp}, ~\ref{fig: isoTurb_tend4_IC0_modK_xpalpha}, and~\ref{fig: isoTurb_tend4_IC0_modK_upalpha}.
Finally, the standard deviation of the particle temperature $\sigma_{T_p}$ and second moment of the random coefficient and temperature $\overline{\alpha^\prime T_p^\prime}$ are presented in Figures~\ref{fig: isoTurb_tend4_IC0_sigma_Tp} and~\ref{fig: isoTurb_tend4_IC0_Tpalpha}.
In this manner, the twelve scalars plotted in Figure~\ref{fig: isoTurb_moments} combine all $42$ moments of the particle phase (including the moments that relate to the random coefficient $\alpha$) and it so admits a  comparison of MC-PSIC and SPARSE results for the complete statistical description of the particle clouds.
Note that for the dF case $k_{\alpha x_p}$, $k_{\alpha u_p}$ and $\overline{\alpha^\prime T_p^\prime}$ are not shown because $\alpha^\prime=0$.


The first moments of the particle cloud (Figures~\ref{fig: isoTurb_tend4_IC0_mean_xpypzp}--\ref{fig: isoTurb_tend4_IC0_mean_Tp}) show that
the the mean cloud motion for the rF and dF case are in close agreement within 3\%.
For an analogous reason that was used to explain that stagnation flow solutions are identical for the dF and the rF test case in the symmetrical vertical direction, the first moments solutions for dF and rF are similar because of the near-isotropy in this turbulent flow also. Note while the carrier-flow is isotropic in the periodic box, the cloud is initialized in only a portion of it, and the carrier-phase in this portion is not precisely isotropic. This means that the uncertainty in the forcing can and does propagate in the dispersed solution, as witnessed by the small deference in dF and rF. 
Specifically, fast and slow responding,  randomly forced particle trajectories   average such that the resulting two moments match those of the dF solution with the average time response of the rF case.
The velocity average (Figure~\ref{fig: isoTurb_tend4_IC0_mean_upvpwp}) reaches a maximum in an initial acceleration of the cloud to reach the decreasing carrier phase velocity field.
After that, a decrease in velocity of the particle cloud follows the decay of turbulence in the flow and the inertia in the particle phase somewhat smoothens the response  as compared with fluid tracers (that move at the flow velocity).
The average temperature of the cloud oscillates around the initial and average value during the entire simulation (unity).
The maximum relative error in the first moments of the particle phase is $0.9\%$ and $0.7\%$ for the average modulus of the particle velocity $|\overline{\boldsymbol{u}}_p|$ for the rF and dF cases respectively.


The second moment trends are shown in Figures~\ref{fig: isoTurb_tend4_IC0_detK_xp}--\ref{fig: isoTurb_tend4_IC0_sigma_Tp}.
Figure~\ref{fig: isoTurb_tend4_IC0_detK_xp} plots the determinant, $|K_{x_p}|$, which is equal to product of the eigenvalues of $K_{x_p}$. This metric  can thus be interpreted a measure of the    spatial spread of the particle cloud because it is equal to  the  product of principle strains (eigenvalues)  of the cloud. It shows that the turbulent field mixes (or diffuses) the dispersed phase which leads to a increase in the  cloud's footprint in space. This spatial  growth is  similar for the dF and rF case because of the near isotropic flow conditions. At later times the cloud occupies a larger portion of the computational and thus the conditions are more isotropic then, and the dF and rF moments are mostly in closer agreement.
The  trends of the determinants in  Figures~\ref{fig: isoTurb_tend4_IC0_detK_up}--\ref{fig: isoTurb_tend4_IC0_sigma_Tp} for the remaining second moments exhibit a rather complex behavior. Its discussion is beyond the scope of the current article that aims to present and validate the rF point-cloud model. It can be concluded, however, that all moments show a very good comparison between rF SPARSE and MC-PSIC.
The maximum relative error  in  $|K_{x_pu_p}|$
for the rF and dF case are $4.5\%$ and $8\%$ respectively. Note that the accuracy for the governing equations of the moments of the point-cloud tracer in  $K_{x_pu_p}$ are directly dependent on truncated statistical moments of order greater than two.
In Figure~\ref{fig: isoTurb_errors_mu1mu2} we plot the convergence of the SPARSE errors with respect to the level of splitting $M$, confirming once again the theoretical third order convergence rate.

\begin{figure}[h!]
	\centering
	\subfloat[]{
		\label{fig: isoTurb_fxp_MC}
		\includegraphics[width=0.32\textwidth]{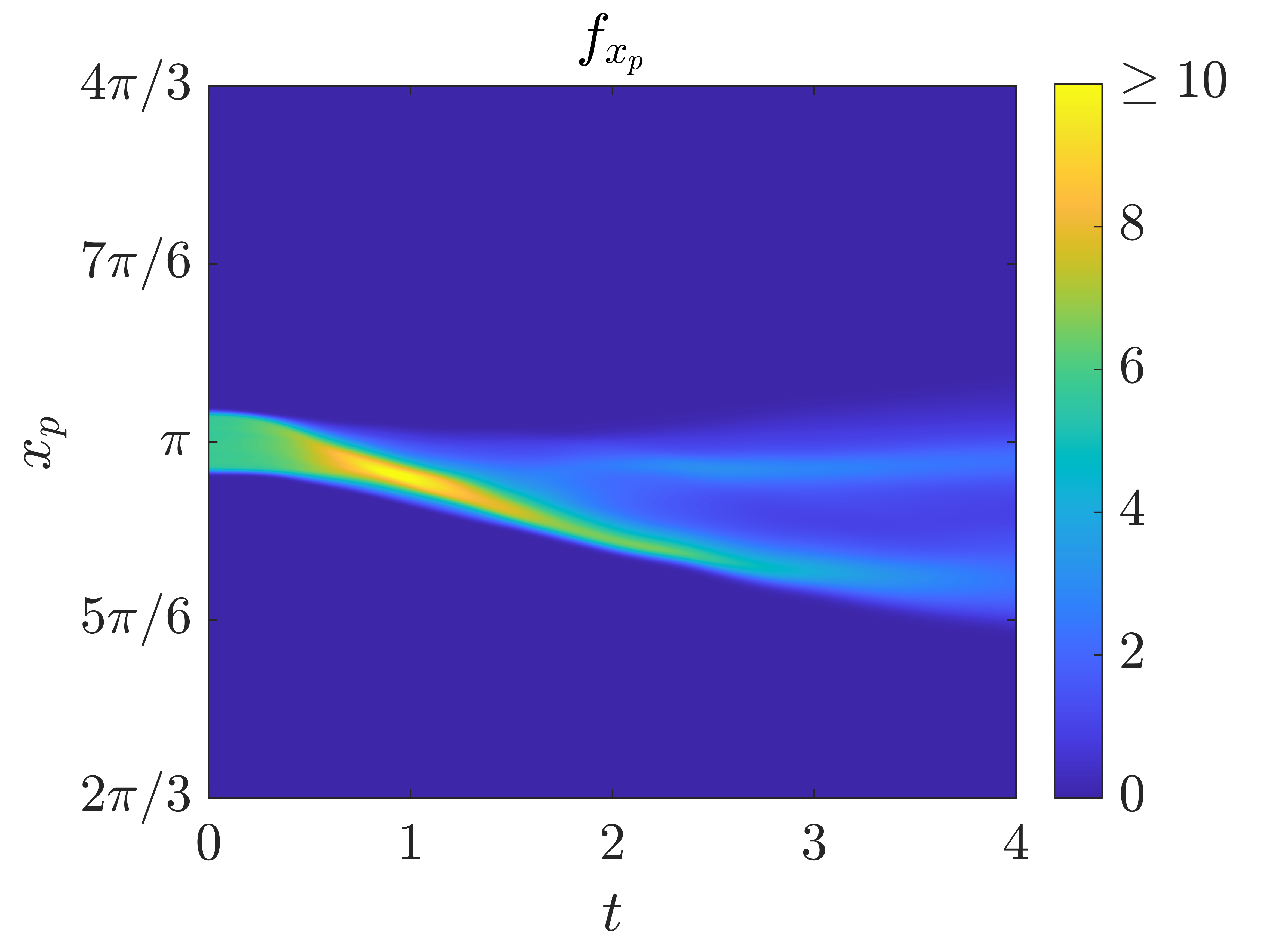}}
        \hfill	
	\subfloat[]{
		\label{fig: isoTurb_fxp_SPARSE}
		\includegraphics[width=0.32\textwidth]{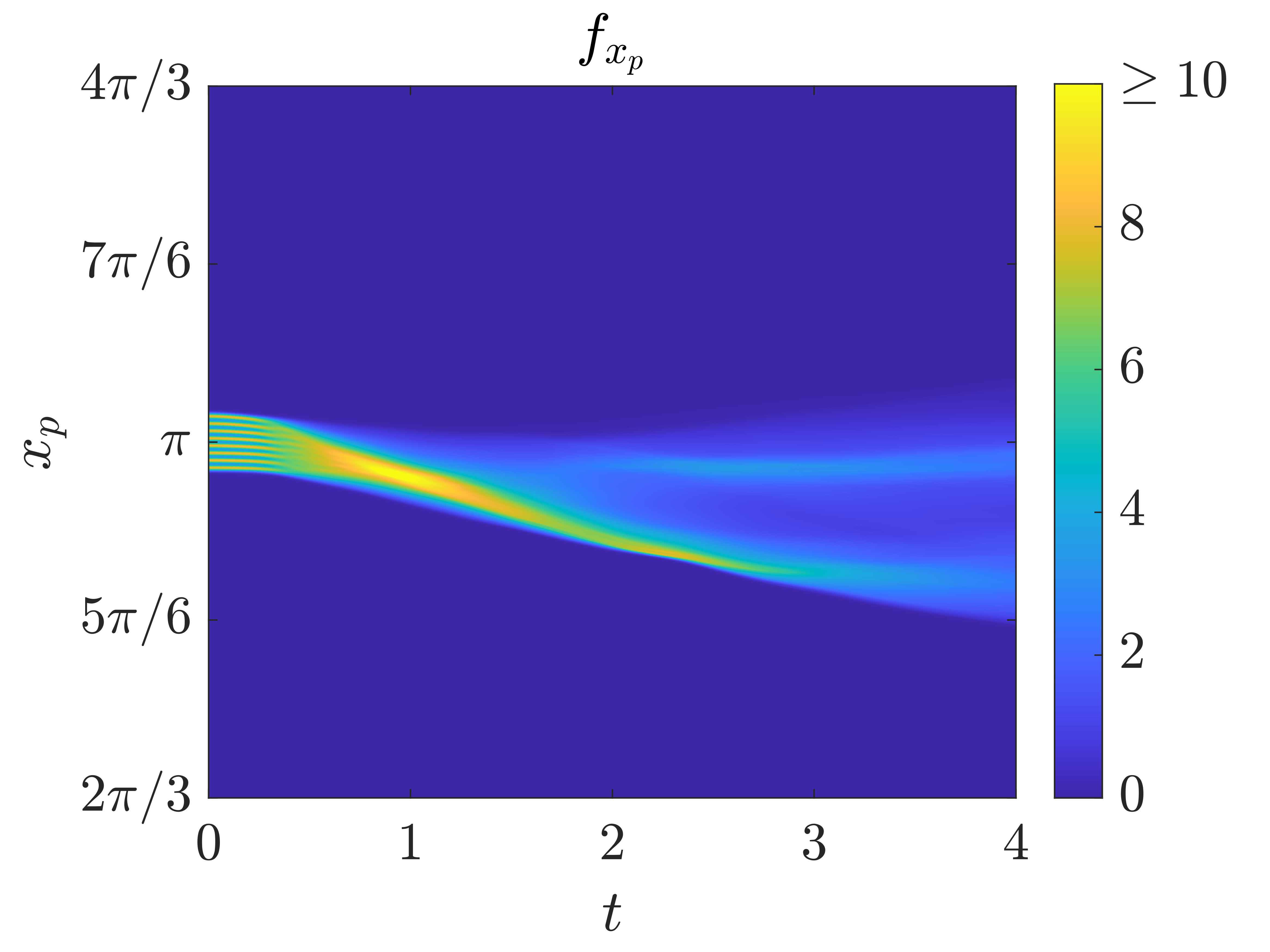}} 
		\hfill
	\subfloat[]{
		\label{fig: isoTurb_fxp_times}
		\includegraphics[width=0.32\textwidth]{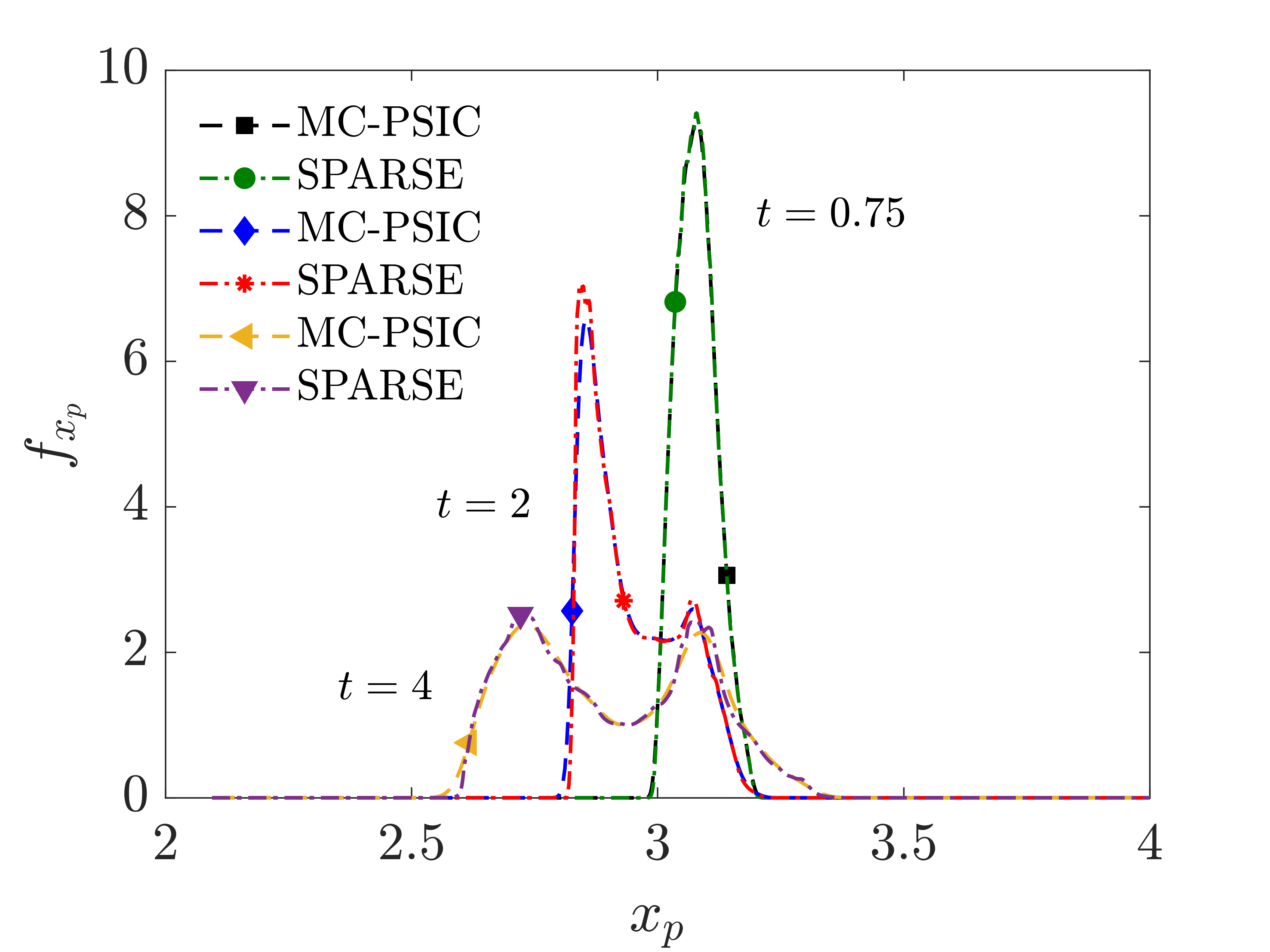}} \\
	\subfloat[]{
		\label{fig: isoTurb_fyp_MC}
		\includegraphics[width=0.32\textwidth]{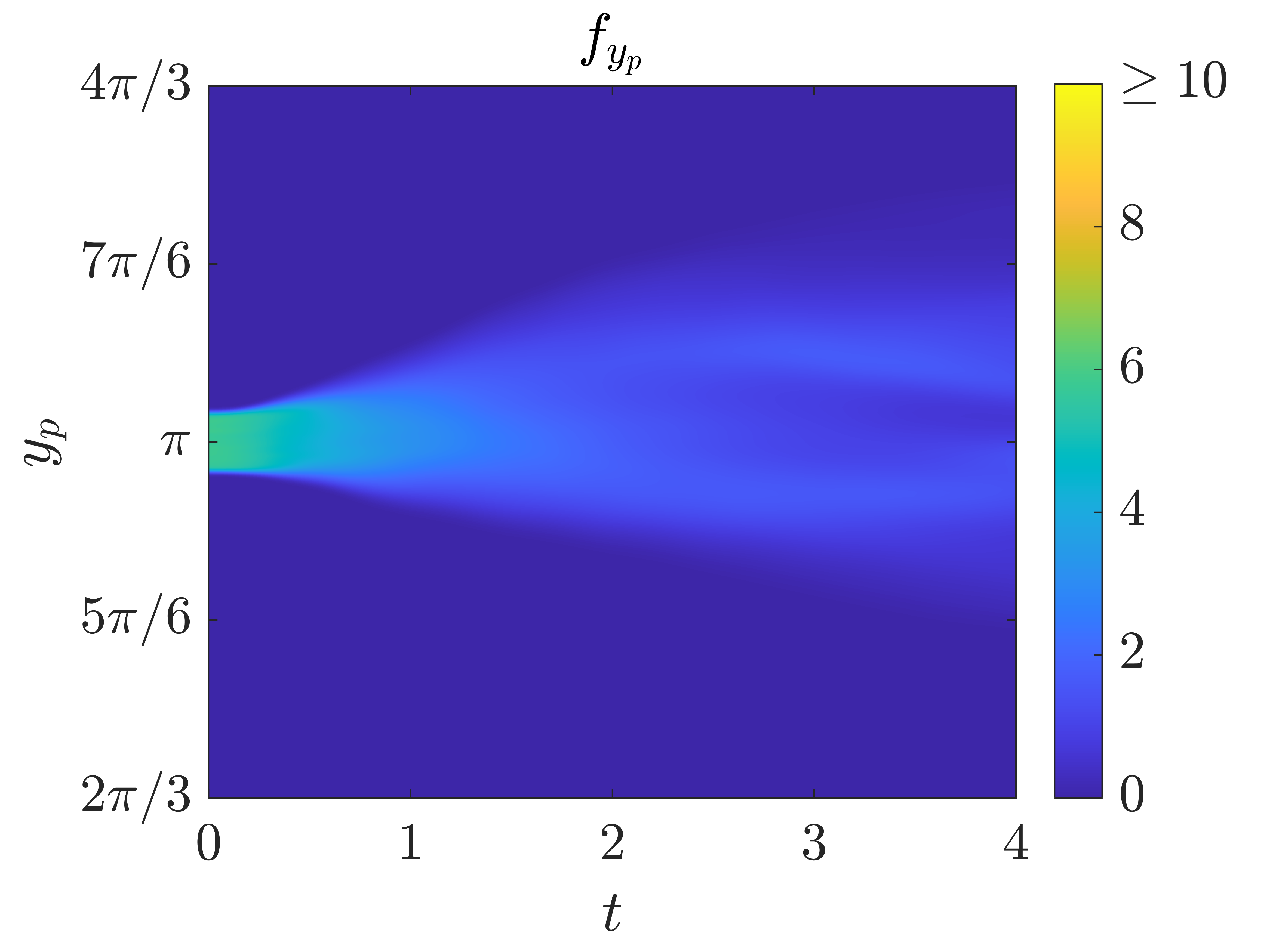}}
        \hfill	
	\subfloat[]{
		\label{fig: isoTurb_fyp_SPARSE}
		\includegraphics[width=0.32\textwidth]{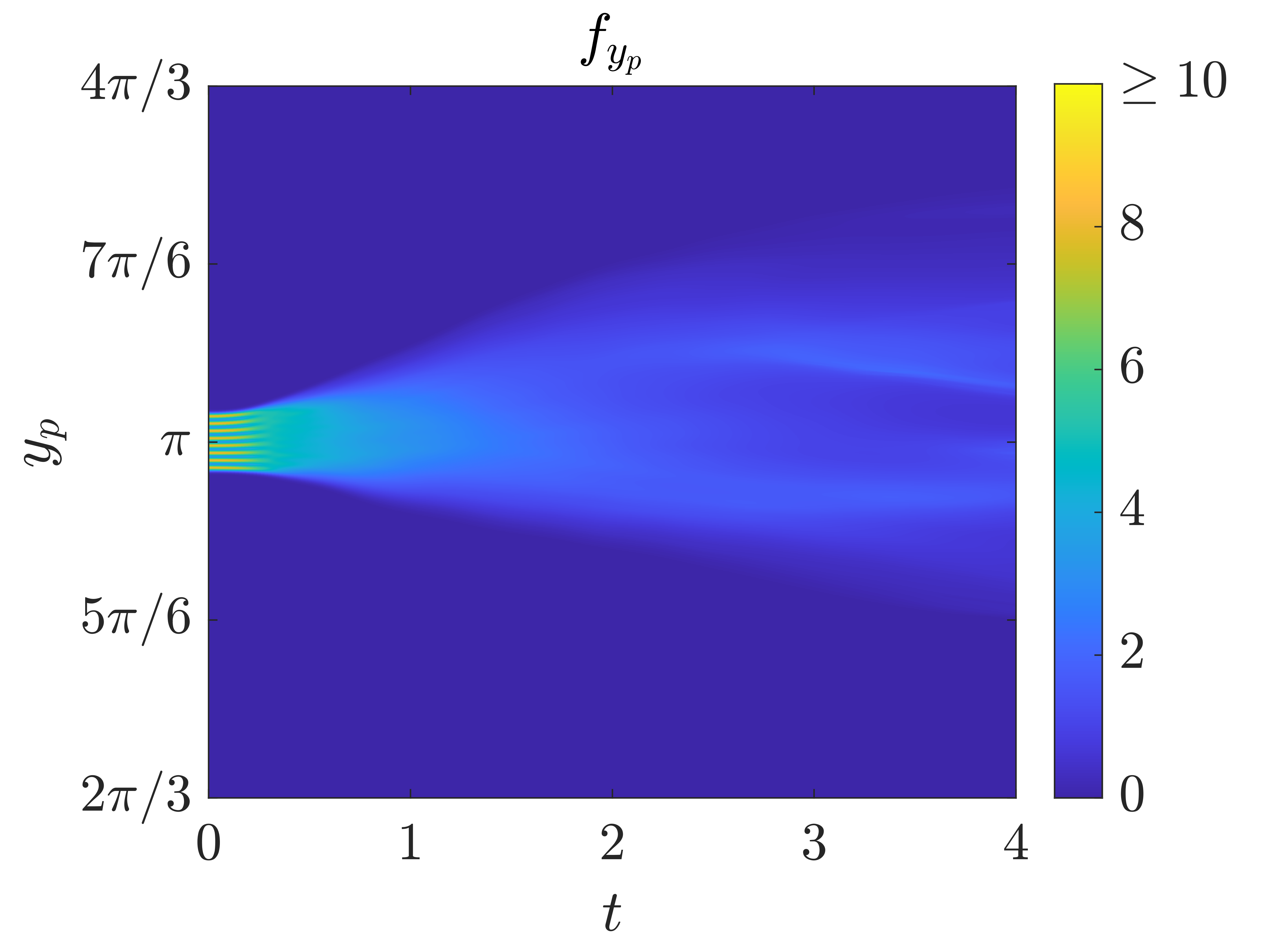}} 
		\hfill
	\subfloat[]{
		\label{fig: isoTurb_fyp_times}
		\includegraphics[width=0.32\textwidth]{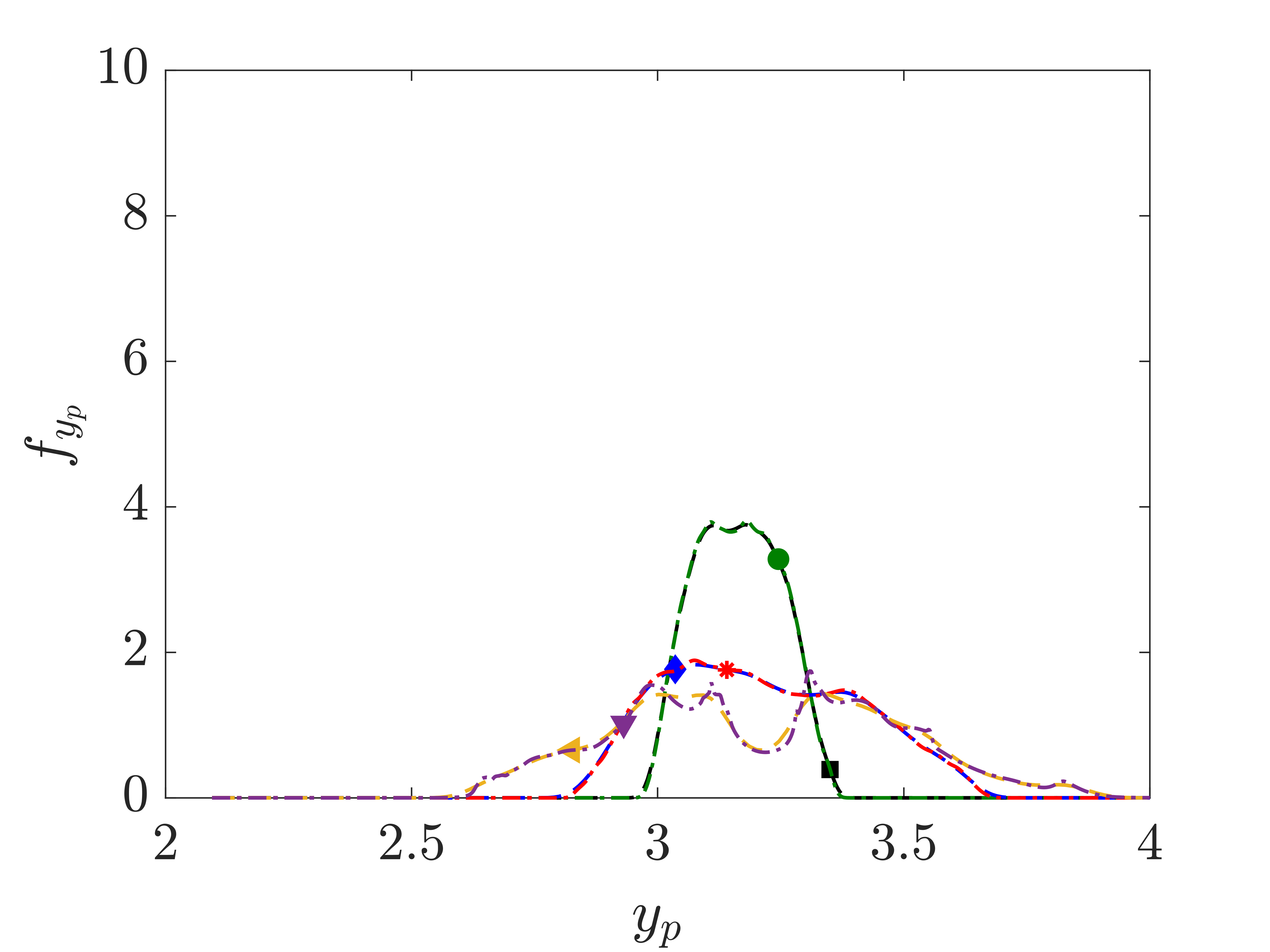}} \\
    \subfloat[]{
		\label{fig: isoTurb_fzp_MC}
		\includegraphics[width=0.32\textwidth]{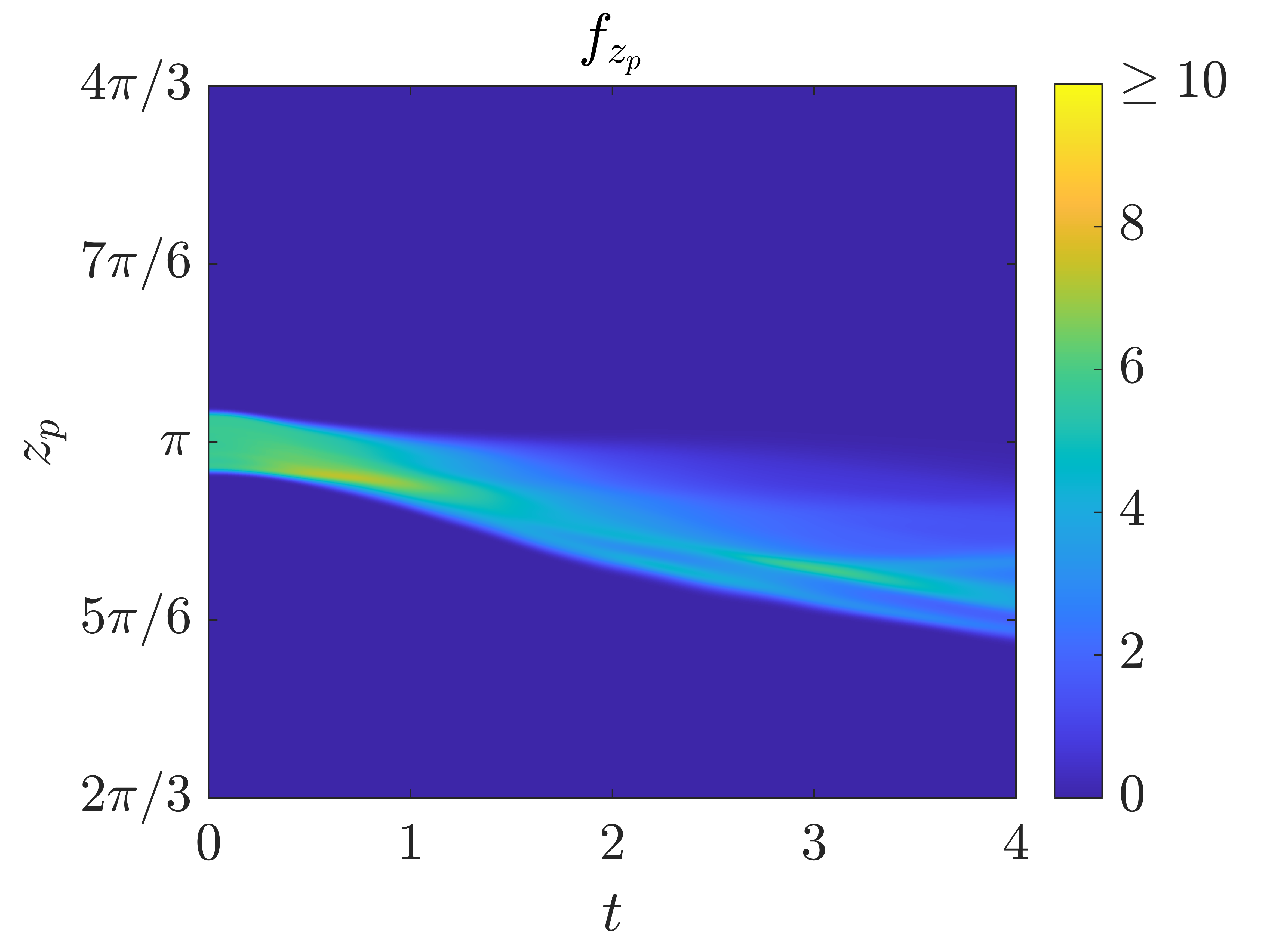}}
		\hfill
	\subfloat[]{
		\label{fig: isoTurb_fzp_SPARSE}
		\includegraphics[width=0.32\textwidth]{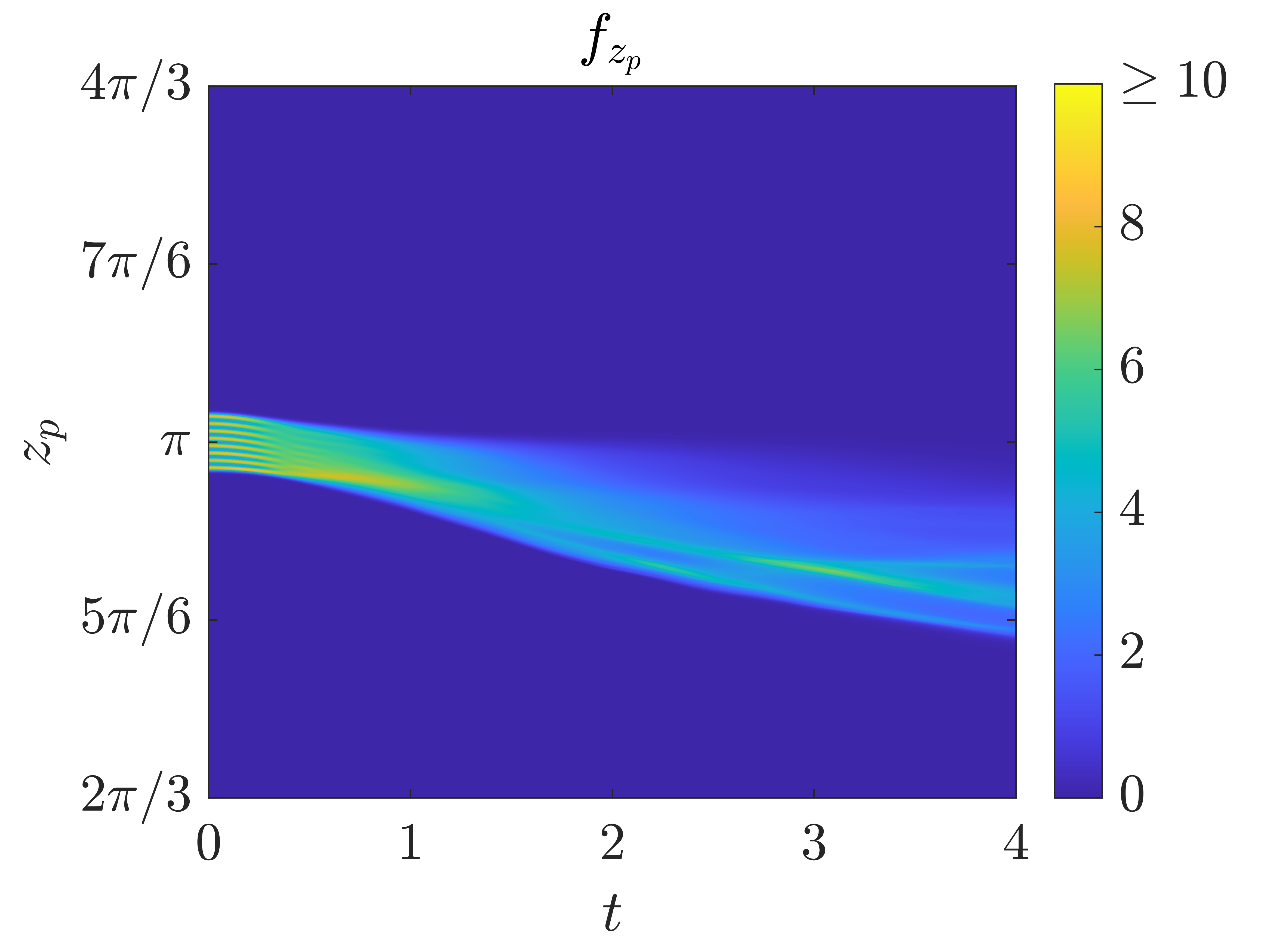}} 
		\hfill
	\subfloat[]{
		\label{fig: isoTurb_fzp_times}
		\includegraphics[width=0.32\textwidth]{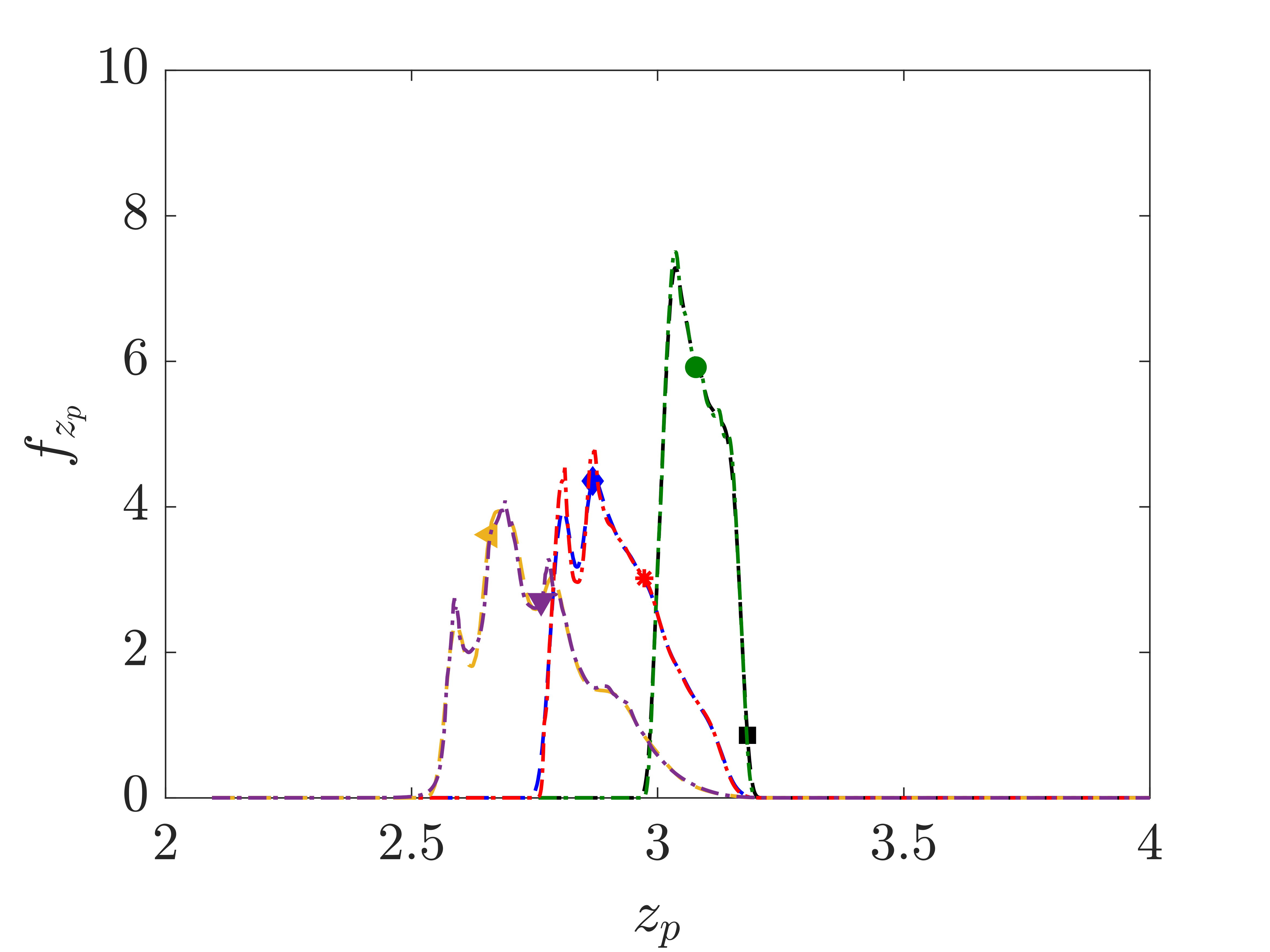}}
	\caption[]{PDFs of the components of the particle locations $f_{x_p}$, $f_{y_p}$ and $f_{z_p}$ for the rF case computed with MC-PSIC and SPARSE methods. The Figure shows: PDF of $x_p$ computed with (a) MC-PSIC and (b) SPARSE and (c) comparison for different times and the corresponding result for $y_p$ in (d)--(f) and for $z_p$ in (g)--(i). The legend in (c) also corresponds to (f) and (i) where the solution is plotted at times $t=[0.75, \ 2, \ 4]$ for both methods.}
	\label{fig: isoTurb_PDF_xpypzp}
\end{figure}

\begin{figure}[h!]
	\centering
	\subfloat[]{
		\label{fig: isoTurb_fup_MC}
		\includegraphics[width=0.32\textwidth]{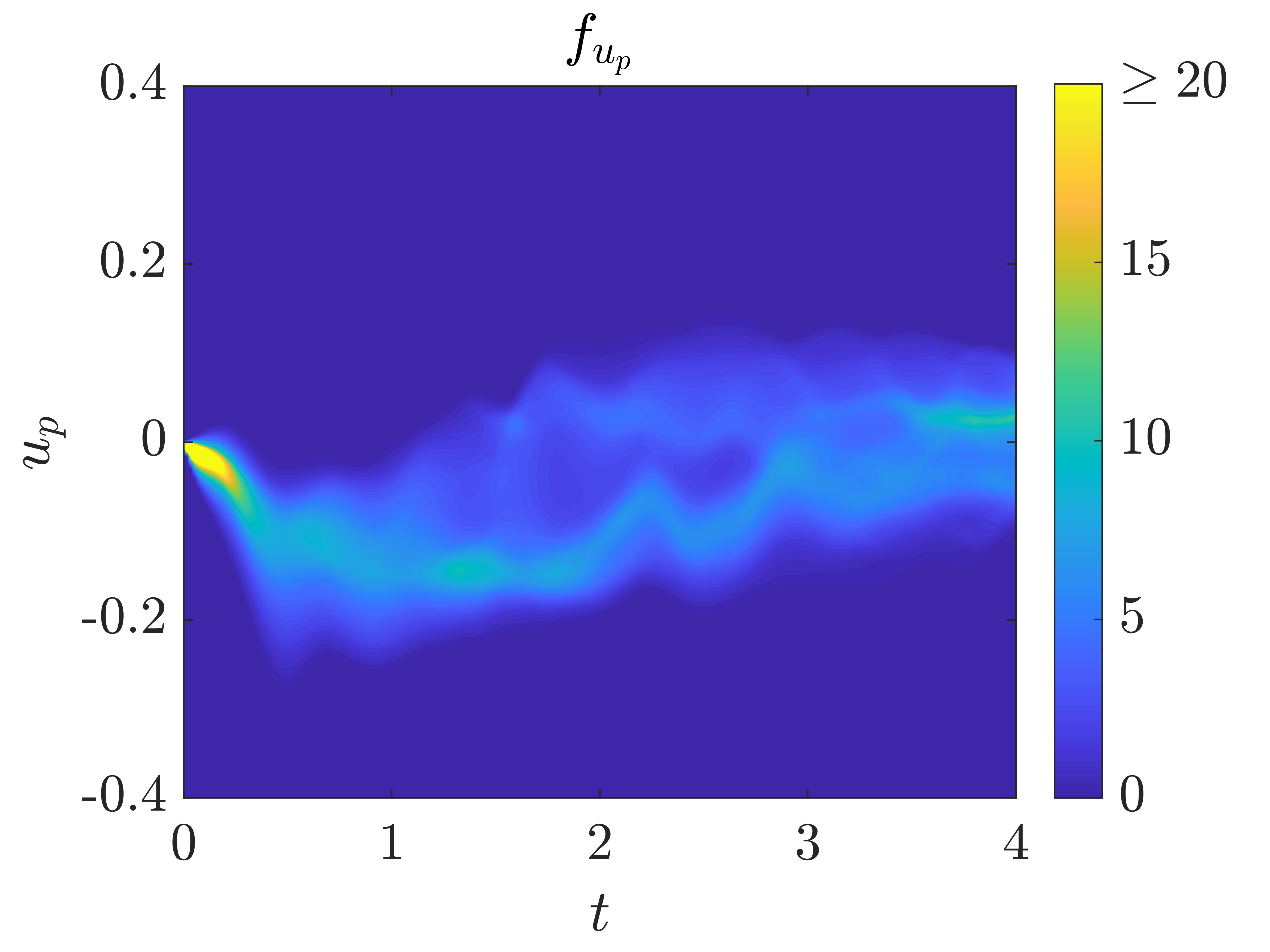}}
        \hfill	
	\subfloat[]{
		\label{fig: isoTurb_fup_SPARSE}
		\includegraphics[width=0.32\textwidth]{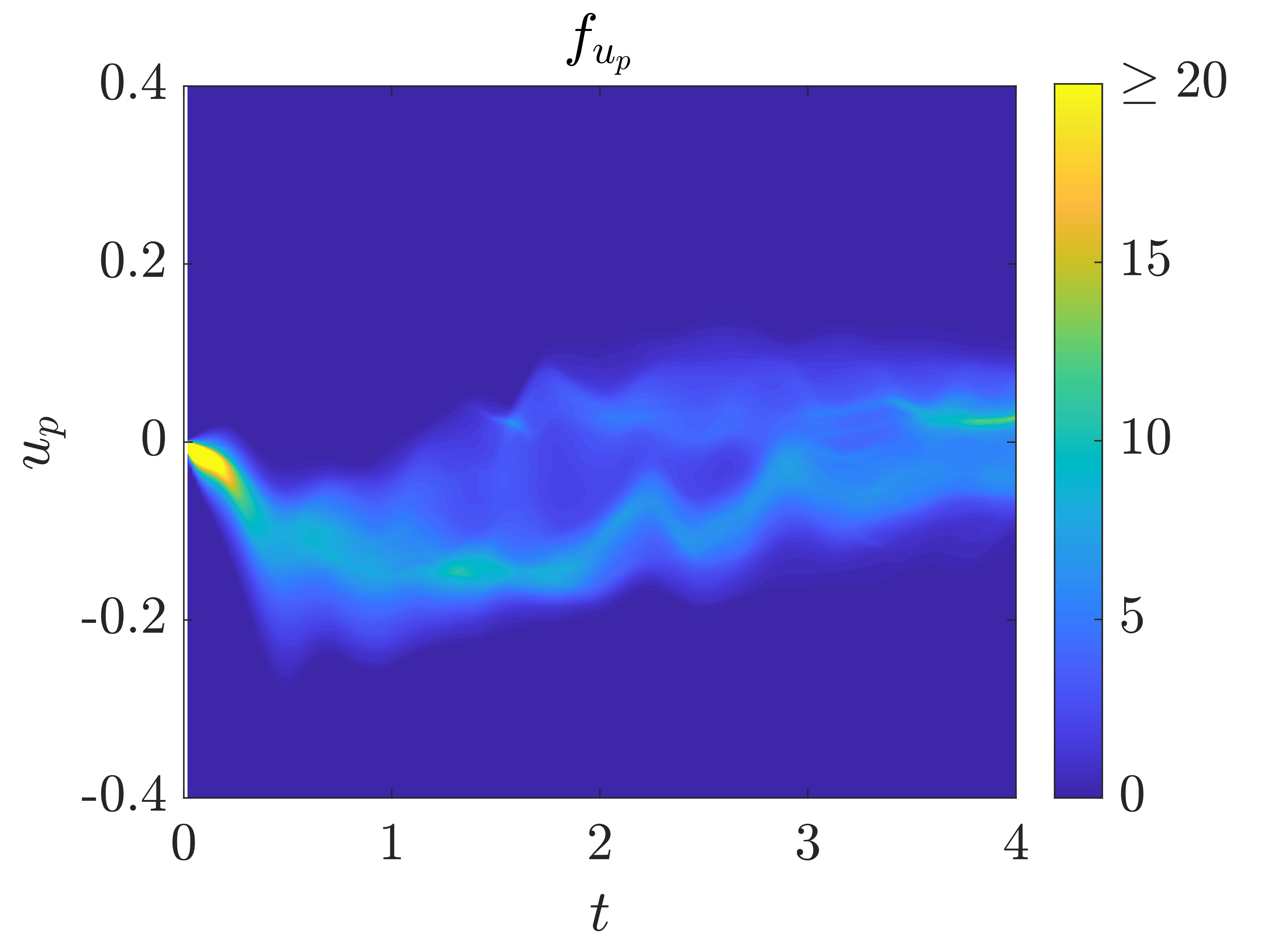}} 
		\hfill
	\subfloat[]{
		\label{fig: isoTurb_fup_times}
		\includegraphics[width=0.32\textwidth]{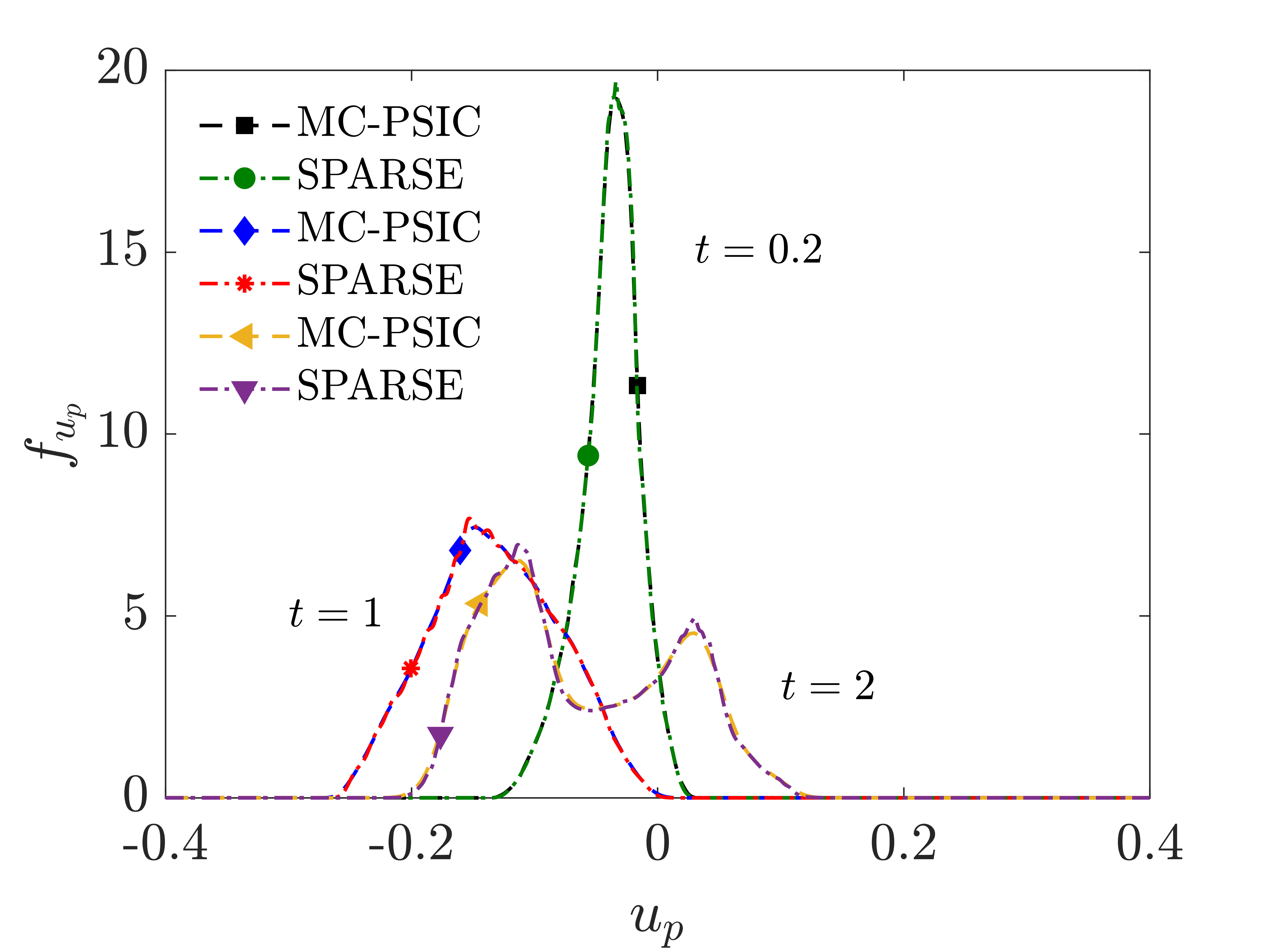}} \\
	\subfloat[]{
		\label{fig: isoTurb_fvp_MC}
		\includegraphics[width=0.32\textwidth]{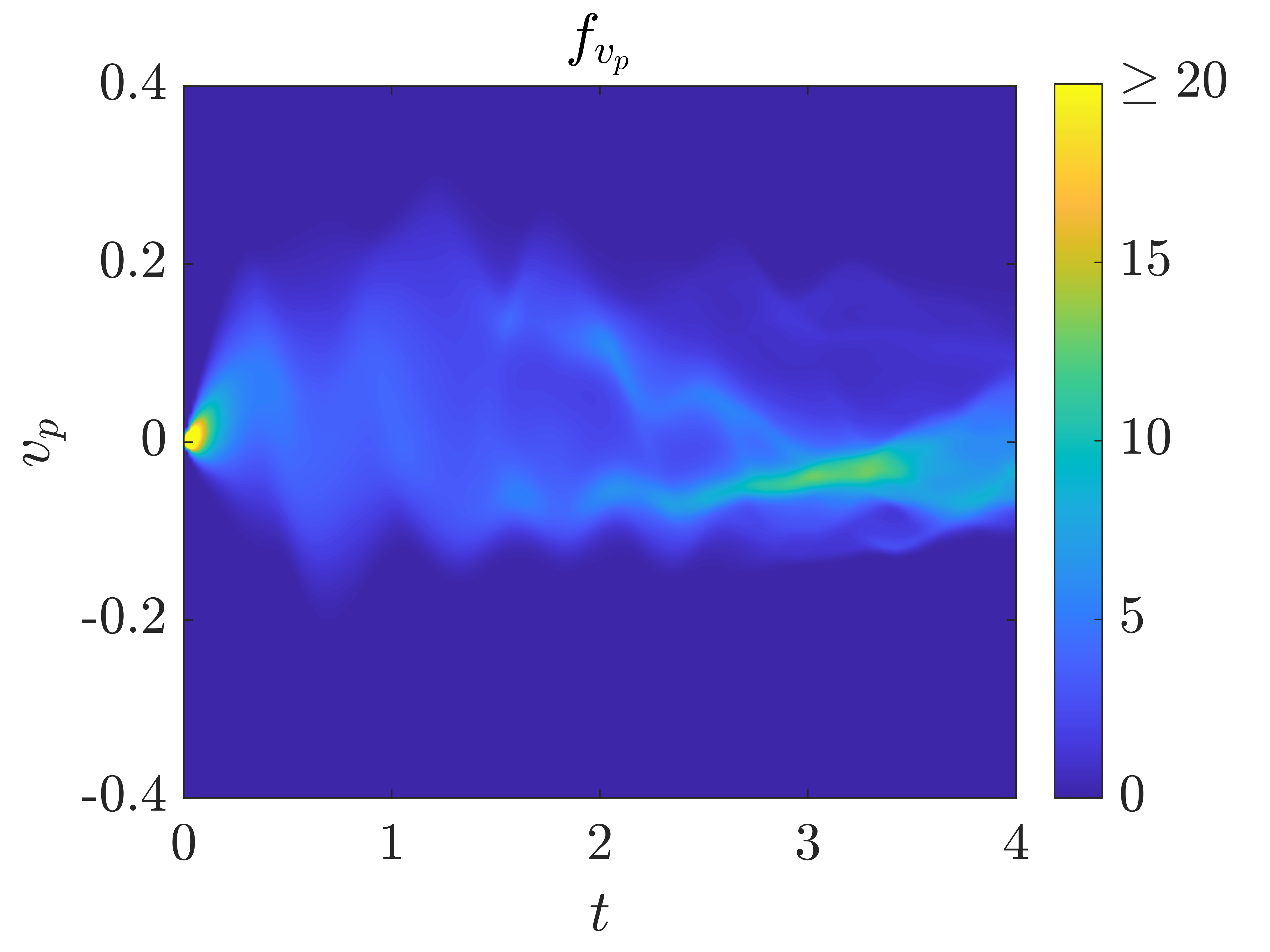}}
        \hfill	
	\subfloat[]{
		\label{fig: isoTurb_fvp_SPARSE}
		\includegraphics[width=0.32\textwidth]{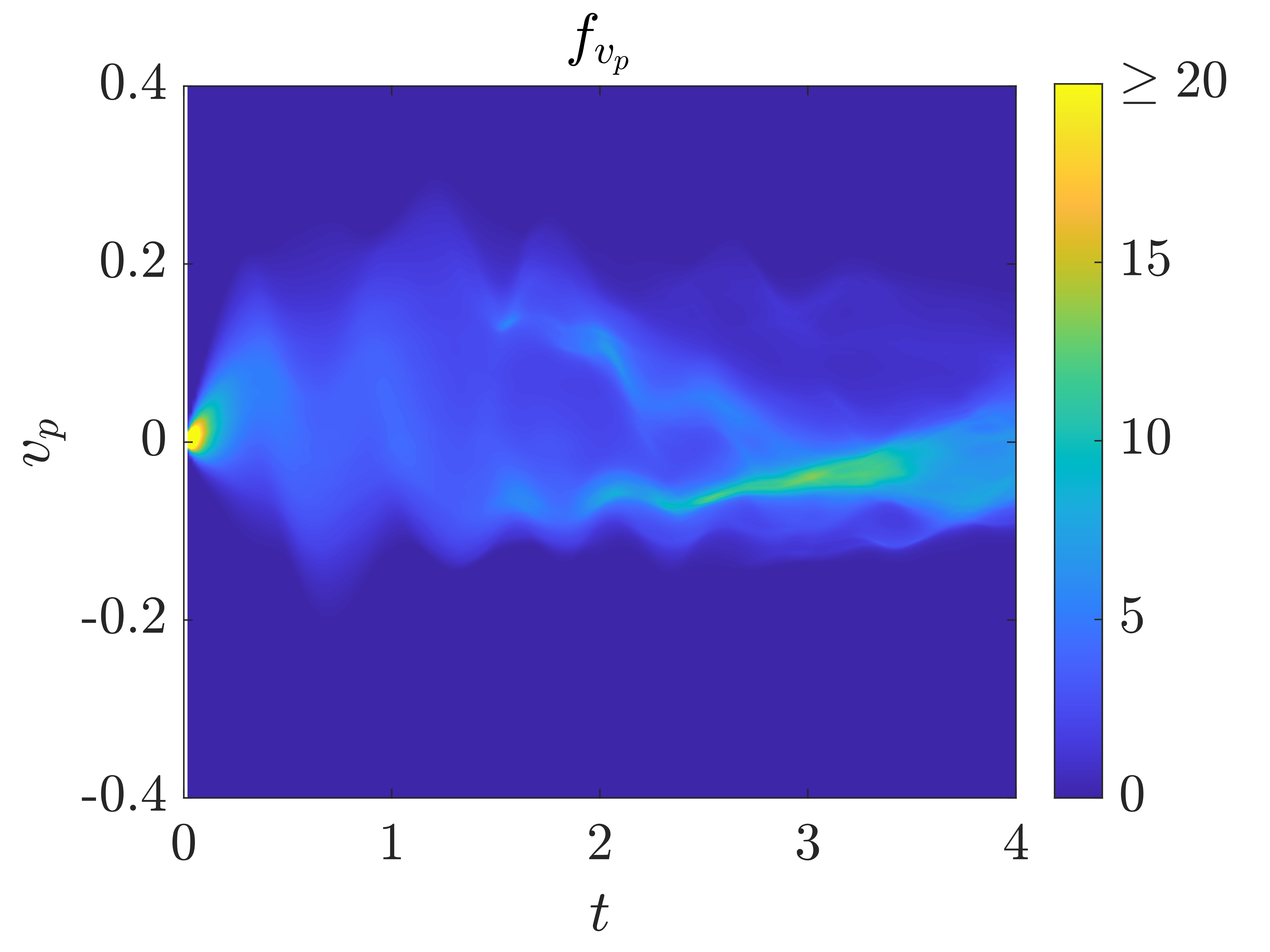}} 
		\hfill
	\subfloat[]{
		\label{fig: isoTurb_fvp_times}
		\includegraphics[width=0.32\textwidth]{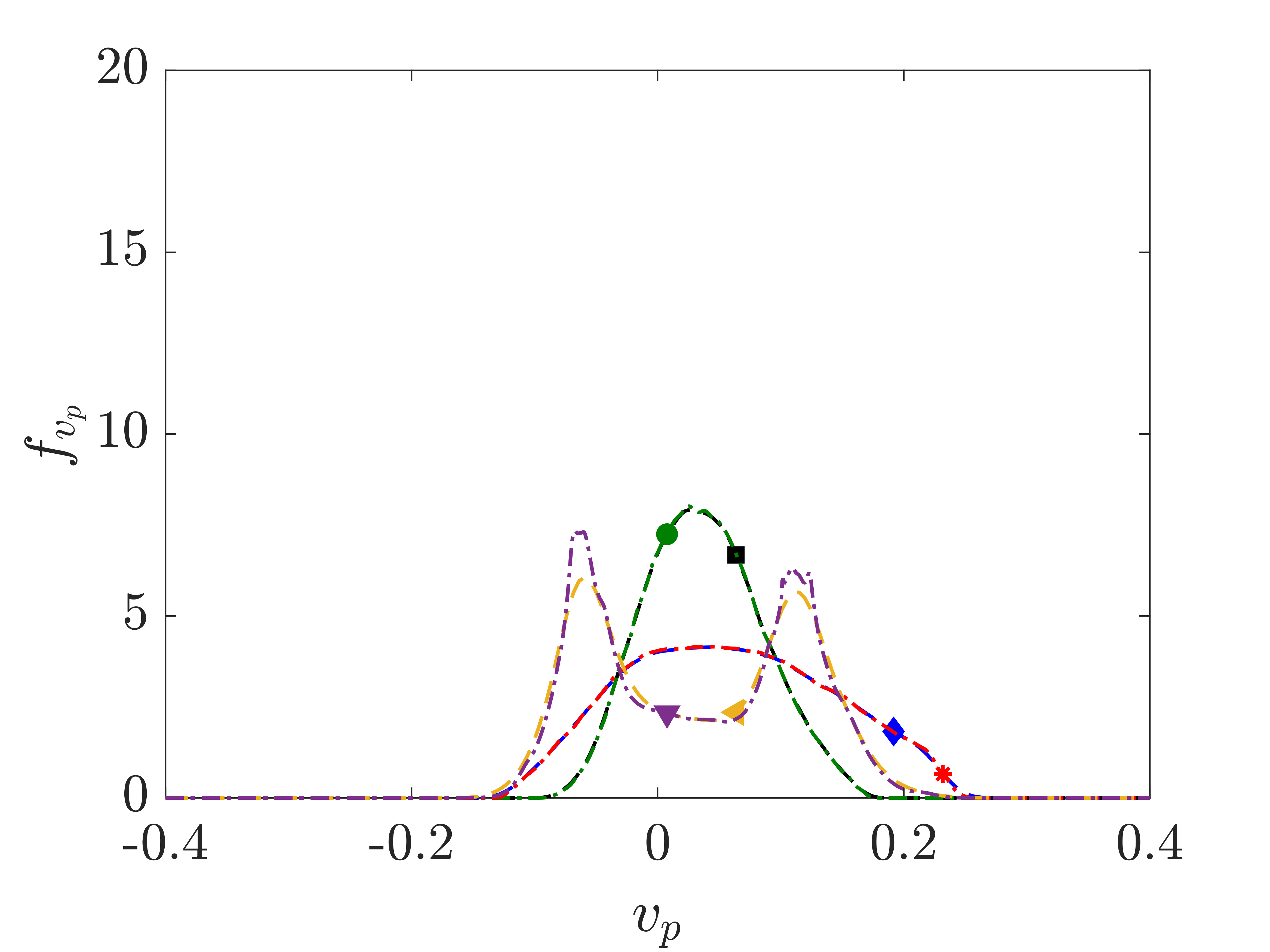}} \\
    \subfloat[]{
		\label{fig: isoTurb_fwp_MC}
		\includegraphics[width=0.32\textwidth]{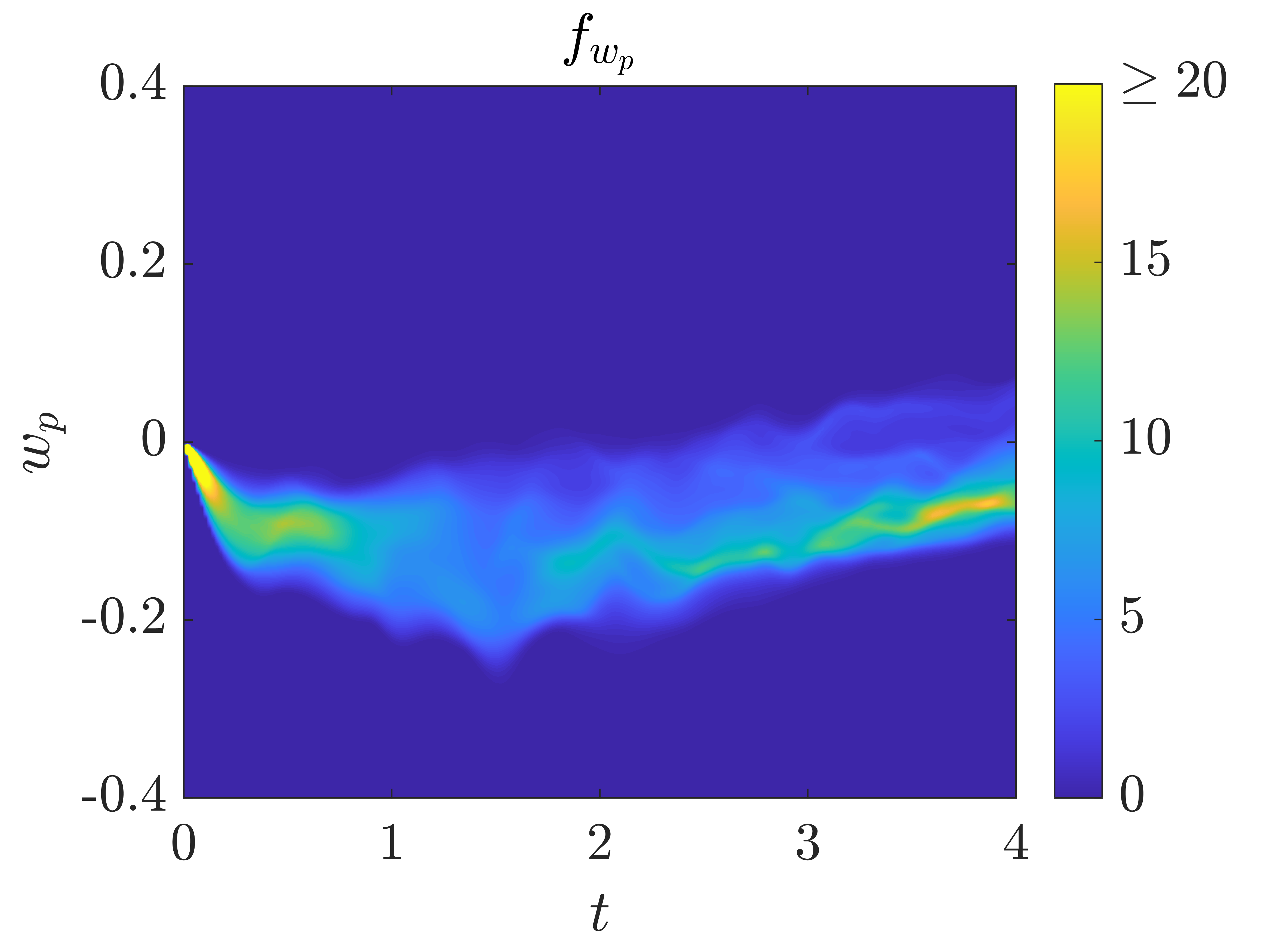}}
		\hfill
	\subfloat[]{
		\label{fig: isoTurb_fwp_SPARSE}
		\includegraphics[width=0.32\textwidth]{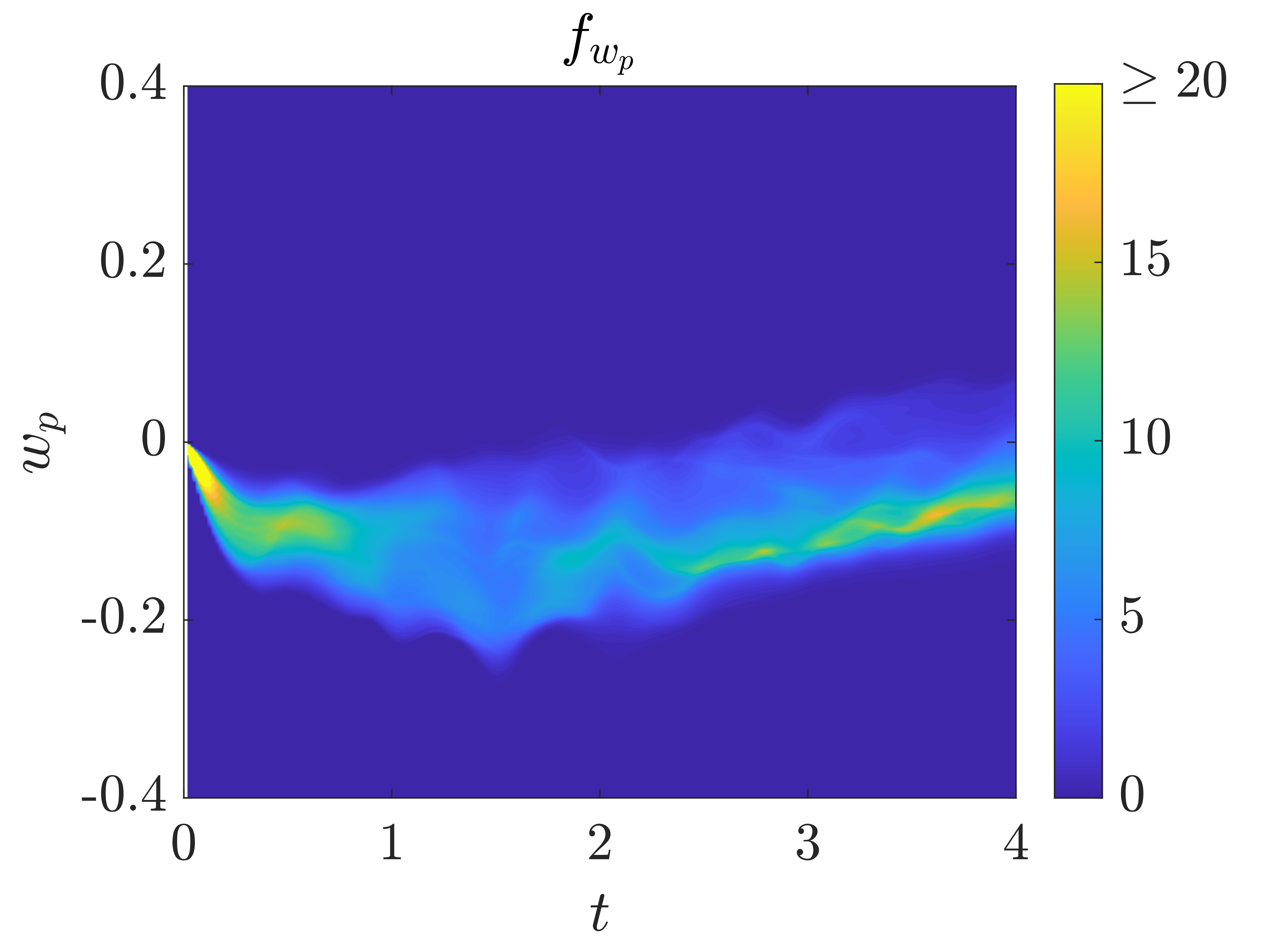}} 
		\hfill
	\subfloat[]{
		\label{fig: isoTurb_fwp_times}
		\includegraphics[width=0.32\textwidth]{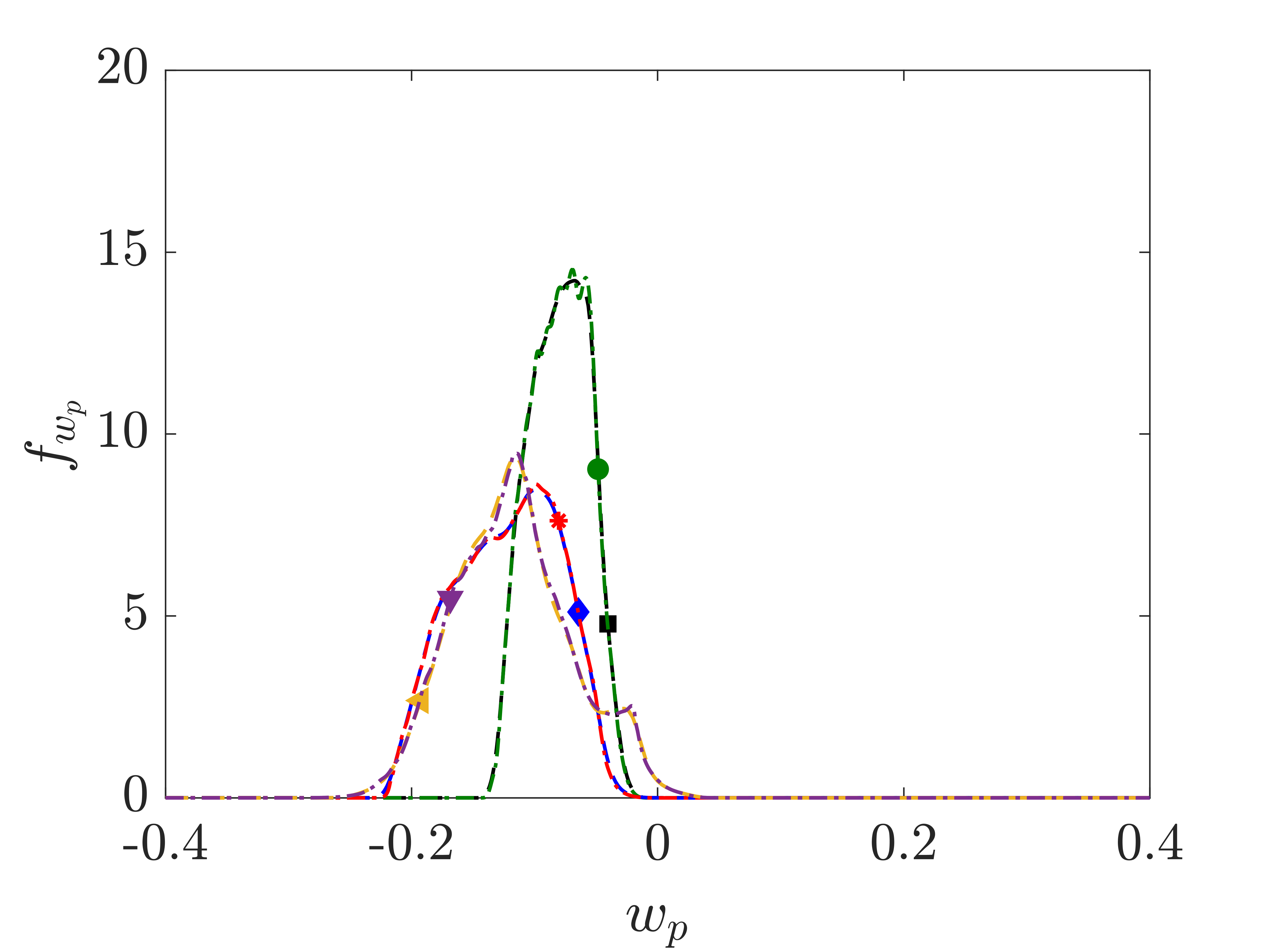}} \\
	\subfloat[]{
		\label{fig: isoTurb_fTp_MC}
		\includegraphics[width=0.32\textwidth]{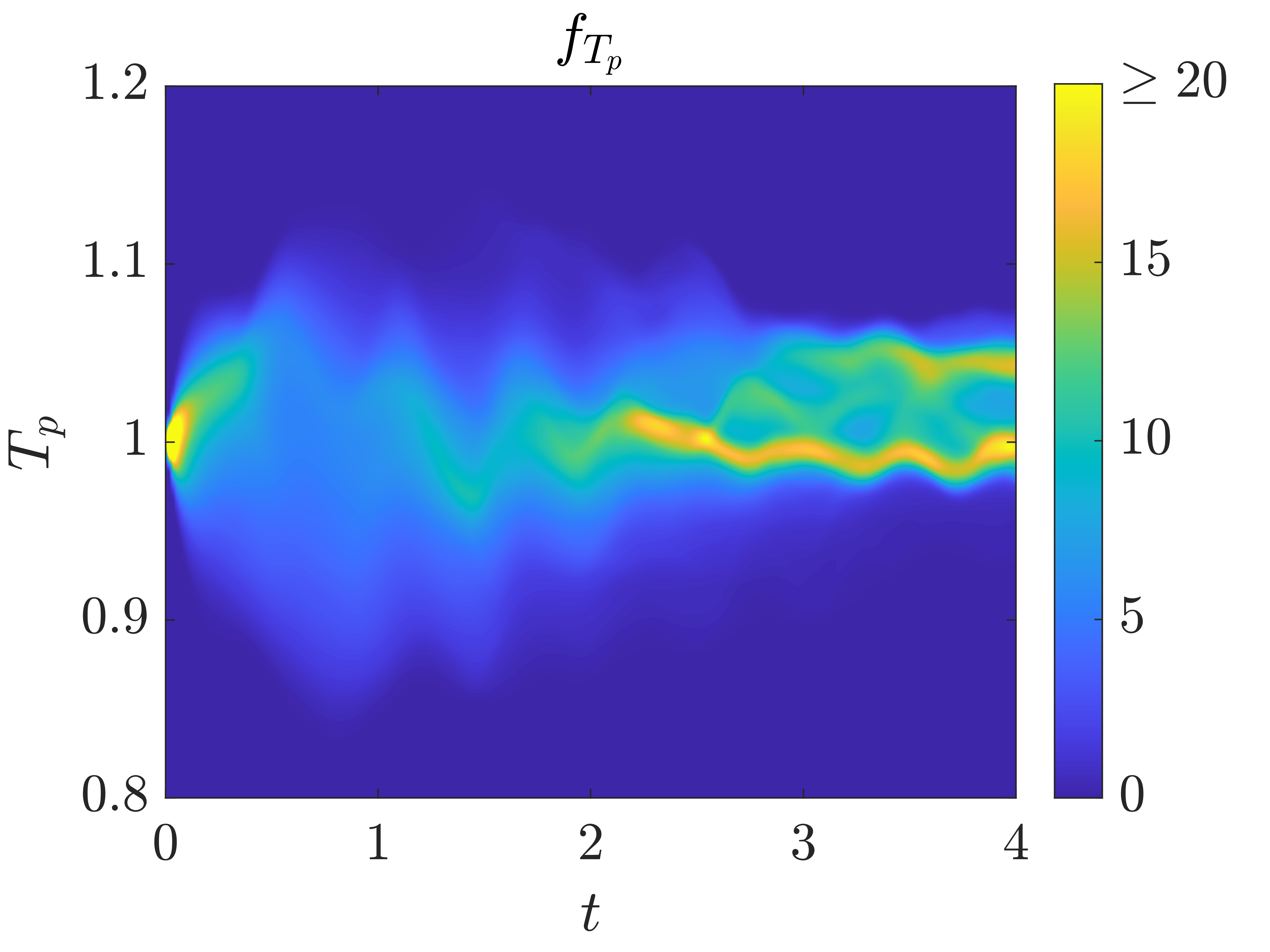}}
		\hfill
	\subfloat[]{
		\label{fig: isoTurb_fTp_SPARSE}
		\includegraphics[width=0.32\textwidth]{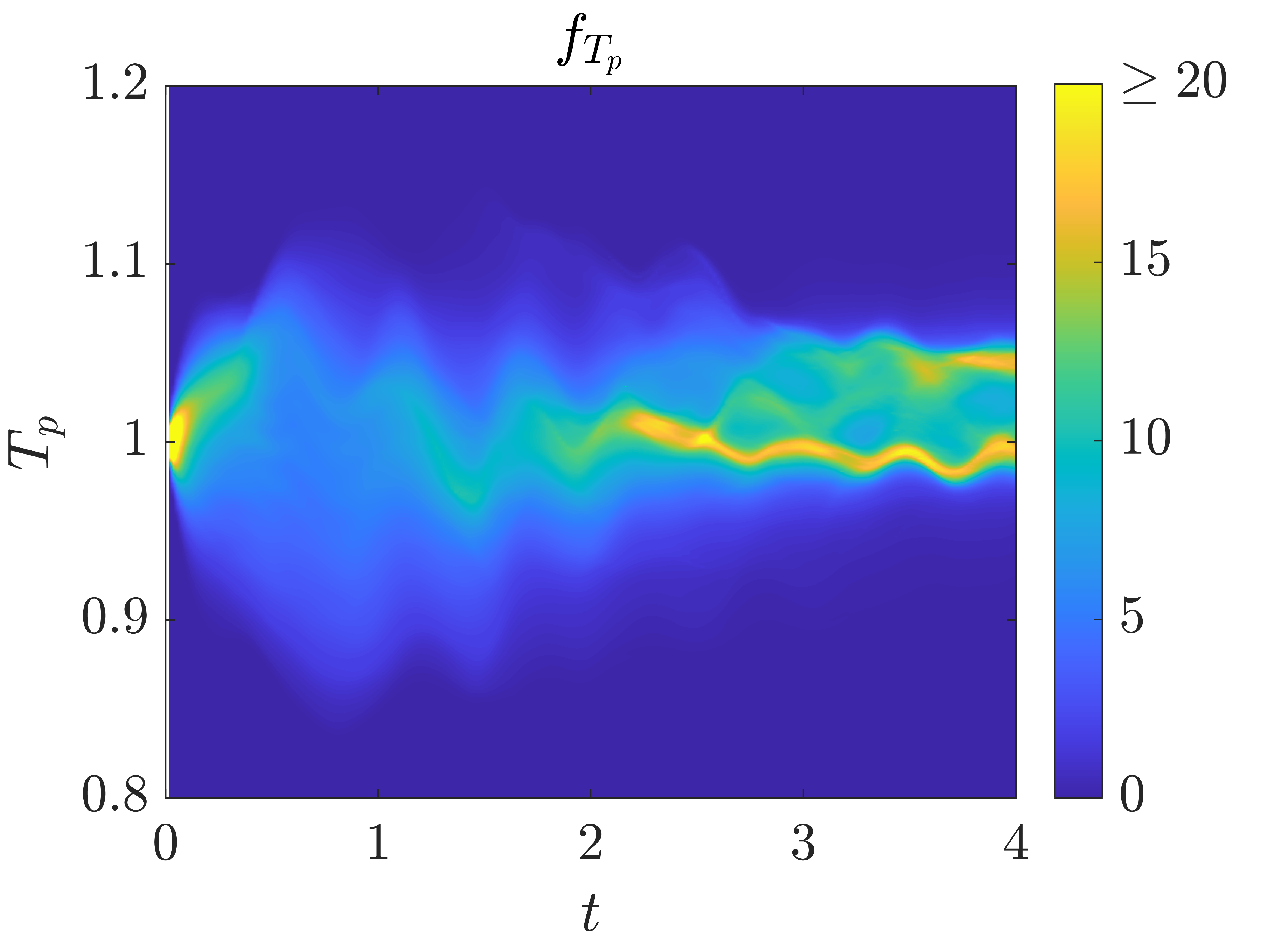}} 
		\hfill
	\subfloat[]{
		\label{fig: isoTurb_fTp_times}
		\includegraphics[width=0.32\textwidth]{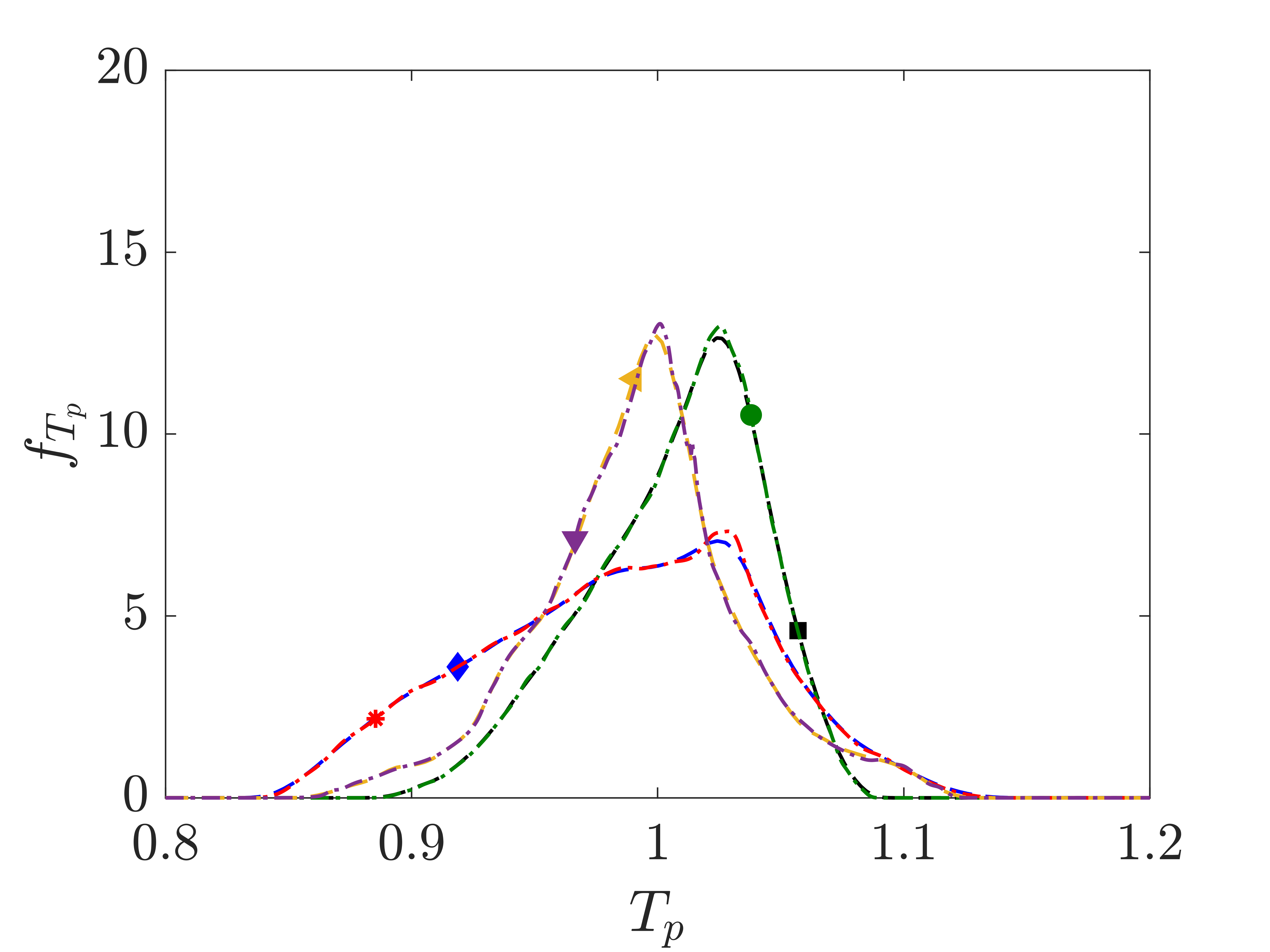}} 
	\caption[]{PDFs of the particle velocity components $f_{u_p}$, $f_{v_p}$ and $f_{w_p}$ and temperature $f_{T_p}$ for the rF case computed with MC-PSIC and SPARSE methods. The Figure shows: PDF of $u_p$ computed with (a) MC-PSIC and (b) SPARSE and (c) comparison for different times and the corresponding result for $v_p$ in (d)--(f), for $w_p$ in (g)--(i) and for $T_p$ in (j)--(l). The legend in (c) also corresponds to (f), (i) and (l) where the solution is plotted at times $t=[0.2, \ 1, \ 2]$ for both methods.}
	\label{fig: isoTurb_PDF_upvpwpTp}
\end{figure}

As a final validation and illustration of the capability of SPARSE, we present    PDFs of the Cumulative Cloud in Figures~\ref{fig: isoTurb_PDF_xpypzp} and~\ref{fig: isoTurb_PDF_upvpwpTp}.
We compare the evolution  of the PDF contours computed with MC-PSIC with the SPARSE method.
For three specific instants of time $t=[0.75, \ 2, \ 4]$, the PDFs  are extracted along lines and plotted versus the random variable only.
 These figures illustrate that SPARSE predicts the PDF with same level of detail as MC-PSIC. For example, the shift of the PDF from a  single peak  to a double peak in $f_{x_p}$ is captured at time $t=4$. 

\section{Conclusions and future work} \label{sec: conclusions}

A closed-form Lagrangian point-cloud model is coined that determines the first and second statistical moments of groups of  randomly forced particles within a cloud region  at a single (singular) point in a computationally efficient manner. 
The union of several  point-clouds can determine the PDF  and higher order statistical moments of the union, which we dub a  Cumulative Cloud. 
The model propagates the uncertainty in the  forcing function, as described by polynomial chaos expansion, that can be either the confidence interval in data-driven and/or empirical forcing models or the stochastic force variance  on the sub-cloud scales, into its  solution. 


Like its deterministically forced sibling, the model is derived using the SPARSE approach, which  combines  a truncated method of moments and a truncated Taylor series expansion of the forcing function and the carrier-flow variable at the points-cloud's mean location and relative velocity. The randomly forced SPARSE-R method is also closed, i.e., the formulation is self-contained, independent of MC results to close higher order correlations of variables in principle unknown.
The convergence of these truncated terms is proportional to the standard deviation of the particle variables to the third power.
This third-order convergence rate is verified in numerical tests.

The  second moments of the variables in the Taylor expanded cloud and its influence on the acceleration of the cloud are relatively small as compared to  the first moments. As a result a point-cloud that is deterministically forced with the mean of the  random forcing function provides a reasonable estimate of the mean of the randomly forced point-cloud  for the three test cases considered.  

The difference of the second moments determined with SPARSE-R approach and the deterministic SPARSE solution provides an accurate estimate of propagation of the uncertainty of the forcing function into confidence intervals of the solution. 

A virtual stress that correlates the random coefficients of the polynomial chaos expansion, with the dispersed phase fluctuations mixes the samples of the randomly forced particle solution in phase space. 
These stresses are correctly and accurately modeled by SPARSE-R and thus SPARSE-R gives direct insight in the fundamental driving mechanics of uncertainty propagation in a solution by inspection and comparison of the magnitudes of the moments  on the rights hand side of the closed systems of dynamic, SPARSE-R ordinary differential equations.

The virtual stress is negligible in directions of symmetry or in isotropic flows, because correlations with respect to the fluctuations forcing coefficients multiply with zero mean field solutions.  Then  randomness does not affect the footprint of the clouds because fast and slow responding particles cancel each other's motions. This  was illustrated in stagnation flow and an isotropic turbulence.

Future developments of the SPARSE-R method include a two-way coupled formulation that extends the existent work by~\cite{taverniers2019two} to include randomness in the forcing.
Additional improvements include an algorithm to split and join subclouds as needed depending on the error of the SPARSE approximations.

\section*{Acknowledgments}

This work was supported by the Air Force Office of Scientific Research under Award No FA9550-19-1-0387 and a San Diego State University Graduate Fellowship. 

\section*{CRediT authorship contribution statement}

\textbf{Daniel Dom\'inguez-V\'azquez:} Conceptualization, Data curation, Formal analysis, Investigation, Methodology, Resources, Software, Validation, Visualization, Writing the original draft, Review and editing. \textbf{Gustaaf B. Jacobs:} Conceptualization, Funding acquisition, Project administration, Supervision, Review and editing.

\appendix


\section{Expected convergence of SPARSE for randomly forced particle clouds}\label{app: convergence}

Here, the expected convergence rate of the SPARSE formulation with respect to the size of the initial cloud is derived.
Consider the Taylor expansion of the forcing function $f_1$ around the average values of the particle cloud in a one-dimensional setting, we can find the next term omitted in the SPARSE method by tracking the leading order term of the Taylor series that were neglected previously in the SPARSE formulation~\eqref{eq: MoM} and~\eqref{eq: closure_means}--\eqref{eq: closure_alphaeta}.
Looking at the approximation of the average forcing $f_1$ given in~\eqref{eq: closure_mean_f1} and retaining terms on the order to fluctuations to the third power one has
\begin{align}
    \overline{f_1(\alpha,a)} &\approx \overline{ A_{00} + A_{10} \alpha^\prime + A_{01}a^\prime + A_{20}{\alpha^\prime}^2 + A_{11}\alpha^\prime a^\prime + A_{02}{a^\prime}^2 + A_{30}{\alpha^\prime}^3 + A_{21}{\alpha^\prime}^2a^\prime + A_{12}\alpha^\prime{a^\prime}^2 + A_{03}{\alpha^\prime}^3  } \nonumber \\
    &= A_{00} + A_{11}\overline{\alpha^\prime a^\prime} + A_{02}\overline{{a^\prime}^2} + A_{12}\overline{\alpha^\prime{a^\prime}^2} + A_{03}\overline{{a^\prime}^3}
\end{align}
where we have considered that $\partial^2 f_1/\partial \alpha^2=\partial^3 f_1/\partial \alpha^3=0$ and the constants $A_{i j}$ are defined by
\begin{align}
    A_{ij} &= \left.\frac{\partial^{(i+j)}f_1}{\partial \alpha^{(i)} \partial a^{(j)}}\right|_{\overline{\alpha},\overline{a}},
    \label{eq: leading_constants_A}
\end{align}
where the superscript in parenthesis indicates the order of the partial derivative.

\clearpage
Then, using the definition of the relative velocity $\boldsymbol{a}=\boldsymbol{u}-\boldsymbol{u}_p$ we arrive to
\begin{align}
    \overline{f\left(\alpha,a \right) } &\approx A_{00} + A_{11}\left( \overline{\alpha^\prime u^\prime}-\overline{\alpha^\prime u_p^\prime} \right) + A_{02}\left( \overline{{u^\prime}^2} -2\overline{u^\prime u_p^\prime}+\overline{{u_p^\prime}^2}\right) +A_{12}\left( \overline{\alpha^\prime{u^\prime}^2} -2\overline{\alpha^\prime u^\prime u_p^\prime}+\overline{\alpha^\prime{u_p^\prime}^2} \right) \nonumber \\
    &+ A_{03}\left(\overline{{u^\prime}^3}+3\overline{u^\prime{u_p^\prime}^2}-3\overline{{u^\prime}^2{u_p^\prime}}-\overline{{u_p^\prime}^3}  \right),
\end{align}
that can be related to variables of the particle phase using the closure~\eqref{eq: closure_means}--\eqref{eq: closure_alphaeta}
\begin{align}
\begin{split}
    \overline{f\left(\alpha,a \right) } &\approx A_{00} + A_{11}\left[ \left( B_1\overline{\alpha^\prime x_p^\prime}+B_2\overline{\alpha^\prime{x_p^\prime}^2} \right) -\overline{\alpha^\prime u_p^\prime} \right] \\ 
    &+ A_{02}\left[ \left( B_1^2\overline{{x_p^\prime}^2} + 2B_1 B_2\overline{{x_p^\prime}^3} \right) - 2\left(B_1\overline{x_p^\prime{u_p^\prime}} + B_2\overline{x_p^\prime{u_p^\prime}^2} \right) +\overline{{u_p^\prime}^2}\right]  \\ 
    &+ A_{12}\left[ B_1^2\overline{\alpha^\prime{x_p^\prime}^2} - 2B_1 \overline{\alpha^\prime x_p^\prime u_p^\prime} + \overline{\alpha^\prime{u_p^\prime}^2} \right]  \\ 
    &+ A_{03}\left(B_1^3\overline{{x_p^\prime}^3}+3B_1\overline{x_p^\prime{u_p^\prime}^2}-3B_1^2\overline{{x_p^\prime}^2{u_p^\prime}}-\overline{{u_p^\prime}^3}  \right),
\end{split}
\end{align}
where the constants are $B_1$ and $B_2$ are defined as
\begin{align}
\begin{split}
    B_1 = \left.\frac{\partial u}{\partial x} \right|_{\overline{x}_p}, \ \ \ \ \ 
    B_2 = \left.\frac{\partial^2 u}{\partial x^2} \right|_{\overline{x}_p},
\end{split}
\label{eq: leading_constants_B}
\end{align}
and reordering one has
\begin{align}
\begin{split}
    \overline{f\left(\alpha,a \right) } &\approx \overbrace{ A_{00} + A_{11}\left(B_1\overline{\alpha^\prime x_p^\prime} -\overline{\alpha^\prime u_p^\prime}\right) + A_{02}\left( B_1^2\overline{{x_p^\prime}^2} -2B_1\overline{x_p^\prime{u_p^\prime}} + \overline{{u_p^\prime}^2}\right)  }^{\text{Terms included in SPARSE}} + A_{11}B_2\overline{\alpha^\prime{x_p^\prime}^2}  \\ 
    &+ 2A_{02}\left( B_1 B_2 \overline{{x_p^\prime}^3} -B_2\overline{x_p^\prime{u_p^\prime}^2}  \right) + A_{12}\left( B_1^2\overline{\alpha^\prime{x_p^\prime}^2} - 2B_1 \overline{\alpha^\prime x_p^\prime u_p^\prime} + \overline{\alpha^\prime{u_p^\prime}^2} \right)  \\ 
    &+ A_{03}\left(B_1^3\overline{{x_p^\prime}^3}+3B_1\overline{x_p^\prime{u_p^\prime}^2}-3B_1^2\overline{{x_p^\prime}^2{u_p^\prime}}-\overline{{u_p^\prime}^3}  \right) .
\end{split}
\end{align}
The leading order term of the error is proportional to third order moments of the particle phase that can be related to the standard deviations of the particle phase variables by making use of the coefficient of skewness $\gamma$ of a given third order central moment.
Additionally, the terms included in the SPARSE formulation (up to second moments) can be also related to the standard deviation of the particle phase variables by the use of the Pearson's coefficient $\rho$ of two given magnitudes.
In this way, one can write the average forcing as
\begin{align}
\begin{split}
    \overline{f_1(\alpha,a)} &\approx \overbrace{A_{00} + A_{02}B_1^2 \sigma_{x_p}^2 + A_{02}\sigma_{u_p}^2 - 2A_{02}B_1\rho_{{x_p u_p}}\sigma_{x_p}\sigma_{u_p} + A_{11}B_1\rho_{\alpha x_p}\sigma_{\alpha}\sigma_{x_p} -A_{11}\rho_{\alpha u_p}\sigma_{\alpha}\sigma_{u_p} }^{\text{Terms included in SPARSE}} \\ 
    &+ 2A_{02}B_1 B_2 \gamma_{x_p^3}\sigma_{x_p}^3 - A_{03}\gamma_{u_p^3}\sigma_{u_p}^3 +\left(3A_{03}B_1-2  A_{02}B_2\right)\gamma_{x_p u_p^2}\sigma_{x_p}\sigma_{u_p}^2 - 3A_{03}B_1^2\gamma_{x_p^2 u_p}\sigma_{x_p}^2\sigma_{u_p} \\
    &+\left( A_{11}{B_2} +A_{12}B_1^2\right) \gamma_{\alpha x_p^2}\sigma_{\alpha}\sigma_{x_p}^2 +A_{12}\gamma_{\alpha u_p^2}\sigma_{\alpha}\sigma_{u_p}^2 -2A_{12}B_1\gamma_{\alpha x_p u_p}\sigma_{\alpha}\sigma_{x_p}\sigma_{u_p},
\end{split}
\label{eq: error_gammas}
\end{align}
where we can see that the leading order term of the error is proportional to the standard deviations of the particle phase variables $\alpha$, $x_p$ and $u_p$ in combinations of order three with the form $\sigma_{\alpha}^{n_\alpha}\sigma_{x_p}^{n_{x_p}}\sigma_{u_p}^{n_{u_p}}$ where the integer exponents are $n_\alpha+n_{x_p}+n_{u_p}=3$ with $0 \leq n_\alpha \leq 3$, $0 \leq n_{x_p}\leq 3$ and $0 \leq n_{u_p} \leq 3$.
Then, when using splitting, i.e., dividing the domain of the cloud in uniform sets that define subclouds (see Appendix~\ref{app: splitting_algorithm}), the number of divisions can be expressed as $M_p=M_p^{\alpha}M_p^{x_p}M_p^{u_p}=M^3$ for a one-dimensional case where the same number of divisions $M$ per dimension ($\alpha$, $x_p$ and $u_p$) is considered.
Every subcloud $k$, with $k=1,\dots,M^3$ is solved with the SPARSE equations~\eqref{eq: MoM} jointly with the closure~\eqref{eq: closure_means}--\eqref{eq: closure_etaf1}.
Each subcloud has a reduced standard deviation in the particle phase magnitudes such that for the $k$-th subcloud ${\sigma_{x_p}}_k \sim \sigma_{x_p}/M$ and similarly for $\alpha$ and $u_p$.
Also, the constants $A_{i j}$ and $B_i$ in~\eqref{eq: leading_constants_A} and \eqref{eq: leading_constants_B} for each subcloud differ for the ones of the cloud without splitting in that they are evaluated at average locations of the different subclouds, for example ${B_1}_k=\left.\frac{\partial u}{\partial x} \right|_{{\overline{x}_p}_k}$, is evaluated at ${\overline{x}_p}_k$ instead of the average location of the total cloud with no splitting ${\overline{x}_p}$.
However, because the cloud has to be small enough for the Taylor series to be consider accurate, we can assume that the constants remain on the same order of magnitude ${A_{i j}}_k \sim A_{i j}$ and ${B_i}_k \sim B_i$ as a result of assuming ${\overline{x}_p}_k \simeq {\overline{x}_p}$.
Additionally, and also based on the same assumption of having a small enough cloud, i.e., having smooth gradients within the cloud, the average coefficient of skewness of the different subclouds remains on the same order of magnitude that the one of the cloud without splitting.
This is $\sum_{k=1}^{M_p}w_k{\gamma_{x_p^3}}_k \sim \gamma_{x_p^3}$ and similarly for other third moments, where $w_k$ is the weight of the $k$-th subcloud (see Algorithm~\ref{alg: spliting}).

With these considerations, the average forcing when using splitting can be written as
\begin{align}
\begin{split}
    \overline{f_1(\alpha,a)} &= \sum_{k=1}^{M^3}w_k \overline{f_1(\alpha,a)}_k ,
\end{split}
\label{eq: error_summ_macroparticles}
\end{align}
where the leading order terms of the error are
\begin{subequations} \label{eq: errors_order3}
\begin{align}
    &2\sum_{k=1}^{M^3}w_k {A_{02}}_k {B_1}_k {B_2}_k {\gamma_{x_p^3}}_k {\sigma_{x_p}^3}_k \sim 2 {A_{02}} {B_1} {B_2} {\gamma_{x_p^3}} \left( \frac{\sigma_{x_p}}{M}\right)^3  ,\\ 
    &\sum_{k=1}^{M^3}w_k {A_{03}}_k{\gamma_{u_p^3}}_k{\sigma_{u_p}^3}_k  \sim  A_{03}\gamma_{u_p^3}\left( \frac{\sigma_{u_p}}{M}\right)^3    ,\\
    &\sum_{k=1}^{M^3}w_k\left(3{A_{03}}_k {B_1}_k - 2 {A_{02}}_k {B_2}_k \right){\gamma_{x_p u_p^2}}_k {\sigma_{x_p}}_k {\sigma_{u_p}^2}_k \sim \left(3A_{03}B_1-2  A_{02}B_2\right)\gamma_{x_p u_p^2} \left( \frac{\sigma_{x_p}}{M}\right) \left( \frac{\sigma_{u_p}}{M}\right)^2 , \\ 
     &3\sum_{k=1}^{M^3}w_k {A_{03}}_k {B_1^2}_k {\gamma_{x_p^2 u_p}}_k {\sigma_{x_p}^2}_k {\sigma_{u_p}}_k \sim  3A_{03}B_1^2\gamma_{x_p^2 u_p}\left( \frac{\sigma_{x_p}}{M}\right)^2 \left( \frac{\sigma_{u_p}}{M}\right) , \\
     &\sum_{k=1}^{M^3}w_k\left( {A_{11}}_k {B_2}_k +{A_{12}}_k{B_1^2}_k\right) {\gamma_{\alpha x_p^2}}_k{\sigma_{\alpha}}_k{\sigma_{x_p}^2}_k \sim \left( A_{11}{B_2} +A_{12}B_1^2\right) \gamma_{\alpha x_p^2}\left( \frac{\sigma_{\alpha}}{M}\right) \left( \frac{\sigma_{x_p}}{M}\right)^2 , \\
     &\sum_{k=1}^{M^3}w_k {A_{12}}_k{\gamma_{\alpha u_p^2}}_k{\sigma_{\alpha}}_k{\sigma_{u_p}^2}_k \sim A_{12}\gamma_{\alpha u_p^2}\left( \frac{\sigma_{\alpha}}{M}\right) \left( \frac{\sigma_{u_p}}{M}\right)^2 ,\\
     &\sum_{k=1}^{M^3}w_k {A_{12}}_k {B_1}_k {\gamma_{\alpha x_p u_p}}_k {\sigma_{\alpha}}_k {\sigma_{x_p}}_k {\sigma_{u_p}}_k \sim A_{12}B_1\gamma_{\alpha x_p u_p}\left( \frac{\sigma_{\alpha}}{M}\right)\left( \frac{\sigma_{x_p}}{M}\right)\left( \frac{\sigma_{u_p}}{M}\right)
\end{align}
\end{subequations}
that shows the proportionality of the errors with $M^{-3}$.
The SPARSE method is therefore expected to converge with a third order rate with the number of divisions per dimension $M$ or level of splitting.
Note that the analysis performed for the term $\overline{f_1(\alpha,a)}$ can be extended to all the terms included in the closure~\eqref{eq: closure_means}--\eqref{eq: closure_etaf1}.
We refer the reader to~\cite{dominguez2023closed} where a similar derivation was performed for other terms in the closure~\eqref{eq: closure_means}--\eqref{eq: closure_etaf1}.

\section{Splitting algorithm} \label{app: splitting_algorithm}

Here, we present the algorithm that has been used to classify the point-particles in a given cloud into sets that define subclouds of point-particles distributed uniformly. 
To define the initial condition for the SPARSE equations for each subcloud, we need to compute the first two moments of each set of point-particles.
This is illustrated in Figure~\ref{fig: IC} where point-particles are visualized by points and subclouds by ellipsoids and colors identify the different divisions of the initial cloud (meaning the cloud without splitting, the total cloud).
The routine to subdivide the initial condition (initial cloud of point-particles) for a one-dimensional case where a single mode is considered is outlined in the Algorithm~\ref{alg: spliting}.
The algorithm takes as inputs the number of divisions along the random parameter $\alpha$, particle location $x_p$ and velocity $u_p$ and loops over the point-particle variables classifying and counting them.
In this manner, the moments of each subcloud are computed as well as the weights $w_k$.

\begin{algorithm2e}
\SetAlgoLined
\DontPrintSemicolon 
\KwIn{Velocities $u_p$, locations $x_p$ and values of the random coefficient $\alpha$ of the $N_p$ point-particles at $t=0$ and number of divisions of the cloud $M_p^{x_p}$, $M_p^{u_p}$ and $M_p^{\alpha}$}
\KwOut{Weights $w$ and first two moments $\overline{x}_p$, $\overline{u}_p$, $\overline{\alpha}$, $\overline{{x_p^\prime}^2}$, $\overline{{u_p^\prime}^2}$, $\overline{{\alpha^\prime}^2}$, $\overline{x_p^\prime u_p^\prime}$, $\overline{\alpha^\prime x_p^\prime}$ and $\overline{\alpha^\prime u_p^\prime}$ of the subclouds at $t=0$.}
Initialize the counter, obtain the size of the subclouds and loop over $x_p$, $u_p$ and $\alpha$ \;
$k \gets 1$ \;
$\delta x_p \gets (max(x_p)-min(x_p))/M_p^{x_p}$ \;
$\delta u_p \gets (max(u_p)-min(u_p))/M_p^{u_p}$ \;
$\delta \alpha \gets (max(\alpha)-min(\alpha))/M_p^{\alpha}$ \;
\For{$p \gets 1$ \textbf{to} $M_p^{x_p}$} {
\For{$q \gets 1$ \textbf{to} $M_p^{u_p}$} {
\For{$r \gets 1$ \textbf{to} $M_p^{\alpha}$} {
    Obtain the limits of the $k$-th subcloud \;
    $\{{x_p}^{left}_k, \ {x_p}^{right}_k \}  \gets \{min(x_p)+(p-1)\delta x_p, \ min(x_p)+p\delta x_p  \} $ \;
    $\{{u_p}^{left}_k, \ {u_p}^{right}_k \}  \gets \{min(u_p)+(q-1)\delta u_p, \ min(u_p)+q\delta u_p  \} $ \;
    $\{{\alpha}^{left}_k, \ {\alpha}^{right}_k \}  \gets \{ min(\alpha)+(r-1)\delta \alpha, \ min(\alpha)+r\delta \alpha  \} $ \;
    Find the point-particles contained in the $k$-th subcloud \;
    $i \gets 1$\;
    \For{$j \gets 1$ \textbf{to} $N_p$} {
        \If{${x_p}^{left}_k \leq {x_p}_j \leq {x_p}^{right}_k \ \& \ {u_p}^{left}_k \leq {u_p}_j \leq {u_p}^{right}_k \ \& \ {\alpha}^{left}_k \leq {\alpha}_j \leq {\alpha}^{right}_k $}{
            ${x_p^m}_i \gets {x_p}_j $ , \ \ \  ${u_p^m}_i \gets {u_p}_j $ , \ \ \ ${\alpha^m_i} \gets {\alpha}_j $ \;
            $i \gets i+1 $ \;
        }
    }
    Compute the weight and first two moments of the $k$-th subcloud \;
    $ w_k \gets  (i-1)/N_p$ \;
    $ {\overline{x}_p}_k \gets  mean(x_p^m)$, \ \ \ $ {\overline{u}_p}_k \gets  mean(u_p^m)$, \ \ \ $ {\overline{\alpha}}_k \gets  mean(\alpha^m)$ \;
    $ {\overline{{x_p^\prime}^2}}_k \gets  mean( (x_p^m - mean(x_p^m))^2)$ \;
    $ {\overline{{u_p^\prime}^2}}_k \gets  mean( (u_p^m - mean(u_p^m))^2)$ \;
    $ {\overline{{\alpha^\prime}^2}}_k \gets  mean( (\alpha^m - mean(\alpha^m))^2)$ \;
    $ {\overline{{x_p^\prime}{u_p^\prime}}}_k \gets  mean( (x_p^m - mean(x_p^m))(u_p^m - mean(u_p^m)))$ \;
    $ {\overline{{\alpha^\prime}{x_p^\prime}}}_k \gets  mean( (\alpha^m - mean(\alpha^m))(x_p^m - mean(x_p^m)))$ \;
    $ {\overline{{\alpha^\prime}{u_p^\prime}}}_k \gets  mean( (\alpha^m - mean(\alpha^m))(u_p^m - mean(u_p^m)))$ \;
    $k \gets k+1$ \;
}}}
\caption{Splitting algorithm}
\label{alg: spliting}
\end{algorithm2e}

\section{SPARSE equations for the stagnation flow} \label{app: rF_equations}

The closed SPARSE equations for the stagnation flow~\eqref{eq: SF_flow} when the Stokes' law is considered ($f_1=\alpha$ and $g_1=1$) are presented here.
The interpolation of the flow velocity at the average particle location is simplified to $\overline{u}= -k \overline{x}_p$ and $\overline{v}= k \overline{v}_p$.
The SPARSE equations for $x-$direction read as
\begin{subequations}\label{eq: SF_equations_x}
\begin{align}
    \frac{\text{d} \overline{x}_p}{\text{d}t} &= \overline{u}_p , 
    \label{eq: SFeqs_mean_xp} \\ 
    St\frac{\text{d} \overline{u}_p}{\text{d}t} &= \overline{\alpha}\left(-k\overline{x}_p-\overline{u}_p \right)-k\overline{\alpha^\prime x_p^\prime}-\overline{\alpha^\prime u_p^\prime} , 
    \label{eq: SFeqs_mean_up} \\ 
    \frac{\text{d} \overline{{x_p^\prime}^2}}{\text{d}t} &= 2\overline{x_p^\prime u_p^\prime} ,
    \label{eq: SFeqs_cm2_xp} \\ 
    St\frac{\text{d} \overline{{u_p^\prime}^2}}{\text{d}t} &= 2\overline{\alpha}\left( -k\overline{x_p^\prime u_p^\prime}-\overline{{u_p^\prime}^2} \right)+2\overline{\alpha^\prime u_p^\prime}\left( -k\overline{x}_p-\overline{u}_p\right) , 
    \label{eq: SFeqs_cm2_up} \\
    \frac{\text{d}}{\text{d}t}\left(\overline{x_p^\prime u_p^\prime}\right) &= \overline{{u_p^\prime}^2}+\frac{1}{St}\left[ \overline{\alpha}\left( -k\overline{{x_p^\prime}^2}-\overline{x_p^\prime u_p^\prime} \right)+\overline{\alpha^\prime x_p^\prime}\left( -k\overline{x}_p-\overline{u}_p\right)  \right], 
    \label{eq: SFeqs_xpup} \\
    \frac{\text{d}}{\text{d}t}\left(\overline{\alpha^\prime x_p^\prime}\right) &=\overline{\alpha^\prime u_p^\prime} ,
    \label{eq: SFeqs_xpa} \\
    St\frac{\text{d}}{\text{d}t}\left(\overline{\alpha^\prime u_p^\prime}\right) &=\overline{\alpha}\left(  -k\overline{\alpha^\prime x_p^\prime}-\overline{\alpha^\prime u_p^\prime}\right)+\overline{{\alpha^\prime}^2}\left( -k\overline{x}_p-\overline{u}_p \right) ,
    \label{eq: SFeqs_upa}
\end{align}
\end{subequations}
for the $y-$direction
\begin{subequations}\label{eq: SF_equations_y}
\begin{align}
    \frac{\text{d} \overline{y}_p}{\text{d}t} &= \overline{v}_p , 
    \label{eq: SFeqs_mean_yp} \\ 
    St\frac{\text{d} \overline{v}_p}{\text{d}t} &= \overline{\alpha}\left(k\overline{y}_p-\overline{v}_p \right)+k\overline{\alpha^\prime y_p^\prime}-\overline{\alpha^\prime v_p^\prime} , 
    \label{eq: SFeqs_mean_vp} \\ 
    \frac{\text{d} \overline{{y_p^\prime}^2}}{\text{d}t} &= 2\overline{y_p^\prime v_p^\prime} ,
    \label{eq: SFeqs_cm2_yp} \\ 
    St\frac{\text{d} \overline{{v_p^\prime}^2}}{\text{d}t} &= 2\overline{\alpha}\left( k\overline{y_p^\prime v_p^\prime}-\overline{{v_p^\prime}^2} \right)+2\overline{\alpha^\prime v_p^\prime}\left( k\overline{y}_p-\overline{v}_p\right) , 
    \label{eq: SFeqs_cm2_vp} \\
    \frac{\text{d}}{\text{d}t}\left(\overline{y_p^\prime v_p^\prime}\right) &= \overline{{v_p^\prime}^2}+\frac{1}{St}\left[ \overline{\alpha}\left( k\overline{{y_p^\prime}^2}-\overline{y_p^\prime v_p^\prime} \right)+\overline{\alpha^\prime y_p^\prime}\left( k\overline{y}_p-\overline{v}_p\right)  \right], 
    \label{eq: SFeqs_ypvp} \\
    \frac{\text{d}}{\text{d}t}\left(\overline{\alpha^\prime y_p^\prime}\right) &=\overline{\alpha^\prime v_p^\prime} ,
    \label{eq: SFeqs_ypa} \\
    St\frac{\text{d}}{\text{d}t}\left(\overline{\alpha^\prime v_p^\prime}\right) &=\overline{\alpha}\left(  k\overline{\alpha^\prime y_p^\prime}-\overline{\alpha^\prime v_p^\prime}\right)+ \overline{{\alpha^\prime}^2}\left( k\overline{y}_p-\overline{v}_p \right) ,
    \label{eq: SFeqs_vpa}
\end{align}
\end{subequations}
and for the variables that combine directions one has
\begin{subequations}\label{eq: SF_equations_xy}
\begin{align}
    \frac{\text{d}}{\text{d}t}\left(\overline{{x_p^\prime}{y_p^\prime}}\right) &= \overline{{x_p^\prime}{v_p^\prime}}+\overline{{y_p^\prime}{u_p^\prime}}
    \label{eq: SFeqs_xpyp} \\ 
    \frac{\text{d}}{\text{d}t}\left(\overline{{x_p^\prime}{v_p^\prime}}\right) &= \overline{{u_p^\prime}{v_p^\prime}}+\frac{1}{St} \left[ \overline{\alpha}\left( k\overline{{x_p^\prime}{y_p^\prime}}-\overline{{x_p^\prime}{v_p^\prime}}\right) +\overline{\alpha^\prime x_p^\prime}\left(k\overline{y}_p-\overline{v}_p \right)  \right] , 
    \label{eq: SFeqs_xpvp} \\
    \frac{\text{d}}{\text{d}t}\left(\overline{{y_p^\prime}{u_p^\prime}}\right) &= \overline{{u_p^\prime}{v_p^\prime}}-\frac{1}{St} \left[ \overline{\alpha} \left( k\overline{{x_p^\prime}{y_p^\prime}} + \overline{{y_p^\prime}{u_p^\prime}} \right) +\overline{\alpha^\prime y_p^\prime}\left(k\overline{x}_p+\overline{u}_p \right)  \right] , 
    \label{eq: SFeqs_ypup} \\
    St\frac{\text{d}}{\text{d}t}\left(\overline{{u_p^\prime}{v_p^\prime}}\right) &= \overline{\alpha}\left(-k\overline{x_p^\prime v_p^\prime}+k\overline{y_p^\prime u_p^\prime} -2\overline{u_p^\prime v_p^\prime}\right)+\overline{\alpha^\prime u_p^\prime}\left( -k\overline{x}_p-\overline{u}_p\right)+\overline{\alpha^\prime v_p^\prime}\left( k\overline{y}_p-\overline{v}_p\right),
    \label{eq: SFeqs_upvp}
\end{align}
\end{subequations}
where the directions are decouple and the horizontal magnitudes can be computed independently of the vertical ones and vice versa.
The moments that combine magnitudes of different directions (equations in~\eqref{eq: SF_equations_xy}) can be solved after system~\eqref{eq: SF_equations_x} and~\eqref{eq: SF_equations_y}.

Considering that the variables $\boldsymbol{x}_p$, $\boldsymbol{u}_p$ and $\alpha$ are uncorrelated at the initial time $t=0$, for infinite samples $N_p\rightarrow \infty$, the initial conditions~\eqref{eq: SF_IC} set all moments of the particle variables to be zero initially but the initial average location ${\overline{x}_p}=-1$ and standard deviations of the particle location $\sigma_{x_p}=\sigma_{y_p}=0.3$ as described in Section~\ref{sec: test_SF}.
This means that the only average relative velocity different from zero at the initial time $t=0$ is on the horizontal direction $\overline{a}_x=\overline{u}-\overline{u}_p=k$ with zero in the vertical direction $\overline{a}_y=\overline{v}-\overline{v}_p=0$.
Note that under this circumstance, the input of uncertainty in the vertical equations~\eqref{eq: SF_equations_y} occurs in the term $\overline{{\alpha^\prime}^2}\left(k\overline{y}_p-\overline{v}_p  \right)$ that is zero because the relative velocity initially is set to zero $\overline{a}_y=0$.
Taking into account this initial condition, some moments remain zero for the entire simulation since the right hand side terms of its corresponding equations are all zero for any later time. 
In particular, the SPARSE equations in the horizontal components are the same as in~\eqref{eq: SF_equations_x}.
However, in the vertical component the system~\eqref{eq: SF_equations_y} simplifies to
\begin{subequations}\label{eq: SF_equations_y_ICsimplified}
\begin{align}
    \frac{\text{d} \overline{{y_p^\prime}^2}}{\text{d}t} &= 2\overline{y_p^\prime v_p^\prime} ,
    \label{eq: SFeqs_cm2_yp_ICsimplified} \\ 
    \frac{\text{d} \overline{{v_p^\prime}^2}}{\text{d}t} &= \frac{2\overline{\alpha}}{St}\left( k\overline{y_p^\prime v_p^\prime}-\overline{{v_p^\prime}^2} \right) , 
    \label{eq: SFeqs_cm2_vp_ICsimplified} \\
    \frac{\text{d}}{\text{d}t}\left(\overline{y_p^\prime v_p^\prime}\right) &= \overline{{v_p^\prime}^2}+\frac{\overline{\alpha}}{St} \left( k\overline{{y_p^\prime}^2}-\overline{y_p^\prime v_p^\prime} \right)  , 
    \label{eq: SFeqs_ypvp_ICsimplified} \\
    \overline{y}_p &= \overline{v}_p = \overline{\alpha^\prime y_p^\prime}=\overline{\alpha^\prime v_p^\prime}=0,
    \label{eq: SFeqs_moments0_ICsimplified}
\end{align}
\end{subequations}
and the equations for the moments combining directions~\eqref{eq: SF_equations_xy} simplify to
\begin{subequations}\label{eq: SF_equations_xy_ICsimplified}
\begin{align}
    \frac{\text{d}}{\text{d}t}\left(\overline{{x_p^\prime}{y_p^\prime}}\right) &= \overline{{x_p^\prime}{v_p^\prime}}+\overline{{y_p^\prime}{u_p^\prime}}
    \label{eq: SFeqs_xpyp_ICsimplified} \\ 
    \frac{\text{d}}{\text{d}t}\left(\overline{{x_p^\prime}{v_p^\prime}}\right) &= \overline{{u_p^\prime}{v_p^\prime}}+\frac{2\overline{\alpha}}{St}  \overline{{x_p^\prime}{y_p^\prime}} , 
    \label{eq: SFeqs_xpvp_ICsimplified} \\
    \frac{\text{d}}{\text{d}t}\left(\overline{{y_p^\prime}{u_p^\prime}}\right) &= \overline{{u_p^\prime}{v_p^\prime}}-\frac{2\overline{\alpha}}{St}  \overline{{x_p^\prime}{y_p^\prime}}  , 
    \label{eq: SFeqs_ypup_ICsimplified} \\
    St\frac{\text{d}}{\text{d}t}\left(\overline{{u_p^\prime}{v_p^\prime}}\right) &= \overline{\alpha}\left(-k\overline{x_p^\prime v_p^\prime}+k\overline{y_p^\prime u_p^\prime} -2\overline{u_p^\prime v_p^\prime}\right)+\overline{\alpha^\prime u_p^\prime}\left( -k\overline{x}_p-\overline{u}_p\right).
    \label{eq: SFeqs_upvp_ICsimplified}
\end{align}
\end{subequations}

It is important to notice that the solution of the moments in the vertical component~\eqref{eq: SF_equations_y_ICsimplified} when the average relative velocity is zero $\overline{a}_y=0$ and the variables are uncorrelated between them initially $\overline{\alpha^\prime y_p^\prime} = \overline{\alpha^\prime v_p^\prime}=0$, that the solution simplifies to the deterministic case, i.e., the deterministic SPARSE equations considering $\alpha=\overline{\alpha}$, with $\alpha^\prime=0$.
This can be seen in Figures~\ref{fig: sFcenter_sigma_xpyp},~\ref{fig: sFcenter_sigma_upvp} and~\ref{fig: sFcenter_xpup_ypvp} where the moments $\overline{{y_p^\prime}^2}$, $\overline{{v_p^\prime}^2}$ and $\overline{{y_p^\prime}{v_p^\prime}}$ are plotted for the rF and dF coinciding for the computed time interval.

\bibliographystyle{unsrtnat}
\bibliography{BIB}  

\end{document}